%% file: main.tex
\documentclass[10pt,a4paper]{scrartcl}

\usepackage[margin=2.3cm]{geometry}
\usepackage{amsmath}
\usepackage{amssymb}
\usepackage{amsfonts}
\usepackage{graphicx}
\usepackage{bm}
\usepackage{multirow}
\usepackage{lmodern}
\usepackage{float}
\usepackage[T1]{fontenc}
\usepackage{booktabs}
\usepackage{textcomp}
\usepackage{authblk}
\usepackage[section]{placeins}
\usepackage[format=plain]{caption}
\usepackage[table]{xcolor}
\usepackage{colortbl}
\usepackage{hyperref}
\hypersetup{
	colorlinks=true,
	linkcolor=blue,
	citecolor=blue,
	filecolor=blue,
	urlcolor=blue}
\usepackage{multicol}
\usepackage{multirow}
\usepackage{listings}
\usepackage{makecell}
\usepackage{threeparttable}
\usepackage{tikz}
\usepackage{enumitem}
\setlist{noitemsep}
\graphicspath{{./figs/}}
\let\pgfimage=\includegraphics
\usetikzlibrary{shapes,positioning,arrows,patterns}
\usetikzlibrary{external}
\usepackage{microtype}
\usepackage[backend=biber,style=apa,natbib=true,bibencoding=utf8]{biblatex}
\bibliography{references.bib}

\DeclareMathOperator{\normal}{normal}
\DeclareMathOperator{\uniform}{uniform}
\DeclareMathOperator{\Bernoulli}{Bernoulli}
\DeclareMathOperator{\gammadist}{gamma}
\DeclareMathOperator{\expdist}{exponential}
\DeclareMathOperator{\negbinom}{negBinomial}
\DeclareMathOperator{\lognormal}{logNormal}
\DeclareMathOperator{\dirichlet}{Dirichlet}
\DeclareMathOperator{\betadist}{beta}
\DeclareMathOperator{\Cauchy}{Cauchy}
\DeclareMathOperator{\GP}{GP}

\DeclareMathOperator{\cjs}{CJS}

\title{Detecting and diagnosing prior and likelihood sensitivity with
  power-scaling}

\author[1]{Noa Kallioinen}
\author[1]{Topi Paananen}
\author[2]{Paul-Christian B\"{u}rkner}
\author[1]{Aki Vehtari}
\affil[1]{Department of Computer Science, Aalto University, Espoo, Finland}
\affil[2]{Cluster of Excellence SimTech, University of Stuttgart, Stuttgart, Germany}

\begin{document}

\maketitle

\input{abstract.tex}

\input{body.tex}

\section{Acknowledgements}
\input{./acknowledgements.tex}

 \printbibliography

\newpage
\section*{Appendix}
\appendix

\input{./appendix.tex}

\end{document}

%% file: abstract.tex
\begin{abstract}
  Determining the sensitivity of the posterior to perturbations of the prior and likelihood is an important part of the
  Bayesian workflow. We introduce a practical and computationally efficient sensitivity analysis approach using importance sampling to estimate properties of posteriors resulting from power-scaling the prior or likelihood. On this basis, we suggest a diagnostic that can indicate the presence of prior-data conflict or likelihood noninformativity and discuss limitations to this power-scaling approach. The approach can be easily included in Bayesian workflows with minimal effort by the model builder and we present an implementation in our new R package \texttt{priorsense}. We further demonstrate
  the workflow on case studies of real data using models varying in complexity from simple linear models to Gaussian process models.
\end{abstract}

%% file: body.tex
\section{Introduction}
\label{sec:intro}

Bayesian inference is characterised by the derivation of a posterior
from a prior and a likelihood. As the posterior is dependent on the
specification of these two components, investigating its sensitivity
to perturbations of the prior and likelihood is a critical step in the Bayesian
workflow~\citep{gelmanBayesianWorkflow2020,depaoliImportancePriorSensitivity2020,lopesConfrontingPriorConvictions2011}. Along
with indicating the robustness of an inference in general, such sensitivity
is related to issues of \textit{prior-data
conflict}~\citep{evansCheckingPriordataConflict2006,allabadiOptimalRobustnessResults2017,reimherrPriorSampleSize2020}
and \textit{likelihood noninformativity}~\citep{gelmanPriorCanOften2017,poirierRevisingBeliefsNonidentified1998}.
Historically, sensitivity analysis has been an important topic in
Bayesian methods
research~\citep[e.g.][]{canavosBayesianEstimationSensitivity1975,skeneBayesianModellingSensitivity1986,bergerRobustBayesianAnalysis1990,bergerOverviewRobustBayesian1994,hillSensitivityBayesianAnalysis1994}.
However, the amount of research on the topic has
diminished~\citep{watsonApproximateModelsRobust2016,bergerBayesianRobustness2000}
and results from sensitivity analyses are seldom reported in empirical
studies employing Bayesian
methods~\citep{vandeschootSystematicReviewBayesian2017}. We
suggest that a reason for this is the lack of
sensitivity analysis approaches that are easily incorporated into existing
modelling workflows.

In this work, we present a sensitivity analysis approach that fits into workflows in which modellers use probabilistic programming languages, such as Stan~\citep{standevelopmentteamStanModellingLanguage2021} or PyMC~\citep{pymc3}, and employ Markov chain Monte Carlo (MCMC) methods to estimate posteriors via posterior draws~\citep[e.g.\ workflows described in][]{grinsztajnBayesianWorkflowDisease2021,gelmanBayesianWorkflow2020,schadPrincipledBayesianWorkflow2021}. The number of active users of such frameworks is currently estimated to be over a hundred thousand~\citep{carpenter100K10Years2022}. We provide examples with models that are commonly used by this community, but the general principles are not tied to any specific model or prior families. Furthermore, as the approach focuses on MCMC-based workflows, analytical derivations that would rely on conjugate priors or specific model families are not the focus and are not presented here.

A common workflow is to begin with a base model with template or `default' priors, and iteratively build more complex models~\citep{gelmanBayesianWorkflow2020}. Recommended template priors, and default priors in higher-level interfaces to Stan and PyMC, such as \texttt{rstanarm}~\citep{rstanarm}, \texttt{brms}~\citep{burknerBrmsPackageBayesian2017}, and \texttt{bambi}~\citep{Capretto2022}, are designed to be weakly informative and should work well when the data is highly informative so that the likelihood dominates. However, the presence of prior and likelihood sensitivity should still be checked, as no prior can be universally applicable. Considering the prevalence of default priors, a tool that assists in checking for prior (and likelihood) sensitivity is a valuable contribution to the community.

User-guided sensitivity analysis can be performed by fitting models with different specified
perturbations to the prior or likelihood~\citep{spiegelhalterBayesianApproachesClinical}, but this can require substantial amounts of both user and computing time~\citep{perezMCMCbasedLocalParametric2006,jacobiAutomatedSensitivityAnalysis2018}. Using more computationally efficient methods can reduce the computation time, but existing methods, while useful in many circumstances, are not always applicable. They are focused on particular types of
models~\citep{Roos2021,hunanyan2021quantification}
or inference
mechanisms~\citep{roosSensitivityAnalysisBayesian2015}, rely on
manual specification of
perturbations~\citep{mccartanAdjustrStanModel2021},
require substantial or technically complex changes to the model code that hinder
widespread
use~\citep{giordanoCovariancesRobustnessVariational2018,jacobiAutomatedSensitivityAnalysis2018}, or may still require substantial amounts of computation time~\citep{hoGlobalRobustBayesian2020,bornnEfficientComputationalApproach2010}.

We present a complementary sensitivity analysis approach that
aims to
\begin{itemize}
    \item be computationally efficient;
    \item be applicable to a wide range of models; 
    \item provide automated diagnostics;
    \item require minimal changes to existing model code and workflows.
\end{itemize}

We emphasise that the approach should not be used for repeated tuning of priors until diagnostic warnings no longer appear. Instead, the approach should be considered as a diagnostic to detect accidentally misspecified priors (for example default priors) and unexpected sensitivities or conflicts. The reaction to diagnostic warnings should always involve careful consideration about domain expertise, priors, and model specification.

\begin{figure}[ht]
  \centering
  \input{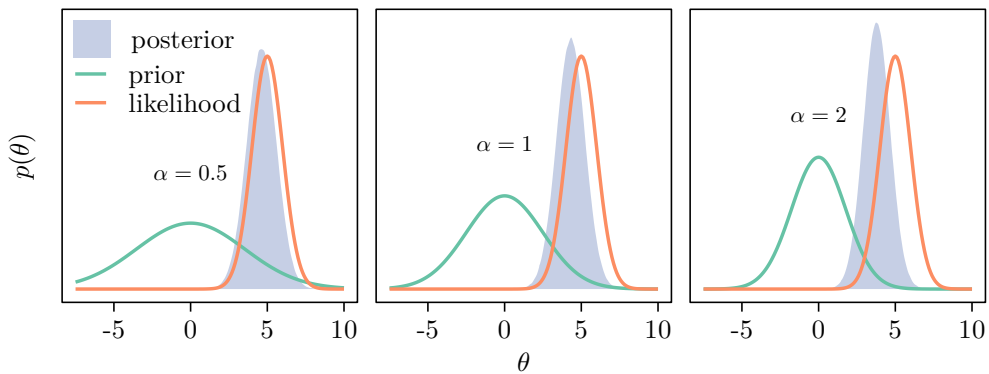}
  \caption{Example of our power-scaling sensitivity approach. Here, the prior is power-scaled, and the effect on the posterior is shown. In this case the prior
    is \(\normal(0, 2.5)\) and the likelihood is equivalent to
    \(\normal(5, 1)\). Power-scaling the prior by different
    \(\alpha\) values (in this case 0.5 and 2.0) shifts the posterior
    (shaded for emphasis), indicating prior sensitivity.}%
  \label{fig:scaling-example}
\end{figure}

\newpage

\begin{figure}[bt]
  \centering
  \small
  
  \begin{tikzpicture}[
    >=stealth,
    node distance = 0.5cm
    ]    

  \node[draw,
  align = center,
  ] (basefit) {(1) Run posterior inference};

  \node[draw,
  align = center,
  below left = 1cm,
  ] (perturbprior) {(2) Estimate properties of\\ posteriors with\\perturbed \textit{prior}};

  \node[draw,
  align = center,
  below right = 1cm,
  ] (perturblik) {(3) Estimate properties of\\ posteriors with\\perturbed \textit{likelihood}};

  \node[draw,
  align = center,
  below = of perturbprior,
  ] (checkprior) {(4) Check \textit{prior} \\sensitivity};

  \node[draw,
  align = center,
  below = of perturblik,
  ] (checklik) {(5) Check \textit{likelihood} \\sensitivity};

  \node[draw,
  align = center,
  below = 4cm,
  ] (diagnose) {(6) Diagnose sensitivity};
  
    \node[draw,
  align = center,
  below left = of diagnose,
  ] (adjust) {(7) Adjust model};

  \node[draw,
  align = center,
  below = of diagnose,
  ] (use) {(8) Use model};

  \draw[-stealth] (basefit) -- (perturbprior);
  \draw[-stealth] (basefit) -- (perturblik);
  \draw[-stealth] (perturbprior) -- (checkprior);
  \draw[-stealth] (perturblik) -- (checklik);
  \draw[-stealth] (checkprior) -- (diagnose);
  \draw[-stealth] (checklik) -- (diagnose);
  \draw[-stealth] (diagnose) -- (use);
  \draw[-stealth] (diagnose) -- (adjust);
  \draw[-stealth] (adjust) -| +(-2,0) |-  (basefit);
  
\end{tikzpicture}
\caption{Workflow of our proposed sensitivity analysis approach.}
\label{fig:workflow}
\end{figure}
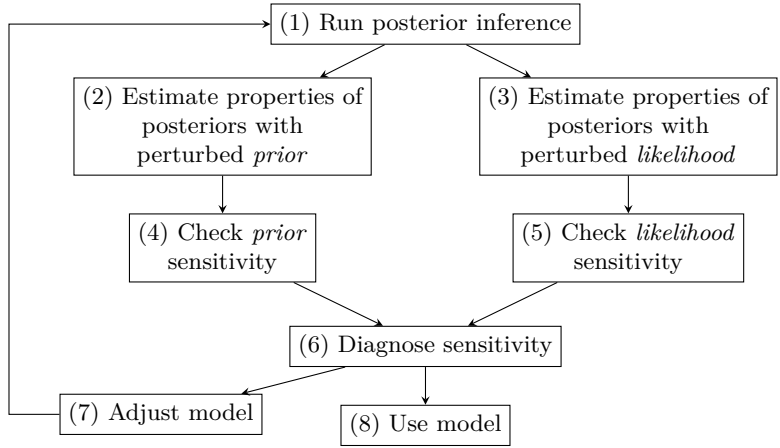

Our proposed approach uses importance sampling to estimate properties of perturbed
posteriors that result from power-scaling (exponentiating by some \(\alpha > 0\)) the prior or likelihood (see Figure~\ref{fig:scaling-example}). We use a variant of importance sampling (Pareto smoothed importance sampling; PSIS) that is self-diagnosing and alerts the user when estimates are untrustworthy~\citep{vehtariParetoSmoothedImportance2019}. We
propose a diagnostic, based on the the change to the posterior induced by this perturbation, that can indicate the
presence of prior-data conflict or likelihood noninformativity. Importantly, as long as the changes to the priors or likelihood induced by power-scaling are not too substantial, the procedure does not require refitting the model, which drastically increases its efficiency. The
envisioned workflow is as follows (also see
Figure~\ref{fig:workflow}):
\begin{enumerate}
\item[(1)] Fit a base model (either a template
  model or a manually specified model) to data, resulting in a base
  posterior distribution.
\item[(2, 3)] Estimate properties of perturbed posteriors that result from separately
  power-scaling the prior and likelihood.
\item[(4, 5)] Evaluate the extent the perturbed posteriors differ from the
  base posterior numerically and visually.
\item[(6)] Diagnose based on the pattern of prior and likelihood sensitivity.
\item[(7)] Reevaluate the assumptions implied by the base model and
  potentially modify it (and repeat (1)--(6)).
\item[(8)] Continue with use of the model for its intended purpose.
\end{enumerate}

\section{Details of the approach}

\subsection{Power-scaling perturbations}
\label{sec:method}

The proposed sensitivity analysis approach relies on separately perturbing the prior or likelihood through
power-scaling (exponentiating by some \(\alpha > 0\) close to 1). This power-scaling is
a controlled, distribution-agnostic method of modifying a probability
distribution. Intuitively, it can be considered to weaken (when \(\alpha < 1\)) or strengthen (when \(\alpha > 1\)) the component being power-scaled in relation to the other. Although power-scaling changes the normalising constant, this is not a concern when using Monte Carlo approaches for estimating posteriors via posterior draws. Furthermore, while the posterior can become improper when \(\alpha\) approaches 0, this is not an issue as we only consider values close to 1.

For all non-uniform distributions, as \(\alpha\)
diverges from \(1\), the shape of the distribution
changes. However, it retains the  support of the base distribution
(if the density at a point in the base distribution is zero,
raising it to any power will still result in zero; likewise any
nonzero density will remain nonzero). In the context of prior perturbations, these properties are desirable as slight
perturbations from power-scaling
result in distributions that likely represent similar implied
assumptions to those of the base distribution. A set of slightly perturbed priors can thus be considered a reasonable class of distributions for prior sensitivity
analysis~\citep[see][]{bergerRobustBayesianAnalysis1990,bergerOverviewRobustBayesian1994}. For the likelihood, power-scaling acts as an approximation for decreasing or increasing the number of (conditionally independent) observations, akin to data cloning~\citep{lele2007DataCloning} and likelihood weighting~\citep{greco_robust_2008,agostinelli_weighted_2013}. 

\begin{figure}[tb]
  \centering
  \input{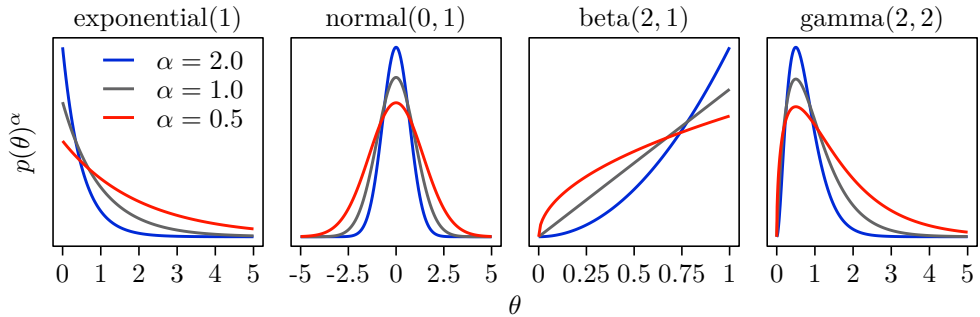}
  \caption{The effect of power-scaling on exponential family distributions commonly
    used as priors. In each case, the resulting distributions can be
    expressed in the same form as the base distribution with modified
    parameters. The power-scaling approach is not tied to specific distribution families, and these formulations are provided for intuition.}%
  \label{fig:example-dists}
\end{figure}

The power-scaling approach is not dependent on the form of the distribution family and will work providing that the distribution family is non-uniform (distributions with parameters controlling the support will only be power-scaled with respect to the base support). To provide intuition, we present analytically how power-scaling affects several exponential family distributions commonly used as priors (Figure~\ref{fig:example-dists} and Table~\ref{tab:dists}). For instance, a normal distribution,
\(\normal(\theta \mid \mu, \sigma) \propto \exp{(-\frac{1}{2}
  (\frac{\theta - \mu}{\sigma})^2)}\), when power-scaled by some
\(\alpha > 0\) simply scales the \(\sigma\) parameter by \(\alpha^{-1/2}\), thus
\(\normal(\theta \mid \mu, \sigma)^\alpha \propto \normal(\theta \mid
\mu, \alpha^{-1/2}\sigma )\)

Power-scaling, while effective, is only able to perturb a distribution in a particular manner. For example, it is not possible to directly shift the location of a distribution via power-scaling, without also changing other aspects. Like most diagnostics, when power-scaling sensitivity analysis does not indicate sensitivity, this only means that it could not detect sensitivity to power-scaling, not that the model is certainly well-behaved or insensitive to other types of perturbations. Nevertheless, power-scaling remains an intuitive perturbation as it mirrors increasing or decreasing the strength of prior beliefs or amount of data.

\begin{table}[tb]
  \centering
  \caption{Forms of power-scaled distributions for common
    distributions.}%
  \label{tab:dists}
  \begin{tabular}{ll}
    \toprule
    Base & Power-scaled \\
    \midrule
    \(\expdist(\theta \mid \lambda)\) & \( \propto \expdist(\theta \mid \alpha \lambda)\)\\
    \(\normal(\theta \mid \mu, \sigma)\) & \( \propto \normal(\theta \mid \mu, \alpha^{-1/2} \sigma)\) \\
    \(\betadist(\theta \mid s_1, s_2)\) & \( \propto \betadist(\theta \mid \alpha s_1 -
                                            \alpha + 1, \alpha s_2 - \alpha + 1)\) \\
    \(\gammadist(\theta \mid s_1, s_2)\) & \( \propto \gammadist(\theta \mid \alpha s_1 -
                                           \alpha + 1, \alpha s_2)\) \\
    \bottomrule
  \end{tabular}
\end{table}

\FloatBarrier

\subsection{Power-scaling priors}
\label{sec:prior}

In order for the sensitivity analysis approach to be independent of the number of parameters in the model, generally all priors are power-scaled simultaneously. However, in some cases, certain priors should be excluded from this set or others selectively power-scaled. For example, in hierarchical models, power-scaling both top- and intermediate-level priors can lead to unintended results. To illustrate this, consider two forms of prior, a non-hierarchical prior with two independent parameters \(p(\theta) \, p(\phi)\) and a hierarchical prior of the form \(p(\theta\mid\phi) \, p(\phi)\). In the first case, the appropriate power-scaling for the prior is  \(p(\phi)^\alpha  \, p(\theta)^\alpha\), while in the second, only the top level prior should be power-scaled, that is, \(p(\theta\mid\phi) \, p(\phi)^\alpha \). If the prior \(p(\theta\mid\phi)\) is also power-scaled, \(\theta\) will be affected by the power-scaling twice, directly and indirectly, perhaps even in opposite directions depending on the parameterisation.

\subsection{Estimating properties of perturbed posteriors}

As the normalizing constant for the posterior distribution can rarely
be computed analytically in real-world analyses, our approach assumes
that the base posterior is approximated using (Markov chain) Monte Carlo draws
(workflow step 1, see Figure~\ref{fig:workflow}). These draws are used
to estimate properties of the perturbed posteriors via importance sampling (workflow steps 2 and 3, see
Figure~\ref{fig:workflow}). Importance
sampling is a method to
estimate expectations of a target distribution by weighting draws from a proposal distribution~\citep{robert2004}. After computing these weights, there are several possibilities
for evaluating sensitivity.
For example, different summaries of perturbed posteriors can be computed
directly, or resampled draws can be generated
using importance resampling~\citep{rubin1988SIR}.

Importance sampling as a method for efficient sensitivity analysis has
been previously described by
\citet{bergerOverviewRobustBayesian1994,besagBayesianComputationStochastic1995,oneillImportanceSamplingBayesian2009,tsaiInfluenceMeasuresRobust2011}. However, one
limitation of importance sampling is that it can be unreliable when
the variance of importance weights is large or infinite. Hence,
as described by \citet{bergerOverviewRobustBayesian1994}, relying
on importance sampling to estimate a posterior resulting from a
perturbed prior or likelihood, without controlling the width of the
perturbation class (e.g. through a continuous parameter to control the
amount of perturbation, \(\alpha\) in our case) is likely to lead to unstable estimates.

To further alleviate issues with importance sampling, we
use Pareto smoothed importance sampling~\citep[PSIS;][]{vehtariParetoSmoothedImportance2019}, which stabilises the importance weights in an efficient,
self-diagnosing and trustworthy manner by modelling the upper tail of the importance weights with a generalised Pareto distribution. In cases where PSIS does not perform adequately, weights are adapted with
importance weighted moment matching~\citep[IWMM;][]{paananenImplicitlyAdaptiveImportance2021}, which is
a generic adaptive importance sampling algorithm that improves
the implicit proposal distribution by iterative weighted moment matching. The combination of using a continuous parameter to control the amount of perturbation, along with PSIS and IWMM, allows for a reliable and self-diagnosing method of estimating properties of perturbed posteriors.

\subsubsection{Calculating importance weights for power-scaling perturbations}

 Consider an expectation
of a function \(h\) of parameters \({\theta}\), which come from a
target distribution \(f({\theta})\):
\begin{equation}  \mathbb{E}_f[h({\theta})] = \int h({\theta})f({\theta}) d{\theta}.
\end{equation}
In cases when draws can be generated from the target distribution, the
simple Monte Carlo estimate can be calculated from a sequence of \(S\)
draws from \(f({\theta})\):
\begin{equation}\mathbb{E}_f[h({\theta})] \approx \frac{1}{S} \sum_{s=1}^Sh({\theta}^{(s)}), \text{where }
  {\theta}^{(s)} \sim f({\theta}).
\end{equation}
As an alternative to calculating the expectation directly with draws from \(f({\theta})\), the
importance sampling estimate instead uses draws from a proposal
distribution \(g({\theta})\) and the ratio between the target and
proposal densities, known as the importance weights \(w\). The self-normalised importance sampling estimate does not require known normalising constants of the target or proposal. Thus, it is well suited for use in the context of probabilistic programming languages, which do not calculate these:
\begin{equation} \label{eq:snis}  \mathbb{E}_f[h({\theta})] \approx \frac{\sum_{s=1}^S
    h(\theta^{(s)}) \frac{f({\theta}^{(s)})}{g({\theta}^{(s)})}}{\sum_{s=1}^S \frac{f({\theta}^{(s)})}{g({\theta}^{(s)})}} =
  \frac{\sum_{s=1}^S h(\theta^{(s)}) w^{(s)}}{\sum_{s=1}^S w^{(s)}}, \text{where } \theta^{(s)} \sim g(\theta).
\end{equation}

In the context of power-scaling perturbations, the proposal
distribution is the base posterior and the target distribution is
a perturbed posterior resulting from power-scaling. If the proposal and target
distributions are expressed as the products of the prior \(p(\theta)\) and
likelihood \(p(y \mid {\theta})\), with one of these components raised to the power of
\(\alpha\), the importance sampling weights only
depend on the density of the component being power-scaled. For prior power-scaling, the importance weights are
\begin{equation} \label{eq:wprior}
  w_{\alpha_\text{pr}}^{(s)} = \frac{p({\theta}^{(s)})^\alpha p(y \mid {\theta}^{(s)})}{p({\theta}^{(s)})p(y \mid {\theta}^{(s)})}\\
  = p({\theta}^{(s)})^{\alpha - 1}.
\end{equation}
Analogously, the importance weights for likelihood power-scaling
are
\begin{equation} \label{eq:wlik}
  w_{\alpha_\text{lik}}^{(s)} = p(y \mid {\theta}^{(s)})^{\alpha - 1}.
\end{equation}
As the importance weights are only dependent on the density of the
power-scaled component at the location of the proposal draws, they are easy to compute for a range of \(\alpha\) values.
See Appendix~\ref{sec:software} for practical implementation details about
computing the weights.

\subsection{Measuring sensitivity}
\label{sec:measuring-sens}

There are different ways to evaluate the effect of power-scaling perturbations on
a posterior (workflow steps 4 and 5, see
Figure~\ref{fig:workflow}). Here we present two options: first, a method
that investigates changes in specific posterior quantities of interest
(e.g.\ mean and standard deviation), and second, a
method based on the distances between the base marginal posteriors and the
perturbed marginal posteriors. These methods should not be considered competing, but rather allow for different levels of sensitivity
analysis, and depending on the context and what the modeller is interested in, one may be more useful than the other. Importantly, the proposed power-scaling approach is not tied to any
particular method of evaluating sensitivity. These methods are our
suggestions, but once quantities or weighted draws from perturbed posteriors
are computed, a multitude of comparisons to the base posterior and other posteriors
can be performed.

\subsubsection{Quantity-based sensitivity}
\label{sec:quantity}

In some cases it can be most useful to investigate
sensitivity of particular quantities of interest. Expectations of interest for a perturbed
posterior can be calculated from the base draws and the importance
weights using Equation~(\ref{eq:snis}). Other quantities that are not expectations (such as the median and quantiles) can be derived from the weighted empirical cumulative distribution function~(ECDF). Computed quantities can then be compared based on the specific interests of the modeller, or local sensitivity can be quantified by derivatives with respect to the perturbation parameter \(\alpha\).

\subsubsection{Distance-based sensitivity}
\label{sec:distance}

We can investigate the sensitivity of marginal posteriors using a distance-based approach. Here, we follow previous work which has quantified sensitivity based on the distance between the base and perturbed posteriors~\citep{ohaganHSSS,allabadiMeasuringBayesianRobustness2021,kurtekBayesianSensitivityAnalysis2015}.
In principle, many different divergence or distance measures can be
used, although there may be slight
differences in interpretation~\citep[see, for example][]{lekHowChoiceDistance2019,chaComprehensiveSurveyDistance2007}, however, the cumulative Jensen-Shannon divergence~\citep[CJS;][]{nguyenNonparametricJensenShannonDivergence2015} has two properties that make it appropriate for our use case. First, its symmetrised form is upper-bounded, like the standard Jensen-Shannon divergence~\citep{linDivergenceMeasuresBased1991}, which aids interpretation. Second, instead of comparing probability density functions (PDFs) or empirical kernel density estimates, as the standard Jensen-Shannon divergence does, it compares cumulative distribution functions (CDFs) or ECDFs, which can be efficiently estimated from Monte Carlo draws. Although PDFs could be estimated from the draws using kernel density estimation and then the standard Jensen-Shannon distance used, this relies on smoothness assumptions and may require substantially more draws to be accurate, and lead to artefacts otherwise~\citep[for further discussion of the benefits of ECDFs, see, for example][]{sailynoja2021graphical}.

Given two CDFs \(P(\theta)\) and \(Q(\theta)\),
\begin{equation}
\label{eq:cjs}
    \cjs(P(\theta) \| Q(\theta)) = \int P(\theta)\log_2 \left(\frac{2 P(\theta)}{P(\theta) + Q(\theta)}\right) \mathop{d\theta} + \frac{1}{2\ln(2)} \int Q(\theta) - P(\theta) \mathop{d\theta}.
\end{equation}
As a distance measure, we use the symmetrised and metric (square root) version of CJS, normalised with respect to its upper bound, such that it is bounded on the 0 to 1 interval~\citep[for further details see][]{nguyenNonparametricJensenShannonDivergence2015}:
\begin{equation}
\label{eq:cjsdist}
    \cjs_{\text{dist}}(P(\theta) \| Q(\theta)) = \sqrt{\frac{\cjs(P(\theta) \| Q(\theta)) + \cjs(Q(\theta) \| P(\theta))}{\int P(\theta) + Q(\theta) \mathop{d\theta}}}.
\end{equation}
As \(\cjs\) is not invariant to the sign of the parameter values, \(\cjs(P(\theta)\|Q(\theta)) \neq \cjs(P(-\theta)\|Q(-\theta))\), we use \(\max(\cjs_{\text{dist}}(P(\theta)\|Q(\theta)),\cjs_{\text{dist}}(P(-\theta)\|Q(-\theta)))\) to account for this and ensure applicability regardless of possible transformations applied to posterior draws that may change the sign.

In our approach, we compare the ECDFs of the base posterior to the perturbed posteriors. The ECDF of the base posterior is estimated from the base posterior draws, whereas the ECDFs of the perturbed posteriors are estimated by first weighting the base draws with the importance weights. The ECDF is a step-function derived from the draws. In an unweighted ECDF, the heights of each step are all equal to \(1/S\), where \(S\) is the number of draws. In the weighted ECDF, the heights of the steps are equal to the normalised importance weights of each draw. As described by \citet{nguyenNonparametricJensenShannonDivergence2015}, when using ECDFs, the integrals in Equations~(\ref{eq:cjs}) and (\ref{eq:cjsdist}) reduce to sums, which allows for efficient computation. 

\subsubsection{Local sensitivity}
\label{sec:local}
Both distance-based and quantity-based sensitivity can be evaluated
for any \(\alpha\) value.
It is also possible to obtain an overall estimate of sensitivity
at \(\alpha = 1\) by differentiation.
This follows previous work which defines
the local sensitivity as the derivative with respect to the
perturbation parameter~\citep{gustafsonLocalRobustnessBayesian2000,maroufyLocalGlobalRobustness2015,sivaganesanRobustBayesianDiagnostics1993,giordanoCovariancesRobustnessVariational2018}. For power-scaling, we suggest considering the derivative with respect to \(\log_2(\alpha)\) as it captures the symmetry of power-scaling around \(\alpha = 1\) and provides values on a natural scale in relation to halving or doubling the log density of the component.

Because of the simplicity of the power-scaling procedure, local sensitivity at \(\alpha = 1\) can be computed analytically with importance sampling for certain quantities, such as the mean and variance, without knowing the analytical form of the posterior.
This allows for a highly computationally efficient method to probe for sensitivity
in common quantities before performing further sensitivity diagnostics. For quantities that are computed as an expectation of some function \(h\), the derivative at \(\alpha = 1\)
can be computed as follows. We denote the power-scaling importance weights as \(p_{\text{ps}}(\theta^{(s)})^{\alpha - 1}\), where \(p_{\text{ps}}(\theta^{(s)})\) is the density of the power-scaled component, which can be either the prior or likelihood depending on the type of scaling. Then
\begin{align*}
\left . \frac{\sum_{s=1}^S h(\theta^{(s)}) p_{\text{ps}}(\theta^{(s)})^{\alpha - 1} }{\sum_{s=1}^S p_{\text{ps}}(\theta^{(s)})^{\alpha - 1}}
\frac{\partial}{\partial \log_2(\alpha)}\right|_{\alpha = 1}\\
= \left . \frac{\left(\sum_{s=1}^S \alpha \ln(2) h(\theta^{(s)}) p_{\text{ps}}(\theta^{(s)})^{\alpha - 1} \ln(p_{\text{ps}}(\theta^{(s)}))\right)\left(\sum_{s=1}^S p_{\text{ps}}(\theta^{(s)})^{\alpha - 1}\right)}{\left(\sum_{s=1}^S p_{\text{ps}}(\theta^{(s)})^{\alpha - 1}\right)^2} \right|_{\alpha = 1}\\
- \left . \frac{\left(\sum_{s=1}^S h(\theta^{(s)}) p_{\text{ps}}(\theta^{(s)})^{\alpha - 1}\right)\left(\sum_{s=1}^S \alpha \ln(2) p_{\text{ps}}(\theta^{(s)})^{\alpha - 1} \ln(p_{\text{ps}}(\theta^{(s)}))\right)}{\left(\sum_{s=1}^S p_{\text{ps}}(\theta^{(s)})^{\alpha - 1}\right)^2} \right|_{\alpha = 1}\\
= \ln(2) \left(\frac{1}{S} \sum_{s=1}^S \ln(p_{\text{ps}}(\theta^{(s)})) h( \theta^{(s)}) - \left (\frac{1}{S} \sum_{s=1}^S h(\theta^{(s)}) \right) \left (\frac{1}{S} \sum_{s=1}^S \ln(p_{\text{ps}}(\theta^{(s)})) \right )\right).
\end{align*}

%%%% ORIGINAL (not log)
% \being{align*}
% \left . \frac{\sum_{s=1}^S h(\theta^{(s)}) (w_{\text{ps}}^{(s)})^{\alpha - 1} }{\sum_{s=1}^S (w_{\text{ps}}^{(s)})^{\alpha - 1}}
% \frac{\partial}{\partial \alpha}\right|_{\alpha = 1}\\
% = \left . \frac{(\sum_{s=1}^S h(\theta^{(s)}) (w_{\text{ps}}^{(s)})^{\alpha - 1} \log (w_{\text{ps}}^{(s)}))(\sum_{s=1}^S (w_{\text{ps}}^{(s)})^{\alpha - 1})}{(\sum_{s=1}^S (w_{\text{ps}}^{(s)})^{\alpha - 1})^2} \right|_{\alpha = 1}\\
% - \left . \frac{(\sum_{s=1}^S h(\theta^{(s)}) (w_{\text{ps}}^{(s)})^{\alpha - 1})(\sum_{s=1}^S (w_{\text{ps}}^{(s)})^{\alpha - 1} \log (w_{\text{ps}}^{(s)}))}{(\sum_{s=1}^S (w_{\text{ps}}^{(s)})^{\alpha - 1})^2} \right|_{\alpha = 1}\\
% = \frac{1}{S} \sum_{s=1}^S \log (w_{\text{ps}}^{(s)}) h( \theta^{(s)}) - \left (\frac{1}{S} \sum_{s=1}^S h(\theta^{(s)}) \right) \left (\frac{1}{S} \sum_{s=1}^S \log (w_{\text{ps}}^{(s)}) \right ) .
% \end{align*}

Consider for example that we are interested in the sensitivity of the posterior
mean of the parameters \(\theta\). For prior scaling,
the derivative of the mean with respect to \(\log_2(\alpha)\) at \(\alpha = 1\) is then
\begin{align}
 D_{\text{mean}} &= \ln(2) \left( \frac{1}{S} \sum_{s=1}^S \ln (p(\theta^{(s)}))  \theta^{(s)} - \left (\frac{1}{S} \sum_{s=1}^S \theta^{(s)} \right) \left (\frac{1}{S} \sum_{s=1}^S \ln (p(\theta^{(s)})) \right ) \right).
\end{align}

As with quantity-based sensitivity, distance-based sensitivity can also be quantified
by taking the corresponding derivative.
\(\cjs_{\text{dist}}\) increases from 0 as \(\alpha\) diverges from 1 (approximately linearly in log scale) and its derivative is discontinuous at \(\alpha = 1\). As a measure of local power-scaling sensitivity, we take the average of the absolute derivatives of the divergence in the negative and positive \(\alpha\) directions, with respect to \(\log_2(\alpha)\). We approximate this numerically from the ECDFs with finite differences:
\begin{align*}
    D &= \frac{1}{2}\left(\left|\frac{f(x - \Delta x) - f(x)}{-\Delta x} \right| + \left| \frac{f(x + \Delta x) - f(x)}{\Delta x} \right| \right) \\
    % D_{\cjs} &= \left(\left| \frac{\cjs_{\text{dist}}(\hat{P}_1(\theta) \| \hat{P}_{1/(1 + \delta)}(\theta)) - \cjs_{\text{dist}}(\hat{P}_1(\theta) \| \hat{P}_{1}(\theta))}{\log_2(1 + \delta)}\right| + \left|\frac{\cjs_{\text{dist}}(\hat{P}_1(\theta) \| \hat{P}_{1 + \delta}(\theta)) - \cjs_{\text{dist}}(\hat{P}_1(\theta)}{-\Delta x} \right|\right)/2 \\
    D_{\cjs} &= \frac{\cjs_{\text{dist}}(\hat{P}_1(\theta) \| \hat{P}_{1/(1 + \delta)}(\theta)) + \cjs_{\text{dist}}(\hat{P}_1(\theta) \| \hat{P}_{1 + \delta}(\theta))}{2\log_2(1+\delta)},
\end{align*}
where \(\hat{P}_1(\theta)\) is the ECDF of the base posterior (when \(\alpha = 1\)), \(\hat{P}_{1/(1+\delta)}(\theta)\) is the weighted ECDF when \(\alpha = 1/(1+\delta)\) and \(\hat{P}_{1+\delta}(\theta)\) is the weighted ECDF when \(\alpha = 1 + \delta\). For implementation we use \(\delta = 0.01\).

\subsubsection{Diagnostic threshold}
\label{sec:threshold}
We consider \(D_{\cjs} \geq 0.05\) to be a reasonable indication of sensitivity. Distance metrics (and corresponding sensitivity diagnostics) can be calibrated and transformed with respect to perturbing a known distribution, such as a standard normal~\citep[e.g.][]{roosSensitivityAnalysisBayesian2015}. While we do not transform the value of \(D_{\cjs}\) in such a way, a comparison with the normal distribution can aid interpretation: For a standard normal, a \(D_{\cjs}\) of 0.05 corresponds to the mean being shifted by more than approximately 0.3 standard deviations, or the standard deviation differing by a factor greater than approximately 0.3, when the power-scaling \(\alpha\) is changed by a factor of two. A graphical depiction of this distance is shown in Figure~\ref{fig:threshold-example} in Appendix~\ref{sec:threshold-appendix}. However, depending on how concerned a modeller is with sensitivity, this threshold can be adapted to reflect what constitutes a meaningful change in the specific model.

\FloatBarrier

\subsection{Diagnosing sensitivity}
\label{sec:examples}

Sensitivity can be diagnosed by comparing the amount of exhibited prior and likelihood sensitivity (workflow step 6, see Table~\ref{tab:sensitivity-types}).
When a model is well-behaved, it is expected that there will be likelihood sensitivity, as power-scaling the likelihood is analogous to changing the number of (conditionally independent) observations. In hierarchical models, it is important to recognise that this is analogous to changing the number of observations within each group, rather than the number of groups. As such, in hierarchical models, lack of likelihood sensitivity based on power-scaling does not necessarily indicate that the likelihood is weak overall. As there can be relations between parameters, the pattern of sensitivity for a single parameter should be considered in the context of others. Cases in which the posterior is insensitive to both prior and likelihood power-scaling (i.e.\ uninformative prior with likelihood noninformativity) will likely be detectable from model fitting issues, and are not further addressed by our approach.

\begin{table}[tb]
  \centering
  \caption{The interplay between prior sensitivity and
    likelihood sensitivity can be used to diagnose the cause.}%
  \label{tab:sensitivity-types}
  \begin{tabular}{cccc}
    \toprule
    & & \multicolumn{2}{c}{\makecell{Prior\\ sensitivity}} \\
    & & No & Yes \\
    \midrule
    \multirowcell{2}[-0.5em]{Likelihood\\ sensitivity} & No & & \makecell[l]{Likelihood\\ noninformativity} \\
    \cmidrule{2-4}
    & Yes & \makecell[l]{Likelihood\\ domination} & \makecell{Prior-data conflict} \\
    \bottomrule
  \end{tabular}
\end{table}

\begin{figure}[tb]
\centering
    \input{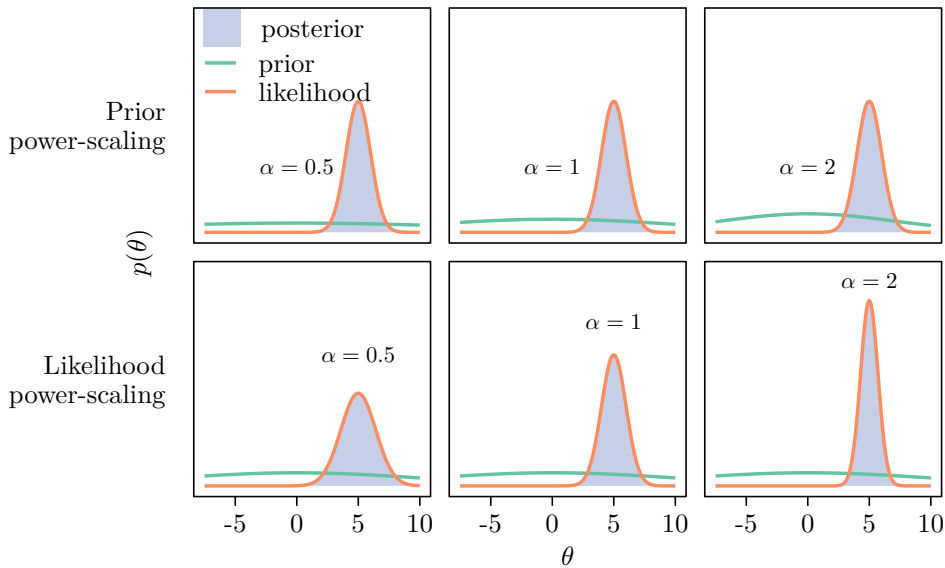}
  
  \caption{A weakly informative \(\normal(0, 10)\) prior  and a
    well-behaving \(\normal(5, 1)\) likelihood lead to likelihood domination. This is indicated by little to no
    prior sensitivity and expected likelihood sensitivity.  This is the outcome that many default priors aim for, as the prior has little influence on the posterior. Top row: the prior is power-scaled; bottom row: the likelihood is power-scaled. Note that
    in the figure the likelihood and posterior densities are almost
    completely overlapping.}%
  \label{fig:weaklyinf}
\end{figure}

Likelihood domination (the combination of a weakly informative
or diffuse prior combined with a well-behaving and informative likelihood) will result
in likelihood sensitivity but no prior sensitivity. This indicates that the posterior is mostly reliant on the data and likelihood rather than the prior (see Figure~\ref{fig:weaklyinf}). This is the outcome that default priors aim for, as the prior has little influence on the posterior.

\begin{figure}
\centering
  \input{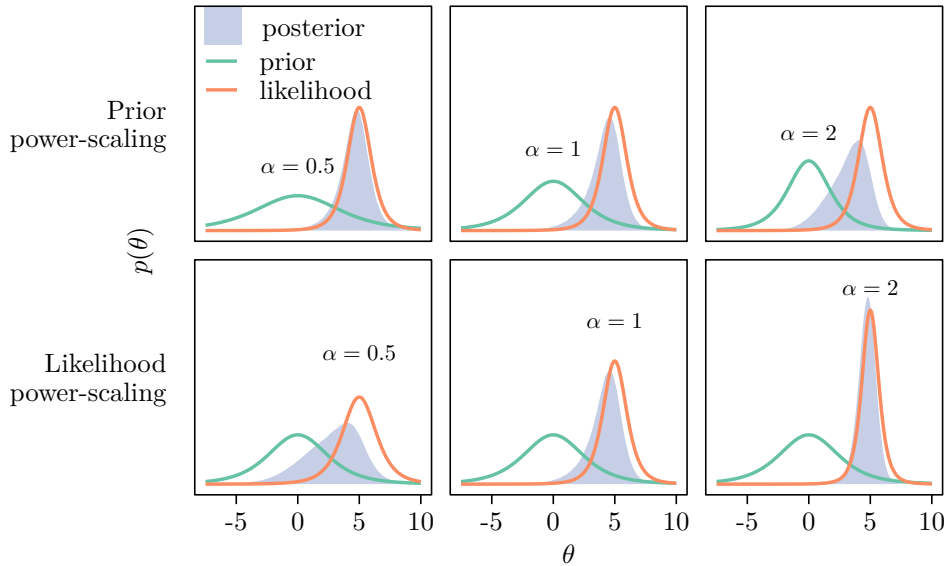}
  \caption{Conflict between \(t_4(0, 2.5)\) prior and \(t_4(5, 1)\) likelihood, results in the posterior (shaded for emphasis)
    being sensitive to both prior and likelihood power-scaling.  Top row: the prior is power-scaled; bottom row: the likelihood is power-scaled.}
      \label{fig:conflict-2}
  \end{figure}
In contrast, prior sensitivity can result from two primary causes, both of which are indications that the model may have an issue: 1)
\textit{prior-data conflict} and 2) \textit{likelihood noninformativity}. In the case of prior-data conflict, the posterior will exhibit both prior and likelihood sensitivity, whereas in the case of likelihood noninformativity (in relation to the prior) there
will be some marginal posteriors which are not as sensitive to
likelihood power-scaling as they are to prior power-scaling (or not at all sensitive to
likelihood power-scaling).

Prior-data conflict~\citep{walter2009,evansCheckingPriordataConflict2006,nottCheckingPriorDataConflict2020} can
arise due to intentionally or unintentionally informative priors
disagreeing with, but not being dominated by, the likelihood. When this is the case, the posterior will
be sensitive to power-scaling both the prior and the likelihood, as
 illustrated in Figure~\ref{fig:conflict-2}. When prior-data conflict has been detected, the modeller may wish to modify the model by using a less informative prior~\citep[e.g.,][]{evansWeakInformativityInformation2011,nottUsingPriorExpansions2020} or using heavy-tailed distributions~\citep[e.g.,][]{gagnon2021robustness,ohaganBayesianHeavytailedModels2012}.

The presence of prior sensitivity but relatively low (or no) likelihood sensitivity is an indication that the likelihood is weakly informative (or noninformative) in relation to the prior. This can occur, for example, when there is complete
separation in a logistic regression. The simplest case of complete
separation occurs when there are observations of only one class. For
example, suppose a researcher is attempting to identify the occurrence
rate of a rare event in a new population. Based on previous research,
it is believed that the rate is close to 1 out of 1000. The researcher
has since collected 100 observations from the new population, all of
which are negative. As the data are only of one class, the posterior
will then exhibit prior sensitivity as the likelihood is relatively weak. In the case of weakly informative or noninformative likelihood, the choice of prior will have a direct impact on the posterior and is therefore of a greater importance and should be considered carefully.
In some cases, the likelihood (or the data) may not be problematic in
and of itself, but if the chosen prior is highly informative and dominates the likelihood, the
posterior may be relatively insensitive to power-scaling the
likelihood. As such, when interpreting sensitivity it is important to consider both the prior and the likelihood and the interplay between them~\citep[see related discussion by][]{gelmanPriorCanOften2017}.

%\FloatBarrier

\subsubsection{Sensitivity for parameter combinations and other quantities}

As discussed, sensitivity can be evaluated for each marginal distribution separately in a relatively automated manner. This approach may lead to interpretation issues when individual parameters are by definition not informed by the likelihood, or are not readily interpretable. In the case when the likelihood may be informative for a combination of parameters, but not any of the parameters individually, it can be useful to perform a whitening transformation (such as principle component analysis)~\citep{kessy2018Sphere} on the posterior draws and then investigate sensitivity in the compressed parameter space. This can indicate which parameter combinations are sensitive to likelihood perturbations, indicating that they are jointly informed by the likelihood, and which are not.

This whitening approach works when there are few parameters, but as the number of parameters grows, the compressed components can be more difficult to interpret. Instead, in more complex cases, we suggest the modeller focus on target quantities of interest. For example, in the case of Gaussian process regression or models specifically focused on predictions, it can be more useful to investigate the sensitivity of predictive distributions~\citep{paananen2021RSense,paananen2019VariableSelection} than posterior distributions of model parameters.

\section{Software implementation}
\label{sec:implementation}

Our approach for power-scaling sensitivity analysis is implemented in
\texttt{priorsense} (\url{https://github.com/n-kall/priorsense}), our new
R~\citep{rcoreteamLanguageEnvironmentStatistical2020} package for
prior sensitivity diagnostics. The implementation focuses on models
fit with
Stan~\citep{standevelopmentteamStanModellingLanguage2021}, but it can
be extended to work with other probabilistic programming frameworks
that provide similar functionality. The package includes numerical
diagnostics and graphical representations of changes in
posteriors. These are available for both distance- and quantity-based
sensitivity. Further details on the usage and implementation are included in Appendix~\ref{sec:software}.

\section{Simulations}

Here we present two simulations demonstrating how the diagnostic \(D_{\cjs}\) performs in two scenarios: (a) when the likelihood corresponds to the true model, but the data realisation may weakly inform some parameters, and (b) when the prior is changed to be in increasing conflict with the likelihood. We show that the diagnostic can detect these two cases.

\subsection{Separation simulation}
\label{sec:separation-sim}

We generated 1000 data realisations of \(N = 25\) observations, with the following structure:
\begin{equation*}
    x_{1,i} \sim \uniform(-1, 1),\quad
    x_{2,i} \sim \uniform(-1, 1),
\end{equation*}
\begin{equation*}
  Y_{i} \sim \Bernoulli(p_{i}),\quad
  \log\left(\frac{p_{i}}{1-p_{i}}\right) = x_{1,i} + x_{2,i}.
\end{equation*}
We then fit a Bernoulli logit model to each realisation as follows:
\begin{equation*}
  Y_{i} \sim \Bernoulli(p_{i}),\quad
  \log\left(\frac{p_{i}}{1-p_{i}}\right) = \beta_0 + \sum_{k=1}^2{\beta_k x_{k,i}}
  \end{equation*}
  \begin{equation*}
  \beta_0 \sim \normal(0, 10),\quad
  \beta_k \sim \normal(0, 2.5).
\end{equation*}
The model is correctly specified and matches the data generating process. However, a single realisation of 25 observations may be weakly informative due to complete or near-complete separation.

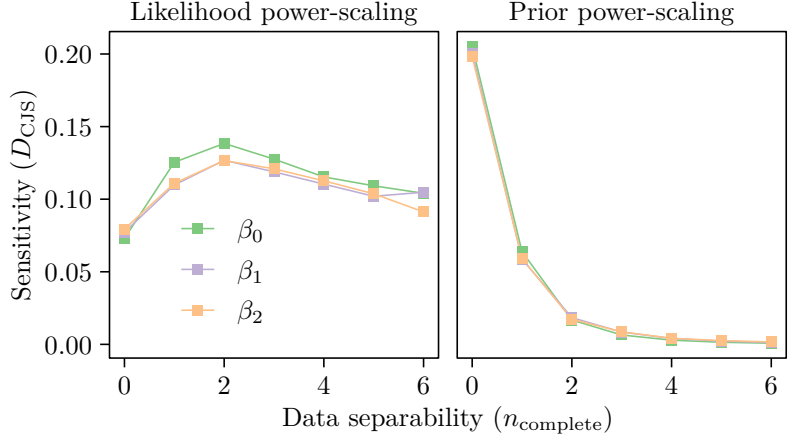
\begin{figure}
    \centering
    \input{./figs/separation_example}
    \caption{Relationship between data separability and sensitivity in the separation simulation. \(n_{\text{complete}}\) is the minumum number of observations that need to be removed to result in complete separation. Each point represents the mean over the data realisations for which the \(n_{\text{complete}}\) were equal.}
    \label{fig:separation-sim}
\end{figure}

For each data realisation, we compare the a measure of separation \(n_{\text{complete}}\)~\citep{christmannMeasuringOverlapBinary2001},  to the power-scaling sensitivity diagnostic \(D_{\text{CJS}}\). \(n_{\text{complete}}\) is defined as the minimum number of observations that need to be removed to result in complete separation. As shown in Figure~\ref{fig:separation-sim}, separability induced high sensitivity. When the data is completely or nearly separable, the prior sensitivity is high and when the data is far from completely separable, the prior sensitivity is low.

\FloatBarrier

\subsection{Conflict simulation}

We generate 100 data realisations of \(N = 25\) observations, with the following structure, for \(k \in {1,2,3,4}\):
{\small
\begin{equation*}
  x_{k,i} \sim \normal(0, 1),\quad
  y_i \sim \normal(\mu_i, 1),\quad
  \mu_i = 0.25 x_{1,i} + 0.25 x_{2,i} + 0.25 x_{3,i} + 0.25 x_{4,i}.
\end{equation*}
}

We then transform each data realisation such that \(x_{1,i} \leftarrow x_{1,i}/c\) and \(x_{2,i} \leftarrow x_{2,i}/c\), for \(c \in \{1, 2, 4, 6, 8\}\) to change the scale of the \(x_1\) and \(x_2\) variables and the corresponding coefficients, but not the values of \(y\). We then fit the following model to each transformed data set:
{\small
\begin{equation*}
  y_i \sim \normal(\mu_i, 1),\quad
  \mu_i = \beta_0 + \sum_{k=1}^{4}x_{k,i}\beta_k,
  \end{equation*}
  \begin{equation*}
  \beta_0 \sim t_3(0, 2.5),\quad
  \beta_k \sim \normal(0, 1).
\end{equation*}
}
The model is well specified in the sense that the parameter space of the model includes the parameter value of the data generating process.

 As \(c\) is increased, the priors on \(\beta_1\) and \(\beta_2\) will begin to conflict with the likelihood from finite data. We investigate the effect of this increase on the power-scaling sensitivity diagnostic \(D_{\cjs}\) for each regression coefficient.

\begin{figure}
\centering
    \input{./figs/covariate_scaling_example}
    \caption{Relation between scaling the covariates and the sensitivity. Each point represents the mean of 100 model fits (using different data realisations).}
    \label{fig:conflict-sim}
\end{figure}
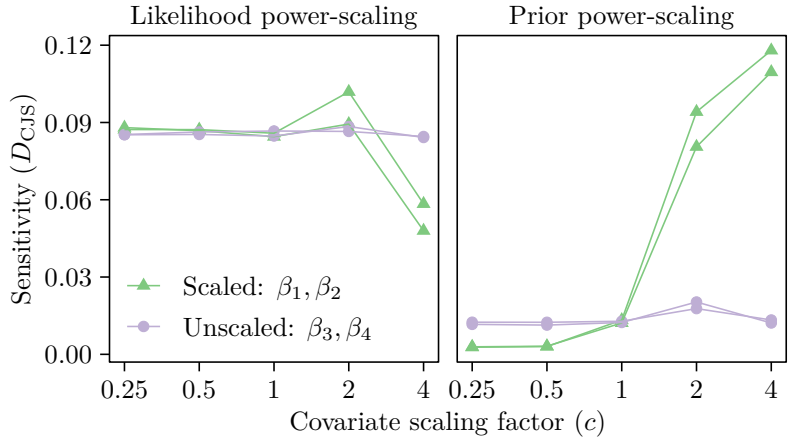

As shown in Figure~\ref{fig:conflict-sim}, the coefficients for the scaled predictors (\(\beta_1, \beta_2\)) exhibit different degrees of sensitivity depending on the degree of scaling. Prior sensitivity increases as the scaling factor increases, indicating prior-data conflict. Importantly, likelihood sensitivity decreases when \(c = 4\), indicating that the prior is beginning to dominate the likelihood. As expected, the other coefficients (\(\beta_3, \beta_4\)), do not exhibit sensitivity or changes in sensitivity.

\FloatBarrier

\section{Case studies}

In this section, we show how \texttt{priorsense} can be used in a Bayesian model building workflow to detect and diagnose prior sensitivity
in realistic models fit to real data (corresponding data and code are available at \url{https://github.com/n-kall/powerscaling-sensitivity}). We present a variety of models and show sensitivity diagnostics for different quantities, including regression coefficients (Sections~\ref{sec:bodyfat} and \ref{sec:banknotes}), scale parameters (Sections~\ref{sec:bodyfat}, \ref{sec:bacteria}, \ref{sec:motorcycle}), model fit (Section~\ref{sec:uscrime}), and posterior predictions (Sections~\ref{sec:motorcycle} and \ref{sec:covid}).

We use the \texttt{brms} package ~\citep{burknerBrmsPackageBayesian2017}, which is a high-level R interface to Stan, to specify and fit the simpler regression
models and Stan directly for the more complex models. Unless further specified, we
use Stan to generate posterior draws using the default settings (4
chains, 2000 iterations per chain, half discarded as warm-up). Convergence diagnostics and effective sample sizes are checked for all model fits~\citep{vehtariRankNormalizationFoldingLocalization2021}, and sampling parameters are adjusted to relieve any identified issues before proceeding with sensitivity analysis. As the quantitative indication of sensitivity, we use \(D_{\text{CJS}}\) and the threshold of 0.05 as described in Section~\ref{sec:measuring-sens}, but we also present graphical checks.

\subsection{Body fat (linear regression)}
\label{sec:bodyfat}

This case study shows a situation in which \textit{prior-data conflict}
can be detected by power-scaling sensitivity analysis. This
conflict results from choosing priors that are not of appropriate
scales for some predictors. For this case study, we use the
\texttt{bodyfat} data
set~\citep{johnsonFittingPercentageBody1996}, which has
previously been the focus of variable selection
experiments~\citep{pavoneUsingReferenceModels2020,heinzeVariableSelectionReview2018}. The
aim of the analysis is to predict an expensive and cumbersome water immersion measurement of body
fat percentage from a set of thirteen easier to measure
characteristics, including age, height, weight, and circumferences of
various body parts.

We begin with a linear regression model to predict body fat percentage
from the aforementioned variables. By default, in \texttt{brms} the
\(\beta_0\) (intercept) and \(\sigma\) parameters are given data-derived weakly
informative priors, and the regression coefficients are given improper
flat priors. Power-scaling will not affect flat priors, so we specify proper priors for the regression coefficients. We specify the
same prior for all coefficients, \(\normal(0, 1)\), which does not
seem unreasonable based on preliminary prior-predictive checks. We arrive at the following model:
{\small
\begin{equation*}
  y_i \sim \normal(\mu_i, \sigma),\quad
  \mu_i = \beta_0 + \sum_{k=1}^{13}   \beta_k x_{k,i},
  \end{equation*}
  \begin{equation*}
  \beta_0 \sim t_3(0, 9.2),\quad
  \beta_k \sim \normal(0, 1),\quad
  \sigma \sim t_3^+(0, 9.2).
    \end{equation*}
}

From the marginal posteriors, there do not appear to be issues, and all estimates are in reasonable ranges (Figure~\ref{fig:bodyfat-posterior}). Power-scaling sensitivity
analysis, performed with the \texttt{powerscale\_sensitivity} function, however, shows that there is both prior sensitivity and likelihood sensitivity for one of the parameters, \(\beta_{\text{wrist}}\)
(Table~\ref{tab:bodyfat-sense}). This
indicates that there may be prior-data conflict.

\begin{table}[ht]
\centering
\begin{threeparttable} 
  \caption{Sensitivity diagnostic values for the body fat case study.}
  \label{tab:bodyfat-sense}
  \small
  \begin{tabular}{lllll}
    \toprule
    & \multicolumn{2}{c}{Original prior} & \multicolumn{2}{c}{Adjusted prior} \\ 
    Parameter & \makecell[l]{Prior\\ sensitivity} & \makecell[l]{Likelihood\\ sensitivity} & \makecell[l]{Prior\\ sensitivity} & \makecell[l]{Likelihood\\ sensitivity} \\
    \midrule
 \(\beta_{\text{wrist}}\) &   \textbf{0.12}    &   0.09 & 0.00    &   0.08 \\
   \(\beta_{\text{weight}}\) &   0.02    &   0.12 &   0.00    &   0.09 \\
  \(\beta_{\text{thigh}}\) &    0.01      & 0.08 &    0.00      & 0.10 \\
   \(\beta_{\text{neck}}\) &     0.01   &    0.11 &     0.00   &    0.09 \\
 \(\beta_{\text{knee}}\) &     0.01     &  0.1  &     0.00     &  0.08  \\
 \(\beta_{\text{hip}}\) &      0.01     &  0.11 &  0.00     &  0.09 \\
 \(\beta_{\text{height}}\) &   0.00      &    0.09 &   0.00      &    0.08 \\
 \(\beta_{\text{forearm}}\) &  0.02     &  0.12 &  0.00     &  0.09 \\
 \(\beta_{\text{chest}}\) &    0.01  &     0.08  &    0.00  &     0.09 \\
 \(\beta_{\text{biceps}}\) &   0.01     &  0.09 &   0.00     &  0.08 \\
 \(\beta_{\text{ankle}}\) &    0.02     &  0.1  &    0.00     &  0.09  \\
 \(\beta_{\text{age}}\) &      0.03     &  0.12  &      0.00     &  0.08 \\
  \(\beta_{\text{abdomen}}\) &  0.00    &      0.09 &  0.00    &      0.10 \\
  \(\beta_{\text{intercept}}\) & 0.00 & 0.07 & 0.00 & 0.10\\
  \(\sigma\) & 0.00 & 0.19 & 0.00 & 0.20 \\
    \bottomrule
  \end{tabular}
  \begin{tablenotes}
\item{Higher sensitivity values indicate greater sensitivity.}
\item{Prior sensitivity above \(0.05\) indicates informative prior (bold).}
\item{Likelihood sensitivity below \(0.05\) indicates weak or noninformative likelihood.}
\end{tablenotes}
\end{threeparttable}
\end{table}

\begin{figure}[tb]
    \centering
    \includegraphics{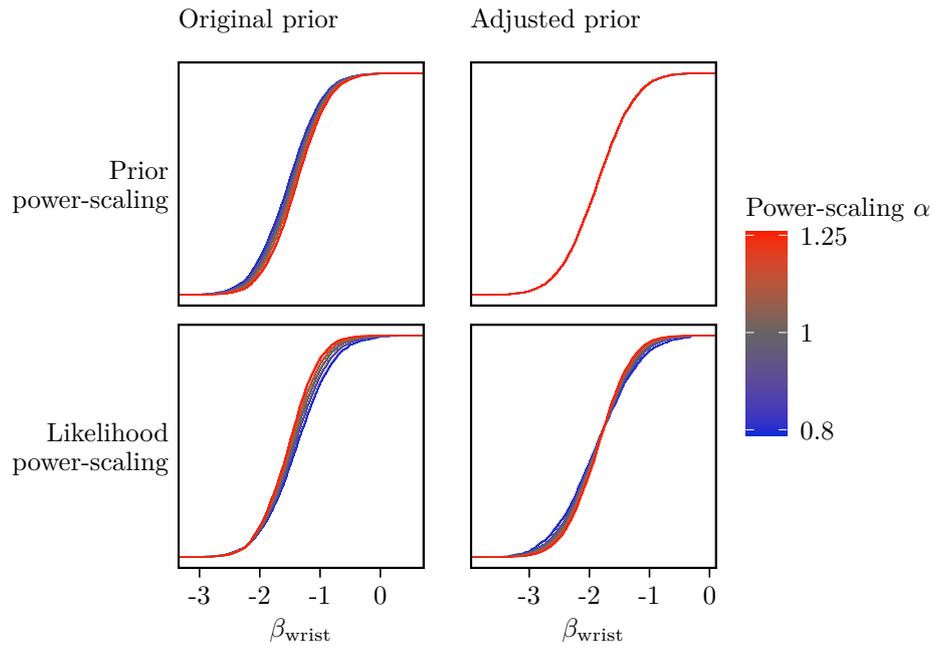}
    \caption{Power-scaling diagnostic plot of marginal ECDFs
      for posterior \(\beta_{\text{wrist}}\) in the body fat case
      study. (Left) Original prior; There is both prior and likelihood sensitivity, as the ECDFs are not overlapping. (Right) Adjusted prior; There is now no prior sensitivity, as the ECDFs are overlapping, whereas there is still likelihood sensitivity.}%
    \label{fig:bodyfat-both-sens}
  \end{figure}

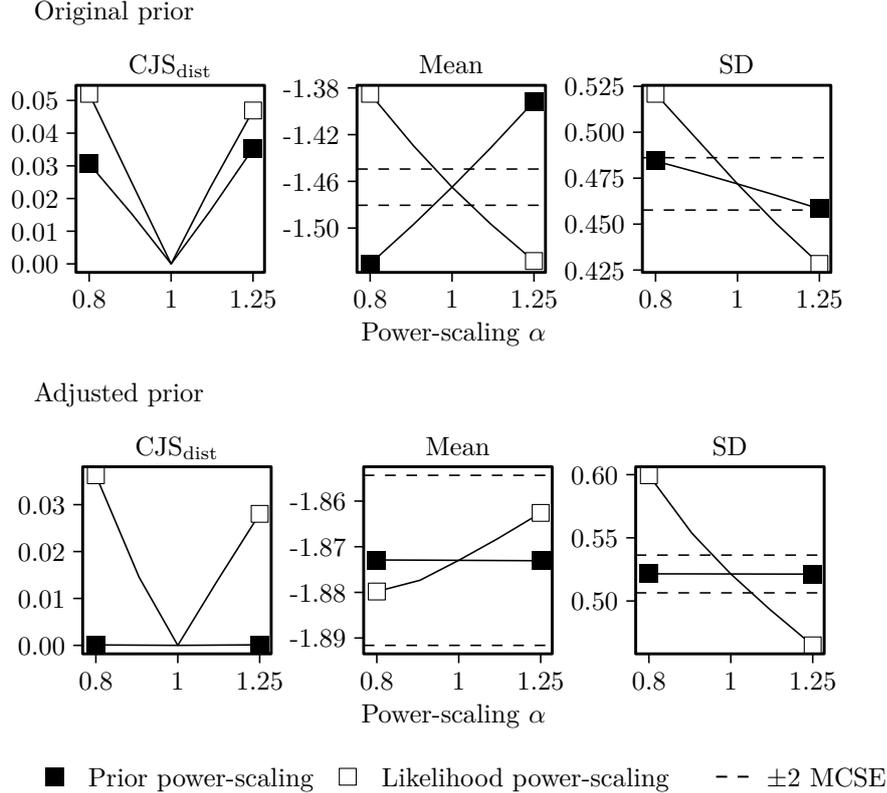
\begin{figure}[tb]
    \centering
    \input{./figs/bodyfat_quantities_joint.tex}
    \caption{Posterior quantities of \(\beta_{\text{wrist}}\) as a function of power-scaling for the body fat case study. With this plot, we can compare the effect of prior and likelihood power-scaling on specific quantities. Shown as dashed lines are \(\pm2\) Monte Carlo standard errors (MCSE) of the base posterior quantity, as guides to whether an observed change is meaningful. Top: original prior; The pattern of the change in the mean indicates prior-data conflict, as power-scaling the prior and likelihood have opposite directional effects on the posterior mean. Bottom: adjusted prior; there is no longer prior or likelihood sensitivity for the mean, indicating no prior-data conflict. Likelihood sensitivity for the posterior standard deviation remains, indicating that the likelihood is informative.}
    \label{fig:bodyfat-summary-sens}
\end{figure}

We then check how the ECDF of the posterior is affected by power-scaling of the prior and likelihood. In \texttt{priorsense}, this is done creating a sequence of weighted draws (for a sequence of \(\alpha\) values) using \texttt{powerscale\_sequence}, and then plotting this sequence with \texttt{powerscale\_plot\_ecdf} (Figure~\ref{fig:bodyfat-both-sens}, left). We see that the posterior is sensitive to both prior and likelihood power-scaling, and that it shifts right (towards zero) as the prior is strengthened, and left (away from zero) as the likelihood is strengthened. This is an indication of prior-data conflict,
which can be further seen by plotting the change in quantities using \texttt{powerscale\_plot\_quantities} (Figure~\ref{fig:bodyfat-summary-sens}). Prior-data conflict is evident by the `X' shape of the mean plot, as the mean is shifting in opposite directions. As there is prior sensitivity arising from prior-data conflict, which
is unexpected and unintentional as our priors were chosen to be weakly informative, we consider modifying the priors. On inspecting the raw
data, we see that although the predictor variables are all measured on
similar scales, the variances of the variables differ
substantially. For example, the variance of wrist circumference is
0.83, while the variance of abdomen is 102.65. This leads to our
chosen prior to be unintentionally informative for some of the
regression coefficients, including wrist, while being weakly
informative for others. To account for this, we refit the model with
priors empirically scaled to the data, \(\beta_k \sim \normal(0, 2.5 s_y/s_{x_k})\), where \(s_y\) is the standard deviation of \(y\) and \(s_{x_k}\) is the standard deviation of predictor variable \(x_k\). This corresponds to the default priors used for regression models in the \texttt{rstanarm} package~\citep{rstanarm}, as described in \citet{gelmanRegressionOtherStories2020} and \citet{gabry2020PriorDistributions}. We refit the model and see that the
posterior mean for \(\beta_{\text{wrist}}\) changes from -1.45 to -1.86,
indicating that the base prior was indeed unintentionally
informative and in conflict with the data, pulling the estimate towards zero. Power-scaling sensitivity
analysis on the adjusted model fit shows that there is no longer prior sensitivity, and there is appropriate likelihood sensitivity (Table~\ref{tab:bodyfat-sense}, Figure~\ref{fig:bodyfat-both-sens} right).

This is a clear example of how power-scaling sensitivity analysis
can detect and diagnose prior-data conflict. Unintentionally informative priors resulted in the conflict, which
could not be detected by only inspecting the posterior estimates of the base model. Once detected and diagnosed, the model could be adjusted and analysis could proceed. It is important to emphasise that the model was modified as the original priors were \emph{unintentionally} informative. If the original priors had been manually specified based on prior knowledge, it may not have been appropriate to modify the priors after observing the sensitivity, as the precise prior specification would be an inherent part of the model.

\FloatBarrier
\subsection{Banknotes (logistic regression)}
\label{sec:banknotes}

This case study is an example of using power-scaling sensitivity analysis to detect and diagnose \textit{likelihood noninformativity}. We use
the \texttt{banknote} data set~\citep{Flury1988} available from the \texttt{mclust} package~\citep{mclust},
which contains measurements of six properties of 100 genuine (\(Y = 0\)) and 100 counterfeit (\(Y = 1\)) Swiss banknotes. We fit a logistic regression on the status of a note based on these measurements. For priors, we
use the template priors
\(\normal(0, 10)\) for the intercept and \(\normal(0, 2.5/s_{x_k})\) for the regression coefficients, where \(s_{x_k}\) is the standard deviation of predictor \(k\). The model is then
{\small
\begin{equation*}
  Y_{i} \sim \Bernoulli(p_{i}),\quad
  \log\left(\frac{p_{i}}{1-p_{i}}\right) = \beta_0 + \sum_{k=1}^6{\beta_k x_{k,i}},
  \end{equation*}
  \begin{equation*}
  \beta_0 \sim \normal(0, 10),\quad
  \beta_k \sim \normal(0, 2.5/s_{x_k}).
\end{equation*}
}

Power-scaling sensitivity analysis indicates prior sensitivity for all predictor coefficients (Table~\ref{tab:bank-sens}). Furthermore, most exhibit low likelihood sensitivity, indicating a weak likelihood. In a Bernoulli model, this may arise if the binary outcome is completely separable by the predictors. This can be confirmed using the \texttt{detectseparation} package~\citep{kosmidis2021}, which detects infinite maximum likelihood estimates (caused by separation) in binary outcome regression models without fitting the model. Indeed, according to this method, the data set is completely separable and the prior sensitivity will remain, regardless of choice of prior. As shown in the simulation study in Section~\ref{sec:separation-sim}, this is not necessarily an indication that the model is misspecified or problematic, but rather the complete separation in the data realisation may be causing issues for estimating the regression coefficients.

\begin{table}[ht]
 \centering
 \begin{threeparttable}
 \caption{Sensitivity diagnostic values for the bank notes case study.}
   \label{tab:bank-sens}
 \begin{tabular}{llll}
 \toprule
   Parameter & \makecell[l]{Prior\\ sensitivity} & \makecell[l]{Likelihood\\ sensitivity} & Comment\\
   \midrule
 \(\beta_{\text{length}}\) & \textbf{0.07} & \textbf{0.02} & weak likelihood \\
 \(\beta_{\text{left}}\) & \textbf{0.10} & \textbf{0.01} & weak likelihood \\
 \(\beta_{\text{right}}\) & \textbf{0.08} & \textbf{0.02} & weak likelihood \\
 \(\beta_{\text{bottom}}\) & \textbf{0.25} & 0.11 & prior-data conflict \\
 \(\beta_{\text{top}}\) & \textbf{0.18} & \textbf{0.04} & weak likelihood \\
 \(\beta_{\text{diagonal}}\) & \textbf{0.13} & 0.05 & prior-data conflict \\
    \bottomrule
 \end{tabular}
 \begin{tablenotes}
\item{Higher sensitivity values indicate greater sensitivity.}
\item{Prior sensitivity above \(0.05\) indicates informative prior (bold).}
\item{Likelihood sensitivity below \(0.05\) indicates weak or noninformative likelihood (bold).}
 \end{tablenotes}
 \end{threeparttable}
 \end{table}

\FloatBarrier

\subsection{Bacteria treatment (hierarchical logistic regression)}
\label{sec:bacteria}

 Here, we use the \texttt{bacteria} data set, available from the \texttt{MASS} package~\citep{mass} to demonstrate power-scaling sensitivity analysis in hierarchical models. This data has previously been used by 
\citet{kurtekBayesianSensitivityAnalysis2015} in a sensitivity analysis comparing posteriors resulting from different priors. We use the same model structure and similar priors and arrive at matching conclusions. Importantly, we show that the problematic prior can be detected from the resulting posterior, without the need to compare to other posteriors (and without the need for multiple fits). The data set contains 220 observations of the effect of a treatment (placebo, drug with low compliance, drug with high compliance) on 50 children with middle ear infection over 5 time points (week). The outcome variable is the presence (\(Y = 1\)) or absence (\(Y = 0\)) of the bacteria targeted by the drug. We fit the same generalised linear multilevel model on the data as \citet{kurtekBayesianSensitivityAnalysis2015}, based on an example from \citet{brownMCMCGeneralizedLinear2010}:
{\small
\begin{equation*}
  Y_{ij} \sim \Bernoulli(p_{ij}),\quad
  \log\left(\frac{p_{ij}}{1-p_{ij}}\right) = \mu + \sum_{k=1}^3{x_{kij}\beta_k + V_i},
  \end{equation*}
  \begin{equation*}
  \mu \sim \normal(0, 10),\quad
  \beta_k \sim \normal(0, 10),\quad
  V_i \sim \normal(0, \sigma),\quad \tau = \frac{1}{\sigma^2} \sim \gammadist(0.01, 0.01).
  \end{equation*}
}

\begin{figure}[ht]
  \centering
  \input{./figs/bacteria_priors.tex}
  \caption{Priors for the hyperparameter \(\tau\) in the bacteria case study. Priors considered reasonable for this application are shown on the left while priors considered unreasonable are shown on the right.}%
  \label{fig:bacteria-priors}
\end{figure}
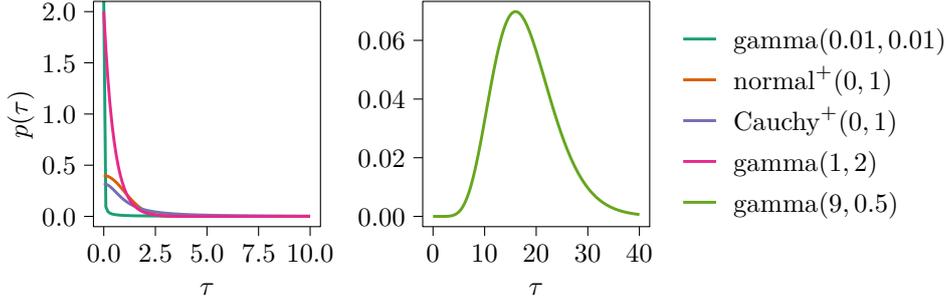

We try different priors for the precision
hyperparameter \(\tau\).
We compare the sensitivity of the base model, with prior \(\tau \sim \gammadist(0.01, 0.01)\), to the comparison
priors. Three of which are considered reasonable, \(\tau \sim \normal^+(0, 10), \Cauchy^+(0, 100), \gammadist(1, 2)\), and one is considered unreasonable, \(\tau \sim \gammadist(9, 0.5)\). These priors are shown in Figure~\ref{fig:bacteria-priors}. We fit each model with four chains of 10000 iterations (2000 discarded as warmup) and perform power-scaling sensitivity analysis on each.
As discussed in Section~\ref{sec:prior}, 
only the top-level parameters in the hierarchical prior are power-scaled (i.e.\ the prior on \(V_i\) is not power-scaled). Posterior quantities and sensitivity diagnostics for all models are shown in Appendix~\ref{sec:bacteria-appendix}. It is
apparent that the \(\tau\) parameter is sensitive to the prior when using the \(\gammadist(9, 0.5)\) prior. This indicates that the prior may be inappropriately informative. Although there is no indication of power-scaling sensitivity for the \(\mu\) and \(\beta\) parameters, comparing the posteriors for the models indicates differences in these parameters for the unreasonable \(\tau\) prior compared to the other priors. This is an important observation, and highlights that power-scaling is a local perturbation and may not influence the model strongly enough to change all quantities, yet can indicate the presence of potential issues.

\FloatBarrier
\subsection{Motorcycle crash (Gaussian process regression)}
\label{sec:motorcycle}
 
Here, we demonstrate power-scaling sensitivity analysis on model without readily interpretable model parameters. We fit a Gaussian process regression to the \texttt{mcycle} data set, also available in the \texttt{MASS} package and show the sensitivity of predictions to perturbations of the prior and likelihood. For a primer on Gaussian process regression, see~\citet{seegerGaussianProcessesMachine2004}.

The data set contains 133 measurements of head acceleration at different time points during a simulated motorcycle crash. It is further described by \citet{Silverman1985}. We fit a Gaussian process regression to the data, predicting the head acceleration (\(y\)) from the time (\(x\)). We use two Gaussian processes; one for the mean and one for the standard deviation of the residuals. The model is
\begin{equation*}
    y \sim \normal(f(x), \exp(g(x))), 
\end{equation*}
\begin{equation*}
 f \sim \GP(0, K_1(x, x^\prime, \rho_f, \sigma_f)),\quad
    g \sim \GP(0, K_2(x, x^\prime, \rho_g, \sigma_g)), 
\end{equation*}
\begin{equation*}
    \rho_f \sim \normal^+(0, 1),\quad
    \rho_g \sim \normal^+(0, 1), 
\end{equation*}
\begin{equation*}
 \sigma_f \sim \normal^+(0, 0.05),\quad
    \sigma_g \sim \normal^+(0, 0.5).
\end{equation*}

For \(K_1\) and \(K_2\) we use Mat\'ern covariance functions with \(\nu = 3/2\). These functions are controlled by the \(\rho\) and \(\sigma\) parameters. The \(\rho\) parameters are the length-scales of the processes and define how close two points \(x\) and \(x^\prime\) must be to influence each other. The \(\sigma\) parameters define the standard deviations of the noise. For efficient sampling with Stan, we use Hilbert space approximate Gaussian processes~\citep{solin2020HilbertSpace,ruitortmayol2022PracticalHilbert}.
The number of basis functions (\(m_f = m_g = 40\)) and the proportional extension factor (\(c_f = c_g = 1.5\)) are adapted such that the posterior length-scale estimates \(\hat\rho_f\) and \(\hat\rho_g\) are above the threshold of that which can be accurately approximated~\citep[see][]{ruitortmayol2022PracticalHilbert}.
We can then focus on the choice of priors for the length-scale parameters (\(\rho_f, \rho_g\)) and the marginal standard deviation parameters (\(\sigma_f, \sigma_g\)). It is known that for a Gaussian process, the \(\rho\) and \(\sigma\) parameters are not well informed independently~\citep{diggle2007Geostatistics}, so the sensitivity of the marginals may not be properly representative as there may be prior sensitivity no matter the choice of prior. We first demonstrate the sensitivity of the marginals before proceeding with a focus on the sensitivity of the model predictions, in accordance with \citet{paananen2021RSense}.

\begin{table}[t]
\caption{Prior and likelihood sensitivity in the motorcycle crash case study using the original prior.}
  \label{tab:motorcycle-base-sens}
  \centering
  \begin{tabular}{lllll}
     \toprule
     & \multicolumn{2}{c}{Original prior} & \multicolumn{2}{c}{Adjusted prior} \\
     Parameter & \makecell[l]{Prior\\ sensitivity} & \makecell[l]{Likelihood\\ sensitivity} &  \makecell[l]{Prior\\ sensitivity} & \makecell[l]{Likelihood\\ sensitivity}\\
     \midrule
     \(\rho_f\) & \textbf{0.52} & 1.62 & \textbf{0.12} & 0.13\\
     \(\rho_g\) & \textbf{0.18} & 0.06 & \textbf{0.15} & 0.25\\
     \(\sigma_f\) & \textbf{0.92} & 2.09 & \textbf{0.35} & 0.20 \\
     \(\sigma_g\) & \textbf{0.14} & 0.18 & \textbf{0.26} & 0.09 \\
     \bottomrule
  \end{tabular}
  
%   Prior sensitivity values above the threshold \((\geq 0.05)\) indicate possible informative prior (bold). Likelihood sensitivity values below the threshold (\(< 0.05\)) indicate possible weak or noninformative likelihood (bold). There is clear sensitivity in the posterior parameter marginals, however, as the parameters are difficult to interpret, it is unclear to what extent this poses an issue. Although alleviated, sensitivity remains in the posterior parameter marginals.
 \end{table}

\begin{figure}[htb]
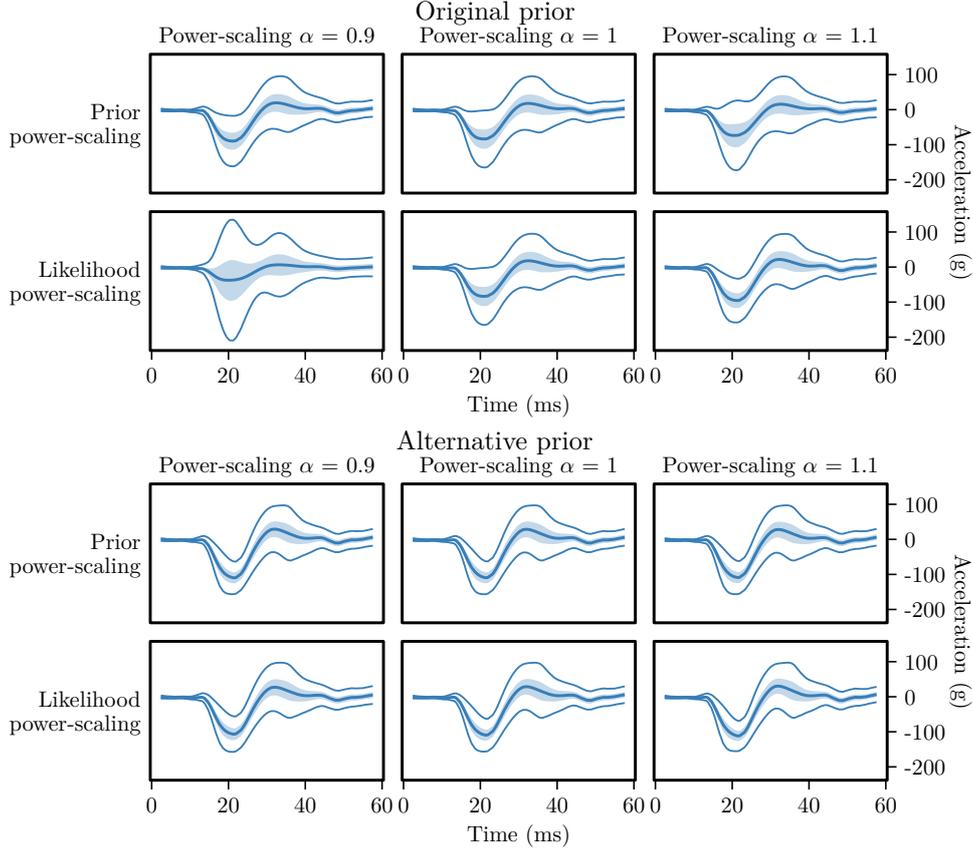

\centering
Original prior
  
  \input{./figs/motorcycle_sense_plot.tex}

Alternative prior

    \input{./figs/motorcycle_sense_adjust_plot.tex}
  \caption{Sensitivity of posterior predictions to prior and likelihood power-scaling in the motorcycle case study. Shown in the plots are the mean, 50\% and 95\% credible intervals for the posterior predictions. Top: original prior \(\sigma_f \sim \normal(0, 0.05)\). There is clear prior and likelihood sensitivity in the predictions around 20~ms after the crash. Bottom: alternative prior \(\sigma_f \sim \normal(0, 0.1)\). There is now no prior sensitivity and minimal likelihood sensitivity for the predictions.}%
  \label{fig:motorcycle-sense}
\end{figure}

As expected, there is prior sensitivity in the marginals (Table~\ref{tab:motorcycle-base-sens}). The prior and likelihood sensitivity for the parameters is high, which may be an indication of an issue, however it is difficult to determine based on the parameter marginals alone. Instead we follow up by plotting how the predictions are affected by power-scaling. As shown in Figure~\ref{fig:motorcycle-sense} (top), the predictions around 20~ms exhibit sensitivity to both prior and likelihood power-scaling. The prediction interval widens as the prior is strengthened (\(\alpha > 1\)), and narrows as it is weakened (\(\alpha < 1\)). Likelihood power-scaling has the opposite effect. This indicates potential prior-data conflict from an unintentionally informative prior. Widening the prior on \(\sigma_f\) from \(\normal(0, 0.05)\) to \(\normal(0, 0.1)\) alleviates the conflict such that it is no longer apparent in the predictions~(Figure~\ref{fig:motorcycle-sense}, bottom). Plotting the predictions with the raw data indicates a good fit (Figure~\ref{fig:motorcycle-data}). However, there remains sensitivity in the parameters, although it is lessened (Table~\ref{tab:motorcycle-base-sens}). This further demonstrates that depending on the model, prior sensitivity may be present, but is not necessarily an issue. We advise modellers to pay attention to specific quantities and properties of interest, particularly when performing sensitivity analyses on more complex models, rather than focusing on parameters without clear interpretations.

\FloatBarrier

\subsection{US Crime (linear regression with shrinkage prior)}
\label{sec:uscrime}

Here, we show how sensitivity can be analysed with respect to model fit. We fit a regression to the \texttt{UScrime} data set, available from the \texttt{MASS}~\citep{mass} package, and use a joint prior on the regression coefficients based on a prior on the model fit, Bayesian \(R^2\)~\citep{gelmanRsquaredBayesianRegression2019}. Such a prior structure can be used to specify a weakly informative prior on the model fit to prevent overfitting~\citep{gelmanRegressionOtherStories2020}. We use the R2-D2 prior~\citep{zhangBayesianRegressionUsing2022} as implemented in \texttt{brms} and check for sensitivity of the posterior \(R^2\) to changes to the prior on \(R^2\).

The data has observations from 47 US states in the year 1960. See \citet{clydeIntroductionBayesianThinking} for further details on the data set.
We model the crime rate \(y\) from 15 predictors \(x_k\) using a logNormal observation model.  All continuous predictors are log transformed, following \citet{mass}. We use the \texttt{brms} default weakly informative priors on the intercept \(\beta_0\) and residual standard deviation \(\sigma\).

The full model, including the R2-D2 prior is specified as
\begin{equation*}
  y_i \sim \lognormal(\mu_i, \sigma),\quad
  \mu_i = \beta_0 + \sum_{k=1}^{15}x_{k,i}\beta_k,
\end{equation*}
\begin{equation*}
  \beta_0 \sim t_3(6.7, 2.5),\quad
  \sigma \sim t_3^+(0, 2.5),\quad
  \beta_k \sim \normal(0, (\frac{\sigma^2}{s^2_{x_k}} \phi_k \tau^2)^{1/2}), 
  \end{equation*}
\begin{equation*}
 \phi \sim \dirichlet(1,\dots,1),\quad
  \tau^2 = \frac{R^2}{1-R^2},\quad
  R^2 \sim \betadist(s_1, s_2),
\end{equation*} where \(s^2_{x_k}\) is the sample variance of predictor \(x_k\).

\begin{figure}
    \centering
    \input{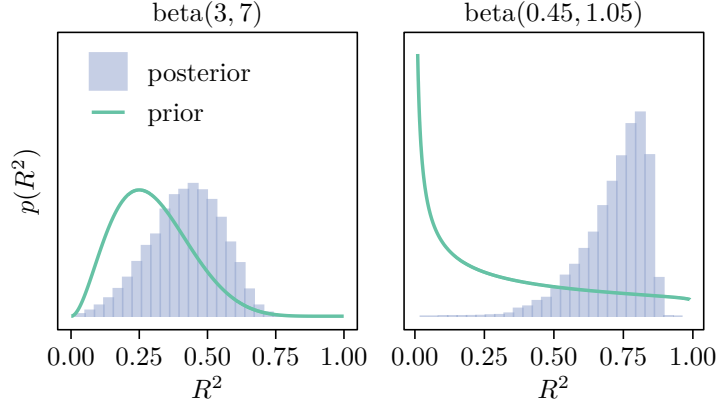}
    \caption{The priors and corresponding posteriors for the model fit (\(R^2\)) in the US Crime case study.}
    \label{fig:r2priors}
\end{figure}

We contrast two prior specifications, prior 1: \(R^2 \sim \betadist(3, 7)\) and prior 2: \(R^2 \sim \betadist(0.45, 1.05)\), shown in Figure~\ref{fig:r2priors}. The sensitivity analysis indicates that prior 1 may be informative and affecting the posterior. Indeed, the posterior for \(R^2\) is lower with prior 1 than prior 2 (Table~\ref{tab:uscrimesens}).

To follow up this, we perform leave-one-out cross validation on both models to compare predictive performance using the \(\text{elpd}_{\text{loo}}\) metric~\citep{vehtariLooEfficientLeaveoneout2020}. The results, also shown in Table~\ref{tab:uscrimesens}, indicate that prior 1 leads to lower predictive performance than prior 2, and induces a lower effective number of parameters (\(p_\text{loo}\)). This further corroborates the results of the sensitivity analysis, and shows that the power-scaling sensitivity diagnostic can be used as an early indication of issues that can influence predictive performance.

\begin{table}[ht]
\centering
    \caption{Power-scaling sensitivity and predictive model performance for prior specifications in the US crime case study.}
    \label{tab:uscrimesens}
    \small
    \begin{tabular}{llllll}
    \toprule
    Prior on \(R^2\) & \makecell[l]{Prior\\ sensitivity\\ \(D_{\text{CJS}}\)} & \makecell[l]{Likelihood\\ sensitivity\\ \(D_{\text{CJS}}\)} & \makecell[l]{Posterior \(R^2\) \\ Median (SE)} & \makecell[l]{Predictive \\ performance\\ \(\text{elpd}_{\text{loo}}\)} &  \makecell[l]{Effective\\number of\\parameters \\ \(p_{\text{loo}}\)}\\ 
    \midrule
      \(\betadist(3, 7)\)  & \textbf{0.17} & 0.57 & 0.42 (0.14) & -326.7 (4.4) & 5.2 (1.0)\\
      \(\betadist(0.45, 1.05)\)  & 0.02  & 1.32 & 0.72 (0.12) & -318.9 (4.4) & 11.6 (2.0)  \\
      \bottomrule
    \end{tabular}
\end{table}

\FloatBarrier

\subsection{COVID-19 interventions (infections and deaths model)}
\label{sec:covid}

In this case study, we evaluate the prior and likelihood sensitivity in a model of deaths from the COVID-19 pandemic~\citep{flaxmanReport13Estimating2020}. The Stan code and data for this model are available from PosteriorDB~\citep{Magnusson_posteriordb_a_set_2021}. 
We focus on the effects of power-scaling the priors on three parameters of the model: \(\tau\), \(\phi\) and \(\kappa\). Due to the complexity of the model, we separately power-scale each prior to determine their individual effects. We evaluate the sensitivity of predictions (expected number of deaths due to COVID-19) in 14 countries over 100 days.

For a full description of the model, see~\citet{flaxmanReport13Estimating2020}.
The parts of the model which we focus on are as follows.
The prior on \(\phi\) which partially controls the variance of the negative binomial likelihood of observed daily deaths \(D_{t,m}\), modelled from the expected deaths due to the virus \(d_{t,m}\) for a given day \(t\) and country \(m\):
\begin{equation*}
    D_{t, m} \sim \negbinom(d_{t,m}, d_{t,m} + d_{t,m}^2/\phi),\quad
    \phi \sim \normal^+(0, 5).
    \end{equation*} \
\(d_{t,m}\) is a function of \(R_{0,m}\) and \(c_{1,m} \dots c_{6,m}\) (among other parameters). The prior on \(\kappa\) which controls the variance of the baseline reproductive number \(R_0\) of the virus for each country \(m\):
\begin{equation*}
    R_{0, m} \sim \normal^+(3.28, \kappa),\quad
    \kappa \sim \normal^+(0, 0.5),
    \end{equation*}
and the prior on \(\tau\) which affects the number of seed infections (infections in the six days following the beginning of the seed period, which is defined as the 30 days before a country observes a total of ten or more deaths):
\begin{equation*}
    c_{1,m}, ..., c_{6, m} \sim \expdist(1/\tau),\quad
    \tau \sim \expdist(0.03).
\end{equation*}

\begin{figure}[ht]
    \centering    \input{./figs/covid_liksubsens}
    \caption{Likelihood sensitivity of posterior predictions (expected deaths due to COVID-19) for four countries. Vertical lines indicate the onset of major governmental intervention. The dotted lines indicate the sensitivity threshold of 0.05, above which we consider sensitivity to be present.}
    \label{fig:covidsubliksens}
\end{figure}
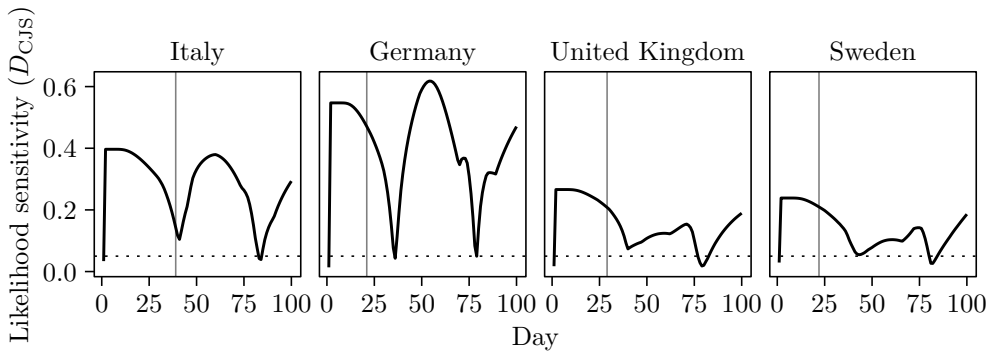

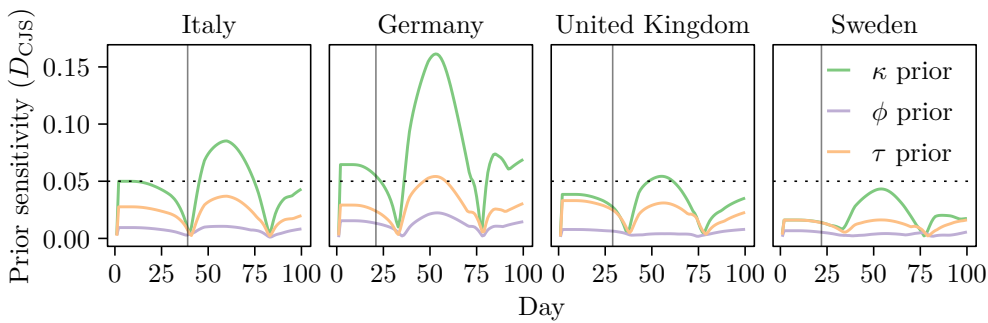
\begin{figure}[ht]
    \centering    \input{./figs/covid_priorsubsens}
    \caption{Prior sensitivity of posterior predictions (expected deaths due to COVID-19) for four countries. The vertical lines indicate the onset of major governmental intervention. The dotted lines indicate the sensitivity threshold of 0.05, above which we consider sensitivity to be present.}
    \label{fig:covidsubpriorsens}
\end{figure}

\begin{figure}[ht]
    \centering    \input{./figs/covid_quants}
    \caption{Prior and likelihood sensitivity of posterior predictions (expected deaths due to COVID-19) for day 50 in Germany. Only the \(\kappa\) prior is power-scaled.}
    \label{fig:covidquants}
\end{figure}
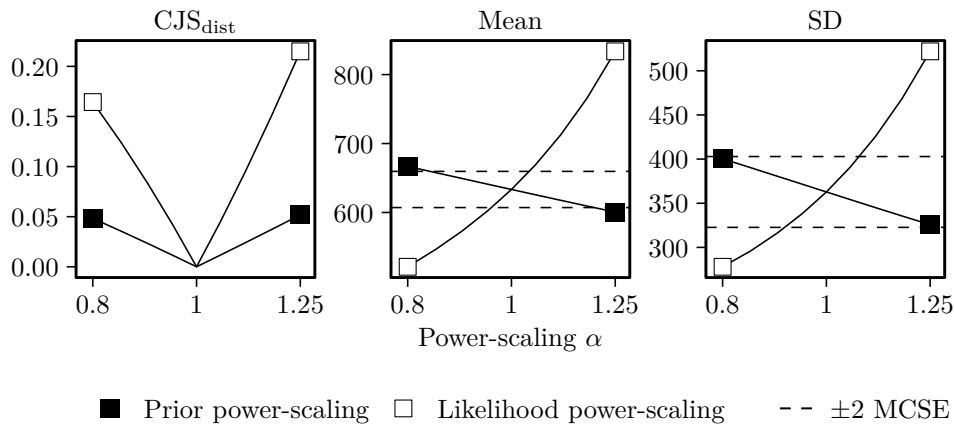

Here we focus on a subset of four countries, but results for all 14 countries are presented in Appendix~\ref{sec:covid-appendix}.
The results shown in Figure~\ref{fig:covidsubliksens} indicate that there is likelihood sensitivity throughout the time period, indicating the data is informative, as seen in Figure~\ref{fig:covidsubpriorsens}. Furthermore, there is clear sensitivity to the \(\kappa\) prior, and some sensitivity to the \(\tau\) prior. This is most pronounced in the predictions of deaths from day 30 to 70, shortly after the first major governmental interventions. Sensitivity is particularly high for the predictions for Germany.  Following this up by plotting the sensitivity of predictions on day 50 in Germany (Figure~\ref{fig:covidquants}), there is an indication that the prior is in conflict with the data, as the mean is shifted in opposite directions by prior and likelihood scaling. 

These results are an indication that the chosen prior on \(\kappa\) may be informative and in conflict with the data, and the justification for this prior should be carefully considered. As the prior on \(R_0\) for each country is centred around a specific value, 3.28, based on previous literature~\citep{liuReproductiveNumberCOVID192020}, some sensitivity to the prior on \(\kappa\) may be expected, however the finding that it may be in conflict with the data is nevertheless important and may warrant further attention. This is an example for how a more complex model can be checked for prior and likelihood sensitivity by selectively perturbing priors and focusing on predictions.

\section{Discussion}
 \label{sec:discussion}

We have introduced an approach and corresponding workflow for prior and likelihood sensitivity
 analysis using power-scaling perturbations of the prior and
 likelihood. The proposed approach is computationally efficient and applicable to
 a wide range of models with minor changes to existing model code. This will allow automated prior sensitivity diagnostics for probabilistic programming languages such as Stan and PyMC, and higher-level interfaces like \texttt{brms}, \texttt{rstanarm} and \texttt{bambi}, and make the use of default priors safer as potential problems can be detected and warnings presented to users. The approach can also be used to identify which priors may need more careful specification. The use
 of PSIS and IWMM ensures that the approach is reliable 
 while being computationally efficient. These
 properties were demonstrated in several simulated examples and case studies of real data, and our sensitivity analysis
 workflow easily fits into a larger Bayesian workflow
 involving model checking and model iteration.

 Rather than fixing the power-scaling \(\alpha\) values, it could be possible to include the \(\alpha\) parameters in the model and place hyperpriors on them. However, this naturally complicates the model by adding additional levels of hierarchy. In addition, the question of sensitivity to the choice of hyperprior would then be raised, which may require further sensitivity analysis, or additional levels of hierarchy, the parameters of which become less and less informed by the data~\citep{goelInformationHyperparamtersHierarchical1981}. Instead, the power-scaling sensitivity approach can be seen as a controlled method for automatically comparing alternative priors, which are interpretable by the modeller.
 
 We have demonstrated checking the
 presence of sensitivity based on the derivative of the cumulative Jensen-Shannon
 distance between the base and perturbed priors with respect to the
 power-scaling factor. While this is a useful diagnostic,
 power-scaling sensitivity analysis is a general approach with multiple
 valid variants. Future work could include further developing quantity-based sensitivity to identify meaningful changes in quantities and predictions
 with respect to power-scaling, and working towards automated guidance on safe model adjustment after sensitivity has been detected and diagnosed. Other extensions include developing additional perturbations that affect different aspects of distributions, such as inducing a mean shift via exponential tilting~\citep{siegmundImportanceSamplingMonte1976}. Additionally, it is possible to use the same framework to investigate the influence of specific observations, by power-scaling the likelihood contribution of a single or subset of observations.

Finally, we emphasise that the presence of prior sensitivity or the absence of likelihood sensitivity are not issues in and of themselves. Rather, context and intention of the model builder need to be taken into account. We suggest that the model builder pay particular attention when the pattern of sensitivity is unexpected or surprising, as this may indicate the model is not behaving as anticipated. We again emphasise that the approach should be coupled with thoughtful consideration of the model specification and not be used for repeated tuning of the priors until diagnostic warnings disappear.

%% file: figs/separation_example.tex
% Created by tikzDevice version 0.12.3.1 on 2022-12-08 14:41:43
% !TEX encoding = UTF-8 Unicode
\begin{tikzpicture}[x=1pt,y=1pt]
\definecolor{fillColor}{RGB}{255,255,255}
\begin{scope}
\definecolor{drawColor}{RGB}{127,201,127}

\path[draw=drawColor,line width= 0.6pt,line join=round] ( 85.13, 79.82) --
	(103.79,108.53) --
	(122.46,115.65) --
	(141.12,109.72) --
	(159.78,103.01) --
	(178.45, 99.66) --
	(197.11, 96.83);
\definecolor{drawColor}{RGB}{190,174,212}

\path[draw=drawColor,line width= 0.6pt,line join=round] ( 85.13, 82.14) --
	(103.79, 99.96) --
	(122.46,109.29) --
	(141.12,104.92) --
	(159.78,100.29) --
	(178.45, 95.71) --
	(197.11, 97.22);
\definecolor{drawColor}{RGB}{253,192,134}

\path[draw=drawColor,line width= 0.6pt,line join=round] ( 85.13, 83.36) --
	(103.79,100.59) --
	(122.46,109.20) --
	(141.12,106.06) --
	(159.78,101.53) --
	(178.45, 96.65) --
	(197.11, 89.81);
\definecolor{fillColor}{RGB}{127,201,127}

\path[fill=fillColor] ( 83.17, 77.86) --
	( 87.09, 77.86) --
	( 87.09, 81.78) --
	( 83.17, 81.78) --
	cycle;

\path[fill=fillColor] (101.83,106.57) --
	(105.76,106.57) --
	(105.76,110.49) --
	(101.83,110.49) --
	cycle;

\path[fill=fillColor] (120.49,113.69) --
	(124.42,113.69) --
	(124.42,117.62) --
	(120.49,117.62) --
	cycle;

\path[fill=fillColor] (139.16,107.76) --
	(143.08,107.76) --
	(143.08,111.68) --
	(139.16,111.68) --
	cycle;

\path[fill=fillColor] (157.82,101.05) --
	(161.74,101.05) --
	(161.74,104.98) --
	(157.82,104.98) --
	cycle;

\path[fill=fillColor] (176.48, 97.70) --
	(180.41, 97.70) --
	(180.41,101.62) --
	(176.48,101.62) --
	cycle;

\path[fill=fillColor] (195.15, 94.86) --
	(199.07, 94.86) --
	(199.07, 98.79) --
	(195.15, 98.79) --
	cycle;
\definecolor{fillColor}{RGB}{190,174,212}

\path[fill=fillColor] ( 83.17, 80.18) --
	( 87.09, 80.18) --
	( 87.09, 84.10) --
	( 83.17, 84.10) --
	cycle;

\path[fill=fillColor] (101.83, 98.00) --
	(105.76, 98.00) --
	(105.76,101.93) --
	(101.83,101.93) --
	cycle;

\path[fill=fillColor] (120.49,107.33) --
	(124.42,107.33) --
	(124.42,111.25) --
	(120.49,111.25) --
	cycle;

\path[fill=fillColor] (139.16,102.96) --
	(143.08,102.96) --
	(143.08,106.89) --
	(139.16,106.89) --
	cycle;

\path[fill=fillColor] (157.82, 98.33) --
	(161.74, 98.33) --
	(161.74,102.25) --
	(157.82,102.25) --
	cycle;

\path[fill=fillColor] (176.48, 93.75) --
	(180.41, 93.75) --
	(180.41, 97.67) --
	(176.48, 97.67) --
	cycle;

\path[fill=fillColor] (195.15, 95.26) --
	(199.07, 95.26) --
	(199.07, 99.19) --
	(195.15, 99.19) --
	cycle;
\definecolor{fillColor}{RGB}{253,192,134}

\path[fill=fillColor] ( 83.17, 81.40) --
	( 87.09, 81.40) --
	( 87.09, 85.32) --
	( 83.17, 85.32) --
	cycle;

\path[fill=fillColor] (101.83, 98.63) --
	(105.76, 98.63) --
	(105.76,102.55) --
	(101.83,102.55) --
	cycle;

\path[fill=fillColor] (120.49,107.24) --
	(124.42,107.24) --
	(124.42,111.16) --
	(120.49,111.16) --
	cycle;

\path[fill=fillColor] (139.16,104.10) --
	(143.08,104.10) --
	(143.08,108.03) --
	(139.16,108.03) --
	cycle;

\path[fill=fillColor] (157.82, 99.57) --
	(161.74, 99.57) --
	(161.74,103.49) --
	(157.82,103.49) --
	cycle;

\path[fill=fillColor] (176.48, 94.68) --
	(180.41, 94.68) --
	(180.41, 98.61) --
	(176.48, 98.61) --
	cycle;

\path[fill=fillColor] (195.15, 87.85) --
	(199.07, 87.85) --
	(199.07, 91.77) --
	(195.15, 91.77) --
	cycle;
\definecolor{drawColor}{RGB}{0,0,0}

\path[draw=drawColor,line width= 0.6pt,line join=round,line cap=round] ( 79.53, 34.67) rectangle (202.71,157.84);
\end{scope}
\begin{scope}
\definecolor{drawColor}{RGB}{127,201,127}

\path[draw=drawColor,line width= 0.6pt,line join=round] (215.31,152.24) --
	(233.97, 74.82) --
	(252.63, 49.02) --
	(271.30, 43.37) --
	(289.96, 41.42) --
	(308.62, 40.63) --
	(327.28, 40.27);
\definecolor{drawColor}{RGB}{190,174,212}

\path[draw=drawColor,line width= 0.6pt,line join=round] (215.31,149.75) --
	(233.97, 72.01) --
	(252.63, 49.91) --
	(271.30, 44.48) --
	(289.96, 42.06) --
	(308.62, 41.07) --
	(327.28, 40.53);
\definecolor{drawColor}{RGB}{253,192,134}

\path[draw=drawColor,line width= 0.6pt,line join=round] (215.31,148.32) --
	(233.97, 72.19) --
	(252.63, 49.31) --
	(271.30, 44.56) --
	(289.96, 42.13) --
	(308.62, 41.27) --
	(327.28, 40.77);
\definecolor{fillColor}{RGB}{127,201,127}

\path[fill=fillColor] (213.34,150.28) --
	(217.27,150.28) --
	(217.27,154.21) --
	(213.34,154.21) --
	cycle;

\path[fill=fillColor] (232.01, 72.86) --
	(235.93, 72.86) --
	(235.93, 76.78) --
	(232.01, 76.78) --
	cycle;

\path[fill=fillColor] (250.67, 47.06) --
	(254.59, 47.06) --
	(254.59, 50.98) --
	(250.67, 50.98) --
	cycle;

\path[fill=fillColor] (269.33, 41.41) --
	(273.26, 41.41) --
	(273.26, 45.34) --
	(269.33, 45.34) --
	cycle;

\path[fill=fillColor] (288.00, 39.45) --
	(291.92, 39.45) --
	(291.92, 43.38) --
	(288.00, 43.38) --
	cycle;

\path[fill=fillColor] (306.66, 38.67) --
	(310.58, 38.67) --
	(310.58, 42.60) --
	(306.66, 42.60) --
	cycle;

\path[fill=fillColor] (325.32, 38.30) --
	(329.25, 38.30) --
	(329.25, 42.23) --
	(325.32, 42.23) --
	cycle;
\definecolor{fillColor}{RGB}{190,174,212}

\path[fill=fillColor] (213.34,147.78) --
	(217.27,147.78) --
	(217.27,151.71) --
	(213.34,151.71) --
	cycle;

\path[fill=fillColor] (232.01, 70.05) --
	(235.93, 70.05) --
	(235.93, 73.98) --
	(232.01, 73.98) --
	cycle;

\path[fill=fillColor] (250.67, 47.95) --
	(254.59, 47.95) --
	(254.59, 51.87) --
	(250.67, 51.87) --
	cycle;

\path[fill=fillColor] (269.33, 42.52) --
	(273.26, 42.52) --
	(273.26, 46.44) --
	(269.33, 46.44) --
	cycle;

\path[fill=fillColor] (288.00, 40.10) --
	(291.92, 40.10) --
	(291.92, 44.03) --
	(288.00, 44.03) --
	cycle;

\path[fill=fillColor] (306.66, 39.11) --
	(310.58, 39.11) --
	(310.58, 43.03) --
	(306.66, 43.03) --
	cycle;

\path[fill=fillColor] (325.32, 38.57) --
	(329.25, 38.57) --
	(329.25, 42.49) --
	(325.32, 42.49) --
	cycle;
\definecolor{fillColor}{RGB}{253,192,134}

\path[fill=fillColor] (213.34,146.36) --
	(217.27,146.36) --
	(217.27,150.29) --
	(213.34,150.29) --
	cycle;

\path[fill=fillColor] (232.01, 70.22) --
	(235.93, 70.22) --
	(235.93, 74.15) --
	(232.01, 74.15) --
	cycle;

\path[fill=fillColor] (250.67, 47.35) --
	(254.59, 47.35) --
	(254.59, 51.27) --
	(250.67, 51.27) --
	cycle;

\path[fill=fillColor] (269.33, 42.59) --
	(273.26, 42.59) --
	(273.26, 46.52) --
	(269.33, 46.52) --
	cycle;

\path[fill=fillColor] (288.00, 40.17) --
	(291.92, 40.17) --
	(291.92, 44.09) --
	(288.00, 44.09) --
	cycle;

\path[fill=fillColor] (306.66, 39.30) --
	(310.58, 39.30) --
	(310.58, 43.23) --
	(306.66, 43.23) --
	cycle;

\path[fill=fillColor] (325.32, 38.81) --
	(329.25, 38.81) --
	(329.25, 42.73) --
	(325.32, 42.73) --
	cycle;
\definecolor{drawColor}{RGB}{0,0,0}

\path[draw=drawColor,line width= 0.6pt,line join=round,line cap=round] (209.71, 34.67) rectangle (332.88,157.84);
\end{scope}
\begin{scope}
\definecolor{drawColor}{RGB}{0,0,0}

\node[text=drawColor,anchor=base,inner sep=0pt, outer sep=0pt, scale=  1.00] at (141.12,162.31) {Likelihood power-scaling};
\end{scope}
\begin{scope}
\definecolor{drawColor}{RGB}{0,0,0}

\node[text=drawColor,anchor=base,inner sep=0pt, outer sep=0pt, scale=  1.00] at (271.30,162.31) {Prior power-scaling};
\end{scope}
\begin{scope}
\definecolor{drawColor}{RGB}{0,0,0}

\path[draw=drawColor,line width= 0.6pt,line join=round] ( 85.13, 31.17) --
	( 85.13, 34.67);

\path[draw=drawColor,line width= 0.6pt,line join=round] (122.46, 31.17) --
	(122.46, 34.67);

\path[draw=drawColor,line width= 0.6pt,line join=round] (159.78, 31.17) --
	(159.78, 34.67);

\path[draw=drawColor,line width= 0.6pt,line join=round] (197.11, 31.17) --
	(197.11, 34.67);
\end{scope}
\begin{scope}
\definecolor{drawColor}{RGB}{0,0,0}

\node[text=drawColor,anchor=base,inner sep=0pt, outer sep=0pt, scale=  1.00] at ( 85.13, 21.28) {0};

\node[text=drawColor,anchor=base,inner sep=0pt, outer sep=0pt, scale=  1.00] at (122.46, 21.28) {2};

\node[text=drawColor,anchor=base,inner sep=0pt, outer sep=0pt, scale=  1.00] at (159.78, 21.28) {4};

\node[text=drawColor,anchor=base,inner sep=0pt, outer sep=0pt, scale=  1.00] at (197.11, 21.28) {6};
\end{scope}
\begin{scope}
\definecolor{drawColor}{RGB}{0,0,0}

\path[draw=drawColor,line width= 0.6pt,line join=round] (215.31, 31.17) --
	(215.31, 34.67);

\path[draw=drawColor,line width= 0.6pt,line join=round] (252.63, 31.17) --
	(252.63, 34.67);

\path[draw=drawColor,line width= 0.6pt,line join=round] (289.96, 31.17) --
	(289.96, 34.67);

\path[draw=drawColor,line width= 0.6pt,line join=round] (327.28, 31.17) --
	(327.28, 34.67);
\end{scope}
\begin{scope}
\definecolor{drawColor}{RGB}{0,0,0}

\node[text=drawColor,anchor=base,inner sep=0pt, outer sep=0pt, scale=  1.00] at (215.31, 21.28) {0};

\node[text=drawColor,anchor=base,inner sep=0pt, outer sep=0pt, scale=  1.00] at (252.63, 21.28) {2};

\node[text=drawColor,anchor=base,inner sep=0pt, outer sep=0pt, scale=  1.00] at (289.96, 21.28) {4};

\node[text=drawColor,anchor=base,inner sep=0pt, outer sep=0pt, scale=  1.00] at (327.28, 21.28) {6};
\end{scope}
\begin{scope}
\definecolor{drawColor}{RGB}{0,0,0}

\node[text=drawColor,anchor=base east,inner sep=0pt, outer sep=0pt, scale=  1.00] at ( 73.03, 36.39) {0.00};

\node[text=drawColor,anchor=base east,inner sep=0pt, outer sep=0pt, scale=  1.00] at ( 73.03, 63.78) {0.05};

\node[text=drawColor,anchor=base east,inner sep=0pt, outer sep=0pt, scale=  1.00] at ( 73.03, 91.16) {0.10};

\node[text=drawColor,anchor=base east,inner sep=0pt, outer sep=0pt, scale=  1.00] at ( 73.03,118.55) {0.15};

\node[text=drawColor,anchor=base east,inner sep=0pt, outer sep=0pt, scale=  1.00] at ( 73.03,145.93) {0.20};
\end{scope}
\begin{scope}
\definecolor{drawColor}{RGB}{0,0,0}

\path[draw=drawColor,line width= 0.6pt,line join=round] ( 76.03, 39.84) --
	( 79.53, 39.84);

\path[draw=drawColor,line width= 0.6pt,line join=round] ( 76.03, 67.22) --
	( 79.53, 67.22);

\path[draw=drawColor,line width= 0.6pt,line join=round] ( 76.03, 94.61) --
	( 79.53, 94.61);

\path[draw=drawColor,line width= 0.6pt,line join=round] ( 76.03,121.99) --
	( 79.53,121.99);

\path[draw=drawColor,line width= 0.6pt,line join=round] ( 76.03,149.38) --
	( 79.53,149.38);
\end{scope}
\begin{scope}
\definecolor{drawColor}{RGB}{0,0,0}

\node[text=drawColor,anchor=base,inner sep=0pt, outer sep=0pt, scale=  1.00] at (206.21,  8.94) {Data separability ($n_{\text{complete}}$)};
\end{scope}
\begin{scope}
\definecolor{drawColor}{RGB}{0,0,0}

\node[text=drawColor,rotate= 90.00,anchor=base,inner sep=0pt, outer sep=0pt, scale=  1.00] at ( 49.81, 96.25) {Sensitivity ($D_{\text{CJS}}$)};
\end{scope}
\begin{scope}
\definecolor{drawColor}{RGB}{127,201,127}

\path[draw=drawColor,line width= 0.6pt,line join=round] (106.41, 83.52) -- (118.73, 83.52);
\end{scope}
\begin{scope}
\definecolor{fillColor}{RGB}{127,201,127}

\path[fill=fillColor] (110.60, 81.56) --
	(114.53, 81.56) --
	(114.53, 85.48) --
	(110.60, 85.48) --
	cycle;
\end{scope}
\begin{scope}
\definecolor{drawColor}{RGB}{190,174,212}

\path[draw=drawColor,line width= 0.6pt,line join=round] (106.41, 68.12) -- (118.73, 68.12);
\end{scope}
\begin{scope}
\definecolor{fillColor}{RGB}{190,174,212}

\path[fill=fillColor] (110.60, 66.16) --
	(114.53, 66.16) --
	(114.53, 70.08) --
	(110.60, 70.08) --
	cycle;
\end{scope}
\begin{scope}
\definecolor{drawColor}{RGB}{253,192,134}

\path[draw=drawColor,line width= 0.6pt,line join=round] (106.41, 52.72) -- (118.73, 52.72);
\end{scope}
\begin{scope}
\definecolor{fillColor}{RGB}{253,192,134}

\path[fill=fillColor] (110.60, 50.76) --
	(114.53, 50.76) --
	(114.53, 54.68) --
	(110.60, 54.68) --
	cycle;
\end{scope}
\begin{scope}
\definecolor{drawColor}{RGB}{0,0,0}

\node[text=drawColor,anchor=base west,inner sep=0pt, outer sep=0pt, scale=  1.00] at (127.27, 80.07) {$\beta_0$};
\end{scope}
\begin{scope}
\definecolor{drawColor}{RGB}{0,0,0}

\node[text=drawColor,anchor=base west,inner sep=0pt, outer sep=0pt, scale=  1.00] at (127.27, 64.67) {$\beta_1$};
\end{scope}
\begin{scope}
\definecolor{drawColor}{RGB}{0,0,0}

\node[text=drawColor,anchor=base west,inner sep=0pt, outer sep=0pt, scale=  1.00] at (127.27, 49.27) {$\beta_2$};
\end{scope}
\end{tikzpicture}

%% file: figs/covariate_scaling_example.tex
% Created by tikzDevice version 0.12.3.1 on 2022-12-08 11:18:17
% !TEX encoding = UTF-8 Unicode
\begin{tikzpicture}[x=1pt,y=1pt]
\definecolor{fillColor}{RGB}{255,255,255}
\begin{scope}
\definecolor{fillColor}{RGB}{127,201,127}

\path[fill=fillColor] ( 85.13,126.17) --
	( 87.77,121.59) --
	( 82.49,121.59) --
	cycle;

\path[fill=fillColor] ( 85.13,125.44) --
	( 87.77,120.86) --
	( 82.49,120.86) --
	cycle;
\definecolor{drawColor}{RGB}{190,174,212}
\definecolor{fillColor}{RGB}{190,174,212}

\path[draw=drawColor,line width= 0.4pt,line join=round,line cap=round,fill=fillColor] ( 85.13,120.41) circle (  1.96);

\path[draw=drawColor,line width= 0.4pt,line join=round,line cap=round,fill=fillColor] ( 85.13,120.50) circle (  1.96);
\definecolor{fillColor}{RGB}{127,201,127}

\path[fill=fillColor] (113.13,125.00) --
	(115.77,120.42) --
	(110.48,120.42) --
	cycle;

\path[fill=fillColor] (113.13,125.41) --
	(115.77,120.83) --
	(110.48,120.83) --
	cycle;
\definecolor{fillColor}{RGB}{190,174,212}

\path[draw=drawColor,line width= 0.4pt,line join=round,line cap=round,fill=fillColor] (113.13,120.53) circle (  1.96);

\path[draw=drawColor,line width= 0.4pt,line join=round,line cap=round,fill=fillColor] (113.13,121.48) circle (  1.96);
\definecolor{fillColor}{RGB}{127,201,127}

\path[fill=fillColor] (141.12,122.76) --
	(143.76,118.18) --
	(138.48,118.18) --
	cycle;

\path[fill=fillColor] (141.12,124.05) --
	(143.76,119.48) --
	(138.48,119.48) --
	cycle;
\definecolor{fillColor}{RGB}{190,174,212}

\path[draw=drawColor,line width= 0.4pt,line join=round,line cap=round,fill=fillColor] (141.12,119.95) circle (  1.96);

\path[draw=drawColor,line width= 0.4pt,line join=round,line cap=round,fill=fillColor] (141.12,121.84) circle (  1.96);
\definecolor{fillColor}{RGB}{127,201,127}

\path[fill=fillColor] (169.11,127.52) --
	(171.76,122.94) --
	(166.47,122.94) --
	cycle;

\path[fill=fillColor] (169.11,139.66) --
	(171.76,135.09) --
	(166.47,135.09) --
	cycle;
\definecolor{fillColor}{RGB}{190,174,212}

\path[draw=drawColor,line width= 0.4pt,line join=round,line cap=round,fill=fillColor] (169.11,123.54) circle (  1.96);

\path[draw=drawColor,line width= 0.4pt,line join=round,line cap=round,fill=fillColor] (169.11,121.68) circle (  1.96);
\definecolor{fillColor}{RGB}{127,201,127}

\path[fill=fillColor] (197.11, 87.26) --
	(199.75, 82.69) --
	(194.47, 82.69) --
	cycle;

\path[fill=fillColor] (197.11, 97.39) --
	(199.75, 92.81) --
	(194.47, 92.81) --
	cycle;
\definecolor{fillColor}{RGB}{190,174,212}

\path[draw=drawColor,line width= 0.4pt,line join=round,line cap=round,fill=fillColor] (197.11,119.39) circle (  1.96);

\path[draw=drawColor,line width= 0.4pt,line join=round,line cap=round,fill=fillColor] (197.11,119.76) circle (  1.96);
\definecolor{drawColor}{RGB}{127,201,127}

\path[draw=drawColor,line width= 0.6pt,line join=round] ( 85.13,123.12) --
	(113.13,121.95) --
	(141.12,119.71) --
	(169.11,124.46) --
	(197.11, 84.21);

\path[draw=drawColor,line width= 0.6pt,line join=round] ( 85.13,122.39) --
	(113.13,122.36) --
	(141.12,121.00) --
	(169.11,136.61) --
	(197.11, 94.34);
\definecolor{drawColor}{RGB}{190,174,212}

\path[draw=drawColor,line width= 0.6pt,line join=round] ( 85.13,120.41) --
	(113.13,120.53) --
	(141.12,119.95) --
	(169.11,123.54) --
	(197.11,119.39);

\path[draw=drawColor,line width= 0.6pt,line join=round] ( 85.13,120.50) --
	(113.13,121.48) --
	(141.12,121.84) --
	(169.11,121.68) --
	(197.11,119.76);
\definecolor{drawColor}{RGB}{0,0,0}

\path[draw=drawColor,line width= 0.6pt,line join=round,line cap=round] ( 79.53, 34.67) rectangle (202.71,157.84);
\end{scope}
\begin{scope}
\definecolor{fillColor}{RGB}{127,201,127}

\path[fill=fillColor] (215.31, 43.47) --
	(217.95, 38.90) --
	(212.66, 38.90) --
	cycle;

\path[fill=fillColor] (215.31, 43.32) --
	(217.95, 38.74) --
	(212.66, 38.74) --
	cycle;
\definecolor{drawColor}{RGB}{190,174,212}
\definecolor{fillColor}{RGB}{190,174,212}

\path[draw=drawColor,line width= 0.4pt,line join=round,line cap=round,fill=fillColor] (215.31, 48.89) circle (  1.96);

\path[draw=drawColor,line width= 0.4pt,line join=round,line cap=round,fill=fillColor] (215.31, 49.71) circle (  1.96);
\definecolor{fillColor}{RGB}{127,201,127}

\path[fill=fillColor] (243.30, 43.70) --
	(245.94, 39.12) --
	(240.66, 39.12) --
	cycle;

\path[fill=fillColor] (243.30, 43.58) --
	(245.94, 39.00) --
	(240.66, 39.00) --
	cycle;
\definecolor{fillColor}{RGB}{190,174,212}

\path[draw=drawColor,line width= 0.4pt,line join=round,line cap=round,fill=fillColor] (243.30, 48.62) circle (  1.96);

\path[draw=drawColor,line width= 0.4pt,line join=round,line cap=round,fill=fillColor] (243.30, 49.68) circle (  1.96);
\definecolor{fillColor}{RGB}{127,201,127}

\path[fill=fillColor] (271.30, 52.48) --
	(273.94, 47.90) --
	(268.65, 47.90) --
	cycle;

\path[fill=fillColor] (271.30, 53.55) --
	(273.94, 48.98) --
	(268.65, 48.98) --
	cycle;
\definecolor{fillColor}{RGB}{190,174,212}

\path[draw=drawColor,line width= 0.4pt,line join=round,line cap=round,fill=fillColor] (271.30, 49.60) circle (  1.96);

\path[draw=drawColor,line width= 0.4pt,line join=round,line cap=round,fill=fillColor] (271.30, 50.07) circle (  1.96);
\definecolor{fillColor}{RGB}{127,201,127}

\path[fill=fillColor] (299.29,118.93) --
	(301.93,114.36) --
	(296.65,114.36) --
	cycle;

\path[fill=fillColor] (299.29,132.12) --
	(301.93,127.54) --
	(296.65,127.54) --
	cycle;
\definecolor{fillColor}{RGB}{190,174,212}

\path[draw=drawColor,line width= 0.4pt,line join=round,line cap=round,fill=fillColor] (299.29, 57.33) circle (  1.96);

\path[draw=drawColor,line width= 0.4pt,line join=round,line cap=round,fill=fillColor] (299.29, 54.84) circle (  1.96);
\definecolor{fillColor}{RGB}{127,201,127}

\path[fill=fillColor] (327.28,147.12) --
	(329.93,142.54) --
	(324.64,142.54) --
	cycle;

\path[fill=fillColor] (327.28,155.29) --
	(329.93,150.72) --
	(324.64,150.72) --
	cycle;
\definecolor{fillColor}{RGB}{190,174,212}

\path[draw=drawColor,line width= 0.4pt,line join=round,line cap=round,fill=fillColor] (327.28, 49.41) circle (  1.96);

\path[draw=drawColor,line width= 0.4pt,line join=round,line cap=round,fill=fillColor] (327.28, 50.55) circle (  1.96);
\definecolor{drawColor}{RGB}{127,201,127}

\path[draw=drawColor,line width= 0.6pt,line join=round] (215.31, 40.42) --
	(243.30, 40.65) --
	(271.30, 49.43) --
	(299.29,115.88) --
	(327.28,144.07);

\path[draw=drawColor,line width= 0.6pt,line join=round] (215.31, 40.27) --
	(243.30, 40.52) --
	(271.30, 50.50) --
	(299.29,129.07) --
	(327.28,152.24);
\definecolor{drawColor}{RGB}{190,174,212}

\path[draw=drawColor,line width= 0.6pt,line join=round] (215.31, 48.89) --
	(243.30, 48.62) --
	(271.30, 49.60) --
	(299.29, 57.33) --
	(327.28, 49.41);

\path[draw=drawColor,line width= 0.6pt,line join=round] (215.31, 49.71) --
	(243.30, 49.68) --
	(271.30, 50.07) --
	(299.29, 54.84) --
	(327.28, 50.55);
\definecolor{drawColor}{RGB}{0,0,0}

\path[draw=drawColor,line width= 0.6pt,line join=round,line cap=round] (209.71, 34.67) rectangle (332.88,157.84);
\end{scope}
\begin{scope}
\definecolor{drawColor}{RGB}{0,0,0}

\node[text=drawColor,anchor=base,inner sep=0pt, outer sep=0pt, scale=  1.00] at (141.12,162.31) {Likelihood power-scaling};
\end{scope}
\begin{scope}
\definecolor{drawColor}{RGB}{0,0,0}

\node[text=drawColor,anchor=base,inner sep=0pt, outer sep=0pt, scale=  1.00] at (271.30,162.31) {Prior power-scaling};
\end{scope}
\begin{scope}
\definecolor{drawColor}{RGB}{0,0,0}

\path[draw=drawColor,line width= 0.6pt,line join=round] ( 85.13, 31.17) --
	( 85.13, 34.67);

\path[draw=drawColor,line width= 0.6pt,line join=round] (113.13, 31.17) --
	(113.13, 34.67);

\path[draw=drawColor,line width= 0.6pt,line join=round] (141.12, 31.17) --
	(141.12, 34.67);

\path[draw=drawColor,line width= 0.6pt,line join=round] (169.11, 31.17) --
	(169.11, 34.67);

\path[draw=drawColor,line width= 0.6pt,line join=round] (197.11, 31.17) --
	(197.11, 34.67);
\end{scope}
\begin{scope}
\definecolor{drawColor}{RGB}{0,0,0}

\node[text=drawColor,anchor=base,inner sep=0pt, outer sep=0pt, scale=  1.00] at ( 85.13, 21.28) {0.25};

\node[text=drawColor,anchor=base,inner sep=0pt, outer sep=0pt, scale=  1.00] at (113.13, 21.28) {0.5};

\node[text=drawColor,anchor=base,inner sep=0pt, outer sep=0pt, scale=  1.00] at (141.12, 21.28) {1};

\node[text=drawColor,anchor=base,inner sep=0pt, outer sep=0pt, scale=  1.00] at (169.11, 21.28) {2};

\node[text=drawColor,anchor=base,inner sep=0pt, outer sep=0pt, scale=  1.00] at (197.11, 21.28) {4};
\end{scope}
\begin{scope}
\definecolor{drawColor}{RGB}{0,0,0}

\path[draw=drawColor,line width= 0.6pt,line join=round] (215.31, 31.17) --
	(215.31, 34.67);

\path[draw=drawColor,line width= 0.6pt,line join=round] (243.30, 31.17) --
	(243.30, 34.67);

\path[draw=drawColor,line width= 0.6pt,line join=round] (271.30, 31.17) --
	(271.30, 34.67);

\path[draw=drawColor,line width= 0.6pt,line join=round] (299.29, 31.17) --
	(299.29, 34.67);

\path[draw=drawColor,line width= 0.6pt,line join=round] (327.28, 31.17) --
	(327.28, 34.67);
\end{scope}
\begin{scope}
\definecolor{drawColor}{RGB}{0,0,0}

\node[text=drawColor,anchor=base,inner sep=0pt, outer sep=0pt, scale=  1.00] at (215.31, 21.28) {0.25};

\node[text=drawColor,anchor=base,inner sep=0pt, outer sep=0pt, scale=  1.00] at (243.30, 21.28) {0.5};

\node[text=drawColor,anchor=base,inner sep=0pt, outer sep=0pt, scale=  1.00] at (271.30, 21.28) {1};

\node[text=drawColor,anchor=base,inner sep=0pt, outer sep=0pt, scale=  1.00] at (299.29, 21.28) {2};

\node[text=drawColor,anchor=base,inner sep=0pt, outer sep=0pt, scale=  1.00] at (327.28, 21.28) {4};
\end{scope}
\begin{scope}
\definecolor{drawColor}{RGB}{0,0,0}

\node[text=drawColor,anchor=base east,inner sep=0pt, outer sep=0pt, scale=  1.00] at ( 73.03, 34.15) {0.00};

\node[text=drawColor,anchor=base east,inner sep=0pt, outer sep=0pt, scale=  1.00] at ( 73.03, 63.30) {0.03};

\node[text=drawColor,anchor=base east,inner sep=0pt, outer sep=0pt, scale=  1.00] at ( 73.03, 92.46) {0.06};

\node[text=drawColor,anchor=base east,inner sep=0pt, outer sep=0pt, scale=  1.00] at ( 73.03,121.61) {0.09};

\node[text=drawColor,anchor=base east,inner sep=0pt, outer sep=0pt, scale=  1.00] at ( 73.03,150.76) {0.12};
\end{scope}
\begin{scope}
\definecolor{drawColor}{RGB}{0,0,0}

\path[draw=drawColor,line width= 0.6pt,line join=round] ( 76.03, 37.59) --
	( 79.53, 37.59);

\path[draw=drawColor,line width= 0.6pt,line join=round] ( 76.03, 66.75) --
	( 79.53, 66.75);

\path[draw=drawColor,line width= 0.6pt,line join=round] ( 76.03, 95.90) --
	( 79.53, 95.90);

\path[draw=drawColor,line width= 0.6pt,line join=round] ( 76.03,125.05) --
	( 79.53,125.05);

\path[draw=drawColor,line width= 0.6pt,line join=round] ( 76.03,154.21) --
	( 79.53,154.21);
\end{scope}
\begin{scope}
\definecolor{drawColor}{RGB}{0,0,0}

\node[text=drawColor,anchor=base,inner sep=0pt, outer sep=0pt, scale=  1.00] at (206.21,  8.94) {Covariate scaling factor ($c$)};
\end{scope}
\begin{scope}
\definecolor{drawColor}{RGB}{0,0,0}

\node[text=drawColor,rotate= 90.00,anchor=base,inner sep=0pt, outer sep=0pt, scale=  1.00] at ( 49.81, 96.25) {Sensitivity ($D_{\text{CJS}}$)};
\end{scope}
\begin{scope}
\definecolor{fillColor}{RGB}{127,201,127}

\path[fill=fillColor] ( 92.30, 66.55) --
	( 94.94, 61.98) --
	( 89.66, 61.98) --
	cycle;
\end{scope}
\begin{scope}
\definecolor{drawColor}{RGB}{127,201,127}

\path[draw=drawColor,line width= 0.6pt,line join=round] ( 86.14, 63.50) -- ( 98.46, 63.50);
\end{scope}
\begin{scope}
\definecolor{drawColor}{RGB}{190,174,212}
\definecolor{fillColor}{RGB}{190,174,212}

\path[draw=drawColor,line width= 0.4pt,line join=round,line cap=round,fill=fillColor] ( 92.30, 48.10) circle (  1.96);
\end{scope}
\begin{scope}
\definecolor{drawColor}{RGB}{190,174,212}

\path[draw=drawColor,line width= 0.6pt,line join=round] ( 86.14, 48.10) -- ( 98.46, 48.10);
\end{scope}
\begin{scope}
\definecolor{drawColor}{RGB}{0,0,0}

\node[text=drawColor,anchor=base west,inner sep=0pt, outer sep=0pt, scale=  1.00] at (107.00, 60.06) {Scaled: $\beta_1, \beta_2$};
\end{scope}
\begin{scope}
\definecolor{drawColor}{RGB}{0,0,0}

\node[text=drawColor,anchor=base west,inner sep=0pt, outer sep=0pt, scale=  1.00] at (107.00, 44.66) {Unscaled: $\beta_3, \beta_4$};
\end{scope}
\end{tikzpicture}

%% file: figs/bodyfat_quantities_joint.tex
% Created by tikzDevice version 0.12.3.1 on 2022-12-08 13:12:06
% !TEX encoding = UTF-8 Unicode
\begin{tikzpicture}[x=1pt,y=1pt]
\definecolor{fillColor}{RGB}{255,255,255}
\begin{scope}
\definecolor{drawColor}{RGB}{255,255,255}
\definecolor{fillColor}{RGB}{255,255,255}

\path[draw=drawColor,line width= 0.6pt,line join=round,line cap=round,fill=fillColor] (  0.00,  0.00) rectangle (375.80,325.21);
\end{scope}
\begin{scope}
\definecolor{drawColor}{RGB}{0,0,0}

\path[draw=drawColor,line width= 0.6pt,line join=round] ( 69.84,251.45) --
	( 86.02,232.25) --
	(100.51,213.68) --
	(114.99,233.49) --
	(131.17,257.17);

\path[draw=drawColor,line width= 0.6pt,line join=round] ( 69.84,277.84) --
	( 86.02,243.86) --
	(100.51,213.68) --
	(114.99,242.15) --
	(131.17,271.55);
\definecolor{fillColor}{RGB}{0,0,0}

\path[fill=fillColor] ( 66.27,247.89) --
	( 73.41,247.89) --
	( 73.41,255.02) --
	( 66.27,255.02) --
	cycle;

\path[fill=fillColor] (127.60,253.60) --
	(134.74,253.60) --
	(134.74,260.74) --
	(127.60,260.74) --
	cycle;
\definecolor{fillColor}{RGB}{255,255,255}

\path[draw=drawColor,line width= 0.4pt,line join=round,line cap=round,fill=fillColor] ( 66.67,274.68) rectangle ( 73.00,281.00);

\path[draw=drawColor,line width= 0.4pt,line join=round,line cap=round,fill=fillColor] (128.01,268.39) rectangle (134.34,274.72);

\path[draw=drawColor,line width= 1.1pt,line join=round,line cap=round] ( 64.90,210.47) rectangle (135.48,281.05);
\end{scope}
\begin{scope}
\definecolor{drawColor}{RGB}{0,0,0}

\path[draw=drawColor,line width= 0.6pt,line join=round] (175.02,213.68) --
	(191.21,228.52) --
	(205.69,242.63) --
	(220.17,257.47) --
	(236.36,274.97);

\path[draw=drawColor,line width= 0.6pt,line join=round] (175.02,277.84) --
	(191.21,258.29) --
	(205.69,242.63) --
	(220.17,228.71) --
	(236.36,214.77);
\definecolor{fillColor}{RGB}{0,0,0}

\path[fill=fillColor] (171.45,210.11) --
	(178.59,210.11) --
	(178.59,217.25) --
	(171.45,217.25) --
	cycle;

\path[fill=fillColor] (232.79,271.40) --
	(239.93,271.40) --
	(239.93,278.53) --
	(232.79,278.53) --
	cycle;
\definecolor{fillColor}{RGB}{255,255,255}

\path[draw=drawColor,line width= 0.4pt,line join=round,line cap=round,fill=fillColor] (171.86,274.68) rectangle (178.19,281.00);

\path[draw=drawColor,line width= 0.4pt,line join=round,line cap=round,fill=fillColor] (233.20,211.60) rectangle (239.52,217.93);

\path[draw=drawColor,line width= 0.6pt,dash pattern=on 4pt off 4pt ,line join=round] (170.09,235.80) -- (240.66,235.80);

\path[draw=drawColor,line width= 0.6pt,dash pattern=on 4pt off 4pt ,line join=round] (170.09,249.46) -- (240.66,249.46);

\path[draw=drawColor,line width= 1.1pt,line join=round,line cap=round] (170.09,210.47) rectangle (240.66,281.05);
\end{scope}
\begin{scope}
\definecolor{drawColor}{RGB}{0,0,0}

\path[draw=drawColor,line width= 0.6pt,line join=round] (281.88,252.49) --
	(298.06,248.06) --
	(312.54,243.86) --
	(327.02,239.53) --
	(343.21,234.47);

\path[draw=drawColor,line width= 0.6pt,line join=round] (281.88,277.84) --
	(298.06,259.93) --
	(312.54,243.86) --
	(327.02,228.90) --
	(343.21,213.68);
\definecolor{fillColor}{RGB}{0,0,0}

\path[fill=fillColor] (278.31,248.93) --
	(285.44,248.93) --
	(285.44,256.06) --
	(278.31,256.06) --
	cycle;

\path[fill=fillColor] (339.64,230.90) --
	(346.78,230.90) --
	(346.78,238.03) --
	(339.64,238.03) --
	cycle;
\definecolor{fillColor}{RGB}{255,255,255}

\path[draw=drawColor,line width= 0.4pt,line join=round,line cap=round,fill=fillColor] (278.71,274.68) rectangle (285.04,281.00);

\path[draw=drawColor,line width= 0.4pt,line join=round,line cap=round,fill=fillColor] (340.05,210.52) rectangle (346.37,216.84);

\path[draw=drawColor,line width= 0.6pt,dash pattern=on 4pt off 4pt ,line join=round] (276.94,233.99) -- (347.51,233.99);

\path[draw=drawColor,line width= 0.6pt,dash pattern=on 4pt off 4pt ,line join=round] (276.94,253.74) -- (347.51,253.74);

\path[draw=drawColor,line width= 1.1pt,line join=round,line cap=round] (276.94,210.47) rectangle (347.51,281.05);
\end{scope}
\begin{scope}
\definecolor{drawColor}{RGB}{0,0,0}

\node[text=drawColor,anchor=base east,inner sep=0pt, outer sep=0pt, scale=  1.00] at (270.44,207.90) {0.425};

\node[text=drawColor,anchor=base east,inner sep=0pt, outer sep=0pt, scale=  1.00] at (270.44,225.24) {0.450};

\node[text=drawColor,anchor=base east,inner sep=0pt, outer sep=0pt, scale=  1.00] at (270.44,242.57) {0.475};

\node[text=drawColor,anchor=base east,inner sep=0pt, outer sep=0pt, scale=  1.00] at (270.44,259.91) {0.500};

\node[text=drawColor,anchor=base east,inner sep=0pt, outer sep=0pt, scale=  1.00] at (270.44,277.24) {0.525};
\end{scope}
\begin{scope}
\definecolor{drawColor}{RGB}{0,0,0}

\path[draw=drawColor,line width= 0.6pt,line join=round] (273.44,211.35) --
	(276.94,211.35);

\path[draw=drawColor,line width= 0.6pt,line join=round] (273.44,228.68) --
	(276.94,228.68);

\path[draw=drawColor,line width= 0.6pt,line join=round] (273.44,246.02) --
	(276.94,246.02);

\path[draw=drawColor,line width= 0.6pt,line join=round] (273.44,263.35) --
	(276.94,263.35);

\path[draw=drawColor,line width= 0.6pt,line join=round] (273.44,280.69) --
	(276.94,280.69);
\end{scope}
\begin{scope}
\definecolor{drawColor}{RGB}{0,0,0}

\node[text=drawColor,anchor=base east,inner sep=0pt, outer sep=0pt, scale=  1.00] at (163.59,223.75) {-1.50};

\node[text=drawColor,anchor=base east,inner sep=0pt, outer sep=0pt, scale=  1.00] at (163.59,241.37) {-1.46};

\node[text=drawColor,anchor=base east,inner sep=0pt, outer sep=0pt, scale=  1.00] at (163.59,259.00) {-1.42};

\node[text=drawColor,anchor=base east,inner sep=0pt, outer sep=0pt, scale=  1.00] at (163.59,276.62) {-1.38};
\end{scope}
\begin{scope}
\definecolor{drawColor}{RGB}{0,0,0}

\path[draw=drawColor,line width= 0.6pt,line join=round] (166.59,227.20) --
	(170.09,227.20);

\path[draw=drawColor,line width= 0.6pt,line join=round] (166.59,244.82) --
	(170.09,244.82);

\path[draw=drawColor,line width= 0.6pt,line join=round] (166.59,262.44) --
	(170.09,262.44);

\path[draw=drawColor,line width= 0.6pt,line join=round] (166.59,280.06) --
	(170.09,280.06);
\end{scope}
\begin{scope}
\definecolor{drawColor}{RGB}{0,0,0}

\node[text=drawColor,anchor=base east,inner sep=0pt, outer sep=0pt, scale=  1.00] at ( 58.40,210.24) {0.00};

\node[text=drawColor,anchor=base east,inner sep=0pt, outer sep=0pt, scale=  1.00] at ( 58.40,222.57) {0.01};

\node[text=drawColor,anchor=base east,inner sep=0pt, outer sep=0pt, scale=  1.00] at ( 58.40,234.90) {0.02};

\node[text=drawColor,anchor=base east,inner sep=0pt, outer sep=0pt, scale=  1.00] at ( 58.40,247.23) {0.03};

\node[text=drawColor,anchor=base east,inner sep=0pt, outer sep=0pt, scale=  1.00] at ( 58.40,259.56) {0.04};

\node[text=drawColor,anchor=base east,inner sep=0pt, outer sep=0pt, scale=  1.00] at ( 58.40,271.89) {0.05};
\end{scope}
\begin{scope}
\definecolor{drawColor}{RGB}{0,0,0}

\path[draw=drawColor,line width= 0.6pt,line join=round] ( 61.40,213.68) --
	( 64.90,213.68);

\path[draw=drawColor,line width= 0.6pt,line join=round] ( 61.40,226.01) --
	( 64.90,226.01);

\path[draw=drawColor,line width= 0.6pt,line join=round] ( 61.40,238.34) --
	( 64.90,238.34);

\path[draw=drawColor,line width= 0.6pt,line join=round] ( 61.40,250.67) --
	( 64.90,250.67);

\path[draw=drawColor,line width= 0.6pt,line join=round] ( 61.40,263.01) --
	( 64.90,263.01);

\path[draw=drawColor,line width= 0.6pt,line join=round] ( 61.40,275.34) --
	( 64.90,275.34);
\end{scope}
\begin{scope}
\definecolor{drawColor}{RGB}{0,0,0}

\node[text=drawColor,anchor=base,inner sep=0pt, outer sep=0pt, scale=  1.00] at (100.19,285.52) {$\text{CJS}_{\text{dist}}$};
\end{scope}
\begin{scope}
\definecolor{drawColor}{RGB}{0,0,0}

\node[text=drawColor,anchor=base,inner sep=0pt, outer sep=0pt, scale=  1.00] at (205.37,285.52) {Mean};
\end{scope}
\begin{scope}
\definecolor{drawColor}{RGB}{0,0,0}

\node[text=drawColor,anchor=base,inner sep=0pt, outer sep=0pt, scale=  1.00] at (312.23,285.52) {SD};
\end{scope}
\begin{scope}
\definecolor{drawColor}{RGB}{0,0,0}

\path[draw=drawColor,line width= 0.6pt,line join=round] ( 69.84,206.97) --
	( 69.84,210.47);

\path[draw=drawColor,line width= 0.6pt,line join=round] (100.51,206.97) --
	(100.51,210.47);

\path[draw=drawColor,line width= 0.6pt,line join=round] (131.17,206.97) --
	(131.17,210.47);
\end{scope}
\begin{scope}
\definecolor{drawColor}{RGB}{0,0,0}

\node[text=drawColor,anchor=base,inner sep=0pt, outer sep=0pt, scale=  1.00] at ( 69.84,197.09) {0.8};

\node[text=drawColor,anchor=base,inner sep=0pt, outer sep=0pt, scale=  1.00] at (100.51,197.09) {1};

\node[text=drawColor,anchor=base,inner sep=0pt, outer sep=0pt, scale=  1.00] at (131.17,197.09) {1.25};
\end{scope}
\begin{scope}
\definecolor{drawColor}{RGB}{0,0,0}

\path[draw=drawColor,line width= 0.6pt,line join=round] (175.02,206.97) --
	(175.02,210.47);

\path[draw=drawColor,line width= 0.6pt,line join=round] (205.69,206.97) --
	(205.69,210.47);

\path[draw=drawColor,line width= 0.6pt,line join=round] (236.36,206.97) --
	(236.36,210.47);
\end{scope}
\begin{scope}
\definecolor{drawColor}{RGB}{0,0,0}

\node[text=drawColor,anchor=base,inner sep=0pt, outer sep=0pt, scale=  1.00] at (175.02,197.09) {0.8};

\node[text=drawColor,anchor=base,inner sep=0pt, outer sep=0pt, scale=  1.00] at (205.69,197.09) {1};

\node[text=drawColor,anchor=base,inner sep=0pt, outer sep=0pt, scale=  1.00] at (236.36,197.09) {1.25};
\end{scope}
\begin{scope}
\definecolor{drawColor}{RGB}{0,0,0}

\path[draw=drawColor,line width= 0.6pt,line join=round] (281.88,206.97) --
	(281.88,210.47);

\path[draw=drawColor,line width= 0.6pt,line join=round] (312.54,206.97) --
	(312.54,210.47);

\path[draw=drawColor,line width= 0.6pt,line join=round] (343.21,206.97) --
	(343.21,210.47);
\end{scope}
\begin{scope}
\definecolor{drawColor}{RGB}{0,0,0}

\node[text=drawColor,anchor=base,inner sep=0pt, outer sep=0pt, scale=  1.00] at (281.88,197.09) {0.8};

\node[text=drawColor,anchor=base,inner sep=0pt, outer sep=0pt, scale=  1.00] at (312.54,197.09) {1};

\node[text=drawColor,anchor=base,inner sep=0pt, outer sep=0pt, scale=  1.00] at (343.21,197.09) {1.25};
\end{scope}
\begin{scope}
\definecolor{drawColor}{RGB}{0,0,0}

\node[text=drawColor,anchor=base,inner sep=0pt, outer sep=0pt, scale=  1.00] at (206.21,184.75) {Power-scaling $\alpha$};
\end{scope}
\begin{scope}
\definecolor{drawColor}{RGB}{0,0,0}

\node[text=drawColor,anchor=base west,inner sep=0pt, outer sep=0pt, scale=  1.00] at ( 49.11,305.83) {Original prior};
\end{scope}
\begin{scope}
\definecolor{drawColor}{RGB}{0,0,0}

\path[draw=drawColor,line width= 0.6pt,line join=round] ( 72.34, 69.93) --
	( 88.52, 69.85) --
	(103.01, 69.77) --
	(117.49, 69.86) --
	(133.67, 69.97);

\path[draw=drawColor,line width= 0.6pt,line join=round] ( 72.34,133.93) --
	( 88.52, 95.48) --
	(103.01, 69.77) --
	(117.49, 93.63) --
	(133.67,119.39);
\definecolor{fillColor}{RGB}{0,0,0}

\path[fill=fillColor] ( 68.77, 66.36) --
	( 75.91, 66.36) --
	( 75.91, 73.50) --
	( 68.77, 73.50) --
	cycle;

\path[fill=fillColor] (130.10, 66.40) --
	(137.24, 66.40) --
	(137.24, 73.54) --
	(130.10, 73.54) --
	cycle;
\definecolor{fillColor}{RGB}{255,255,255}

\path[draw=drawColor,line width= 0.4pt,line join=round,line cap=round,fill=fillColor] ( 69.17,130.77) rectangle ( 75.50,137.10);

\path[draw=drawColor,line width= 0.4pt,line join=round,line cap=round,fill=fillColor] (130.51,116.22) rectangle (136.84,122.55);

\path[draw=drawColor,line width= 1.1pt,line join=round,line cap=round] ( 67.40, 66.57) rectangle (137.98,137.14);
\end{scope}
\begin{scope}
\definecolor{drawColor}{RGB}{0,0,0}

\path[draw=drawColor,line width= 0.6pt,line join=round] (177.52,101.98) --
	(193.71,101.92) --
	(208.19,101.85) --
	(222.67,101.78) --
	(238.86,101.70);

\path[draw=drawColor,line width= 0.6pt,line join=round] (177.52, 90.09) --
	(193.71, 94.33) --
	(208.19,101.85) --
	(222.67,109.96) --
	(238.86,119.74);
\definecolor{fillColor}{RGB}{0,0,0}

\path[fill=fillColor] (173.95, 98.41) --
	(181.09, 98.41) --
	(181.09,105.55) --
	(173.95,105.55) --
	cycle;

\path[fill=fillColor] (235.29, 98.13) --
	(242.43, 98.13) --
	(242.43,105.27) --
	(235.29,105.27) --
	cycle;
\definecolor{fillColor}{RGB}{255,255,255}

\path[draw=drawColor,line width= 0.4pt,line join=round,line cap=round,fill=fillColor] (174.36, 86.92) rectangle (180.69, 93.25);

\path[draw=drawColor,line width= 0.4pt,line join=round,line cap=round,fill=fillColor] (235.70,116.58) rectangle (242.02,122.91);

\path[draw=drawColor,line width= 0.6pt,dash pattern=on 4pt off 4pt ,line join=round] (172.59, 69.77) -- (243.16, 69.77);

\path[draw=drawColor,line width= 0.6pt,dash pattern=on 4pt off 4pt ,line join=round] (172.59,133.93) -- (243.16,133.93);

\path[draw=drawColor,line width= 1.1pt,line join=round,line cap=round] (172.59, 66.57) rectangle (243.16,137.14);
\end{scope}
\begin{scope}
\definecolor{drawColor}{RGB}{0,0,0}

\path[draw=drawColor,line width= 0.6pt,line join=round] (279.38, 96.75) --
	(295.56, 96.73) --
	(310.04, 96.70) --
	(324.52, 96.67) --
	(340.71, 96.64);

\path[draw=drawColor,line width= 0.6pt,line join=round] (279.38,133.93) --
	(295.56,111.98) --
	(310.04, 96.70) --
	(324.52, 83.44) --
	(340.71, 69.77);
\definecolor{fillColor}{RGB}{0,0,0}

\path[fill=fillColor] (275.81, 93.18) --
	(282.94, 93.18) --
	(282.94,100.32) --
	(275.81,100.32) --
	cycle;

\path[fill=fillColor] (337.14, 93.07) --
	(344.28, 93.07) --
	(344.28,100.21) --
	(337.14,100.21) --
	cycle;
\definecolor{fillColor}{RGB}{255,255,255}

\path[draw=drawColor,line width= 0.4pt,line join=round,line cap=round,fill=fillColor] (276.21,130.77) rectangle (282.54,137.10);

\path[draw=drawColor,line width= 0.4pt,line join=round,line cap=round,fill=fillColor] (337.55, 66.61) rectangle (343.87, 72.94);

\path[draw=drawColor,line width= 0.6pt,dash pattern=on 4pt off 4pt ,line join=round] (274.44, 89.58) -- (345.01, 89.58);

\path[draw=drawColor,line width= 0.6pt,dash pattern=on 4pt off 4pt ,line join=round] (274.44,103.82) -- (345.01,103.82);

\path[draw=drawColor,line width= 1.1pt,line join=round,line cap=round] (274.44, 66.57) rectangle (345.01,137.14);
\end{scope}
\begin{scope}
\definecolor{drawColor}{RGB}{0,0,0}

\node[text=drawColor,anchor=base east,inner sep=0pt, outer sep=0pt, scale=  1.00] at (267.94, 83.07) {0.50};

\node[text=drawColor,anchor=base east,inner sep=0pt, outer sep=0pt, scale=  1.00] at (267.94,106.90) {0.55};

\node[text=drawColor,anchor=base east,inner sep=0pt, outer sep=0pt, scale=  1.00] at (267.94,130.73) {0.60};
\end{scope}
\begin{scope}
\definecolor{drawColor}{RGB}{0,0,0}

\path[draw=drawColor,line width= 0.6pt,line join=round] (270.94, 86.51) --
	(274.44, 86.51);

\path[draw=drawColor,line width= 0.6pt,line join=round] (270.94,110.34) --
	(274.44,110.34);

\path[draw=drawColor,line width= 0.6pt,line join=round] (270.94,134.17) --
	(274.44,134.17);
\end{scope}
\begin{scope}
\definecolor{drawColor}{RGB}{0,0,0}

\node[text=drawColor,anchor=base east,inner sep=0pt, outer sep=0pt, scale=  1.00] at (166.09, 69.10) {-1.89};

\node[text=drawColor,anchor=base east,inner sep=0pt, outer sep=0pt, scale=  1.00] at (166.09, 86.32) {-1.88};

\node[text=drawColor,anchor=base east,inner sep=0pt, outer sep=0pt, scale=  1.00] at (166.09,103.54) {-1.87};

\node[text=drawColor,anchor=base east,inner sep=0pt, outer sep=0pt, scale=  1.00] at (166.09,120.76) {-1.86};
\end{scope}
\begin{scope}
\definecolor{drawColor}{RGB}{0,0,0}

\path[draw=drawColor,line width= 0.6pt,line join=round] (169.09, 72.54) --
	(172.59, 72.54);

\path[draw=drawColor,line width= 0.6pt,line join=round] (169.09, 89.76) --
	(172.59, 89.76);

\path[draw=drawColor,line width= 0.6pt,line join=round] (169.09,106.98) --
	(172.59,106.98);

\path[draw=drawColor,line width= 0.6pt,line join=round] (169.09,124.20) --
	(172.59,124.20);
\end{scope}
\begin{scope}
\definecolor{drawColor}{RGB}{0,0,0}

\node[text=drawColor,anchor=base east,inner sep=0pt, outer sep=0pt, scale=  1.00] at ( 60.90, 66.33) {0.00};

\node[text=drawColor,anchor=base east,inner sep=0pt, outer sep=0pt, scale=  1.00] at ( 60.90, 84.02) {0.01};

\node[text=drawColor,anchor=base east,inner sep=0pt, outer sep=0pt, scale=  1.00] at ( 60.90,101.72) {0.02};

\node[text=drawColor,anchor=base east,inner sep=0pt, outer sep=0pt, scale=  1.00] at ( 60.90,119.41) {0.03};
\end{scope}
\begin{scope}
\definecolor{drawColor}{RGB}{0,0,0}

\path[draw=drawColor,line width= 0.6pt,line join=round] ( 63.90, 69.77) --
	( 67.40, 69.77);

\path[draw=drawColor,line width= 0.6pt,line join=round] ( 63.90, 87.47) --
	( 67.40, 87.47);

\path[draw=drawColor,line width= 0.6pt,line join=round] ( 63.90,105.16) --
	( 67.40,105.16);

\path[draw=drawColor,line width= 0.6pt,line join=round] ( 63.90,122.85) --
	( 67.40,122.85);
\end{scope}
\begin{scope}
\definecolor{drawColor}{RGB}{0,0,0}

\node[text=drawColor,anchor=base,inner sep=0pt, outer sep=0pt, scale=  1.00] at (102.69,141.61) {$\text{CJS}_{\text{dist}}$};
\end{scope}
\begin{scope}
\definecolor{drawColor}{RGB}{0,0,0}

\node[text=drawColor,anchor=base,inner sep=0pt, outer sep=0pt, scale=  1.00] at (207.87,141.61) {Mean};
\end{scope}
\begin{scope}
\definecolor{drawColor}{RGB}{0,0,0}

\node[text=drawColor,anchor=base,inner sep=0pt, outer sep=0pt, scale=  1.00] at (309.73,141.61) {SD};
\end{scope}
\begin{scope}
\definecolor{drawColor}{RGB}{0,0,0}

\path[draw=drawColor,line width= 0.6pt,line join=round] ( 72.34, 63.07) --
	( 72.34, 66.57);

\path[draw=drawColor,line width= 0.6pt,line join=round] (103.01, 63.07) --
	(103.01, 66.57);

\path[draw=drawColor,line width= 0.6pt,line join=round] (133.67, 63.07) --
	(133.67, 66.57);
\end{scope}
\begin{scope}
\definecolor{drawColor}{RGB}{0,0,0}

\node[text=drawColor,anchor=base,inner sep=0pt, outer sep=0pt, scale=  1.00] at ( 72.34, 53.18) {0.8};

\node[text=drawColor,anchor=base,inner sep=0pt, outer sep=0pt, scale=  1.00] at (103.01, 53.18) {1};

\node[text=drawColor,anchor=base,inner sep=0pt, outer sep=0pt, scale=  1.00] at (133.67, 53.18) {1.25};
\end{scope}
\begin{scope}
\definecolor{drawColor}{RGB}{0,0,0}

\path[draw=drawColor,line width= 0.6pt,line join=round] (177.52, 63.07) --
	(177.52, 66.57);

\path[draw=drawColor,line width= 0.6pt,line join=round] (208.19, 63.07) --
	(208.19, 66.57);

\path[draw=drawColor,line width= 0.6pt,line join=round] (238.86, 63.07) --
	(238.86, 66.57);
\end{scope}
\begin{scope}
\definecolor{drawColor}{RGB}{0,0,0}

\node[text=drawColor,anchor=base,inner sep=0pt, outer sep=0pt, scale=  1.00] at (177.52, 53.18) {0.8};

\node[text=drawColor,anchor=base,inner sep=0pt, outer sep=0pt, scale=  1.00] at (208.19, 53.18) {1};

\node[text=drawColor,anchor=base,inner sep=0pt, outer sep=0pt, scale=  1.00] at (238.86, 53.18) {1.25};
\end{scope}
\begin{scope}
\definecolor{drawColor}{RGB}{0,0,0}

\path[draw=drawColor,line width= 0.6pt,line join=round] (279.38, 63.07) --
	(279.38, 66.57);

\path[draw=drawColor,line width= 0.6pt,line join=round] (310.04, 63.07) --
	(310.04, 66.57);

\path[draw=drawColor,line width= 0.6pt,line join=round] (340.71, 63.07) --
	(340.71, 66.57);
\end{scope}
\begin{scope}
\definecolor{drawColor}{RGB}{0,0,0}

\node[text=drawColor,anchor=base,inner sep=0pt, outer sep=0pt, scale=  1.00] at (279.38, 53.18) {0.8};

\node[text=drawColor,anchor=base,inner sep=0pt, outer sep=0pt, scale=  1.00] at (310.04, 53.18) {1};

\node[text=drawColor,anchor=base,inner sep=0pt, outer sep=0pt, scale=  1.00] at (340.71, 53.18) {1.25};
\end{scope}
\begin{scope}
\definecolor{drawColor}{RGB}{0,0,0}

\node[text=drawColor,anchor=base,inner sep=0pt, outer sep=0pt, scale=  1.00] at (206.21, 40.84) {Power-scaling $\alpha$};
\end{scope}
\begin{scope}
\definecolor{drawColor}{RGB}{0,0,0}

\node[text=drawColor,anchor=base west,inner sep=0pt, outer sep=0pt, scale=  1.00] at ( 49.11,161.92) {Adjusted prior};
\end{scope}
\begin{scope}
\definecolor{fillColor}{RGB}{0,0,0}

\path[fill=fillColor] ( 53.23, 16.63) --
	( 60.36, 16.63) --
	( 60.36, 23.77) --
	( 53.23, 23.77) --
	cycle;
\end{scope}
\begin{scope}
\definecolor{drawColor}{RGB}{0,0,0}
\definecolor{fillColor}{RGB}{255,255,255}

\path[draw=drawColor,line width= 0.4pt,line join=round,line cap=round,fill=fillColor] (163.48, 17.04) rectangle (169.80, 23.36);
\end{scope}
\begin{scope}
\definecolor{drawColor}{RGB}{0,0,0}

\node[text=drawColor,anchor=base west,inner sep=0pt, outer sep=0pt, scale=  1.00] at ( 69.49, 16.76) {Prior power-scaling};
\end{scope}
\begin{scope}
\definecolor{drawColor}{RGB}{0,0,0}

\node[text=drawColor,anchor=base west,inner sep=0pt, outer sep=0pt, scale=  1.00] at (179.34, 16.76) {Likelihood power-scaling};
\end{scope}
\begin{scope}
\definecolor{drawColor}{RGB}{0,0,0}

\path[draw=drawColor,line width= 0.6pt,dash pattern=on 4pt off 4pt ,line join=round] (304.60, 20.20) -- (316.92, 20.20);
\end{scope}
\begin{scope}
\definecolor{drawColor}{RGB}{0,0,0}

\path[draw=drawColor,line width= 0.6pt,dash pattern=on 4pt off 4pt ,line join=round] (304.60, 20.20) -- (316.92, 20.20);
\end{scope}
\begin{scope}
\definecolor{drawColor}{RGB}{0,0,0}

\node[text=drawColor,anchor=base west,inner sep=0pt, outer sep=0pt, scale=  1.00] at (323.46, 16.76) {$\pm2$ MCSE};
\end{scope}
\end{tikzpicture}

%% file: figs/bacteria_priors.tex
% Created by tikzDevice version 0.12.3.1 on 2022-12-08 13:08:56
% !TEX encoding = UTF-8 Unicode
\begin{tikzpicture}[x=1pt,y=1pt]
\definecolor{fillColor}{RGB}{255,255,255}
\begin{scope}
\definecolor{drawColor}{RGB}{255,255,255}
\definecolor{fillColor}{RGB}{255,255,255}

\path[draw=drawColor,line width= 0.6pt,line join=round,line cap=round,fill=fillColor] (  0.00,  3.45) rectangle (375.80,141.09);
\end{scope}
\begin{scope}
\definecolor{drawColor}{RGB}{27,158,119}

\path[draw=drawColor,line width= 1.1pt,line join=round] ( 47.97,128.59) --
	( 48.75, 51.10) --
	( 49.52, 49.30) --
	( 50.29, 48.69) --
	( 51.06, 48.39) --
	( 51.84, 48.21) --
	( 52.61, 48.09) --
	( 53.38, 48.00) --
	( 54.15, 47.93) --
	( 54.93, 47.88) --
	( 55.70, 47.84) --
	( 56.47, 47.81) --
	( 57.24, 47.78) --
	( 58.02, 47.76) --
	( 58.79, 47.74) --
	( 59.56, 47.72) --
	( 60.33, 47.70) --
	( 61.11, 47.69) --
	( 61.88, 47.68) --
	( 62.65, 47.67) --
	( 63.42, 47.66) --
	( 64.20, 47.65) --
	( 64.97, 47.64) --
	( 65.74, 47.63) --
	( 66.52, 47.63) --
	( 67.29, 47.62) --
	( 68.06, 47.62) --
	( 68.83, 47.61) --
	( 69.61, 47.61) --
	( 70.38, 47.60) --
	( 71.15, 47.60) --
	( 71.92, 47.59) --
	( 72.70, 47.59) --
	( 73.47, 47.59) --
	( 74.24, 47.58) --
	( 75.01, 47.58) --
	( 75.79, 47.58) --
	( 76.56, 47.57) --
	( 77.33, 47.57) --
	( 78.10, 47.57) --
	( 78.88, 47.57) --
	( 79.65, 47.56) --
	( 80.42, 47.56) --
	( 81.19, 47.56) --
	( 81.97, 47.56) --
	( 82.74, 47.56) --
	( 83.51, 47.55) --
	( 84.28, 47.55) --
	( 85.06, 47.55) --
	( 85.83, 47.55) --
	( 86.60, 47.55) --
	( 87.37, 47.55) --
	( 88.15, 47.54) --
	( 88.92, 47.54) --
	( 89.69, 47.54) --
	( 90.46, 47.54) --
	( 91.24, 47.54) --
	( 92.01, 47.54) --
	( 92.78, 47.54) --
	( 93.55, 47.54) --
	( 94.33, 47.53) --
	( 95.10, 47.53) --
	( 95.87, 47.53) --
	( 96.65, 47.53) --
	( 97.42, 47.53) --
	( 98.19, 47.53) --
	( 98.96, 47.53) --
	( 99.74, 47.53) --
	(100.51, 47.53) --
	(101.28, 47.53) --
	(102.05, 47.53) --
	(102.83, 47.52) --
	(103.60, 47.52) --
	(104.37, 47.52) --
	(105.14, 47.52) --
	(105.92, 47.52) --
	(106.69, 47.52) --
	(107.46, 47.52) --
	(108.23, 47.52) --
	(109.01, 47.52) --
	(109.78, 47.52) --
	(110.55, 47.52) --
	(111.32, 47.52) --
	(112.10, 47.52) --
	(112.87, 47.52) --
	(113.64, 47.52) --
	(114.41, 47.52) --
	(115.19, 47.52) --
	(115.96, 47.51) --
	(116.73, 47.51) --
	(117.50, 47.51) --
	(118.28, 47.51) --
	(119.05, 47.51) --
	(119.82, 47.51) --
	(120.59, 47.51) --
	(121.37, 47.51) --
	(122.14, 47.51) --
	(122.91, 47.51) --
	(123.68, 47.51) --
	(124.46, 47.51) --
	(125.23, 47.51);
\definecolor{drawColor}{RGB}{217,95,2}

\path[draw=drawColor,line width= 1.1pt,line join=round] ( 47.97, 62.89) --
	( 48.75, 62.81) --
	( 49.52, 62.58) --
	( 50.29, 62.21) --
	( 51.06, 61.70) --
	( 51.84, 61.07) --
	( 52.61, 60.35) --
	( 53.38, 59.54) --
	( 54.15, 58.67) --
	( 54.93, 57.75) --
	( 55.70, 56.82) --
	( 56.47, 55.89) --
	( 57.24, 54.98) --
	( 58.02, 54.09) --
	( 58.79, 53.26) --
	( 59.56, 52.48) --
	( 60.33, 51.76) --
	( 61.11, 51.11) --
	( 61.88, 50.52) --
	( 62.65, 50.01) --
	( 63.42, 49.56) --
	( 64.20, 49.17) --
	( 64.97, 48.85) --
	( 65.74, 48.57) --
	( 66.52, 48.34) --
	( 67.29, 48.15) --
	( 68.06, 48.00) --
	( 68.83, 47.88) --
	( 69.61, 47.78) --
	( 70.38, 47.71) --
	( 71.15, 47.65) --
	( 71.92, 47.60) --
	( 72.70, 47.57) --
	( 73.47, 47.54) --
	( 74.24, 47.52) --
	( 75.01, 47.51) --
	( 75.79, 47.50) --
	( 76.56, 47.49) --
	( 77.33, 47.49) --
	( 78.10, 47.48) --
	( 78.88, 47.48) --
	( 79.65, 47.48) --
	( 80.42, 47.48) --
	( 81.19, 47.48) --
	( 81.97, 47.48) --
	( 82.74, 47.48) --
	( 83.51, 47.48) --
	( 84.28, 47.48) --
	( 85.06, 47.48) --
	( 85.83, 47.48) --
	( 86.60, 47.48) --
	( 87.37, 47.48) --
	( 88.15, 47.48) --
	( 88.92, 47.48) --
	( 89.69, 47.48) --
	( 90.46, 47.48) --
	( 91.24, 47.48) --
	( 92.01, 47.48) --
	( 92.78, 47.48) --
	( 93.55, 47.48) --
	( 94.33, 47.48) --
	( 95.10, 47.48) --
	( 95.87, 47.48) --
	( 96.65, 47.48) --
	( 97.42, 47.48) --
	( 98.19, 47.48) --
	( 98.96, 47.48) --
	( 99.74, 47.48) --
	(100.51, 47.48) --
	(101.28, 47.48) --
	(102.05, 47.48) --
	(102.83, 47.48) --
	(103.60, 47.48) --
	(104.37, 47.48) --
	(105.14, 47.48) --
	(105.92, 47.48) --
	(106.69, 47.48) --
	(107.46, 47.48) --
	(108.23, 47.48) --
	(109.01, 47.48) --
	(109.78, 47.48) --
	(110.55, 47.48) --
	(111.32, 47.48) --
	(112.10, 47.48) --
	(112.87, 47.48) --
	(113.64, 47.48) --
	(114.41, 47.48) --
	(115.19, 47.48) --
	(115.96, 47.48) --
	(116.73, 47.48) --
	(117.50, 47.48) --
	(118.28, 47.48) --
	(119.05, 47.48) --
	(119.82, 47.48) --
	(120.59, 47.48) --
	(121.37, 47.48) --
	(122.14, 47.48) --
	(122.91, 47.48) --
	(123.68, 47.48) --
	(124.46, 47.48) --
	(125.23, 47.48);
\definecolor{drawColor}{RGB}{117,112,179}

\path[draw=drawColor,line width= 1.1pt,line join=round] ( 47.97, 59.77) --
	( 48.75, 59.65) --
	( 49.52, 59.30) --
	( 50.29, 58.76) --
	( 51.06, 58.07) --
	( 51.84, 57.31) --
	( 52.61, 56.52) --
	( 53.38, 55.73) --
	( 54.15, 54.97) --
	( 54.93, 54.27) --
	( 55.70, 53.62) --
	( 56.47, 53.04) --
	( 57.24, 52.51) --
	( 58.02, 52.05) --
	( 58.79, 51.63) --
	( 59.56, 51.26) --
	( 60.33, 50.93) --
	( 61.11, 50.64) --
	( 61.88, 50.37) --
	( 62.65, 50.14) --
	( 63.42, 49.93) --
	( 64.20, 49.75) --
	( 64.97, 49.58) --
	( 65.74, 49.43) --
	( 66.52, 49.29) --
	( 67.29, 49.17) --
	( 68.06, 49.06) --
	( 68.83, 48.96) --
	( 69.61, 48.87) --
	( 70.38, 48.78) --
	( 71.15, 48.70) --
	( 71.92, 48.63) --
	( 72.70, 48.57) --
	( 73.47, 48.51) --
	( 74.24, 48.45) --
	( 75.01, 48.40) --
	( 75.79, 48.36) --
	( 76.56, 48.31) --
	( 77.33, 48.27) --
	( 78.10, 48.23) --
	( 78.88, 48.20) --
	( 79.65, 48.17) --
	( 80.42, 48.13) --
	( 81.19, 48.11) --
	( 81.97, 48.08) --
	( 82.74, 48.05) --
	( 83.51, 48.03) --
	( 84.28, 48.01) --
	( 85.06, 47.99) --
	( 85.83, 47.97) --
	( 86.60, 47.95) --
	( 87.37, 47.93) --
	( 88.15, 47.91) --
	( 88.92, 47.90) --
	( 89.69, 47.88) --
	( 90.46, 47.87) --
	( 91.24, 47.86) --
	( 92.01, 47.84) --
	( 92.78, 47.83) --
	( 93.55, 47.82) --
	( 94.33, 47.81) --
	( 95.10, 47.80) --
	( 95.87, 47.79) --
	( 96.65, 47.78) --
	( 97.42, 47.77) --
	( 98.19, 47.76) --
	( 98.96, 47.75) --
	( 99.74, 47.74) --
	(100.51, 47.74) --
	(101.28, 47.73) --
	(102.05, 47.72) --
	(102.83, 47.71) --
	(103.60, 47.71) --
	(104.37, 47.70) --
	(105.14, 47.70) --
	(105.92, 47.69) --
	(106.69, 47.68) --
	(107.46, 47.68) --
	(108.23, 47.67) --
	(109.01, 47.67) --
	(109.78, 47.66) --
	(110.55, 47.66) --
	(111.32, 47.66) --
	(112.10, 47.65) --
	(112.87, 47.65) --
	(113.64, 47.64) --
	(114.41, 47.64) --
	(115.19, 47.64) --
	(115.96, 47.63) --
	(116.73, 47.63) --
	(117.50, 47.63) --
	(118.28, 47.62) --
	(119.05, 47.62) --
	(119.82, 47.62) --
	(120.59, 47.61) --
	(121.37, 47.61) --
	(122.14, 47.61) --
	(122.91, 47.60) --
	(123.68, 47.60) --
	(124.46, 47.60) --
	(125.23, 47.60);
\definecolor{drawColor}{RGB}{231,41,138}

\path[draw=drawColor,line width= 1.1pt,line join=round] ( 47.97,124.73) --
	( 48.75,110.73) --
	( 49.52, 99.26) --
	( 50.29, 89.87) --
	( 51.06, 82.19) --
	( 51.84, 75.90) --
	( 52.61, 70.74) --
	( 53.38, 66.53) --
	( 54.15, 63.07) --
	( 54.93, 60.25) --
	( 55.70, 57.93) --
	( 56.47, 56.04) --
	( 57.24, 54.48) --
	( 58.02, 53.21) --
	( 58.79, 52.17) --
	( 59.56, 51.32) --
	( 60.33, 50.62) --
	( 61.11, 50.05) --
	( 61.88, 49.59) --
	( 62.65, 49.20) --
	( 63.42, 48.89) --
	( 64.20, 48.63) --
	( 64.97, 48.42) --
	( 65.74, 48.25) --
	( 66.52, 48.11) --
	( 67.29, 48.00) --
	( 68.06, 47.90) --
	( 68.83, 47.82) --
	( 69.61, 47.76) --
	( 70.38, 47.71) --
	( 71.15, 47.67) --
	( 71.92, 47.63) --
	( 72.70, 47.60) --
	( 73.47, 47.58) --
	( 74.24, 47.56) --
	( 75.01, 47.55) --
	( 75.79, 47.53) --
	( 76.56, 47.52) --
	( 77.33, 47.51) --
	( 78.10, 47.51) --
	( 78.88, 47.50) --
	( 79.65, 47.50) --
	( 80.42, 47.49) --
	( 81.19, 47.49) --
	( 81.97, 47.49) --
	( 82.74, 47.48) --
	( 83.51, 47.48) --
	( 84.28, 47.48) --
	( 85.06, 47.48) --
	( 85.83, 47.48) --
	( 86.60, 47.48) --
	( 87.37, 47.48) --
	( 88.15, 47.48) --
	( 88.92, 47.48) --
	( 89.69, 47.48) --
	( 90.46, 47.48) --
	( 91.24, 47.48) --
	( 92.01, 47.48) --
	( 92.78, 47.48) --
	( 93.55, 47.48) --
	( 94.33, 47.48) --
	( 95.10, 47.48) --
	( 95.87, 47.48) --
	( 96.65, 47.48) --
	( 97.42, 47.48) --
	( 98.19, 47.48) --
	( 98.96, 47.48) --
	( 99.74, 47.48) --
	(100.51, 47.48) --
	(101.28, 47.48) --
	(102.05, 47.48) --
	(102.83, 47.48) --
	(103.60, 47.48) --
	(104.37, 47.48) --
	(105.14, 47.48) --
	(105.92, 47.48) --
	(106.69, 47.48) --
	(107.46, 47.48) --
	(108.23, 47.48) --
	(109.01, 47.48) --
	(109.78, 47.48) --
	(110.55, 47.48) --
	(111.32, 47.48) --
	(112.10, 47.48) --
	(112.87, 47.48) --
	(113.64, 47.48) --
	(114.41, 47.48) --
	(115.19, 47.48) --
	(115.96, 47.48) --
	(116.73, 47.48) --
	(117.50, 47.48) --
	(118.28, 47.48) --
	(119.05, 47.48) --
	(119.82, 47.48) --
	(120.59, 47.48) --
	(121.37, 47.48) --
	(122.14, 47.48) --
	(122.91, 47.48) --
	(123.68, 47.48) --
	(124.46, 47.48) --
	(125.23, 47.48);
\definecolor{drawColor}{RGB}{0,0,0}

\path[draw=drawColor,line width= 0.4pt,line join=round,line cap=round] ( 44.11, 43.61) rectangle (129.09,128.59);
\end{scope}
\begin{scope}
\definecolor{drawColor}{RGB}{0,0,0}

\node[text=drawColor,anchor=base east,inner sep=0pt, outer sep=0pt, scale=  1.00] at ( 37.61, 44.03) {0.0};

\node[text=drawColor,anchor=base east,inner sep=0pt, outer sep=0pt, scale=  1.00] at ( 37.61, 63.34) {0.5};

\node[text=drawColor,anchor=base east,inner sep=0pt, outer sep=0pt, scale=  1.00] at ( 37.61, 82.66) {1.0};

\node[text=drawColor,anchor=base east,inner sep=0pt, outer sep=0pt, scale=  1.00] at ( 37.61,101.97) {1.5};

\node[text=drawColor,anchor=base east,inner sep=0pt, outer sep=0pt, scale=  1.00] at ( 37.61,121.29) {2.0};
\end{scope}
\begin{scope}
\definecolor{drawColor}{RGB}{0,0,0}

\path[draw=drawColor,line width= 0.4pt,line join=round] ( 40.61, 47.48) --
	( 44.11, 47.48);

\path[draw=drawColor,line width= 0.4pt,line join=round] ( 40.61, 66.79) --
	( 44.11, 66.79);

\path[draw=drawColor,line width= 0.4pt,line join=round] ( 40.61, 86.10) --
	( 44.11, 86.10);

\path[draw=drawColor,line width= 0.4pt,line join=round] ( 40.61,105.42) --
	( 44.11,105.42);

\path[draw=drawColor,line width= 0.4pt,line join=round] ( 40.61,124.73) --
	( 44.11,124.73);
\end{scope}
\begin{scope}
\definecolor{drawColor}{RGB}{0,0,0}

\node[text=drawColor,rotate= 90.00,anchor=base,inner sep=0pt, outer sep=0pt, scale=  1.00] at ( 19.39, 86.10) {$p(\tau)$};
\end{scope}
\begin{scope}
\definecolor{drawColor}{RGB}{0,0,0}

\path[draw=drawColor,line width= 0.4pt,line join=round] ( 47.97, 40.11) --
	( 47.97, 43.61);

\path[draw=drawColor,line width= 0.4pt,line join=round] ( 67.29, 40.11) --
	( 67.29, 43.61);

\path[draw=drawColor,line width= 0.4pt,line join=round] ( 86.60, 40.11) --
	( 86.60, 43.61);

\path[draw=drawColor,line width= 0.4pt,line join=round] (105.92, 40.11) --
	(105.92, 43.61);

\path[draw=drawColor,line width= 0.4pt,line join=round] (125.23, 40.11) --
	(125.23, 43.61);
\end{scope}
\begin{scope}
\definecolor{drawColor}{RGB}{0,0,0}

\node[text=drawColor,anchor=base,inner sep=0pt, outer sep=0pt, scale=  1.00] at ( 47.97, 30.22) {0.0};

\node[text=drawColor,anchor=base,inner sep=0pt, outer sep=0pt, scale=  1.00] at ( 67.29, 30.22) {2.5};

\node[text=drawColor,anchor=base,inner sep=0pt, outer sep=0pt, scale=  1.00] at ( 86.60, 30.22) {5.0};

\node[text=drawColor,anchor=base,inner sep=0pt, outer sep=0pt, scale=  1.00] at (105.92, 30.22) {7.5};

\node[text=drawColor,anchor=base,inner sep=0pt, outer sep=0pt, scale=  1.00] at (125.23, 30.22) {10.0};
\end{scope}
\begin{scope}
\definecolor{drawColor}{RGB}{0,0,0}

\node[text=drawColor,anchor=base,inner sep=0pt, outer sep=0pt, scale=  1.00] at ( 86.60, 17.89) {$\tau$};
\end{scope}
\begin{scope}
\definecolor{drawColor}{RGB}{102,166,30}

\path[draw=drawColor,line width= 1.1pt,line join=round] (171.23, 47.48) --
	(171.43, 47.48) --
	(171.62, 47.48) --
	(171.81, 47.48) --
	(172.01, 47.48) --
	(172.20, 47.48) --
	(172.39, 47.48) --
	(172.59, 47.48) --
	(172.78, 47.48) --
	(172.97, 47.48) --
	(173.16, 47.48) --
	(173.36, 47.48) --
	(173.55, 47.48) --
	(173.74, 47.48) --
	(173.94, 47.48) --
	(174.13, 47.48) --
	(174.32, 47.48) --
	(174.52, 47.48) --
	(174.71, 47.48) --
	(174.90, 47.48) --
	(175.10, 47.48) --
	(175.29, 47.48) --
	(175.48, 47.48) --
	(175.68, 47.49) --
	(175.87, 47.49) --
	(176.06, 47.50) --
	(176.25, 47.51) --
	(176.45, 47.51) --
	(176.64, 47.53) --
	(176.83, 47.54) --
	(177.03, 47.55) --
	(177.22, 47.57) --
	(177.41, 47.59) --
	(177.61, 47.62) --
	(177.80, 47.65) --
	(177.99, 47.68) --
	(178.19, 47.73) --
	(178.38, 47.77) --
	(178.57, 47.82) --
	(178.77, 47.88) --
	(178.96, 47.95) --
	(179.15, 48.03) --
	(179.35, 48.11) --
	(179.54, 48.21) --
	(179.73, 48.31) --
	(179.92, 48.43) --
	(180.12, 48.55) --
	(180.31, 48.69) --
	(180.50, 48.85) --
	(180.70, 49.01) --
	(180.89, 49.19) --
	(181.08, 49.39) --
	(181.28, 49.60) --
	(181.47, 49.83) --
	(181.66, 50.08) --
	(181.86, 50.35) --
	(182.05, 50.63) --
	(182.24, 50.93) --
	(182.44, 51.25) --
	(182.63, 51.60) --
	(182.82, 51.96) --
	(183.01, 52.34) --
	(183.21, 52.75) --
	(183.40, 53.18) --
	(183.59, 53.63) --
	(183.79, 54.10) --
	(183.98, 54.60) --
	(184.17, 55.11) --
	(184.37, 55.66) --
	(184.56, 56.22) --
	(184.75, 56.81) --
	(184.95, 57.42) --
	(185.14, 58.06) --
	(185.33, 58.71) --
	(185.53, 59.40) --
	(185.72, 60.10) --
	(185.91, 60.83) --
	(186.11, 61.58) --
	(186.30, 62.35) --
	(186.49, 63.14) --
	(186.68, 63.95) --
	(186.88, 64.79) --
	(187.07, 65.64) --
	(187.26, 66.51) --
	(187.46, 67.41) --
	(187.65, 68.32) --
	(187.84, 69.24) --
	(188.04, 70.19) --
	(188.23, 71.15) --
	(188.42, 72.13) --
	(188.62, 73.12) --
	(188.81, 74.12) --
	(189.00, 75.14) --
	(189.20, 76.16) --
	(189.39, 77.20) --
	(189.58, 78.25) --
	(189.77, 79.31) --
	(189.97, 80.37) --
	(190.16, 81.45) --
	(190.35, 82.52) --
	(190.55, 83.60) --
	(190.74, 84.69) --
	(190.93, 85.78) --
	(191.13, 86.87) --
	(191.32, 87.96) --
	(191.51, 89.05) --
	(191.71, 90.13) --
	(191.90, 91.22) --
	(192.09, 92.30) --
	(192.29, 93.38) --
	(192.48, 94.45) --
	(192.67, 95.51) --
	(192.86, 96.57) --
	(193.06, 97.62) --
	(193.25, 98.65) --
	(193.44, 99.68) --
	(193.64,100.70) --
	(193.83,101.70) --
	(194.02,102.69) --
	(194.22,103.66) --
	(194.41,104.62) --
	(194.60,105.57) --
	(194.80,106.50) --
	(194.99,107.41) --
	(195.18,108.30) --
	(195.38,109.17) --
	(195.57,110.03) --
	(195.76,110.86) --
	(195.96,111.67) --
	(196.15,112.47) --
	(196.34,113.24) --
	(196.53,113.98) --
	(196.73,114.71) --
	(196.92,115.41) --
	(197.11,116.09) --
	(197.31,116.74) --
	(197.50,117.37) --
	(197.69,117.97) --
	(197.89,118.55) --
	(198.08,119.11) --
	(198.27,119.63) --
	(198.47,120.14) --
	(198.66,120.61) --
	(198.85,121.06) --
	(199.05,121.49) --
	(199.24,121.89) --
	(199.43,122.26) --
	(199.62,122.60) --
	(199.82,122.92) --
	(200.01,123.22) --
	(200.20,123.48) --
	(200.40,123.72) --
	(200.59,123.94) --
	(200.78,124.12) --
	(200.98,124.29) --
	(201.17,124.42) --
	(201.36,124.54) --
	(201.56,124.62) --
	(201.75,124.68) --
	(201.94,124.72) --
	(202.14,124.73) --
	(202.33,124.72) --
	(202.52,124.68) --
	(202.72,124.62) --
	(202.91,124.54) --
	(203.10,124.44) --
	(203.29,124.31) --
	(203.49,124.16) --
	(203.68,123.99) --
	(203.87,123.79) --
	(204.07,123.58) --
	(204.26,123.35) --
	(204.45,123.09) --
	(204.65,122.82) --
	(204.84,122.53) --
	(205.03,122.22) --
	(205.23,121.89) --
	(205.42,121.54) --
	(205.61,121.18) --
	(205.81,120.79) --
	(206.00,120.40) --
	(206.19,119.98) --
	(206.38,119.56) --
	(206.58,119.11) --
	(206.77,118.66) --
	(206.96,118.18) --
	(207.16,117.70) --
	(207.35,117.20) --
	(207.54,116.69) --
	(207.74,116.17) --
	(207.93,115.64) --
	(208.12,115.10) --
	(208.32,114.54) --
	(208.51,113.98) --
	(208.70,113.41) --
	(208.90,112.82) --
	(209.09,112.23) --
	(209.28,111.63) --
	(209.48,111.03) --
	(209.67,110.41) --
	(209.86,109.79) --
	(210.05,109.17) --
	(210.25,108.54) --
	(210.44,107.90) --
	(210.63,107.26) --
	(210.83,106.61) --
	(211.02,105.96) --
	(211.21,105.30) --
	(211.41,104.65) --
	(211.60,103.98) --
	(211.79,103.32) --
	(211.99,102.66) --
	(212.18,101.99) --
	(212.37,101.32) --
	(212.57,100.65) --
	(212.76, 99.98) --
	(212.95, 99.30) --
	(213.14, 98.63) --
	(213.34, 97.96) --
	(213.53, 97.29) --
	(213.72, 96.62) --
	(213.92, 95.95) --
	(214.11, 95.28) --
	(214.30, 94.61) --
	(214.50, 93.95) --
	(214.69, 93.29) --
	(214.88, 92.63) --
	(215.08, 91.97) --
	(215.27, 91.31) --
	(215.46, 90.66) --
	(215.66, 90.01) --
	(215.85, 89.37) --
	(216.04, 88.72) --
	(216.23, 88.09) --
	(216.43, 87.45) --
	(216.62, 86.82) --
	(216.81, 86.20) --
	(217.01, 85.58) --
	(217.20, 84.96) --
	(217.39, 84.35) --
	(217.59, 83.74) --
	(217.78, 83.14) --
	(217.97, 82.54) --
	(218.17, 81.95) --
	(218.36, 81.36) --
	(218.55, 80.78) --
	(218.75, 80.21) --
	(218.94, 79.64) --
	(219.13, 79.08) --
	(219.33, 78.52) --
	(219.52, 77.97) --
	(219.71, 77.42) --
	(219.90, 76.88) --
	(220.10, 76.35) --
	(220.29, 75.82) --
	(220.48, 75.30) --
	(220.68, 74.78) --
	(220.87, 74.27) --
	(221.06, 73.77) --
	(221.26, 73.27) --
	(221.45, 72.78) --
	(221.64, 72.30) --
	(221.84, 71.82) --
	(222.03, 71.35) --
	(222.22, 70.89) --
	(222.42, 70.43) --
	(222.61, 69.98) --
	(222.80, 69.53) --
	(222.99, 69.10) --
	(223.19, 68.66) --
	(223.38, 68.24) --
	(223.57, 67.82) --
	(223.77, 67.40) --
	(223.96, 67.00) --
	(224.15, 66.60) --
	(224.35, 66.20) --
	(224.54, 65.81) --
	(224.73, 65.43) --
	(224.93, 65.05) --
	(225.12, 64.68) --
	(225.31, 64.32) --
	(225.51, 63.96) --
	(225.70, 63.61) --
	(225.89, 63.26) --
	(226.09, 62.92) --
	(226.28, 62.59) --
	(226.47, 62.26) --
	(226.66, 61.94) --
	(226.86, 61.62) --
	(227.05, 61.31) --
	(227.24, 61.00) --
	(227.44, 60.70) --
	(227.63, 60.41) --
	(227.82, 60.12) --
	(228.02, 59.83) --
	(228.21, 59.55) --
	(228.40, 59.28) --
	(228.60, 59.01) --
	(228.79, 58.75) --
	(228.98, 58.49) --
	(229.18, 58.24) --
	(229.37, 57.99) --
	(229.56, 57.74) --
	(229.75, 57.51) --
	(229.95, 57.27) --
	(230.14, 57.04) --
	(230.33, 56.82) --
	(230.53, 56.60) --
	(230.72, 56.38) --
	(230.91, 56.17) --
	(231.11, 55.96) --
	(231.30, 55.76) --
	(231.49, 55.56) --
	(231.69, 55.36) --
	(231.88, 55.17) --
	(232.07, 54.99) --
	(232.27, 54.80) --
	(232.46, 54.62) --
	(232.65, 54.45) --
	(232.85, 54.28) --
	(233.04, 54.11) --
	(233.23, 53.95) --
	(233.42, 53.79) --
	(233.62, 53.63) --
	(233.81, 53.47) --
	(234.00, 53.32) --
	(234.20, 53.18) --
	(234.39, 53.03) --
	(234.58, 52.89) --
	(234.78, 52.76) --
	(234.97, 52.62) --
	(235.16, 52.49) --
	(235.36, 52.36) --
	(235.55, 52.24) --
	(235.74, 52.12) --
	(235.94, 52.00) --
	(236.13, 51.88) --
	(236.32, 51.77) --
	(236.51, 51.65) --
	(236.71, 51.55) --
	(236.90, 51.44) --
	(237.09, 51.34) --
	(237.29, 51.23) --
	(237.48, 51.14) --
	(237.67, 51.04) --
	(237.87, 50.94) --
	(238.06, 50.85) --
	(238.25, 50.76) --
	(238.45, 50.68) --
	(238.64, 50.59) --
	(238.83, 50.51) --
	(239.03, 50.43) --
	(239.22, 50.35) --
	(239.41, 50.27) --
	(239.60, 50.19) --
	(239.80, 50.12) --
	(239.99, 50.05) --
	(240.18, 49.98) --
	(240.38, 49.91) --
	(240.57, 49.84) --
	(240.76, 49.78) --
	(240.96, 49.72) --
	(241.15, 49.65) --
	(241.34, 49.59) --
	(241.54, 49.54) --
	(241.73, 49.48) --
	(241.92, 49.42) --
	(242.12, 49.37) --
	(242.31, 49.32) --
	(242.50, 49.26) --
	(242.70, 49.21) --
	(242.89, 49.17) --
	(243.08, 49.12) --
	(243.27, 49.07) --
	(243.47, 49.03) --
	(243.66, 48.98) --
	(243.85, 48.94) --
	(244.05, 48.90) --
	(244.24, 48.86) --
	(244.43, 48.82) --
	(244.63, 48.78) --
	(244.82, 48.74) --
	(245.01, 48.71) --
	(245.21, 48.67) --
	(245.40, 48.64) --
	(245.59, 48.60) --
	(245.79, 48.57) --
	(245.98, 48.54) --
	(246.17, 48.51) --
	(246.36, 48.48) --
	(246.56, 48.45) --
	(246.75, 48.42) --
	(246.94, 48.39) --
	(247.14, 48.37) --
	(247.33, 48.34) --
	(247.52, 48.32) --
	(247.72, 48.29) --
	(247.91, 48.27) --
	(248.10, 48.24) --
	(248.30, 48.22) --
	(248.49, 48.20);
\definecolor{drawColor}{RGB}{0,0,0}

\path[draw=drawColor,line width= 0.4pt,line join=round,line cap=round] (167.37, 43.61) rectangle (252.35,128.59);
\end{scope}
\begin{scope}
\definecolor{drawColor}{RGB}{0,0,0}

\node[text=drawColor,anchor=base east,inner sep=0pt, outer sep=0pt, scale=  1.00] at (160.87, 44.03) {0.00};

\node[text=drawColor,anchor=base east,inner sep=0pt, outer sep=0pt, scale=  1.00] at (160.87, 66.17) {0.02};

\node[text=drawColor,anchor=base east,inner sep=0pt, outer sep=0pt, scale=  1.00] at (160.87, 88.31) {0.04};

\node[text=drawColor,anchor=base east,inner sep=0pt, outer sep=0pt, scale=  1.00] at (160.87,110.45) {0.06};
\end{scope}
\begin{scope}
\definecolor{drawColor}{RGB}{0,0,0}

\path[draw=drawColor,line width= 0.4pt,line join=round] (163.87, 47.48) --
	(167.37, 47.48);

\path[draw=drawColor,line width= 0.4pt,line join=round] (163.87, 69.61) --
	(167.37, 69.61);

\path[draw=drawColor,line width= 0.4pt,line join=round] (163.87, 91.75) --
	(167.37, 91.75);

\path[draw=drawColor,line width= 0.4pt,line join=round] (163.87,113.89) --
	(167.37,113.89);
\end{scope}
\begin{scope}
\definecolor{drawColor}{RGB}{0,0,0}

\path[draw=drawColor,line width= 0.4pt,line join=round] (171.23, 40.11) --
	(171.23, 43.61);

\path[draw=drawColor,line width= 0.4pt,line join=round] (190.55, 40.11) --
	(190.55, 43.61);

\path[draw=drawColor,line width= 0.4pt,line join=round] (209.86, 40.11) --
	(209.86, 43.61);

\path[draw=drawColor,line width= 0.4pt,line join=round] (229.18, 40.11) --
	(229.18, 43.61);

\path[draw=drawColor,line width= 0.4pt,line join=round] (248.49, 40.11) --
	(248.49, 43.61);
\end{scope}
\begin{scope}
\definecolor{drawColor}{RGB}{0,0,0}

\node[text=drawColor,anchor=base,inner sep=0pt, outer sep=0pt, scale=  1.00] at (171.23, 30.22) {0};

\node[text=drawColor,anchor=base,inner sep=0pt, outer sep=0pt, scale=  1.00] at (190.55, 30.22) {10};

\node[text=drawColor,anchor=base,inner sep=0pt, outer sep=0pt, scale=  1.00] at (209.86, 30.22) {20};

\node[text=drawColor,anchor=base,inner sep=0pt, outer sep=0pt, scale=  1.00] at (229.18, 30.22) {30};

\node[text=drawColor,anchor=base,inner sep=0pt, outer sep=0pt, scale=  1.00] at (248.49, 30.22) {40};
\end{scope}
\begin{scope}
\definecolor{drawColor}{RGB}{0,0,0}

\node[text=drawColor,anchor=base,inner sep=0pt, outer sep=0pt, scale=  1.00] at (209.86, 17.89) {$\tau$};
\end{scope}
\begin{scope}
\definecolor{drawColor}{RGB}{27,158,119}

\path[draw=drawColor,line width= 1.1pt,line join=round] (264.89,114.40) -- (277.21,114.40);
\end{scope}
\begin{scope}
\definecolor{drawColor}{RGB}{217,95,2}

\path[draw=drawColor,line width= 1.1pt,line join=round] (264.89, 99.00) -- (277.21, 99.00);
\end{scope}
\begin{scope}
\definecolor{drawColor}{RGB}{117,112,179}

\path[draw=drawColor,line width= 1.1pt,line join=round] (264.89, 83.60) -- (277.21, 83.60);
\end{scope}
\begin{scope}
\definecolor{drawColor}{RGB}{231,41,138}

\path[draw=drawColor,line width= 1.1pt,line join=round] (264.89, 68.20) -- (277.21, 68.20);
\end{scope}
\begin{scope}
\definecolor{drawColor}{RGB}{102,166,30}

\path[draw=drawColor,line width= 1.1pt,line join=round] (264.89, 52.80) -- (277.21, 52.80);
\end{scope}
\begin{scope}
\definecolor{drawColor}{RGB}{0,0,0}

\node[text=drawColor,anchor=base west,inner sep=0pt, outer sep=0pt, scale=  1.00] at (283.75,110.96) {$\gammadist(0.01, 0.01)$};
\end{scope}
\begin{scope}
\definecolor{drawColor}{RGB}{0,0,0}

\node[text=drawColor,anchor=base west,inner sep=0pt, outer sep=0pt, scale=  1.00] at (283.75, 95.56) {$\normal^+(0, 1)$};
\end{scope}
\begin{scope}
\definecolor{drawColor}{RGB}{0,0,0}

\node[text=drawColor,anchor=base west,inner sep=0pt, outer sep=0pt, scale=  1.00] at (283.75, 80.16) {$\Cauchy^+(0, 1)$};
\end{scope}
\begin{scope}
\definecolor{drawColor}{RGB}{0,0,0}

\node[text=drawColor,anchor=base west,inner sep=0pt, outer sep=0pt, scale=  1.00] at (283.75, 64.76) {$\gammadist(1, 2)$};
\end{scope}
\begin{scope}
\definecolor{drawColor}{RGB}{0,0,0}

\node[text=drawColor,anchor=base west,inner sep=0pt, outer sep=0pt, scale=  1.00] at (283.75, 49.36) {$\gammadist(9, 0.5)$};
\end{scope}
\end{tikzpicture}

%% file: figs/covid_liksubsens.tex
% Created by tikzDevice version 0.12.3.1 on 2022-12-08 15:20:21
% !TEX encoding = UTF-8 Unicode
\begin{tikzpicture}[x=1pt,y=1pt]
\definecolor{fillColor}{RGB}{255,255,255}
\begin{scope}
\definecolor{drawColor}{RGB}{0,0,0}

\path[draw=drawColor,line width= 1.1pt,line join=round] ( 42.12, 81.54) --
	( 42.83,123.73) --
	( 43.54,123.73) --
	( 44.25,123.73) --
	( 44.96,123.73) --
	( 45.67,123.73) --
	( 46.38,123.73) --
	( 47.09,123.73) --
	( 47.80,123.72) --
	( 48.51,123.68) --
	( 49.22,123.60) --
	( 49.93,123.47) --
	( 50.64,123.27) --
	( 51.35,123.01) --
	( 52.06,122.68) --
	( 52.77,122.30) --
	( 53.48,121.85) --
	( 54.19,121.36) --
	( 54.90,120.81) --
	( 55.61,120.23) --
	( 56.32,119.60) --
	( 57.03,118.94) --
	( 57.74,118.25) --
	( 58.45,117.53) --
	( 59.16,116.79) --
	( 59.87,116.02) --
	( 60.58,115.24) --
	( 61.29,114.43) --
	( 62.00,113.60) --
	( 62.71,112.69) --
	( 63.42,111.53) --
	( 64.13,110.27) --
	( 64.84,108.71) --
	( 65.55,106.84) --
	( 66.26,104.79) --
	( 66.97,102.53) --
	( 67.68,100.02) --
	( 68.39, 97.23) --
	( 69.10, 94.21) --
	( 69.81, 91.20) --
	( 70.52, 89.72) --
	( 71.23, 92.96) --
	( 71.94, 96.43) --
	( 72.65, 99.49) --
	( 73.36,102.08) --
	( 74.07,105.79) --
	( 74.78,109.65) --
	( 75.49,112.83) --
	( 76.20,114.63) --
	( 76.91,116.07) --
	( 77.61,117.24) --
	( 78.32,118.23) --
	( 79.03,119.05) --
	( 79.74,119.75) --
	( 80.45,120.33) --
	( 81.16,120.81) --
	( 81.87,121.20) --
	( 82.58,121.48) --
	( 83.29,121.67) --
	( 84.00,121.74) --
	( 84.71,121.49) --
	( 85.42,121.13) --
	( 86.13,120.67) --
	( 86.84,120.11) --
	( 87.55,119.44) --
	( 88.26,118.65) --
	( 88.97,117.76) --
	( 89.68,116.76) --
	( 90.39,115.66) --
	( 91.10,114.46) --
	( 91.81,113.15) --
	( 92.52,111.75) --
	( 93.23,110.23) --
	( 93.94,109.07) --
	( 94.65,108.29) --
	( 95.36,107.16) --
	( 96.07,105.56) --
	( 96.78,102.97) --
	( 97.49, 99.47) --
	( 98.20, 95.19) --
	( 98.91, 90.20) --
	( 99.62, 86.26) --
	(100.33, 82.50) --
	(101.04, 82.12) --
	(101.75, 86.07) --
	(102.46, 89.65) --
	(103.17, 92.22) --
	(103.88, 94.24) --
	(104.59, 95.84) --
	(105.30, 97.13) --
	(106.01, 98.48) --
	(106.72,100.45) --
	(107.43,102.26) --
	(108.14,103.91) --
	(108.85,105.44) --
	(109.56,106.86) --
	(110.27,108.19) --
	(110.98,109.43) --
	(111.69,110.61) --
	(112.40,111.71);

\path[draw=drawColor,line width= 0.6pt,dash pattern=on 1pt off 3pt ,line join=round] ( 38.61, 83.41) -- (115.91, 83.41);
\definecolor{drawColor}{RGB}{0,0,0}

\path[draw=drawColor,draw opacity=0.50,line width= 0.6pt,line join=round] ( 69.10, 75.67) -- ( 69.10,152.97);
\definecolor{drawColor}{RGB}{0,0,0}

\path[draw=drawColor,line width= 0.6pt,line join=round,line cap=round] ( 38.61, 75.67) rectangle (115.91,152.97);
\end{scope}
\begin{scope}
\definecolor{drawColor}{RGB}{0,0,0}

\path[draw=drawColor,line width= 1.1pt,line join=round] (126.42, 79.19) --
	(127.13,141.23) --
	(127.84,141.23) --
	(128.55,141.23) --
	(129.26,141.23) --
	(129.97,141.23) --
	(130.68,141.23) --
	(131.39,141.22) --
	(132.10,141.20) --
	(132.81,141.12) --
	(133.52,140.95) --
	(134.23,140.65) --
	(134.94,140.21) --
	(135.65,139.64) --
	(136.36,138.92) --
	(137.07,138.08) --
	(137.78,137.12) --
	(138.49,136.05) --
	(139.20,134.89) --
	(139.91,133.65) --
	(140.62,132.35) --
	(141.33,130.98) --
	(142.04,129.57) --
	(142.75,128.02) --
	(143.46,126.33) --
	(144.17,124.54) --
	(144.88,122.63) --
	(145.59,120.56) --
	(146.30,118.27) --
	(147.01,115.57) --
	(147.72,111.88) --
	(148.43,107.38) --
	(149.14,101.75) --
	(149.85, 94.72) --
	(150.56, 86.53) --
	(151.27, 82.60) --
	(151.98, 90.69) --
	(152.69, 99.94) --
	(153.40,108.16) --
	(154.11,114.42) --
	(154.82,119.65) --
	(155.52,124.12) --
	(156.23,128.06) --
	(156.94,131.55) --
	(157.65,134.71) --
	(158.36,137.59) --
	(159.07,140.21) --
	(159.78,142.59) --
	(160.49,144.73) --
	(161.20,146.22) --
	(161.91,147.48) --
	(162.62,148.46) --
	(163.33,149.14) --
	(164.04,149.46) --
	(164.75,149.41) --
	(165.46,148.98) --
	(166.17,148.18) --
	(166.88,147.02) --
	(167.59,145.54) --
	(168.30,143.77) --
	(169.01,141.74) --
	(169.72,139.49) --
	(170.43,137.06) --
	(171.14,134.47) --
	(171.85,131.73) --
	(172.56,128.87) --
	(173.27,125.89) --
	(173.98,122.82) --
	(174.69,119.65) --
	(175.40,117.90) --
	(176.11,119.93) --
	(176.82,120.40) --
	(177.53,120.30) --
	(178.24,118.53) --
	(178.95,113.63) --
	(179.66,106.95) --
	(180.37, 98.16) --
	(181.08, 87.29) --
	(181.79, 83.38) --
	(182.50, 93.92) --
	(183.21,102.15) --
	(183.92,107.38) --
	(184.63,111.03) --
	(185.34,113.72) --
	(186.05,114.84) --
	(186.76,114.93) --
	(187.47,114.84) --
	(188.18,114.66) --
	(188.89,114.38) --
	(189.60,116.50) --
	(190.31,118.49) --
	(191.02,120.36) --
	(191.73,122.11) --
	(192.44,123.78) --
	(193.14,125.36) --
	(193.85,126.88) --
	(194.56,128.33) --
	(195.27,129.71) --
	(195.98,131.03) --
	(196.69,132.30);

\path[draw=drawColor,line width= 0.6pt,dash pattern=on 1pt off 3pt ,line join=round] (122.91, 83.41) -- (200.21, 83.41);
\definecolor{drawColor}{RGB}{0,0,0}

\path[draw=drawColor,draw opacity=0.50,line width= 0.6pt,line join=round] (140.62, 75.67) -- (140.62,152.97);
\definecolor{drawColor}{RGB}{0,0,0}

\path[draw=drawColor,line width= 0.6pt,line join=round,line cap=round] (122.91, 75.67) rectangle (200.21,152.97);
\end{scope}
\begin{scope}
\definecolor{drawColor}{RGB}{0,0,0}

\path[draw=drawColor,line width= 1.1pt,line join=round] (210.72, 79.61) --
	(211.43,108.54) --
	(212.14,108.54) --
	(212.85,108.54) --
	(213.56,108.54) --
	(214.27,108.54) --
	(214.98,108.54) --
	(215.69,108.54) --
	(216.40,108.53) --
	(217.11,108.50) --
	(217.82,108.45) --
	(218.53,108.35) --
	(219.24,108.22) --
	(219.95,108.04) --
	(220.66,107.82) --
	(221.37,107.56) --
	(222.08,107.26) --
	(222.79,106.94) --
	(223.50,106.59) --
	(224.21,106.22) --
	(224.92,105.83) --
	(225.63,105.43) --
	(226.34,105.02) --
	(227.05,104.59) --
	(227.76,104.06) --
	(228.47,103.52) --
	(229.18,102.97) --
	(229.89,102.38) --
	(230.60,101.79) --
	(231.31,101.17) --
	(232.02,100.33) --
	(232.73, 99.39) --
	(233.43, 98.40) --
	(234.14, 97.32) --
	(234.85, 96.09) --
	(235.56, 94.68) --
	(236.27, 92.99) --
	(236.98, 90.90) --
	(237.69, 88.28) --
	(238.40, 86.14) --
	(239.11, 86.65) --
	(239.82, 87.09) --
	(240.53, 87.47) --
	(241.24, 87.82) --
	(241.95, 88.15) --
	(242.66, 88.43) --
	(243.37, 88.68) --
	(244.08, 89.14) --
	(244.79, 89.67) --
	(245.50, 90.14) --
	(246.21, 90.54) --
	(246.92, 90.89) --
	(247.63, 91.20) --
	(248.34, 91.46) --
	(249.05, 91.67) --
	(249.76, 91.84) --
	(250.47, 91.94) --
	(251.18, 92.01) --
	(251.89, 92.04) --
	(252.60, 92.02) --
	(253.31, 91.95) --
	(254.02, 91.83) --
	(254.73, 92.14) --
	(255.44, 92.58) --
	(256.15, 93.05) --
	(256.86, 93.54) --
	(257.57, 94.03) --
	(258.28, 94.51) --
	(258.99, 94.95) --
	(259.70, 95.32) --
	(260.41, 95.52) --
	(261.12, 95.05) --
	(261.83, 94.19) --
	(262.54, 92.55) --
	(263.25, 89.45) --
	(263.96, 86.15) --
	(264.67, 83.02) --
	(265.38, 80.92) --
	(266.09, 79.69) --
	(266.80, 79.94) --
	(267.51, 81.25) --
	(268.22, 82.51) --
	(268.93, 84.06) --
	(269.64, 85.54) --
	(270.35, 86.92) --
	(271.05, 88.20) --
	(271.76, 89.36) --
	(272.47, 90.46) --
	(273.18, 91.51) --
	(273.89, 92.49) --
	(274.60, 93.40) --
	(275.31, 94.26) --
	(276.02, 95.07) --
	(276.73, 95.84) --
	(277.44, 96.56) --
	(278.15, 97.25) --
	(278.86, 97.89) --
	(279.57, 98.49) --
	(280.28, 99.07) --
	(280.99, 99.62);

\path[draw=drawColor,line width= 0.6pt,dash pattern=on 1pt off 3pt ,line join=round] (207.21, 83.41) -- (284.51, 83.41);
\definecolor{drawColor}{RGB}{0,0,0}

\path[draw=drawColor,draw opacity=0.50,line width= 0.6pt,line join=round] (230.60, 75.67) -- (230.60,152.97);
\definecolor{drawColor}{RGB}{0,0,0}

\path[draw=drawColor,line width= 0.6pt,line join=round,line cap=round] (207.21, 75.67) rectangle (284.51,152.97);
\end{scope}
\begin{scope}
\definecolor{drawColor}{RGB}{0,0,0}

\path[draw=drawColor,line width= 1.1pt,line join=round] (295.02, 81.00) --
	(295.73,105.34) --
	(296.44,105.34) --
	(297.15,105.34) --
	(297.86,105.34) --
	(298.57,105.34) --
	(299.28,105.34) --
	(299.99,105.34) --
	(300.70,105.33) --
	(301.41,105.30) --
	(302.12,105.24) --
	(302.83,105.14) --
	(303.54,105.00) --
	(304.25,104.80) --
	(304.96,104.56) --
	(305.67,104.28) --
	(306.38,103.96) --
	(307.09,103.61) --
	(307.80,103.23) --
	(308.51,102.83) --
	(309.22,102.41) --
	(309.93,101.98) --
	(310.64,101.54) --
	(311.34,101.10) --
	(312.05,100.63) --
	(312.76,100.06) --
	(313.47, 99.47) --
	(314.18, 98.87) --
	(314.89, 98.25) --
	(315.60, 97.61) --
	(316.31, 96.93) --
	(317.02, 96.20) --
	(317.73, 95.50) --
	(318.44, 94.67) --
	(319.15, 93.85) --
	(319.86, 92.84) --
	(320.57, 91.32) --
	(321.28, 89.45) --
	(321.99, 87.88) --
	(322.70, 86.44) --
	(323.41, 85.02) --
	(324.12, 84.25) --
	(324.83, 83.95) --
	(325.54, 84.03) --
	(326.25, 84.33) --
	(326.96, 84.72) --
	(327.67, 85.27) --
	(328.38, 85.83) --
	(329.09, 86.26) --
	(329.80, 86.70) --
	(330.51, 87.21) --
	(331.22, 87.68) --
	(331.93, 88.11) --
	(332.64, 88.50) --
	(333.35, 88.84) --
	(334.06, 89.12) --
	(334.77, 89.33) --
	(335.48, 89.49) --
	(336.19, 89.59) --
	(336.90, 89.63) --
	(337.61, 89.64) --
	(338.32, 89.59) --
	(339.03, 89.51) --
	(339.74, 89.37) --
	(340.45, 89.18) --
	(341.16, 89.06) --
	(341.87, 89.69) --
	(342.58, 90.37) --
	(343.29, 91.11) --
	(344.00, 91.93) --
	(344.71, 92.84) --
	(345.42, 93.79) --
	(346.13, 94.02) --
	(346.84, 94.14) --
	(347.55, 94.14) --
	(348.26, 93.90) --
	(348.96, 92.34) --
	(349.67, 90.03) --
	(350.38, 87.05) --
	(351.09, 83.69) --
	(351.80, 80.72) --
	(352.51, 80.66) --
	(353.22, 81.71) --
	(353.93, 82.93) --
	(354.64, 84.17) --
	(355.35, 85.36) --
	(356.06, 86.49) --
	(356.77, 87.59) --
	(357.48, 88.65) --
	(358.19, 89.69) --
	(358.90, 90.70) --
	(359.61, 91.76) --
	(360.32, 92.77) --
	(361.03, 93.74) --
	(361.74, 94.67) --
	(362.45, 95.61) --
	(363.16, 96.55) --
	(363.87, 97.45) --
	(364.58, 98.35) --
	(365.29, 99.23);

\path[draw=drawColor,line width= 0.6pt,dash pattern=on 1pt off 3pt ,line join=round] (291.51, 83.41) -- (368.80, 83.41);
\definecolor{drawColor}{RGB}{0,0,0}

\path[draw=drawColor,draw opacity=0.50,line width= 0.6pt,line join=round] (309.93, 75.67) -- (309.93,152.97);
\definecolor{drawColor}{RGB}{0,0,0}

\path[draw=drawColor,line width= 0.6pt,line join=round,line cap=round] (291.51, 75.67) rectangle (368.80,152.97);
\end{scope}
\begin{scope}
\definecolor{drawColor}{RGB}{0,0,0}

\node[text=drawColor,anchor=base,inner sep=0pt, outer sep=0pt, scale=  1.00] at ( 77.26,157.44) {Italy};
\end{scope}
\begin{scope}
\definecolor{drawColor}{RGB}{0,0,0}

\node[text=drawColor,anchor=base,inner sep=0pt, outer sep=0pt, scale=  1.00] at (161.56,157.44) {Germany};
\end{scope}
\begin{scope}
\definecolor{drawColor}{RGB}{0,0,0}

\node[text=drawColor,anchor=base,inner sep=0pt, outer sep=0pt, scale=  1.00] at (245.86,157.44) {United Kingdom};
\end{scope}
\begin{scope}
\definecolor{drawColor}{RGB}{0,0,0}

\node[text=drawColor,anchor=base,inner sep=0pt, outer sep=0pt, scale=  1.00] at (330.15,157.44) {Sweden};
\end{scope}
\begin{scope}
\definecolor{drawColor}{RGB}{0,0,0}

\path[draw=drawColor,line width= 0.6pt,line join=round] ( 41.41, 72.17) --
	( 41.41, 75.67);

\path[draw=drawColor,line width= 0.6pt,line join=round] ( 59.16, 72.17) --
	( 59.16, 75.67);

\path[draw=drawColor,line width= 0.6pt,line join=round] ( 76.91, 72.17) --
	( 76.91, 75.67);

\path[draw=drawColor,line width= 0.6pt,line join=round] ( 94.65, 72.17) --
	( 94.65, 75.67);

\path[draw=drawColor,line width= 0.6pt,line join=round] (112.40, 72.17) --
	(112.40, 75.67);
\end{scope}
\begin{scope}
\definecolor{drawColor}{RGB}{0,0,0}

\node[text=drawColor,anchor=base,inner sep=0pt, outer sep=0pt, scale=  1.00] at ( 41.41, 62.28) {0};

\node[text=drawColor,anchor=base,inner sep=0pt, outer sep=0pt, scale=  1.00] at ( 59.16, 62.28) {25};

\node[text=drawColor,anchor=base,inner sep=0pt, outer sep=0pt, scale=  1.00] at ( 76.91, 62.28) {50};

\node[text=drawColor,anchor=base,inner sep=0pt, outer sep=0pt, scale=  1.00] at ( 94.65, 62.28) {75};

\node[text=drawColor,anchor=base,inner sep=0pt, outer sep=0pt, scale=  1.00] at (112.40, 62.28) {100};
\end{scope}
\begin{scope}
\definecolor{drawColor}{RGB}{0,0,0}

\path[draw=drawColor,line width= 0.6pt,line join=round] (125.71, 72.17) --
	(125.71, 75.67);

\path[draw=drawColor,line width= 0.6pt,line join=round] (143.46, 72.17) --
	(143.46, 75.67);

\path[draw=drawColor,line width= 0.6pt,line join=round] (161.20, 72.17) --
	(161.20, 75.67);

\path[draw=drawColor,line width= 0.6pt,line join=round] (178.95, 72.17) --
	(178.95, 75.67);

\path[draw=drawColor,line width= 0.6pt,line join=round] (196.69, 72.17) --
	(196.69, 75.67);
\end{scope}
\begin{scope}
\definecolor{drawColor}{RGB}{0,0,0}

\node[text=drawColor,anchor=base,inner sep=0pt, outer sep=0pt, scale=  1.00] at (125.71, 62.28) {0};

\node[text=drawColor,anchor=base,inner sep=0pt, outer sep=0pt, scale=  1.00] at (143.46, 62.28) {25};

\node[text=drawColor,anchor=base,inner sep=0pt, outer sep=0pt, scale=  1.00] at (161.20, 62.28) {50};

\node[text=drawColor,anchor=base,inner sep=0pt, outer sep=0pt, scale=  1.00] at (178.95, 62.28) {75};

\node[text=drawColor,anchor=base,inner sep=0pt, outer sep=0pt, scale=  1.00] at (196.69, 62.28) {100};
\end{scope}
\begin{scope}
\definecolor{drawColor}{RGB}{0,0,0}

\path[draw=drawColor,line width= 0.6pt,line join=round] (210.01, 72.17) --
	(210.01, 75.67);

\path[draw=drawColor,line width= 0.6pt,line join=round] (227.76, 72.17) --
	(227.76, 75.67);

\path[draw=drawColor,line width= 0.6pt,line join=round] (245.50, 72.17) --
	(245.50, 75.67);

\path[draw=drawColor,line width= 0.6pt,line join=round] (263.25, 72.17) --
	(263.25, 75.67);

\path[draw=drawColor,line width= 0.6pt,line join=round] (280.99, 72.17) --
	(280.99, 75.67);
\end{scope}
\begin{scope}
\definecolor{drawColor}{RGB}{0,0,0}

\node[text=drawColor,anchor=base,inner sep=0pt, outer sep=0pt, scale=  1.00] at (210.01, 62.28) {0};

\node[text=drawColor,anchor=base,inner sep=0pt, outer sep=0pt, scale=  1.00] at (227.76, 62.28) {25};

\node[text=drawColor,anchor=base,inner sep=0pt, outer sep=0pt, scale=  1.00] at (245.50, 62.28) {50};

\node[text=drawColor,anchor=base,inner sep=0pt, outer sep=0pt, scale=  1.00] at (263.25, 62.28) {75};

\node[text=drawColor,anchor=base,inner sep=0pt, outer sep=0pt, scale=  1.00] at (280.99, 62.28) {100};
\end{scope}
\begin{scope}
\definecolor{drawColor}{RGB}{0,0,0}

\path[draw=drawColor,line width= 0.6pt,line join=round] (294.31, 72.17) --
	(294.31, 75.67);

\path[draw=drawColor,line width= 0.6pt,line join=round] (312.05, 72.17) --
	(312.05, 75.67);

\path[draw=drawColor,line width= 0.6pt,line join=round] (329.80, 72.17) --
	(329.80, 75.67);

\path[draw=drawColor,line width= 0.6pt,line join=round] (347.55, 72.17) --
	(347.55, 75.67);

\path[draw=drawColor,line width= 0.6pt,line join=round] (365.29, 72.17) --
	(365.29, 75.67);
\end{scope}
\begin{scope}
\definecolor{drawColor}{RGB}{0,0,0}

\node[text=drawColor,anchor=base,inner sep=0pt, outer sep=0pt, scale=  1.00] at (294.31, 62.28) {0};

\node[text=drawColor,anchor=base,inner sep=0pt, outer sep=0pt, scale=  1.00] at (312.05, 62.28) {25};

\node[text=drawColor,anchor=base,inner sep=0pt, outer sep=0pt, scale=  1.00] at (329.80, 62.28) {50};

\node[text=drawColor,anchor=base,inner sep=0pt, outer sep=0pt, scale=  1.00] at (347.55, 62.28) {75};

\node[text=drawColor,anchor=base,inner sep=0pt, outer sep=0pt, scale=  1.00] at (365.29, 62.28) {100};
\end{scope}
\begin{scope}
\definecolor{drawColor}{RGB}{0,0,0}

\node[text=drawColor,anchor=base east,inner sep=0pt, outer sep=0pt, scale=  1.00] at ( 32.11, 74.15) {0.0};

\node[text=drawColor,anchor=base east,inner sep=0pt, outer sep=0pt, scale=  1.00] at ( 32.11, 97.41) {0.2};

\node[text=drawColor,anchor=base east,inner sep=0pt, outer sep=0pt, scale=  1.00] at ( 32.11,120.68) {0.4};

\node[text=drawColor,anchor=base east,inner sep=0pt, outer sep=0pt, scale=  1.00] at ( 32.11,143.94) {0.6};
\end{scope}
\begin{scope}
\definecolor{drawColor}{RGB}{0,0,0}

\path[draw=drawColor,line width= 0.6pt,line join=round] ( 35.11, 77.60) --
	( 38.61, 77.60);

\path[draw=drawColor,line width= 0.6pt,line join=round] ( 35.11,100.86) --
	( 38.61,100.86);

\path[draw=drawColor,line width= 0.6pt,line join=round] ( 35.11,124.12) --
	( 38.61,124.12);

\path[draw=drawColor,line width= 0.6pt,line join=round] ( 35.11,147.38) --
	( 38.61,147.38);
\end{scope}
\begin{scope}
\definecolor{drawColor}{RGB}{0,0,0}

\node[text=drawColor,anchor=base,inner sep=0pt, outer sep=0pt, scale=  1.00] at (203.71, 49.95) {Day};
\end{scope}
\begin{scope}
\definecolor{drawColor}{RGB}{0,0,0}

\node[text=drawColor,rotate= 90.00,anchor=base,inner sep=0pt, outer sep=0pt, scale=  1.00] at ( 13.89,114.32) {Likelihood sensitivity ($D_{\text{CJS}})$};
\end{scope}
\end{tikzpicture}

%% file: figs/covid_priorsubsens.tex
% Created by tikzDevice version 0.12.3.1 on 2022-12-08 18:43:51
% !TEX encoding = UTF-8 Unicode
\begin{tikzpicture}[x=1pt,y=1pt]
\definecolor{fillColor}{RGB}{255,255,255}
\begin{scope}
\definecolor{drawColor}{RGB}{127,201,127}

\path[draw=drawColor,line width= 1.1pt,line join=round] ( 47.07, 80.58) --
	( 47.77,100.84) --
	( 48.46,100.84) --
	( 49.16,100.84) --
	( 49.86,100.84) --
	( 50.56,100.84) --
	( 51.26,100.84) --
	( 51.96,100.84) --
	( 52.65,100.84) --
	( 53.35,100.82) --
	( 54.05,100.79) --
	( 54.75,100.73) --
	( 55.45,100.65) --
	( 56.15,100.54) --
	( 56.84,100.40) --
	( 57.54,100.23) --
	( 58.24,100.04) --
	( 58.94, 99.83) --
	( 59.64, 99.59) --
	( 60.34, 99.33) --
	( 61.03, 99.04) --
	( 61.73, 98.74) --
	( 62.43, 98.42) --
	( 63.13, 98.07) --
	( 63.83, 97.71) --
	( 64.53, 97.33) --
	( 65.22, 96.92) --
	( 65.92, 96.49) --
	( 66.62, 96.02) --
	( 67.32, 95.50) --
	( 68.02, 94.82) --
	( 68.72, 94.06) --
	( 69.41, 93.12) --
	( 70.11, 91.97) --
	( 70.81, 90.67) --
	( 71.51, 89.16) --
	( 72.21, 87.40) --
	( 72.91, 85.31) --
	( 73.60, 82.88) --
	( 74.30, 81.27) --
	( 75.00, 83.53) --
	( 75.70, 87.64) --
	( 76.40, 92.02) --
	( 77.10, 96.13) --
	( 77.79, 99.92) --
	( 78.49,103.27) --
	( 79.19,106.20) --
	( 79.89,108.67) --
	( 80.59,110.04) --
	( 81.29,111.18) --
	( 81.98,112.14) --
	( 82.68,112.96) --
	( 83.38,113.68) --
	( 84.08,114.30) --
	( 84.78,114.82) --
	( 85.48,115.27) --
	( 86.17,115.62) --
	( 86.87,115.88) --
	( 87.57,116.03) --
	( 88.27,116.07) --
	( 88.97,115.80) --
	( 89.67,115.41) --
	( 90.36,114.91) --
	( 91.06,114.29) --
	( 91.76,113.55) --
	( 92.46,112.69) --
	( 93.16,111.72) --
	( 93.86,110.64) --
	( 94.55,109.46) --
	( 95.25,108.18) --
	( 95.95,106.81) --
	( 96.65,105.33) --
	( 97.35,103.75) --
	( 98.05,102.08) --
	( 98.74,100.30) --
	( 99.44, 98.38) --
	(100.14, 96.29) --
	(100.84, 93.80) --
	(101.54, 91.09) --
	(102.24, 88.31) --
	(102.93, 85.49) --
	(103.63, 82.80) --
	(104.33, 80.74) --
	(105.03, 83.30) --
	(105.73, 86.11) --
	(106.43, 88.56) --
	(107.12, 90.31) --
	(107.82, 91.70) --
	(108.52, 92.81) --
	(109.22, 93.70) --
	(109.92, 94.42) --
	(110.62, 94.79) --
	(111.31, 95.01) --
	(112.01, 95.17) --
	(112.71, 95.38) --
	(113.41, 95.97) --
	(114.11, 96.51) --
	(114.81, 97.02) --
	(115.50, 97.49) --
	(116.20, 97.93);
\definecolor{drawColor}{RGB}{190,174,212}

\path[draw=drawColor,line width= 1.1pt,line join=round] ( 47.07, 80.52) --
	( 47.77, 83.38) --
	( 48.46, 83.38) --
	( 49.16, 83.38) --
	( 49.86, 83.38) --
	( 50.56, 83.38) --
	( 51.26, 83.38) --
	( 51.96, 83.38) --
	( 52.65, 83.38) --
	( 53.35, 83.38) --
	( 54.05, 83.37) --
	( 54.75, 83.35) --
	( 55.45, 83.33) --
	( 56.15, 83.31) --
	( 56.84, 83.27) --
	( 57.54, 83.23) --
	( 58.24, 83.18) --
	( 58.94, 83.13) --
	( 59.64, 83.07) --
	( 60.34, 83.00) --
	( 61.03, 82.94) --
	( 61.73, 82.86) --
	( 62.43, 82.78) --
	( 63.13, 82.70) --
	( 63.83, 82.61) --
	( 64.53, 82.53) --
	( 65.22, 82.43) --
	( 65.92, 82.33) --
	( 66.62, 82.23) --
	( 67.32, 82.11) --
	( 68.02, 81.97) --
	( 68.72, 81.82) --
	( 69.41, 81.65) --
	( 70.11, 81.44) --
	( 70.81, 81.22) --
	( 71.51, 80.99) --
	( 72.21, 80.75) --
	( 72.91, 80.54) --
	( 73.60, 80.59) --
	( 74.30, 80.82) --
	( 75.00, 81.09) --
	( 75.70, 81.40) --
	( 76.40, 81.73) --
	( 77.10, 82.20) --
	( 77.79, 82.61) --
	( 78.49, 82.98) --
	( 79.19, 83.31) --
	( 79.89, 83.56) --
	( 80.59, 83.64) --
	( 81.29, 83.70) --
	( 81.98, 83.75) --
	( 82.68, 83.78) --
	( 83.38, 83.81) --
	( 84.08, 83.83) --
	( 84.78, 83.84) --
	( 85.48, 83.85) --
	( 86.17, 83.85) --
	( 86.87, 83.85) --
	( 87.57, 83.84) --
	( 88.27, 83.82) --
	( 88.97, 83.76) --
	( 89.67, 83.71) --
	( 90.36, 83.64) --
	( 91.06, 83.57) --
	( 91.76, 83.49) --
	( 92.46, 83.39) --
	( 93.16, 83.29) --
	( 93.86, 83.17) --
	( 94.55, 83.04) --
	( 95.25, 82.90) --
	( 95.95, 82.75) --
	( 96.65, 82.59) --
	( 97.35, 82.46) --
	( 98.05, 82.46) --
	( 98.74, 82.38) --
	( 99.44, 82.27) --
	(100.14, 82.12) --
	(100.84, 81.87) --
	(101.54, 81.55) --
	(102.24, 81.13) --
	(102.93, 80.65) --
	(103.63, 80.13) --
	(104.33, 79.77) --
	(105.03, 79.89) --
	(105.73, 80.26) --
	(106.43, 80.59) --
	(107.12, 80.83) --
	(107.82, 81.01) --
	(108.52, 81.18) --
	(109.22, 81.39) --
	(109.92, 81.60) --
	(110.62, 81.80) --
	(111.31, 81.98) --
	(112.01, 82.15) --
	(112.71, 82.29) --
	(113.41, 82.42) --
	(114.11, 82.55) --
	(114.81, 82.67) --
	(115.50, 82.78) --
	(116.20, 82.89);
\definecolor{drawColor}{RGB}{253,192,134}

\path[draw=drawColor,line width= 1.1pt,line join=round] ( 47.07, 80.28) --
	( 47.77, 91.21) --
	( 48.46, 91.21) --
	( 49.16, 91.21) --
	( 49.86, 91.21) --
	( 50.56, 91.21) --
	( 51.26, 91.21) --
	( 51.96, 91.21) --
	( 52.65, 91.21) --
	( 53.35, 91.20) --
	( 54.05, 91.18) --
	( 54.75, 91.15) --
	( 55.45, 91.10) --
	( 56.15, 91.04) --
	( 56.84, 90.97) --
	( 57.54, 90.87) --
	( 58.24, 90.77) --
	( 58.94, 90.65) --
	( 59.64, 90.53) --
	( 60.34, 90.39) --
	( 61.03, 90.24) --
	( 61.73, 90.09) --
	( 62.43, 89.92) --
	( 63.13, 89.75) --
	( 63.83, 89.56) --
	( 64.53, 89.38) --
	( 65.22, 89.18) --
	( 65.92, 88.98) --
	( 66.62, 88.77) --
	( 67.32, 88.53) --
	( 68.02, 88.22) --
	( 68.72, 87.87) --
	( 69.41, 87.44) --
	( 70.11, 86.90) --
	( 70.81, 86.28) --
	( 71.51, 85.57) --
	( 72.21, 84.73) --
	( 72.91, 83.71) --
	( 73.60, 82.48) --
	( 74.30, 81.00) --
	( 75.00, 80.37) --
	( 75.70, 82.07) --
	( 76.40, 84.20) --
	( 77.10, 86.18) --
	( 77.79, 87.96) --
	( 78.49, 89.51) --
	( 79.19, 90.83) --
	( 79.89, 91.93) --
	( 80.59, 92.54) --
	( 81.29, 93.05) --
	( 81.98, 93.48) --
	( 82.68, 93.85) --
	( 83.38, 94.17) --
	( 84.08, 94.44) --
	( 84.78, 94.68) --
	( 85.48, 94.87) --
	( 86.17, 95.02) --
	( 86.87, 95.14) --
	( 87.57, 95.20) --
	( 88.27, 95.22) --
	( 88.97, 95.10) --
	( 89.67, 94.93) --
	( 90.36, 94.71) --
	( 91.06, 94.44) --
	( 91.76, 94.12) --
	( 92.46, 93.74) --
	( 93.16, 93.31) --
	( 93.86, 92.83) --
	( 94.55, 92.30) --
	( 95.25, 91.73) --
	( 95.95, 91.11) --
	( 96.65, 90.46) --
	( 97.35, 89.76) --
	( 98.05, 89.02) --
	( 98.74, 88.23) --
	( 99.44, 87.40) --
	(100.14, 86.67) --
	(100.84, 86.05) --
	(101.54, 85.19) --
	(102.24, 84.12) --
	(102.93, 82.77) --
	(103.63, 81.26) --
	(104.33, 80.20) --
	(105.03, 81.02) --
	(105.73, 82.34) --
	(106.43, 83.53) --
	(107.12, 84.40) --
	(107.82, 85.07) --
	(108.52, 85.60) --
	(109.22, 86.01) --
	(109.92, 86.35) --
	(110.62, 86.52) --
	(111.31, 86.62) --
	(112.01, 86.69) --
	(112.71, 86.73) --
	(113.41, 86.89) --
	(114.11, 87.16) --
	(114.81, 87.42) --
	(115.50, 87.67) --
	(116.20, 87.90);
\definecolor{drawColor}{RGB}{0,0,0}

\path[draw=drawColor,line width= 0.6pt,dash pattern=on 1pt off 3pt ,line join=round] ( 43.61,100.86) -- (119.66,100.86);
\definecolor{drawColor}{RGB}{0,0,0}

\path[draw=drawColor,draw opacity=0.50,line width= 0.6pt,line join=round] ( 73.60, 76.30) -- ( 73.60,152.35);
\definecolor{drawColor}{RGB}{0,0,0}

\path[draw=drawColor,line width= 0.6pt,line join=round,line cap=round] ( 43.61, 76.30) rectangle (119.66,152.35);
\end{scope}
\begin{scope}
\definecolor{drawColor}{RGB}{127,201,127}

\path[draw=drawColor,line width= 1.1pt,line join=round] (130.12, 80.66) --
	(130.81,107.14) --
	(131.51,107.14) --
	(132.21,107.14) --
	(132.91,107.14) --
	(133.61,107.14) --
	(134.31,107.14) --
	(135.00,107.14) --
	(135.70,107.13) --
	(136.40,107.09) --
	(137.10,107.01) --
	(137.80,106.88) --
	(138.50,106.69) --
	(139.19,106.43) --
	(139.89,106.10) --
	(140.59,105.71) --
	(141.29,105.25) --
	(141.99,104.74) --
	(142.69,104.16) --
	(143.38,103.53) --
	(144.08,102.84) --
	(144.78,102.10) --
	(145.48,101.29) --
	(146.18,100.38) --
	(146.88, 99.34) --
	(147.57, 98.18) --
	(148.27, 96.87) --
	(148.97, 95.36) --
	(149.67, 93.58) --
	(150.37, 91.40) --
	(151.07, 88.63) --
	(151.76, 85.68) --
	(152.46, 83.58) --
	(153.16, 87.09) --
	(153.86, 93.39) --
	(154.56,100.25) --
	(155.26,107.25) --
	(155.95,113.97) --
	(156.65,119.63) --
	(157.35,123.58) --
	(158.05,127.03) --
	(158.75,130.09) --
	(159.45,132.87) --
	(160.14,135.46) --
	(160.84,137.88) --
	(161.54,140.14) --
	(162.24,142.26) --
	(162.94,144.22) --
	(163.64,145.98) --
	(164.33,147.12) --
	(165.03,148.02) --
	(165.73,148.63) --
	(166.43,148.89) --
	(167.13,148.75) --
	(167.83,148.18) --
	(168.52,147.18) --
	(169.22,145.77) --
	(169.92,143.98) --
	(170.62,141.86) --
	(171.32,139.45) --
	(172.02,136.81) --
	(172.71,133.97) --
	(173.41,130.99) --
	(174.11,127.89) --
	(174.81,124.71) --
	(175.51,121.46) --
	(176.21,118.15) --
	(176.90,114.79) --
	(177.60,111.36) --
	(178.30,107.86) --
	(179.00,104.25) --
	(179.70,102.00) --
	(180.40,100.95) --
	(181.09, 98.82) --
	(181.79, 94.97) --
	(182.49, 90.07) --
	(183.19, 83.97) --
	(183.89, 83.26) --
	(184.59, 91.44) --
	(185.28, 98.70) --
	(185.98,104.14) --
	(186.68,107.31) --
	(187.38,109.40) --
	(188.08,110.89) --
	(188.78,111.18) --
	(189.47,110.73) --
	(190.17,110.24) --
	(190.87,109.75) --
	(191.57,108.93) --
	(192.27,107.83) --
	(192.97,106.57) --
	(193.66,105.46) --
	(194.36,105.64) --
	(195.06,106.22) --
	(195.76,106.77) --
	(196.46,107.28) --
	(197.16,107.77) --
	(197.85,108.23) --
	(198.55,108.67) --
	(199.25,109.08);
\definecolor{drawColor}{RGB}{190,174,212}

\path[draw=drawColor,line width= 1.1pt,line join=round] (130.12, 79.75) --
	(130.81, 85.96) --
	(131.51, 85.96) --
	(132.21, 85.96) --
	(132.91, 85.96) --
	(133.61, 85.96) --
	(134.31, 85.96) --
	(135.00, 85.96) --
	(135.70, 85.96) --
	(136.40, 85.95) --
	(137.10, 85.93) --
	(137.80, 85.90) --
	(138.50, 85.85) --
	(139.19, 85.79) --
	(139.89, 85.72) --
	(140.59, 85.63) --
	(141.29, 85.52) --
	(141.99, 85.41) --
	(142.69, 85.28) --
	(143.38, 85.15) --
	(144.08, 85.00) --
	(144.78, 84.85) --
	(145.48, 84.69) --
	(146.18, 84.52) --
	(146.88, 84.33) --
	(147.57, 84.12) --
	(148.27, 83.90) --
	(148.97, 83.66) --
	(149.67, 83.39) --
	(150.37, 83.07) --
	(151.07, 82.64) --
	(151.76, 82.14) --
	(152.46, 81.49) --
	(153.16, 80.73) --
	(153.86, 80.21) --
	(154.56, 80.53) --
	(155.26, 81.45) --
	(155.95, 82.51) --
	(156.65, 83.50) --
	(157.35, 84.26) --
	(158.05, 84.89) --
	(158.75, 85.44) --
	(159.45, 85.95) --
	(160.14, 86.40) --
	(160.84, 86.84) --
	(161.54, 87.25) --
	(162.24, 87.63) --
	(162.94, 87.98) --
	(163.64, 88.30) --
	(164.33, 88.52) --
	(165.03, 88.70) --
	(165.73, 88.83) --
	(166.43, 88.91) --
	(167.13, 88.93) --
	(167.83, 88.89) --
	(168.52, 88.79) --
	(169.22, 88.63) --
	(169.92, 88.43) --
	(170.62, 88.19) --
	(171.32, 87.91) --
	(172.02, 87.61) --
	(172.71, 87.28) --
	(173.41, 86.94) --
	(174.11, 86.58) --
	(174.81, 86.20) --
	(175.51, 85.80) --
	(176.21, 85.40) --
	(176.90, 84.99) --
	(177.60, 84.58) --
	(178.30, 84.14) --
	(179.00, 83.84) --
	(179.70, 83.84) --
	(180.40, 83.78) --
	(181.09, 83.55) --
	(181.79, 82.97) --
	(182.49, 82.21) --
	(183.19, 81.25) --
	(183.89, 80.22) --
	(184.59, 80.44) --
	(185.28, 81.57) --
	(185.98, 82.50) --
	(186.68, 83.08) --
	(187.38, 83.48) --
	(188.08, 83.78) --
	(188.78, 83.90) --
	(189.47, 83.90) --
	(190.17, 83.87) --
	(190.87, 83.83) --
	(191.57, 83.74) --
	(192.27, 83.95) --
	(192.97, 84.16) --
	(193.66, 84.36) --
	(194.36, 84.54) --
	(195.06, 84.72) --
	(195.76, 84.88) --
	(196.46, 85.04) --
	(197.16, 85.19) --
	(197.85, 85.34) --
	(198.55, 85.48) --
	(199.25, 85.61);
\definecolor{drawColor}{RGB}{253,192,134}

\path[draw=drawColor,line width= 1.1pt,line join=round] (130.12, 80.40) --
	(130.81, 91.87) --
	(131.51, 91.87) --
	(132.21, 91.87) --
	(132.91, 91.87) --
	(133.61, 91.87) --
	(134.31, 91.87) --
	(135.00, 91.87) --
	(135.70, 91.86) --
	(136.40, 91.84) --
	(137.10, 91.80) --
	(137.80, 91.73) --
	(138.50, 91.62) --
	(139.19, 91.48) --
	(139.89, 91.30) --
	(140.59, 91.09) --
	(141.29, 90.85) --
	(141.99, 90.58) --
	(142.69, 90.27) --
	(143.38, 89.94) --
	(144.08, 89.58) --
	(144.78, 89.20) --
	(145.48, 88.79) --
	(146.18, 88.33) --
	(146.88, 87.81) --
	(147.57, 87.25) --
	(148.27, 86.63) --
	(148.97, 85.94) --
	(149.67, 85.13) --
	(150.37, 84.15) --
	(151.07, 82.88) --
	(151.76, 81.50) --
	(152.46, 80.75) --
	(153.16, 82.60) --
	(153.86, 85.36) --
	(154.56, 88.06) --
	(155.26, 90.61) --
	(155.95, 92.93) --
	(156.65, 94.77) --
	(157.35, 95.94) --
	(158.05, 96.92) --
	(158.75, 97.77) --
	(159.45, 98.53) --
	(160.14, 99.22) --
	(160.84, 99.87) --
	(161.54,100.47) --
	(162.24,101.03) --
	(162.94,101.55) --
	(163.64,102.01) --
	(164.33,102.28) --
	(165.03,102.49) --
	(165.73,102.63) --
	(166.43,102.67) --
	(167.13,102.60) --
	(167.83,102.42) --
	(168.52,102.11) --
	(169.22,101.68) --
	(169.92,101.13) --
	(170.62,100.48) --
	(171.32, 99.73) --
	(172.02, 98.91) --
	(172.71, 98.03) --
	(173.41, 97.09) --
	(174.11, 96.11) --
	(174.81, 95.09) --
	(175.51, 94.05) --
	(176.21, 92.97) --
	(176.90, 91.88) --
	(177.60, 90.75) --
	(178.30, 89.60) --
	(179.00, 89.24) --
	(179.70, 89.23) --
	(180.40, 89.08) --
	(181.09, 88.53) --
	(181.79, 87.31) --
	(182.49, 85.79) --
	(183.19, 83.93) --
	(183.89, 81.72) --
	(184.59, 81.93) --
	(185.28, 84.82) --
	(185.98, 87.11) --
	(186.68, 88.53) --
	(187.38, 89.50) --
	(188.08, 90.18) --
	(188.78, 90.40) --
	(189.47, 90.33) --
	(190.17, 90.22) --
	(190.87, 90.10) --
	(191.57, 89.85) --
	(192.27, 89.49) --
	(192.97, 89.55) --
	(193.66, 89.92) --
	(194.36, 90.26) --
	(195.06, 90.57) --
	(195.76, 90.91) --
	(196.46, 91.25) --
	(197.16, 91.58) --
	(197.85, 91.89) --
	(198.55, 92.20) --
	(199.25, 92.50);
\definecolor{drawColor}{RGB}{0,0,0}

\path[draw=drawColor,line width= 0.6pt,dash pattern=on 1pt off 3pt ,line join=round] (126.66,100.86) -- (202.71,100.86);
\definecolor{drawColor}{RGB}{0,0,0}

\path[draw=drawColor,draw opacity=0.50,line width= 0.6pt,line join=round] (144.08, 76.30) -- (144.08,152.35);
\definecolor{drawColor}{RGB}{0,0,0}

\path[draw=drawColor,line width= 0.6pt,line join=round,line cap=round] (126.66, 76.30) rectangle (202.71,152.35);
\end{scope}
\begin{scope}
\definecolor{drawColor}{RGB}{127,201,127}

\path[draw=drawColor,line width= 1.1pt,line join=round] (213.16, 80.72) --
	(213.86, 95.90) --
	(214.56, 95.90) --
	(215.26, 95.90) --
	(215.96, 95.90) --
	(216.66, 95.90) --
	(217.35, 95.90) --
	(218.05, 95.90) --
	(218.75, 95.90) --
	(219.45, 95.89) --
	(220.15, 95.86) --
	(220.85, 95.81) --
	(221.54, 95.73) --
	(222.24, 95.64) --
	(222.94, 95.51) --
	(223.64, 95.37) --
	(224.34, 95.20) --
	(225.04, 95.02) --
	(225.73, 94.81) --
	(226.43, 94.58) --
	(227.13, 94.34) --
	(227.83, 94.07) --
	(228.53, 93.79) --
	(229.23, 93.48) --
	(229.92, 93.09) --
	(230.62, 92.68) --
	(231.32, 92.21) --
	(232.02, 91.69) --
	(232.72, 91.10) --
	(233.42, 90.44) --
	(234.11, 89.58) --
	(234.81, 88.55) --
	(235.51, 87.36) --
	(236.21, 85.94) --
	(236.91, 84.25) --
	(237.61, 82.36) --
	(238.30, 81.08) --
	(239.00, 83.03) --
	(239.70, 86.01) --
	(240.40, 88.98) --
	(241.10, 91.72) --
	(241.80, 94.22) --
	(242.49, 96.47) --
	(243.19, 97.73) --
	(243.89, 98.67) --
	(244.59, 99.45) --
	(245.29,100.12) --
	(245.99,100.69) --
	(246.68,101.18) --
	(247.38,101.61) --
	(248.08,101.98) --
	(248.78,102.27) --
	(249.48,102.50) --
	(250.18,102.66) --
	(250.87,102.72) --
	(251.57,102.70) --
	(252.27,102.57) --
	(252.97,102.34) --
	(253.67,102.02) --
	(254.37,101.59) --
	(255.06,101.06) --
	(255.76,100.44) --
	(256.46, 99.74) --
	(257.16, 98.95) --
	(257.86, 98.09) --
	(258.56, 97.15) --
	(259.25, 96.09) --
	(259.95, 94.97) --
	(260.65, 93.77) --
	(261.35, 92.47) --
	(262.05, 91.08) --
	(262.75, 89.66) --
	(263.44, 88.75) --
	(264.14, 87.33) --
	(264.84, 85.11) --
	(265.54, 83.22) --
	(266.24, 81.75) --
	(266.94, 81.09) --
	(267.63, 82.88) --
	(268.33, 84.50) --
	(269.03, 85.90) --
	(269.73, 86.91) --
	(270.43, 87.72) --
	(271.13, 88.37) --
	(271.82, 88.89) --
	(272.52, 89.31) --
	(273.22, 89.67) --
	(273.92, 90.15) --
	(274.62, 90.70) --
	(275.32, 91.19) --
	(276.01, 91.63) --
	(276.71, 92.04) --
	(277.41, 92.42) --
	(278.11, 92.78) --
	(278.81, 93.11) --
	(279.51, 93.43) --
	(280.20, 93.72) --
	(280.90, 94.00) --
	(281.60, 94.27) --
	(282.30, 94.52);
\definecolor{drawColor}{RGB}{190,174,212}

\path[draw=drawColor,line width= 1.1pt,line join=round] (213.16, 79.87) --
	(213.86, 82.66) --
	(214.56, 82.66) --
	(215.26, 82.66) --
	(215.96, 82.66) --
	(216.66, 82.66) --
	(217.35, 82.66) --
	(218.05, 82.66) --
	(218.75, 82.66) --
	(219.45, 82.65) --
	(220.15, 82.65) --
	(220.85, 82.64) --
	(221.54, 82.63) --
	(222.24, 82.61) --
	(222.94, 82.58) --
	(223.64, 82.56) --
	(224.34, 82.53) --
	(225.04, 82.49) --
	(225.73, 82.46) --
	(226.43, 82.42) --
	(227.13, 82.38) --
	(227.83, 82.33) --
	(228.53, 82.29) --
	(229.23, 82.24) --
	(229.92, 82.20) --
	(230.62, 82.16) --
	(231.32, 82.13) --
	(232.02, 82.09) --
	(232.72, 82.04) --
	(233.42, 81.97) --
	(234.11, 81.87) --
	(234.81, 81.76) --
	(235.51, 81.65) --
	(236.21, 81.51) --
	(236.91, 81.35) --
	(237.61, 81.14) --
	(238.30, 80.90) --
	(239.00, 80.64) --
	(239.70, 80.51) --
	(240.40, 80.52) --
	(241.10, 80.56) --
	(241.80, 80.59) --
	(242.49, 80.61) --
	(243.19, 80.63) --
	(243.89, 80.64) --
	(244.59, 80.72) --
	(245.29, 80.82) --
	(245.99, 80.88) --
	(246.68, 80.93) --
	(247.38, 80.97) --
	(248.08, 81.00) --
	(248.78, 81.03) --
	(249.48, 81.04) --
	(250.18, 81.05) --
	(250.87, 81.06) --
	(251.57, 81.06) --
	(252.27, 81.05) --
	(252.97, 81.06) --
	(253.67, 81.06) --
	(254.37, 81.06) --
	(255.06, 81.04) --
	(255.76, 81.02) --
	(256.46, 80.98) --
	(257.16, 80.94) --
	(257.86, 80.93) --
	(258.56, 80.94) --
	(259.25, 80.94) --
	(259.95, 80.94) --
	(260.65, 80.92) --
	(261.35, 80.87) --
	(262.05, 80.79) --
	(262.75, 80.61) --
	(263.44, 80.39) --
	(264.14, 80.12) --
	(264.84, 80.08) --
	(265.54, 80.26) --
	(266.24, 80.43) --
	(266.94, 80.56) --
	(267.63, 80.70) --
	(268.33, 80.82) --
	(269.03, 80.95) --
	(269.73, 81.11) --
	(270.43, 81.27) --
	(271.13, 81.41) --
	(271.82, 81.54) --
	(272.52, 81.68) --
	(273.22, 81.80) --
	(273.92, 81.91) --
	(274.62, 82.01) --
	(275.32, 82.09) --
	(276.01, 82.17) --
	(276.71, 82.24) --
	(277.41, 82.31) --
	(278.11, 82.38) --
	(278.81, 82.44) --
	(279.51, 82.50) --
	(280.20, 82.56) --
	(280.90, 82.62) --
	(281.60, 82.67) --
	(282.30, 82.72);
\definecolor{drawColor}{RGB}{253,192,134}

\path[draw=drawColor,line width= 1.1pt,line join=round] (213.16, 80.03) --
	(213.86, 93.55) --
	(214.56, 93.55) --
	(215.26, 93.55) --
	(215.96, 93.55) --
	(216.66, 93.55) --
	(217.35, 93.55) --
	(218.05, 93.55) --
	(218.75, 93.55) --
	(219.45, 93.54) --
	(220.15, 93.51) --
	(220.85, 93.47) --
	(221.54, 93.41) --
	(222.24, 93.34) --
	(222.94, 93.24) --
	(223.64, 93.12) --
	(224.34, 92.99) --
	(225.04, 92.84) --
	(225.73, 92.68) --
	(226.43, 92.50) --
	(227.13, 92.32) --
	(227.83, 92.12) --
	(228.53, 91.91) --
	(229.23, 91.69) --
	(229.92, 91.41) --
	(230.62, 91.11) --
	(231.32, 90.79) --
	(232.02, 90.42) --
	(232.72, 90.02) --
	(233.42, 89.55) --
	(234.11, 88.93) --
	(234.81, 88.19) --
	(235.51, 87.34) --
	(236.21, 86.35) --
	(236.91, 85.17) --
	(237.61, 83.79) --
	(238.30, 82.33) --
	(239.00, 81.38) --
	(239.70, 82.16) --
	(240.40, 83.79) --
	(241.10, 85.45) --
	(241.80, 86.94) --
	(242.49, 88.29) --
	(243.19, 89.12) --
	(243.89, 89.76) --
	(244.59, 90.30) --
	(245.29, 90.75) --
	(245.99, 91.14) --
	(246.68, 91.47) --
	(247.38, 91.76) --
	(248.08, 92.02) --
	(248.78, 92.23) --
	(249.48, 92.41) --
	(250.18, 92.54) --
	(250.87, 92.63) --
	(251.57, 92.67) --
	(252.27, 92.66) --
	(252.97, 92.59) --
	(253.67, 92.47) --
	(254.37, 92.30) --
	(255.06, 92.07) --
	(255.76, 91.80) --
	(256.46, 91.48) --
	(257.16, 91.13) --
	(257.86, 90.74) --
	(258.56, 90.30) --
	(259.25, 89.80) --
	(259.95, 89.27) --
	(260.65, 88.70) --
	(261.35, 88.42) --
	(262.05, 88.48) --
	(262.75, 88.21) --
	(263.44, 87.78) --
	(264.14, 86.98) --
	(264.84, 85.46) --
	(265.54, 83.82) --
	(266.24, 82.49) --
	(266.94, 81.54) --
	(267.63, 80.86) --
	(268.33, 81.56) --
	(269.03, 82.53) --
	(269.73, 83.29) --
	(270.43, 83.90) --
	(271.13, 84.39) --
	(271.82, 84.79) --
	(272.52, 85.12) --
	(273.22, 85.39) --
	(273.92, 85.61) --
	(274.62, 85.88) --
	(275.32, 86.28) --
	(276.01, 86.65) --
	(276.71, 86.99) --
	(277.41, 87.31) --
	(278.11, 87.61) --
	(278.81, 87.90) --
	(279.51, 88.16) --
	(280.20, 88.42) --
	(280.90, 88.66) --
	(281.60, 88.89) --
	(282.30, 89.12);
\definecolor{drawColor}{RGB}{0,0,0}

\path[draw=drawColor,line width= 0.6pt,dash pattern=on 1pt off 3pt ,line join=round] (209.71,100.86) -- (285.76,100.86);
\definecolor{drawColor}{RGB}{0,0,0}

\path[draw=drawColor,draw opacity=0.50,line width= 0.6pt,line join=round] (232.72, 76.30) -- (232.72,152.35);
\definecolor{drawColor}{RGB}{0,0,0}

\path[draw=drawColor,line width= 0.6pt,line join=round,line cap=round] (209.71, 76.30) rectangle (285.76,152.35);
\end{scope}
\begin{scope}
\definecolor{drawColor}{RGB}{127,201,127}

\path[draw=drawColor,line width= 1.1pt,line join=round] (296.21, 81.07) --
	(296.91, 86.23) --
	(297.61, 86.23) --
	(298.31, 86.23) --
	(299.01, 86.23) --
	(299.70, 86.23) --
	(300.40, 86.23) --
	(301.10, 86.23) --
	(301.80, 86.23) --
	(302.50, 86.22) --
	(303.20, 86.21) --
	(303.89, 86.18) --
	(304.59, 86.15) --
	(305.29, 86.10) --
	(305.99, 86.05) --
	(306.69, 85.98) --
	(307.39, 85.89) --
	(308.08, 85.80) --
	(308.78, 85.69) --
	(309.48, 85.57) --
	(310.18, 85.43) --
	(310.88, 85.27) --
	(311.58, 85.10) --
	(312.27, 84.90) --
	(312.97, 84.68) --
	(313.67, 84.40) --
	(314.37, 84.19) --
	(315.07, 84.10) --
	(315.77, 83.97) --
	(316.46, 83.77) --
	(317.16, 83.48) --
	(317.86, 83.15) --
	(318.56, 82.68) --
	(319.26, 83.26) --
	(319.96, 84.30) --
	(320.65, 85.70) --
	(321.35, 87.17) --
	(322.05, 88.67) --
	(322.75, 90.20) --
	(323.45, 91.41) --
	(324.15, 92.19) --
	(324.84, 92.90) --
	(325.54, 93.60) --
	(326.24, 94.24) --
	(326.94, 94.84) --
	(327.64, 95.40) --
	(328.34, 95.93) --
	(329.03, 96.41) --
	(329.73, 96.83) --
	(330.43, 97.20) --
	(331.13, 97.52) --
	(331.83, 97.76) --
	(332.53, 97.90) --
	(333.22, 97.95) --
	(333.92, 97.90) --
	(334.62, 97.75) --
	(335.32, 97.50) --
	(336.02, 97.15) --
	(336.72, 96.71) --
	(337.41, 96.18) --
	(338.11, 95.58) --
	(338.81, 94.90) --
	(339.51, 94.16) --
	(340.21, 93.37) --
	(340.91, 92.56) --
	(341.60, 91.70) --
	(342.30, 90.74) --
	(343.00, 89.70) --
	(343.70, 88.66) --
	(344.40, 87.60) --
	(345.10, 86.49) --
	(345.79, 85.38) --
	(346.49, 84.25) --
	(347.19, 83.39) --
	(347.89, 82.46) --
	(348.59, 81.34) --
	(349.29, 80.22) --
	(349.98, 80.71) --
	(350.68, 82.31) --
	(351.38, 83.99) --
	(352.08, 85.53) --
	(352.78, 86.73) --
	(353.48, 87.55) --
	(354.17, 87.81) --
	(354.87, 87.95) --
	(355.57, 88.00) --
	(356.27, 88.01) --
	(356.97, 87.98) --
	(357.67, 87.94) --
	(358.36, 87.88) --
	(359.06, 87.81) --
	(359.76, 87.65) --
	(360.46, 87.41) --
	(361.16, 87.14) --
	(361.86, 86.88) --
	(362.55, 86.64) --
	(363.25, 86.59) --
	(363.95, 86.67) --
	(364.65, 86.75) --
	(365.35, 86.83);
\definecolor{drawColor}{RGB}{190,174,212}

\path[draw=drawColor,line width= 1.1pt,line join=round] (296.21, 79.92) --
	(296.91, 82.19) --
	(297.61, 82.19) --
	(298.31, 82.19) --
	(299.01, 82.19) --
	(299.70, 82.19) --
	(300.40, 82.19) --
	(301.10, 82.19) --
	(301.80, 82.19) --
	(302.50, 82.19) --
	(303.20, 82.18) --
	(303.89, 82.16) --
	(304.59, 82.13) --
	(305.29, 82.10) --
	(305.99, 82.06) --
	(306.69, 82.01) --
	(307.39, 81.96) --
	(308.08, 81.90) --
	(308.78, 81.84) --
	(309.48, 81.77) --
	(310.18, 81.70) --
	(310.88, 81.62) --
	(311.58, 81.54) --
	(312.27, 81.45) --
	(312.97, 81.37) --
	(313.67, 81.27) --
	(314.37, 81.16) --
	(315.07, 81.06) --
	(315.77, 80.93) --
	(316.46, 80.79) --
	(317.16, 80.65) --
	(317.86, 80.53) --
	(318.56, 80.49) --
	(319.26, 80.42) --
	(319.96, 80.44) --
	(320.65, 80.39) --
	(321.35, 80.38) --
	(322.05, 80.29) --
	(322.75, 80.27) --
	(323.45, 80.33) --
	(324.15, 80.36) --
	(324.84, 80.43) --
	(325.54, 80.46) --
	(326.24, 80.49) --
	(326.94, 80.53) --
	(327.64, 80.61) --
	(328.34, 80.69) --
	(329.03, 80.78) --
	(329.73, 80.85) --
	(330.43, 80.90) --
	(331.13, 80.95) --
	(331.83, 81.01) --
	(332.53, 81.05) --
	(333.22, 81.09) --
	(333.92, 81.11) --
	(334.62, 81.12) --
	(335.32, 81.12) --
	(336.02, 81.11) --
	(336.72, 81.08) --
	(337.41, 81.04) --
	(338.11, 81.01) --
	(338.81, 80.97) --
	(339.51, 80.91) --
	(340.21, 80.86) --
	(340.91, 80.78) --
	(341.60, 80.71) --
	(342.30, 80.61) --
	(343.00, 80.67) --
	(343.70, 80.77) --
	(344.40, 80.89) --
	(345.10, 81.02) --
	(345.79, 81.16) --
	(346.49, 81.24) --
	(347.19, 81.32) --
	(347.89, 81.42) --
	(348.59, 81.50) --
	(349.29, 81.43) --
	(349.98, 81.22) --
	(350.68, 80.81) --
	(351.38, 80.28) --
	(352.08, 80.15) --
	(352.78, 80.12) --
	(353.48, 80.28) --
	(354.17, 80.39) --
	(354.87, 80.47) --
	(355.57, 80.52) --
	(356.27, 80.54) --
	(356.97, 80.54) --
	(357.67, 80.52) --
	(358.36, 80.51) --
	(359.06, 80.53) --
	(359.76, 80.61) --
	(360.46, 80.71) --
	(361.16, 80.83) --
	(361.86, 80.96) --
	(362.55, 81.11) --
	(363.25, 81.25) --
	(363.95, 81.41) --
	(364.65, 81.57) --
	(365.35, 81.72);
\definecolor{drawColor}{RGB}{253,192,134}

\path[draw=drawColor,line width= 1.1pt,line join=round] (296.21, 80.67) --
	(296.91, 86.22) --
	(297.61, 86.22) --
	(298.31, 86.22) --
	(299.01, 86.22) --
	(299.70, 86.22) --
	(300.40, 86.22) --
	(301.10, 86.22) --
	(301.80, 86.21) --
	(302.50, 86.21) --
	(303.20, 86.19) --
	(303.89, 86.16) --
	(304.59, 86.13) --
	(305.29, 86.07) --
	(305.99, 86.01) --
	(306.69, 85.93) --
	(307.39, 85.84) --
	(308.08, 85.73) --
	(308.78, 85.62) --
	(309.48, 85.49) --
	(310.18, 85.36) --
	(310.88, 85.21) --
	(311.58, 85.05) --
	(312.27, 84.88) --
	(312.97, 84.69) --
	(313.67, 84.46) --
	(314.37, 84.20) --
	(315.07, 83.91) --
	(315.77, 83.60) --
	(316.46, 83.25) --
	(317.16, 82.87) --
	(317.86, 82.44) --
	(318.56, 82.02) --
	(319.26, 81.67) --
	(319.96, 81.68) --
	(320.65, 82.04) --
	(321.35, 82.58) --
	(322.05, 83.20) --
	(322.75, 83.90) --
	(323.45, 84.42) --
	(324.15, 84.72) --
	(324.84, 84.91) --
	(325.54, 85.08) --
	(326.24, 85.28) --
	(326.94, 85.48) --
	(327.64, 85.64) --
	(328.34, 85.78) --
	(329.03, 85.90) --
	(329.73, 86.01) --
	(330.43, 86.11) --
	(331.13, 86.18) --
	(331.83, 86.23) --
	(332.53, 86.26) --
	(333.22, 86.27) --
	(333.92, 86.25) --
	(334.62, 86.23) --
	(335.32, 86.17) --
	(336.02, 86.08) --
	(336.72, 85.98) --
	(337.41, 85.86) --
	(338.11, 85.72) --
	(338.81, 85.56) --
	(339.51, 85.38) --
	(340.21, 85.18) --
	(340.91, 84.98) --
	(341.60, 84.76) --
	(342.30, 84.52) --
	(343.00, 84.25) --
	(343.70, 83.95) --
	(344.40, 83.68) --
	(345.10, 83.80) --
	(345.79, 83.89) --
	(346.49, 83.71) --
	(347.19, 83.43) --
	(347.89, 83.04) --
	(348.59, 82.55) --
	(349.29, 81.77) --
	(349.98, 80.94) --
	(350.68, 80.23) --
	(351.38, 80.53) --
	(352.08, 81.20) --
	(352.78, 81.77) --
	(353.48, 82.25) --
	(354.17, 82.58) --
	(354.87, 83.04) --
	(355.57, 83.43) --
	(356.27, 83.78) --
	(356.97, 84.09) --
	(357.67, 84.34) --
	(358.36, 84.57) --
	(359.06, 84.80) --
	(359.76, 85.04) --
	(360.46, 85.30) --
	(361.16, 85.49) --
	(361.86, 85.64) --
	(362.55, 85.79) --
	(363.25, 85.91) --
	(363.95, 85.98) --
	(364.65, 86.12) --
	(365.35, 86.28);
\definecolor{drawColor}{RGB}{0,0,0}

\path[draw=drawColor,line width= 0.6pt,dash pattern=on 1pt off 3pt ,line join=round] (292.76,100.86) -- (368.80,100.86);
\definecolor{drawColor}{RGB}{0,0,0}

\path[draw=drawColor,draw opacity=0.50,line width= 0.6pt,line join=round] (310.88, 76.30) -- (310.88,152.35);
\definecolor{drawColor}{RGB}{0,0,0}

\path[draw=drawColor,line width= 0.6pt,line join=round,line cap=round] (292.76, 76.30) rectangle (368.80,152.35);
\end{scope}
\begin{scope}
\definecolor{drawColor}{RGB}{0,0,0}

\node[text=drawColor,anchor=base,inner sep=0pt, outer sep=0pt, scale=  1.00] at ( 81.64,156.82) {Italy};
\end{scope}
\begin{scope}
\definecolor{drawColor}{RGB}{0,0,0}

\node[text=drawColor,anchor=base,inner sep=0pt, outer sep=0pt, scale=  1.00] at (164.68,156.82) {Germany};
\end{scope}
\begin{scope}
\definecolor{drawColor}{RGB}{0,0,0}

\node[text=drawColor,anchor=base,inner sep=0pt, outer sep=0pt, scale=  1.00] at (247.73,156.82) {United Kingdom};
\end{scope}
\begin{scope}
\definecolor{drawColor}{RGB}{0,0,0}

\node[text=drawColor,anchor=base,inner sep=0pt, outer sep=0pt, scale=  1.00] at (330.78,156.82) {Sweden};
\end{scope}
\begin{scope}
\definecolor{drawColor}{RGB}{0,0,0}

\path[draw=drawColor,line width= 0.6pt,line join=round] ( 46.37, 72.80) --
	( 46.37, 76.30);

\path[draw=drawColor,line width= 0.6pt,line join=round] ( 63.83, 72.80) --
	( 63.83, 76.30);

\path[draw=drawColor,line width= 0.6pt,line join=round] ( 81.29, 72.80) --
	( 81.29, 76.30);

\path[draw=drawColor,line width= 0.6pt,line join=round] ( 98.74, 72.80) --
	( 98.74, 76.30);

\path[draw=drawColor,line width= 0.6pt,line join=round] (116.20, 72.80) --
	(116.20, 76.30);
\end{scope}
\begin{scope}
\definecolor{drawColor}{RGB}{0,0,0}

\node[text=drawColor,anchor=base,inner sep=0pt, outer sep=0pt, scale=  1.00] at ( 46.37, 62.91) {0};

\node[text=drawColor,anchor=base,inner sep=0pt, outer sep=0pt, scale=  1.00] at ( 63.83, 62.91) {25};

\node[text=drawColor,anchor=base,inner sep=0pt, outer sep=0pt, scale=  1.00] at ( 81.29, 62.91) {50};

\node[text=drawColor,anchor=base,inner sep=0pt, outer sep=0pt, scale=  1.00] at ( 98.74, 62.91) {75};

\node[text=drawColor,anchor=base,inner sep=0pt, outer sep=0pt, scale=  1.00] at (116.20, 62.91) {100};
\end{scope}
\begin{scope}
\definecolor{drawColor}{RGB}{0,0,0}

\path[draw=drawColor,line width= 0.6pt,line join=round] (129.42, 72.80) --
	(129.42, 76.30);

\path[draw=drawColor,line width= 0.6pt,line join=round] (146.88, 72.80) --
	(146.88, 76.30);

\path[draw=drawColor,line width= 0.6pt,line join=round] (164.33, 72.80) --
	(164.33, 76.30);

\path[draw=drawColor,line width= 0.6pt,line join=round] (181.79, 72.80) --
	(181.79, 76.30);

\path[draw=drawColor,line width= 0.6pt,line join=round] (199.25, 72.80) --
	(199.25, 76.30);
\end{scope}
\begin{scope}
\definecolor{drawColor}{RGB}{0,0,0}

\node[text=drawColor,anchor=base,inner sep=0pt, outer sep=0pt, scale=  1.00] at (129.42, 62.91) {0};

\node[text=drawColor,anchor=base,inner sep=0pt, outer sep=0pt, scale=  1.00] at (146.88, 62.91) {25};

\node[text=drawColor,anchor=base,inner sep=0pt, outer sep=0pt, scale=  1.00] at (164.33, 62.91) {50};

\node[text=drawColor,anchor=base,inner sep=0pt, outer sep=0pt, scale=  1.00] at (181.79, 62.91) {75};

\node[text=drawColor,anchor=base,inner sep=0pt, outer sep=0pt, scale=  1.00] at (199.25, 62.91) {100};
\end{scope}
\begin{scope}
\definecolor{drawColor}{RGB}{0,0,0}

\path[draw=drawColor,line width= 0.6pt,line join=round] (212.47, 72.80) --
	(212.47, 76.30);

\path[draw=drawColor,line width= 0.6pt,line join=round] (229.92, 72.80) --
	(229.92, 76.30);

\path[draw=drawColor,line width= 0.6pt,line join=round] (247.38, 72.80) --
	(247.38, 76.30);

\path[draw=drawColor,line width= 0.6pt,line join=round] (264.84, 72.80) --
	(264.84, 76.30);

\path[draw=drawColor,line width= 0.6pt,line join=round] (282.30, 72.80) --
	(282.30, 76.30);
\end{scope}
\begin{scope}
\definecolor{drawColor}{RGB}{0,0,0}

\node[text=drawColor,anchor=base,inner sep=0pt, outer sep=0pt, scale=  1.00] at (212.47, 62.91) {0};

\node[text=drawColor,anchor=base,inner sep=0pt, outer sep=0pt, scale=  1.00] at (229.92, 62.91) {25};

\node[text=drawColor,anchor=base,inner sep=0pt, outer sep=0pt, scale=  1.00] at (247.38, 62.91) {50};

\node[text=drawColor,anchor=base,inner sep=0pt, outer sep=0pt, scale=  1.00] at (264.84, 62.91) {75};

\node[text=drawColor,anchor=base,inner sep=0pt, outer sep=0pt, scale=  1.00] at (282.30, 62.91) {100};
\end{scope}
\begin{scope}
\definecolor{drawColor}{RGB}{0,0,0}

\path[draw=drawColor,line width= 0.6pt,line join=round] (295.51, 72.80) --
	(295.51, 76.30);

\path[draw=drawColor,line width= 0.6pt,line join=round] (312.97, 72.80) --
	(312.97, 76.30);

\path[draw=drawColor,line width= 0.6pt,line join=round] (330.43, 72.80) --
	(330.43, 76.30);

\path[draw=drawColor,line width= 0.6pt,line join=round] (347.89, 72.80) --
	(347.89, 76.30);

\path[draw=drawColor,line width= 0.6pt,line join=round] (365.35, 72.80) --
	(365.35, 76.30);
\end{scope}
\begin{scope}
\definecolor{drawColor}{RGB}{0,0,0}

\node[text=drawColor,anchor=base,inner sep=0pt, outer sep=0pt, scale=  1.00] at (295.51, 62.91) {0};

\node[text=drawColor,anchor=base,inner sep=0pt, outer sep=0pt, scale=  1.00] at (312.97, 62.91) {25};

\node[text=drawColor,anchor=base,inner sep=0pt, outer sep=0pt, scale=  1.00] at (330.43, 62.91) {50};

\node[text=drawColor,anchor=base,inner sep=0pt, outer sep=0pt, scale=  1.00] at (347.89, 62.91) {75};

\node[text=drawColor,anchor=base,inner sep=0pt, outer sep=0pt, scale=  1.00] at (365.35, 62.91) {100};
\end{scope}
\begin{scope}
\definecolor{drawColor}{RGB}{0,0,0}

\node[text=drawColor,anchor=base east,inner sep=0pt, outer sep=0pt, scale=  1.00] at ( 37.11, 75.85) {0.00};

\node[text=drawColor,anchor=base east,inner sep=0pt, outer sep=0pt, scale=  1.00] at ( 37.11, 97.42) {0.05};

\node[text=drawColor,anchor=base east,inner sep=0pt, outer sep=0pt, scale=  1.00] at ( 37.11,118.98) {0.10};

\node[text=drawColor,anchor=base east,inner sep=0pt, outer sep=0pt, scale=  1.00] at ( 37.11,140.55) {0.15};
\end{scope}
\begin{scope}
\definecolor{drawColor}{RGB}{0,0,0}

\path[draw=drawColor,line width= 0.6pt,line join=round] ( 40.11, 79.29) --
	( 43.61, 79.29);

\path[draw=drawColor,line width= 0.6pt,line join=round] ( 40.11,100.86) --
	( 43.61,100.86);

\path[draw=drawColor,line width= 0.6pt,line join=round] ( 40.11,122.43) --
	( 43.61,122.43);

\path[draw=drawColor,line width= 0.6pt,line join=round] ( 40.11,143.99) --
	( 43.61,143.99);
\end{scope}
\begin{scope}
\definecolor{drawColor}{RGB}{0,0,0}

\node[text=drawColor,anchor=base,inner sep=0pt, outer sep=0pt, scale=  1.00] at (206.21, 50.58) {Day};
\end{scope}
\begin{scope}
\definecolor{drawColor}{RGB}{0,0,0}

\node[text=drawColor,rotate= 90.00,anchor=base,inner sep=0pt, outer sep=0pt, scale=  1.00] at ( 13.89,114.32) {Prior sensitivity ($D_{\text{CJS}})$};
\end{scope}
\begin{scope}
\definecolor{drawColor}{RGB}{127,201,127}

\path[draw=drawColor,line width= 1.1pt,line join=round] (312.98,142.80) -- (321.65,142.80);
\end{scope}
\begin{scope}
\definecolor{drawColor}{RGB}{190,174,212}

\path[draw=drawColor,line width= 1.1pt,line join=round] (312.98,127.40) -- (321.65,127.40);
\end{scope}
\begin{scope}
\definecolor{drawColor}{RGB}{253,192,134}

\path[draw=drawColor,line width= 1.1pt,line join=round] (312.98,112.00) -- (321.65,112.00);
\end{scope}
\begin{scope}
\definecolor{drawColor}{RGB}{0,0,0}

\node[text=drawColor,anchor=base west,inner sep=0pt, outer sep=0pt, scale=  1.00] at (329.74,139.35) {$\kappa$ prior};
\end{scope}
\begin{scope}
\definecolor{drawColor}{RGB}{0,0,0}

\node[text=drawColor,anchor=base west,inner sep=0pt, outer sep=0pt, scale=  1.00] at (329.74,123.95) {$\phi$ prior};
\end{scope}
\begin{scope}
\definecolor{drawColor}{RGB}{0,0,0}

\node[text=drawColor,anchor=base west,inner sep=0pt, outer sep=0pt, scale=  1.00] at (329.74,108.55) {$\tau$ prior};
\end{scope}
\end{tikzpicture}

%% file: figs/covid_quants.tex
% Created by tikzDevice version 0.12.3.1 on 2022-11-29 12:46:13
% !TEX encoding = UTF-8 Unicode
\begin{tikzpicture}[x=1pt,y=1pt]
\definecolor{fillColor}{RGB}{255,255,255}
\begin{scope}
\definecolor{drawColor}{RGB}{0,0,0}

\path[draw=drawColor,line width= 0.6pt,line join=round] ( 49.86, 92.18) --
	( 60.42, 87.39) --
	( 70.37, 82.76) --
	( 79.78, 78.27) --
	( 88.71, 73.93) --
	( 97.64, 78.34) --
	(107.05, 83.05) --
	(117.00, 88.10) --
	(127.56, 93.50);

\path[draw=drawColor,line width= 0.6pt,line join=round] ( 49.86,136.07) --
	( 60.42,120.91) --
	( 70.37,105.48) --
	( 79.78, 89.82) --
	( 88.71, 73.93) --
	( 97.64, 90.91) --
	(107.05,109.98) --
	(117.00,131.36) --
	(127.56,155.20);
\definecolor{fillColor}{RGB}{0,0,0}

\path[fill=fillColor] ( 46.30, 88.61) --
	( 53.43, 88.61) --
	( 53.43, 95.75) --
	( 46.30, 95.75) --
	cycle;

\path[fill=fillColor] (123.99, 89.93) --
	(131.13, 89.93) --
	(131.13, 97.07) --
	(123.99, 97.07) --
	cycle;
\definecolor{fillColor}{RGB}{255,255,255}

\path[draw=drawColor,line width= 0.4pt,line join=round,line cap=round,fill=fillColor] ( 46.70,132.91) rectangle ( 53.03,139.23);

\path[draw=drawColor,line width= 0.4pt,line join=round,line cap=round,fill=fillColor] (124.40,152.04) rectangle (130.72,158.37);

\path[draw=drawColor,line width= 1.1pt,line join=round,line cap=round] ( 43.61, 69.87) rectangle (133.01,159.27);
\end{scope}
\begin{scope}
\definecolor{drawColor}{RGB}{0,0,0}

\path[draw=drawColor,line width= 0.6pt,line join=round] (167.76,111.81) --
	(178.32,109.44) --
	(188.27,107.21) --
	(197.68,105.09) --
	(206.61,103.10) --
	(215.54,101.11) --
	(224.95, 99.03) --
	(234.90, 96.85) --
	(245.46, 94.56);

\path[draw=drawColor,line width= 0.6pt,line join=round] (167.76, 73.93) --
	(178.32, 80.52) --
	(188.27, 87.52) --
	(197.68, 95.02) --
	(206.61,103.10) --
	(215.54,112.33) --
	(224.95,123.58) --
	(234.90,137.54) --
	(245.46,155.20);
\definecolor{fillColor}{RGB}{0,0,0}

\path[fill=fillColor] (164.19,108.25) --
	(171.33,108.25) --
	(171.33,115.38) --
	(164.19,115.38) --
	cycle;

\path[fill=fillColor] (241.89, 90.99) --
	(249.02, 90.99) --
	(249.02, 98.13) --
	(241.89, 98.13) --
	cycle;
\definecolor{fillColor}{RGB}{255,255,255}

\path[draw=drawColor,line width= 0.4pt,line join=round,line cap=round,fill=fillColor] (164.60, 70.77) rectangle (170.92, 77.09);

\path[draw=drawColor,line width= 0.4pt,line join=round,line cap=round,fill=fillColor] (242.29,152.04) rectangle (248.62,158.37);

\path[draw=drawColor,line width= 0.6pt,dash pattern=on 4pt off 4pt ,line join=round] (161.51, 96.27) -- (250.91, 96.27);

\path[draw=drawColor,line width= 0.6pt,dash pattern=on 4pt off 4pt ,line join=round] (161.51,109.92) -- (250.91,109.92);

\path[draw=drawColor,line width= 1.1pt,line join=round,line cap=round] (161.51, 69.87) rectangle (250.91,159.27);
\end{scope}
\begin{scope}
\definecolor{drawColor}{RGB}{0,0,0}

\path[draw=drawColor,line width= 0.6pt,line join=round] (285.66,114.68) --
	(296.21,111.30) --
	(306.16,108.09) --
	(315.58,105.05) --
	(324.51,102.16) --
	(333.44, 99.28) --
	(342.85, 96.26) --
	(352.80, 93.11) --
	(363.35, 89.82);

\path[draw=drawColor,line width= 0.6pt,line join=round] (285.66, 73.93) --
	(296.21, 79.95) --
	(306.16, 86.68) --
	(315.58, 94.07) --
	(324.51,102.16) --
	(333.44,111.54) --
	(342.85,123.03) --
	(352.80,137.32) --
	(363.35,155.20);
\definecolor{fillColor}{RGB}{0,0,0}

\path[fill=fillColor] (282.09,111.11) --
	(289.23,111.11) --
	(289.23,118.25) --
	(282.09,118.25) --
	cycle;

\path[fill=fillColor] (359.78, 86.25) --
	(366.92, 86.25) --
	(366.92, 93.39) --
	(359.78, 93.39) --
	cycle;
\definecolor{fillColor}{RGB}{255,255,255}

\path[draw=drawColor,line width= 0.4pt,line join=round,line cap=round,fill=fillColor] (282.50, 70.77) rectangle (288.82, 77.09);

\path[draw=drawColor,line width= 0.4pt,line join=round,line cap=round,fill=fillColor] (360.19,152.04) rectangle (366.52,158.37);

\path[draw=drawColor,line width= 0.6pt,dash pattern=on 4pt off 4pt ,line join=round] (279.41, 88.80) -- (368.80, 88.80);

\path[draw=drawColor,line width= 0.6pt,dash pattern=on 4pt off 4pt ,line join=round] (279.41,115.52) -- (368.80,115.52);

\path[draw=drawColor,line width= 1.1pt,line join=round,line cap=round] (279.41, 69.87) rectangle (368.80,159.27);
\end{scope}
\begin{scope}
\definecolor{drawColor}{RGB}{0,0,0}

\node[text=drawColor,anchor=base,inner sep=0pt, outer sep=0pt, scale=  1.00] at ( 88.31,163.74) {$\text{CJS}_{\text{dist}}$};
\end{scope}
\begin{scope}
\definecolor{drawColor}{RGB}{0,0,0}

\node[text=drawColor,anchor=base,inner sep=0pt, outer sep=0pt, scale=  1.00] at (206.21,163.74) {Mean};
\end{scope}
\begin{scope}
\definecolor{drawColor}{RGB}{0,0,0}

\node[text=drawColor,anchor=base,inner sep=0pt, outer sep=0pt, scale=  1.00] at (324.11,163.74) {SD};
\end{scope}
\begin{scope}
\definecolor{drawColor}{RGB}{0,0,0}

\path[draw=drawColor,line width= 0.6pt,line join=round] ( 49.86, 66.37) --
	( 49.86, 69.87);

\path[draw=drawColor,line width= 0.6pt,line join=round] ( 88.71, 66.37) --
	( 88.71, 69.87);

\path[draw=drawColor,line width= 0.6pt,line join=round] (127.56, 66.37) --
	(127.56, 69.87);
\end{scope}
\begin{scope}
\definecolor{drawColor}{RGB}{0,0,0}

\node[text=drawColor,anchor=base,inner sep=0pt, outer sep=0pt, scale=  1.00] at ( 49.86, 56.48) {0.8};

\node[text=drawColor,anchor=base,inner sep=0pt, outer sep=0pt, scale=  1.00] at ( 88.71, 56.48) {1};

\node[text=drawColor,anchor=base,inner sep=0pt, outer sep=0pt, scale=  1.00] at (127.56, 56.48) {1.25};
\end{scope}
\begin{scope}
\definecolor{drawColor}{RGB}{0,0,0}

\path[draw=drawColor,line width= 0.6pt,line join=round] (167.76, 66.37) --
	(167.76, 69.87);

\path[draw=drawColor,line width= 0.6pt,line join=round] (206.61, 66.37) --
	(206.61, 69.87);

\path[draw=drawColor,line width= 0.6pt,line join=round] (245.46, 66.37) --
	(245.46, 69.87);
\end{scope}
\begin{scope}
\definecolor{drawColor}{RGB}{0,0,0}

\node[text=drawColor,anchor=base,inner sep=0pt, outer sep=0pt, scale=  1.00] at (167.76, 56.48) {0.8};

\node[text=drawColor,anchor=base,inner sep=0pt, outer sep=0pt, scale=  1.00] at (206.61, 56.48) {1};

\node[text=drawColor,anchor=base,inner sep=0pt, outer sep=0pt, scale=  1.00] at (245.46, 56.48) {1.25};
\end{scope}
\begin{scope}
\definecolor{drawColor}{RGB}{0,0,0}

\path[draw=drawColor,line width= 0.6pt,line join=round] (285.66, 66.37) --
	(285.66, 69.87);

\path[draw=drawColor,line width= 0.6pt,line join=round] (324.51, 66.37) --
	(324.51, 69.87);

\path[draw=drawColor,line width= 0.6pt,line join=round] (363.35, 66.37) --
	(363.35, 69.87);
\end{scope}
\begin{scope}
\definecolor{drawColor}{RGB}{0,0,0}

\node[text=drawColor,anchor=base,inner sep=0pt, outer sep=0pt, scale=  1.00] at (285.66, 56.48) {0.8};

\node[text=drawColor,anchor=base,inner sep=0pt, outer sep=0pt, scale=  1.00] at (324.51, 56.48) {1};

\node[text=drawColor,anchor=base,inner sep=0pt, outer sep=0pt, scale=  1.00] at (363.35, 56.48) {1.25};
\end{scope}
\begin{scope}
\definecolor{drawColor}{RGB}{0,0,0}

\node[text=drawColor,anchor=base east,inner sep=0pt, outer sep=0pt, scale=  1.00] at (272.91, 77.80) {300};

\node[text=drawColor,anchor=base east,inner sep=0pt, outer sep=0pt, scale=  1.00] at (272.91, 94.46) {350};

\node[text=drawColor,anchor=base east,inner sep=0pt, outer sep=0pt, scale=  1.00] at (272.91,111.12) {400};

\node[text=drawColor,anchor=base east,inner sep=0pt, outer sep=0pt, scale=  1.00] at (272.91,127.77) {450};

\node[text=drawColor,anchor=base east,inner sep=0pt, outer sep=0pt, scale=  1.00] at (272.91,144.43) {500};
\end{scope}
\begin{scope}
\definecolor{drawColor}{RGB}{0,0,0}

\path[draw=drawColor,line width= 0.6pt,line join=round] (275.91, 81.25) --
	(279.41, 81.25);

\path[draw=drawColor,line width= 0.6pt,line join=round] (275.91, 97.90) --
	(279.41, 97.90);

\path[draw=drawColor,line width= 0.6pt,line join=round] (275.91,114.56) --
	(279.41,114.56);

\path[draw=drawColor,line width= 0.6pt,line join=round] (275.91,131.22) --
	(279.41,131.22);

\path[draw=drawColor,line width= 0.6pt,line join=round] (275.91,147.87) --
	(279.41,147.87);
\end{scope}
\begin{scope}
\definecolor{drawColor}{RGB}{0,0,0}

\node[text=drawColor,anchor=base east,inner sep=0pt, outer sep=0pt, scale=  1.00] at (155.01, 90.97) {600};

\node[text=drawColor,anchor=base east,inner sep=0pt, outer sep=0pt, scale=  1.00] at (155.01,116.98) {700};

\node[text=drawColor,anchor=base east,inner sep=0pt, outer sep=0pt, scale=  1.00] at (155.01,142.99) {800};
\end{scope}
\begin{scope}
\definecolor{drawColor}{RGB}{0,0,0}

\path[draw=drawColor,line width= 0.6pt,line join=round] (158.01, 94.42) --
	(161.51, 94.42);

\path[draw=drawColor,line width= 0.6pt,line join=round] (158.01,120.43) --
	(161.51,120.43);

\path[draw=drawColor,line width= 0.6pt,line join=round] (158.01,146.44) --
	(161.51,146.44);
\end{scope}
\begin{scope}
\definecolor{drawColor}{RGB}{0,0,0}

\node[text=drawColor,anchor=base east,inner sep=0pt, outer sep=0pt, scale=  1.00] at ( 37.11, 70.49) {0.00};

\node[text=drawColor,anchor=base east,inner sep=0pt, outer sep=0pt, scale=  1.00] at ( 37.11, 89.39) {0.05};

\node[text=drawColor,anchor=base east,inner sep=0pt, outer sep=0pt, scale=  1.00] at ( 37.11,108.29) {0.10};

\node[text=drawColor,anchor=base east,inner sep=0pt, outer sep=0pt, scale=  1.00] at ( 37.11,127.19) {0.15};

\node[text=drawColor,anchor=base east,inner sep=0pt, outer sep=0pt, scale=  1.00] at ( 37.11,146.09) {0.20};
\end{scope}
\begin{scope}
\definecolor{drawColor}{RGB}{0,0,0}

\path[draw=drawColor,line width= 0.6pt,line join=round] ( 40.11, 73.93) --
	( 43.61, 73.93);

\path[draw=drawColor,line width= 0.6pt,line join=round] ( 40.11, 92.83) --
	( 43.61, 92.83);

\path[draw=drawColor,line width= 0.6pt,line join=round] ( 40.11,111.73) --
	( 43.61,111.73);

\path[draw=drawColor,line width= 0.6pt,line join=round] ( 40.11,130.64) --
	( 43.61,130.64);

\path[draw=drawColor,line width= 0.6pt,line join=round] ( 40.11,149.54) --
	( 43.61,149.54);
\end{scope}
\begin{scope}
\definecolor{drawColor}{RGB}{0,0,0}

\node[text=drawColor,anchor=base,inner sep=0pt, outer sep=0pt, scale=  1.00] at (206.21, 44.15) {Power-scaling $\alpha$};
\end{scope}
\begin{scope}
\definecolor{fillColor}{RGB}{0,0,0}

\path[fill=fillColor] ( 52.74, 16.93) --
	( 59.88, 16.93) --
	( 59.88, 24.07) --
	( 52.74, 24.07) --
	cycle;
\end{scope}
\begin{scope}
\definecolor{drawColor}{RGB}{0,0,0}
\definecolor{fillColor}{RGB}{255,255,255}

\path[draw=drawColor,line width= 0.4pt,line join=round,line cap=round,fill=fillColor] (162.99, 17.34) rectangle (169.32, 23.66);
\end{scope}
\begin{scope}
\definecolor{drawColor}{RGB}{0,0,0}

\node[text=drawColor,anchor=base west,inner sep=0pt, outer sep=0pt, scale=  1.00] at ( 69.01, 17.06) {Prior power-scaling};
\end{scope}
\begin{scope}
\definecolor{drawColor}{RGB}{0,0,0}

\node[text=drawColor,anchor=base west,inner sep=0pt, outer sep=0pt, scale=  1.00] at (178.86, 17.06) {Likelihood power-scaling};
\end{scope}
\begin{scope}
\definecolor{drawColor}{RGB}{0,0,0}

\path[draw=drawColor,line width= 0.6pt,dash pattern=on 4pt off 4pt ,line join=round] (307.12, 20.50) -- (319.44, 20.50);
\end{scope}
\begin{scope}
\definecolor{drawColor}{RGB}{0,0,0}

\path[draw=drawColor,line width= 0.6pt,dash pattern=on 4pt off 4pt ,line join=round] (307.12, 20.50) -- (319.44, 20.50);
\end{scope}
\begin{scope}
\definecolor{drawColor}{RGB}{0,0,0}

\node[text=drawColor,anchor=base west,inner sep=0pt, outer sep=0pt, scale=  1.00] at (325.98, 17.06) {$\pm2$ MCSE};
\end{scope}
\end{tikzpicture}

%% file: acknowledgements.tex
We thank Osvaldo Martin, Nikolas Siccha, Lukas Prediger, Andrew Manderson and Sona Hunanyan for insightful comments on a previous draft, and Cory McCartan for a helpful discussion on implementation details. We acknowledge the computational resources provided by the Aalto Science-IT project and support by the Academy of Finland Flagship programme: Finnish Center for Artificial Intelligence, FCAI. This work was partially funded by Deutsche Forschungsgemeinschaft (DFG, German Research Foundation) under Germany's Excellence Strategy - EXC 2075 -- 390740016.

%% file: appendix.tex
\section{Diagnostic threshold}
\label{sec:threshold-appendix}
\setcounter{figure}{0}

\makeatletter 
\renewcommand{\thefigure}{S\@arabic\c@figure}
\makeatother

\begin{figure}[h]
    \centering    \input{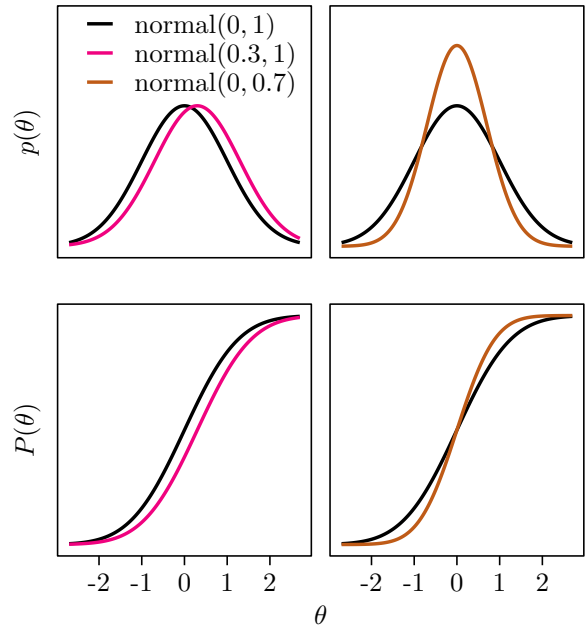}
    \caption{PDFs (top) and CDFs (bottom) of \(\normal(0, 1)\) and perturbed distributions differing by \(\cjs_{\text{dist}} \approx 0.05\) from the \(\normal(0, 1)\). Note that \(\cjs\) quantifies the difference between two CDFs; the corresponding PDFs are shown here to aid interpretation.}
    \label{fig:threshold-example}
\end{figure}

\section{Software implementation}
\label{sec:software}

\subsection{Usage}

Conducting a power-scaling sensitivity analysis with \texttt{priorsense} can be done as follows: given a fitted model object, \texttt{powerscale\_sensitivity} will automatically perform workflow steps 2--5 and return the local sensitivity of each parameter in a model fit (based on numerical derivatives of \(\cjs_\text{dist}\) by default). Follow-up analysis for diagnosing the sensitivity can be performed with \texttt{powerscale\_sequence}, which returns an object containing the base posterior draws along with weights corresponding to each perturbed posterior (or optionally resampled posterior draws). This can be plotted to visualise the change in ECDFs, kernel density etimates, or estimated quantities, with respect to the degree of power-scaling. Sensitivity of posterior quantities such as the mean, median or standard deviation can be assessed with the \texttt{powerscale\_derivative} (for analytical derivatives) and \texttt{powerscale\_gradients} (for numerical derivatives) functions. All functions will provide warnings when estimates derived from PSIS or IWMM may not be trustworthy due to too large differences between the perturbed and base posteriors.

\subsection{Practical implementation details}

In this section, we provide more details for a practical implementation of the approach.
The importance weights for power-scaling the prior or likelihood rely on density evaluations of the power-scaled component. Thus, the following are required for estimating properties of the perturbed posteriors:
\begin{itemize}
    \item posterior draws from the base posterior
    \item (log of) likelihood evaluations at the locations of the posterior draws
    \item (log of) joint prior evaluations (for the priors to be power-scaled) at the locations of the posterior draws 
\end{itemize}

In R, posterior draws can be accessed from the model fit object
directly, while the \texttt{posterior}
package~\citep{burknerPosteriorToolsWorking2020} provides
convenient functions for working with them. Existing R packages interfacing with Stan already make use of the log likelihood evaluations~\citep[e.g.\ the \texttt{loo} package;][]{vehtariLooEfficientLeaveoneout2020}, and the log prior evaluations can be specified in the model code, for example in the
\texttt{generated quantities} block of the Stan code (as shown
in Listing~\ref{lst:example-stan}). In cases where some
priors should be excluded from the power-scaling (such as intermediate
priors in hierarchical models), only the priors to be power-scaled should be included here. Log prior evaluations can also be stored in an array, allowing for selective power-scaling of subsets of priors.

\texttt{priorsense} uses the \texttt{loo} package for PSIS, while IWMM is
currently implemented directly.  \(\cjs_\text{dist}\) is also implemented directly, while other divergence measures can be used from
\texttt{philentropy}~\citep{drostPhilentropyInformationTheory2018}. Functions from \texttt{matrixStats}~\citep{bengtssonMatrixStatsFunctionsThat2020} and \texttt{spatstat}~\citep{spatstat} are used for calculating weighted quantities and weighted ECDFs, respectively. Diagnostics graphics are created using \texttt{ggplot2}~\citep{wickhamGgplot2ElegantGraphics2016}.

 \begin{lstlisting}[
 float,
 basicstyle = \ttfamily,
 caption = Example Stan code with log prior and log likelihood
    specified such that the resulting fitted model can be used with \texttt{priorsense}.,
    label = lst:example-stan
    ]
data {
  int<lower=1> N;
  vector[N] y;
}
parameters {
  real mu;
  real<lower=0> sigma;
}
model {
  // log priors
  target += student_t_lpdf(mu | 4, 0, 10);
  target += exponential_lpdf(sigma | 0.1);
  // log likelihood
  target += normal_lpdf(y | mu, sigma);
}
generated quantities {
  vector[N] log_lik; // log likelihood
  real lprior; // joint log prior
  // log likelihood
  for (n in 1:N) log_lik[n] = normal_lpdf(y[n] | mu, sigma);
  // log prior
  lprior = student_t_lpdf(mu | 4, 0, 10)
            + exponential_lpdf(sigma | 0.1);
}
\end{lstlisting}

\clearpage

\section{Body fat (linear regression)}
\label{sec:bodyfat-appendix}

\begin{figure}[h]
    \centering
    \input{./figs/bodyfat_posterior_base.tex}
    \caption{Marginal posteriors for the body fat case study. Points show means, thick and thin lines correspond to 50\% and 95\% credible intervals.}%
    \label{fig:bodyfat-posterior}
  \end{figure}
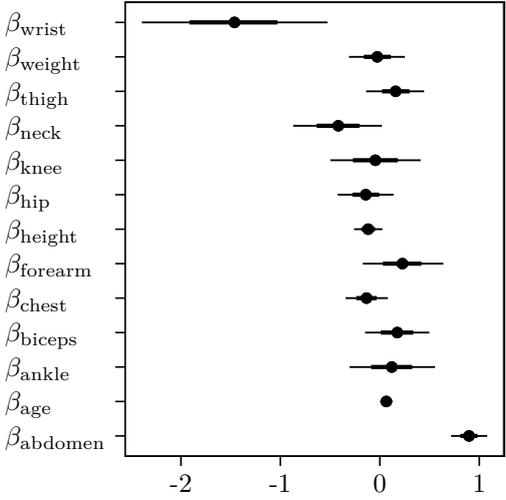

\clearpage

\section{Bacteria (hierarchical logistic regression)}
\label{sec:bacteria-appendix}

 \begin{table}[h]
   \centering
   \caption{Sensitivity diagnostic values for the bacteria case study. }
   \label{tab:bacteria}
   \small
   \begin{tabular}{llllll}
     \toprule
     \makecell[l]{Prior\\(comment)} & Parameter & \makecell[l]{Post.\\mean} & newp\makecell[l]{Post.\\SD} & \makecell[l]{Prior\\ sensitivity} & \makecell[l]{Likelihood\\ sensitivity} \\
     \midrule
     \(\tau \sim \gammadist(0.01, 0.01)\) \\
      & \(\tau\) & 0.36 & 0.3 & 0.02  &    0.10 \\
      & \(\mu\) & 3.8 & 0.76 & 0.03 &     0.18 \\
      & \(\beta_{\text{week}}\) & -0.17 & 0.06 & 0.02 & 0.10 \\
      & \(\beta_{\text{trtDrugP}}\) & -1.01 & 0.92 & 0.02 & 0.10 \\
      & \(\beta_{\text{trtDrug}}\) & -1.55 & 0.92 & 0.02 & 0.10 \\
     \midrule
     \(\tau \sim \normal^+(0, 1)\) \\
     & \(\tau\) & 0.44 & 0.27  & 0.01 & 0.20 \\
     & \(\mu\) & 3.6 & 0.76 & 0.00 & 0.16 \\
     & \(\beta_{\text{week}}\) & -0.16 & 0.05 & 0.00 & 0.11 \\
     & \(\beta_{\text{trtDrugP}}\) & -0.94 & 0.85  & 0.00 & 0.10 \\
     & \(\beta_{\text{trtDrug}}\) & -1.48 & 0.84  & 0.01 & 0.10 \\
     \midrule
     \(\tau \sim \Cauchy^+(0, 1)\) \\
     & \(\tau\) & 0.44  & 0.30 & 0.01 & 0.12 \\
     & \(\mu\) & 3.62 & 0.76 & 0.01 & 0.17  \\
     & \(\beta_{\text{week}}\) & -0.16 & 0.06 & 0.00 & 0.11 \\
     & \(\beta_{\text{trtDrugP}}\) & -0.95 & 0.85 & 0.00 & 0.11 \\
     & \(\beta_{\text{trtDrug}}\) & -1.47 & 0.85 & 0.00 & 0.09 \\
     \midrule
     \(\tau \sim \gammadist(1, 2)\) \\
     & \(\tau\) & 0.37 & 0.21  & 0.02 & 0.17 \\
     & \(\mu\) & 3.73 & 0.8 & 0.02 & 0.17 \\
     & \(\beta_{\text{week}}\) & -0.17 & 0.06 & 0.01 & 0.09 \\
     & \(\beta_{\text{trtDrugP}}\) & -0.99 & 0.90 & 0.01 & 0.09 \\
     & \(\beta_{\text{trtDrug}}\) & -1.53 & 0.88& 0.01 & 0.09 \\
     \midrule
     \(\tau \sim \gammadist(9, 0.5)\) \\
     &\\
     prior-data conflict & \(\tau\) & 13.8 & 5.4 & \textbf{0.10} & 0.13 \\
     posterior differs & \(\mu\) & 2.63 & 0.42 & 0.01 & 0.06  \\
     posterior differs & \(\beta_{\text{week}}\) & -0.12 & 0.05 & 0.00 & 0.08 \\
     posterior differs & \(\beta_{\text{trtDrugP}}\) & -0.66 & 0.46 & 0.01 & 0.07  \\
     posterior differs & \(\beta_{\text{trtDrug}}\) & -1.14 & 0.45 &  0.01 & 0.08\\
     \bottomrule
   \end{tabular}
   %Values derived from numerical derivatives of \(\cjs_\text{dist}\) with respect to \(\log_2(\alpha)\) around \(\alpha = 1\). Higher sensitivity values indicate greater sensitivity. Prior sensitivity values above the threshold \((\geq 0.05)\) indicate potential issues. Likelihood sensitivity values below the threshold (\(< 0.05\)) indicate weak or noninformative likelihood. There appears to be prior-data conflict for the unreasonable prior (\(\tau \sim \text{gamma}(9, 0.5)\)), which leads to drastically differing posterior estimates for all parameters.
 \end{table}

\clearpage

\section{Motorcycle crash (Gaussian process regression)}
\label{sec:motorcycle-appendix}

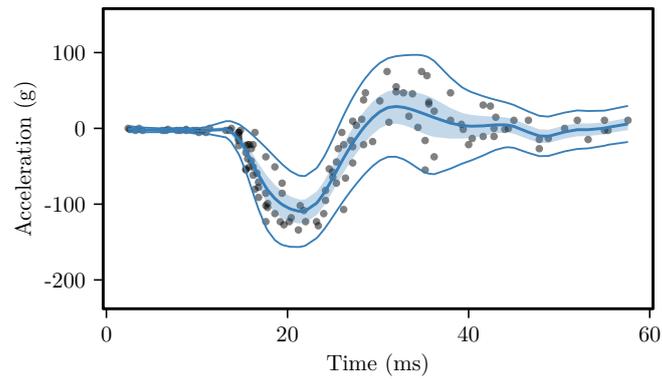
\begin{figure}[h]
  \centering
  \input{./figs/motorcycle_adjust_plot.tex}
  \caption{Prediction plot for the adjusted model with the data superimposed. Shown in the plot are the mean, 50\% and 95\% credible intervals for the posterior predictions. The predictions capture the raw data well, indicating that we have arrived at a reasonable model.}%
  \label{fig:motorcycle-data}
\end{figure}

\clearpage

\section{COVID-19 interventions (infections and deaths model)}
\label{sec:covid-appendix}
\begin{figure}[h]
    \centering    \input{./figs/covid_liksens}
    \caption{Likelihood sensitivity of posterior predictions (expected deaths due to COVID-19) for a sequence of 100 days, for each country. The vertical line indicates the onset of major governmental intervention. The dotted line indicates the sensitivity threshold of 0.05, above which we consider sensitivity to be present.}
    \label{fig:covidliksens}
\end{figure}
\begin{figure}[ht]
    \centering    \input{./figs/covid_sens}
    \caption{Prior sensitivity of posterior predictions (expected deaths due to COVID-19) for a sequence of 100 days, for each country. The vertical lines indicate the onset of major governmental intervention. The dotted lines indicate the sensitivity threshold of 0.05, above which we consider sensitivity to be present.}
    \label{fig:covidsens}
\end{figure}

%% file: figs/bodyfat_posterior_base.tex
% Created by tikzDevice version 0.12.3.1 on 2022-04-28 10:31:58
% !TEX encoding = UTF-8 Unicode
\begin{tikzpicture}[x=1pt,y=1pt]
\definecolor{fillColor}{RGB}{255,255,255}
\path[use as bounding box,fill=fillColor,fill opacity=0.00] (0,0) rectangle (216.81,216.81);
\begin{scope}
\path[clip] ( 67.61, 38.20) rectangle (209.81,209.81);
\definecolor{drawColor}{RGB}{0,0,0}

\path[draw=drawColor,line width= 0.7pt,line join=round] ( 74.08,202.01) -- (143.60,202.01);

\path[draw=drawColor,line width= 0.7pt,line join=round] (144.66,150.01) -- (178.44,150.01);

\path[draw=drawColor,line width= 0.7pt,line join=round] (130.75,163.01) -- (163.93,163.01);

\path[draw=drawColor,line width= 1.6pt,line join=round] ( 91.89,202.01) -- (124.88,202.01);

\path[draw=drawColor,line width= 0.7pt,line join=round] (151.86, 72.00) -- (183.83, 72.00);

\path[draw=drawColor,line width= 0.7pt,line join=round] (156.78,111.00) -- (186.97,111.00);

\path[draw=drawColor,line width= 0.7pt,line join=round] (157.66, 85.00) -- (181.69, 85.00);

\path[draw=drawColor,line width= 0.7pt,line join=round] (158.03,176.01) -- (179.78,176.01);

\path[draw=drawColor,line width= 0.7pt,line join=round] (151.63,189.01) -- (172.56,189.01);

\path[draw=drawColor,line width= 0.7pt,line join=round] (147.40,137.01) -- (168.31,137.01);

\path[draw=drawColor,line width= 1.6pt,line join=round] (153.10,150.01) -- (169.92,150.01);

\path[draw=drawColor,line width= 1.6pt,line join=round] (139.54,163.01) -- (155.57,163.01);

\path[draw=drawColor,line width= 0.7pt,line join=round] (150.37, 98.00) -- (166.13, 98.00);

\path[draw=drawColor,line width= 1.6pt,line join=round] (159.91, 72.00) -- (175.29, 72.00);

\path[draw=drawColor,line width= 1.6pt,line join=round] (164.33,111.00) -- (178.77,111.00);

\path[draw=drawColor,line width= 0.7pt,line join=round] (189.87, 46.00) -- (203.35, 46.00);

\path[draw=drawColor,line width= 1.6pt,line join=round] (163.50, 85.00) -- (175.69, 85.00);

\path[draw=drawColor,line width= 0.7pt,line join=round] (153.52,124.00) -- (164.14,124.00);

\path[draw=drawColor,line width= 1.6pt,line join=round] (164.01,176.01) -- (174.30,176.01);

\path[draw=drawColor,line width= 1.6pt,line join=round] (157.17,189.01) -- (167.31,189.01);

\path[draw=drawColor,line width= 1.6pt,line join=round] (152.93,137.01) -- (162.96,137.01);

\path[draw=drawColor,line width= 1.6pt,line join=round] (154.38, 98.00) -- (161.97, 98.00);

\path[draw=drawColor,line width= 1.6pt,line join=round] (193.35, 46.00) -- (199.73, 46.00);

\path[draw=drawColor,line width= 1.6pt,line join=round] (156.30,124.00) -- (161.41,124.00);

\path[draw=drawColor,line width= 0.7pt,line join=round] (163.31, 59.00) -- (167.88, 59.00);

\path[draw=drawColor,line width= 1.6pt,line join=round] (164.51, 59.00) -- (166.73, 59.00);
\definecolor{fillColor}{RGB}{0,0,0}

\path[draw=drawColor,line width= 0.6pt,line join=round,line cap=round,fill=fillColor] (108.77,202.01) circle (  1.87);

\path[draw=drawColor,line width= 0.6pt,line join=round,line cap=round,fill=fillColor] (161.49,150.01) circle (  1.87);

\path[draw=drawColor,line width= 0.6pt,line join=round,line cap=round,fill=fillColor] (147.62,163.01) circle (  1.87);

\path[draw=drawColor,line width= 0.6pt,line join=round,line cap=round,fill=fillColor] (108.77,202.01) circle (  1.87);

\path[draw=drawColor,line width= 0.6pt,line join=round,line cap=round,fill=fillColor] (167.63, 72.00) circle (  1.87);

\path[draw=drawColor,line width= 0.6pt,line join=round,line cap=round,fill=fillColor] (171.62,111.00) circle (  1.87);

\path[draw=drawColor,line width= 0.6pt,line join=round,line cap=round,fill=fillColor] (169.70, 85.00) circle (  1.87);

\path[draw=drawColor,line width= 0.6pt,line join=round,line cap=round,fill=fillColor] (169.08,176.01) circle (  1.87);

\path[draw=drawColor,line width= 0.6pt,line join=round,line cap=round,fill=fillColor] (162.20,189.01) circle (  1.87);

\path[draw=drawColor,line width= 0.6pt,line join=round,line cap=round,fill=fillColor] (157.90,137.01) circle (  1.87);

\path[draw=drawColor,line width= 0.6pt,line join=round,line cap=round,fill=fillColor] (161.49,150.01) circle (  1.87);

\path[draw=drawColor,line width= 0.6pt,line join=round,line cap=round,fill=fillColor] (147.62,163.01) circle (  1.87);

\path[draw=drawColor,line width= 0.6pt,line join=round,line cap=round,fill=fillColor] (158.19, 98.00) circle (  1.87);

\path[draw=drawColor,line width= 0.6pt,line join=round,line cap=round,fill=fillColor] (167.63, 72.00) circle (  1.87);

\path[draw=drawColor,line width= 0.6pt,line join=round,line cap=round,fill=fillColor] (171.62,111.00) circle (  1.87);

\path[draw=drawColor,line width= 0.6pt,line join=round,line cap=round,fill=fillColor] (196.58, 46.00) circle (  1.87);

\path[draw=drawColor,line width= 0.6pt,line join=round,line cap=round,fill=fillColor] (169.70, 85.00) circle (  1.87);

\path[draw=drawColor,line width= 0.6pt,line join=round,line cap=round,fill=fillColor] (158.81,124.00) circle (  1.87);

\path[draw=drawColor,line width= 0.6pt,line join=round,line cap=round,fill=fillColor] (169.08,176.01) circle (  1.87);

\path[draw=drawColor,line width= 0.6pt,line join=round,line cap=round,fill=fillColor] (162.20,189.01) circle (  1.87);

\path[draw=drawColor,line width= 0.6pt,line join=round,line cap=round,fill=fillColor] (157.90,137.01) circle (  1.87);

\path[draw=drawColor,line width= 0.6pt,line join=round,line cap=round,fill=fillColor] (158.19, 98.00) circle (  1.87);

\path[draw=drawColor,line width= 0.6pt,line join=round,line cap=round,fill=fillColor] (196.58, 46.00) circle (  1.87);

\path[draw=drawColor,line width= 0.6pt,line join=round,line cap=round,fill=fillColor] (158.81,124.00) circle (  1.87);

\path[draw=drawColor,line width= 0.6pt,line join=round,line cap=round,fill=fillColor] (165.59, 59.00) circle (  1.87);

\path[draw=drawColor,line width= 0.6pt,line join=round,line cap=round,fill=fillColor] (165.59, 59.00) circle (  1.87);

\path[draw=drawColor,line width= 1.1pt,line join=round,line cap=round] ( 67.61, 38.20) rectangle (209.81,209.81);
\end{scope}
\begin{scope}
\path[clip] (  0.00,  0.00) rectangle (216.81,216.81);
\definecolor{drawColor}{RGB}{0,0,0}

\node[text=drawColor,anchor=base west,inner sep=0pt, outer sep=0pt, scale=  1.00] at ( 22.87, 42.56) {$\beta_{\text{abdomen}}$};

\node[text=drawColor,anchor=base west,inner sep=0pt, outer sep=0pt, scale=  1.00] at ( 22.87, 55.56) {$\beta_{\text{age}}$};

\node[text=drawColor,anchor=base west,inner sep=0pt, outer sep=0pt, scale=  1.00] at ( 22.87, 68.56) {$\beta_{\text{ankle}}$};

\node[text=drawColor,anchor=base west,inner sep=0pt, outer sep=0pt, scale=  1.00] at ( 22.87, 81.56) {$\beta_{\text{biceps}}$};

\node[text=drawColor,anchor=base west,inner sep=0pt, outer sep=0pt, scale=  1.00] at ( 22.87, 94.56) {$\beta_{\text{chest}}$};

\node[text=drawColor,anchor=base west,inner sep=0pt, outer sep=0pt, scale=  1.00] at ( 22.87,107.56) {$\beta_{\text{forearm}}$};

\node[text=drawColor,anchor=base west,inner sep=0pt, outer sep=0pt, scale=  1.00] at ( 22.87,120.56) {$\beta_{\text{height}}$};

\node[text=drawColor,anchor=base west,inner sep=0pt, outer sep=0pt, scale=  1.00] at ( 22.87,133.56) {$\beta_{\text{hip}}$};

\node[text=drawColor,anchor=base west,inner sep=0pt, outer sep=0pt, scale=  1.00] at ( 22.87,146.56) {$\beta_{\text{knee}}$};

\node[text=drawColor,anchor=base west,inner sep=0pt, outer sep=0pt, scale=  1.00] at ( 22.87,159.56) {$\beta_{\text{neck}}$};

\node[text=drawColor,anchor=base west,inner sep=0pt, outer sep=0pt, scale=  1.00] at ( 22.87,172.56) {$\beta_{\text{thigh}}$};

\node[text=drawColor,anchor=base west,inner sep=0pt, outer sep=0pt, scale=  1.00] at ( 22.87,185.56) {$\beta_{\text{weight}}$};

\node[text=drawColor,anchor=base west,inner sep=0pt, outer sep=0pt, scale=  1.00] at ( 22.87,198.57) {$\beta_{\text{wrist}}$};
\end{scope}
\begin{scope}
\path[clip] (  0.00,  0.00) rectangle (216.81,216.81);
\definecolor{drawColor}{RGB}{0,0,0}

\path[draw=drawColor,line width= 0.6pt,line join=round] ( 64.11, 46.00) --
	( 67.61, 46.00);

\path[draw=drawColor,line width= 0.6pt,line join=round] ( 64.11, 59.00) --
	( 67.61, 59.00);

\path[draw=drawColor,line width= 0.6pt,line join=round] ( 64.11, 72.00) --
	( 67.61, 72.00);

\path[draw=drawColor,line width= 0.6pt,line join=round] ( 64.11, 85.00) --
	( 67.61, 85.00);

\path[draw=drawColor,line width= 0.6pt,line join=round] ( 64.11, 98.00) --
	( 67.61, 98.00);

\path[draw=drawColor,line width= 0.6pt,line join=round] ( 64.11,111.00) --
	( 67.61,111.00);

\path[draw=drawColor,line width= 0.6pt,line join=round] ( 64.11,124.00) --
	( 67.61,124.00);

\path[draw=drawColor,line width= 0.6pt,line join=round] ( 64.11,137.01) --
	( 67.61,137.01);

\path[draw=drawColor,line width= 0.6pt,line join=round] ( 64.11,150.01) --
	( 67.61,150.01);

\path[draw=drawColor,line width= 0.6pt,line join=round] ( 64.11,163.01) --
	( 67.61,163.01);

\path[draw=drawColor,line width= 0.6pt,line join=round] ( 64.11,176.01) --
	( 67.61,176.01);

\path[draw=drawColor,line width= 0.6pt,line join=round] ( 64.11,189.01) --
	( 67.61,189.01);

\path[draw=drawColor,line width= 0.6pt,line join=round] ( 64.11,202.01) --
	( 67.61,202.01);
\end{scope}
\begin{scope}
\path[clip] (  0.00,  0.00) rectangle (216.81,216.81);
\definecolor{drawColor}{RGB}{0,0,0}

\path[draw=drawColor,line width= 0.6pt,line join=round] ( 88.71, 34.70) --
	( 88.71, 38.20);

\path[draw=drawColor,line width= 0.6pt,line join=round] (125.94, 34.70) --
	(125.94, 38.20);

\path[draw=drawColor,line width= 0.6pt,line join=round] (163.17, 34.70) --
	(163.17, 38.20);

\path[draw=drawColor,line width= 0.6pt,line join=round] (200.40, 34.70) --
	(200.40, 38.20);
\end{scope}
\begin{scope}
\path[clip] (  0.00,  0.00) rectangle (216.81,216.81);
\definecolor{drawColor}{RGB}{0,0,0}

\node[text=drawColor,anchor=base,inner sep=0pt, outer sep=0pt, scale=  1.00] at ( 88.71, 24.81) {-2};

\node[text=drawColor,anchor=base,inner sep=0pt, outer sep=0pt, scale=  1.00] at (125.94, 24.81) {-1};

\node[text=drawColor,anchor=base,inner sep=0pt, outer sep=0pt, scale=  1.00] at (163.17, 24.81) {0};

\node[text=drawColor,anchor=base,inner sep=0pt, outer sep=0pt, scale=  1.00] at (200.40, 24.81) {1};
\end{scope}
\end{tikzpicture}

%% file: figs/motorcycle_adjust_plot.tex
% Created by tikzDevice version 0.12.3.1 on 2022-05-04 14:29:14
% !TEX encoding = UTF-8 Unicode
\begin{tikzpicture}[x=1pt,y=1pt]
\definecolor{fillColor}{RGB}{255,255,255}
\begin{scope}
\definecolor{drawColor}{RGB}{0,0,0}
\definecolor{fillColor}{RGB}{0,0,0}

\path[draw=drawColor,draw opacity=0.50,line width= 0.4pt,line join=round,line cap=round,fill=fillColor,fill opacity=0.50] ( 49.18,213.36) circle (  1.21);

\path[draw=drawColor,draw opacity=0.50,line width= 0.4pt,line join=round,line cap=round,fill=fillColor,fill opacity=0.50] ( 49.85,212.98) circle (  1.21);

\path[draw=drawColor,draw opacity=0.50,line width= 0.4pt,line join=round,line cap=round,fill=fillColor,fill opacity=0.50] ( 51.89,212.58) circle (  1.21);

\path[draw=drawColor,draw opacity=0.50,line width= 0.4pt,line join=round,line cap=round,fill=fillColor,fill opacity=0.50] ( 53.24,213.36) circle (  1.21);

\path[draw=drawColor,draw opacity=0.50,line width= 0.4pt,line join=round,line cap=round,fill=fillColor,fill opacity=0.50] ( 54.60,212.58) circle (  1.21);

\path[draw=drawColor,draw opacity=0.50,line width= 0.4pt,line join=round,line cap=round,fill=fillColor,fill opacity=0.50] ( 62.06,212.58) circle (  1.21);

\path[draw=drawColor,draw opacity=0.50,line width= 0.4pt,line join=round,line cap=round,fill=fillColor,fill opacity=0.50] ( 63.41,212.58) circle (  1.21);

\path[draw=drawColor,draw opacity=0.50,line width= 0.4pt,line join=round,line cap=round,fill=fillColor,fill opacity=0.50] ( 64.09,212.98) circle (  1.21);

\path[draw=drawColor,draw opacity=0.50,line width= 0.4pt,line join=round,line cap=round,fill=fillColor,fill opacity=0.50] ( 67.48,212.58) circle (  1.21);

\path[draw=drawColor,draw opacity=0.50,line width= 0.4pt,line join=round,line cap=round,fill=fillColor,fill opacity=0.50] ( 68.84,212.58) circle (  1.21);

\path[draw=drawColor,draw opacity=0.50,line width= 0.4pt,line join=round,line cap=round,fill=fillColor,fill opacity=0.50] ( 70.87,212.98) circle (  1.21);

\path[draw=drawColor,draw opacity=0.50,line width= 0.4pt,line join=round,line cap=round,fill=fillColor,fill opacity=0.50] ( 70.87,212.58) circle (  1.21);

\path[draw=drawColor,draw opacity=0.50,line width= 0.4pt,line join=round,line cap=round,fill=fillColor,fill opacity=0.50] ( 73.59,212.58) circle (  1.21);

\path[draw=drawColor,draw opacity=0.50,line width= 0.4pt,line join=round,line cap=round,fill=fillColor,fill opacity=0.50] ( 74.94,212.58) circle (  1.21);

\path[draw=drawColor,draw opacity=0.50,line width= 0.4pt,line join=round,line cap=round,fill=fillColor,fill opacity=0.50] ( 75.62,211.81) circle (  1.21);

\path[draw=drawColor,draw opacity=0.50,line width= 0.4pt,line join=round,line cap=round,fill=fillColor,fill opacity=0.50] ( 76.98,212.58) circle (  1.21);

\path[draw=drawColor,draw opacity=0.50,line width= 0.4pt,line join=round,line cap=round,fill=fillColor,fill opacity=0.50] ( 78.33,211.81) circle (  1.21);

\path[draw=drawColor,draw opacity=0.50,line width= 0.4pt,line join=round,line cap=round,fill=fillColor,fill opacity=0.50] ( 79.69,213.36) circle (  1.21);

\path[draw=drawColor,draw opacity=0.50,line width= 0.4pt,line join=round,line cap=round,fill=fillColor,fill opacity=0.50] ( 85.79,212.58) circle (  1.21);

\path[draw=drawColor,draw opacity=0.50,line width= 0.4pt,line join=round,line cap=round,fill=fillColor,fill opacity=0.50] ( 87.15,212.58) circle (  1.21);

\path[draw=drawColor,draw opacity=0.50,line width= 0.4pt,line join=round,line cap=round,fill=fillColor,fill opacity=0.50] ( 87.82,213.36) circle (  1.21);

\path[draw=drawColor,draw opacity=0.50,line width= 0.4pt,line join=round,line cap=round,fill=fillColor,fill opacity=0.50] ( 90.54,209.55) circle (  1.21);

\path[draw=drawColor,draw opacity=0.50,line width= 0.4pt,line join=round,line cap=round,fill=fillColor,fill opacity=0.50] ( 90.54,211.81) circle (  1.21);

\path[draw=drawColor,draw opacity=0.50,line width= 0.4pt,line join=round,line cap=round,fill=fillColor,fill opacity=0.50] ( 90.54,211.81) circle (  1.21);

\path[draw=drawColor,draw opacity=0.50,line width= 0.4pt,line join=round,line cap=round,fill=fillColor,fill opacity=0.50] ( 90.54,210.70) circle (  1.21);

\path[draw=drawColor,draw opacity=0.50,line width= 0.4pt,line join=round,line cap=round,fill=fillColor,fill opacity=0.50] ( 90.54,208.78) circle (  1.21);

\path[draw=drawColor,draw opacity=0.50,line width= 0.4pt,line join=round,line cap=round,fill=fillColor,fill opacity=0.50] ( 90.54,206.84) circle (  1.21);

\path[draw=drawColor,draw opacity=0.50,line width= 0.4pt,line join=round,line cap=round,fill=fillColor,fill opacity=0.50] ( 91.21,212.58) circle (  1.21);

\path[draw=drawColor,draw opacity=0.50,line width= 0.4pt,line join=round,line cap=round,fill=fillColor,fill opacity=0.50] ( 93.25,206.84) circle (  1.21);

\path[draw=drawColor,draw opacity=0.50,line width= 0.4pt,line join=round,line cap=round,fill=fillColor,fill opacity=0.50] ( 93.25,204.18) circle (  1.21);

\path[draw=drawColor,draw opacity=0.50,line width= 0.4pt,line join=round,line cap=round,fill=fillColor,fill opacity=0.50] ( 93.25,198.06) circle (  1.21);

\path[draw=drawColor,draw opacity=0.50,line width= 0.4pt,line join=round,line cap=round,fill=fillColor,fill opacity=0.50] ( 93.25,197.66) circle (  1.21);

\path[draw=drawColor,draw opacity=0.50,line width= 0.4pt,line join=round,line cap=round,fill=fillColor,fill opacity=0.50] ( 93.93,201.86) circle (  1.21);

\path[draw=drawColor,draw opacity=0.50,line width= 0.4pt,line join=round,line cap=round,fill=fillColor,fill opacity=0.50] ( 93.93,207.21) circle (  1.21);

\path[draw=drawColor,draw opacity=0.50,line width= 0.4pt,line join=round,line cap=round,fill=fillColor,fill opacity=0.50] ( 94.60,207.21) circle (  1.21);

\path[draw=drawColor,draw opacity=0.50,line width= 0.4pt,line join=round,line cap=round,fill=fillColor,fill opacity=0.50] ( 94.60,198.83) circle (  1.21);

\path[draw=drawColor,draw opacity=0.50,line width= 0.4pt,line join=round,line cap=round,fill=fillColor,fill opacity=0.50] ( 95.28,201.09) circle (  1.21);

\path[draw=drawColor,draw opacity=0.50,line width= 0.4pt,line join=round,line cap=round,fill=fillColor,fill opacity=0.50] ( 95.28,205.69) circle (  1.21);

\path[draw=drawColor,draw opacity=0.50,line width= 0.4pt,line join=round,line cap=round,fill=fillColor,fill opacity=0.50] ( 95.96,207.21) circle (  1.21);

\path[draw=drawColor,draw opacity=0.50,line width= 0.4pt,line join=round,line cap=round,fill=fillColor,fill opacity=0.50] ( 95.96,198.83) circle (  1.21);

\path[draw=drawColor,draw opacity=0.50,line width= 0.4pt,line join=round,line cap=round,fill=fillColor,fill opacity=0.50] ( 95.96,195.72) circle (  1.21);

\path[draw=drawColor,draw opacity=0.50,line width= 0.4pt,line join=round,line cap=round,fill=fillColor,fill opacity=0.50] ( 96.64,211.81) circle (  1.21);

\path[draw=drawColor,draw opacity=0.50,line width= 0.4pt,line join=round,line cap=round,fill=fillColor,fill opacity=0.50] ( 96.64,190.37) circle (  1.21);

\path[draw=drawColor,draw opacity=0.50,line width= 0.4pt,line join=round,line cap=round,fill=fillColor,fill opacity=0.50] ( 97.32,196.49) circle (  1.21);

\path[draw=drawColor,draw opacity=0.50,line width= 0.4pt,line join=round,line cap=round,fill=fillColor,fill opacity=0.50] ( 97.99,193.06) circle (  1.21);

\path[draw=drawColor,draw opacity=0.50,line width= 0.4pt,line join=round,line cap=round,fill=fillColor,fill opacity=0.50] ( 97.99,187.31) circle (  1.21);

\path[draw=drawColor,draw opacity=0.50,line width= 0.4pt,line join=round,line cap=round,fill=fillColor,fill opacity=0.50] ( 97.99,191.14) circle (  1.21);

\path[draw=drawColor,draw opacity=0.50,line width= 0.4pt,line join=round,line cap=round,fill=fillColor,fill opacity=0.50] (100.71,202.63) circle (  1.21);

\path[draw=drawColor,draw opacity=0.50,line width= 0.4pt,line join=round,line cap=round,fill=fillColor,fill opacity=0.50] (100.71,188.88) circle (  1.21);

\path[draw=drawColor,draw opacity=0.50,line width= 0.4pt,line join=round,line cap=round,fill=fillColor,fill opacity=0.50] (100.71,178.16) circle (  1.21);

\path[draw=drawColor,draw opacity=0.50,line width= 0.4pt,line join=round,line cap=round,fill=fillColor,fill opacity=0.50] (100.71,184.22) circle (  1.21);

\path[draw=drawColor,draw opacity=0.50,line width= 0.4pt,line join=round,line cap=round,fill=fillColor,fill opacity=0.50] (101.38,185.02) circle (  1.21);

\path[draw=drawColor,draw opacity=0.50,line width= 0.4pt,line join=round,line cap=round,fill=fillColor,fill opacity=0.50] (101.38,183.51) circle (  1.21);

\path[draw=drawColor,draw opacity=0.50,line width= 0.4pt,line join=round,line cap=round,fill=fillColor,fill opacity=0.50] (104.10,181.19) circle (  1.21);

\path[draw=drawColor,draw opacity=0.50,line width= 0.4pt,line join=round,line cap=round,fill=fillColor,fill opacity=0.50] (104.10,198.83) circle (  1.21);

\path[draw=drawColor,draw opacity=0.50,line width= 0.4pt,line join=round,line cap=round,fill=fillColor,fill opacity=0.50] (106.13,178.16) circle (  1.21);

\path[draw=drawColor,draw opacity=0.50,line width= 0.4pt,line join=round,line cap=round,fill=fillColor,fill opacity=0.50] (106.81,188.88) circle (  1.21);

\path[draw=drawColor,draw opacity=0.50,line width= 0.4pt,line join=round,line cap=round,fill=fillColor,fill opacity=0.50] (106.81,192.69) circle (  1.21);

\path[draw=drawColor,draw opacity=0.50,line width= 0.4pt,line join=round,line cap=round,fill=fillColor,fill opacity=0.50] (107.49,176.99) circle (  1.21);

\path[draw=drawColor,draw opacity=0.50,line width= 0.4pt,line join=round,line cap=round,fill=fillColor,fill opacity=0.50] (109.52,178.16) circle (  1.21);

\path[draw=drawColor,draw opacity=0.50,line width= 0.4pt,line join=round,line cap=round,fill=fillColor,fill opacity=0.50] (110.20,179.65) circle (  1.21);

\path[draw=drawColor,draw opacity=0.50,line width= 0.4pt,line join=round,line cap=round,fill=fillColor,fill opacity=0.50] (112.91,175.05) circle (  1.21);

\path[draw=drawColor,draw opacity=0.50,line width= 0.4pt,line join=round,line cap=round,fill=fillColor,fill opacity=0.50] (113.59,184.22) circle (  1.21);

\path[draw=drawColor,draw opacity=0.50,line width= 0.4pt,line join=round,line cap=round,fill=fillColor,fill opacity=0.50] (114.94,182.36) circle (  1.21);

\path[draw=drawColor,draw opacity=0.50,line width= 0.4pt,line join=round,line cap=round,fill=fillColor,fill opacity=0.50] (115.62,178.16) circle (  1.21);

\path[draw=drawColor,draw opacity=0.50,line width= 0.4pt,line join=round,line cap=round,fill=fillColor,fill opacity=0.50] (119.69,178.16) circle (  1.21);

\path[draw=drawColor,draw opacity=0.50,line width= 0.4pt,line join=round,line cap=round,fill=fillColor,fill opacity=0.50] (120.37,176.62) circle (  1.21);

\path[draw=drawColor,draw opacity=0.50,line width= 0.4pt,line join=round,line cap=round,fill=fillColor,fill opacity=0.50] (122.40,181.19) circle (  1.21);

\path[draw=drawColor,draw opacity=0.50,line width= 0.4pt,line join=round,line cap=round,fill=fillColor,fill opacity=0.50] (123.08,186.17) circle (  1.21);

\path[draw=drawColor,draw opacity=0.50,line width= 0.4pt,line join=round,line cap=round,fill=fillColor,fill opacity=0.50] (123.08,189.97) circle (  1.21);

\path[draw=drawColor,draw opacity=0.50,line width= 0.4pt,line join=round,line cap=round,fill=fillColor,fill opacity=0.50] (124.44,198.06) circle (  1.21);

\path[draw=drawColor,draw opacity=0.50,line width= 0.4pt,line join=round,line cap=round,fill=fillColor,fill opacity=0.50] (125.79,194.94) circle (  1.21);

\path[draw=drawColor,draw opacity=0.50,line width= 0.4pt,line join=round,line cap=round,fill=fillColor,fill opacity=0.50] (125.79,196.89) circle (  1.21);

\path[draw=drawColor,draw opacity=0.50,line width= 0.4pt,line join=round,line cap=round,fill=fillColor,fill opacity=0.50] (127.15,192.69) circle (  1.21);

\path[draw=drawColor,draw opacity=0.50,line width= 0.4pt,line join=round,line cap=round,fill=fillColor,fill opacity=0.50] (127.15,200.69) circle (  1.21);

\path[draw=drawColor,draw opacity=0.50,line width= 0.4pt,line join=round,line cap=round,fill=fillColor,fill opacity=0.50] (127.83,205.69) circle (  1.21);

\path[draw=drawColor,draw opacity=0.50,line width= 0.4pt,line join=round,line cap=round,fill=fillColor,fill opacity=0.50] (129.18,211.81) circle (  1.21);

\path[draw=drawColor,draw opacity=0.50,line width= 0.4pt,line join=round,line cap=round,fill=fillColor,fill opacity=0.50] (129.86,182.74) circle (  1.21);

\path[draw=drawColor,draw opacity=0.50,line width= 0.4pt,line join=round,line cap=round,fill=fillColor,fill opacity=0.50] (129.86,207.21) circle (  1.21);

\path[draw=drawColor,draw opacity=0.50,line width= 0.4pt,line join=round,line cap=round,fill=fillColor,fill opacity=0.50] (130.54,194.60) circle (  1.21);

\path[draw=drawColor,draw opacity=0.50,line width= 0.4pt,line join=round,line cap=round,fill=fillColor,fill opacity=0.50] (132.57,208.78) circle (  1.21);

\path[draw=drawColor,draw opacity=0.50,line width= 0.4pt,line join=round,line cap=round,fill=fillColor,fill opacity=0.50] (133.25,200.32) circle (  1.21);

\path[draw=drawColor,draw opacity=0.50,line width= 0.4pt,line join=round,line cap=round,fill=fillColor,fill opacity=0.50] (133.25,206.44) circle (  1.21);

\path[draw=drawColor,draw opacity=0.50,line width= 0.4pt,line join=round,line cap=round,fill=fillColor,fill opacity=0.50] (133.25,216.07) circle (  1.21);

\path[draw=drawColor,draw opacity=0.50,line width= 0.4pt,line join=round,line cap=round,fill=fillColor,fill opacity=0.50] (134.61,214.50) circle (  1.21);

\path[draw=drawColor,draw opacity=0.50,line width= 0.4pt,line join=round,line cap=round,fill=fillColor,fill opacity=0.50] (136.64,216.79) circle (  1.21);

\path[draw=drawColor,draw opacity=0.50,line width= 0.4pt,line join=round,line cap=round,fill=fillColor,fill opacity=0.50] (137.32,207.21) circle (  1.21);

\path[draw=drawColor,draw opacity=0.50,line width= 0.4pt,line join=round,line cap=round,fill=fillColor,fill opacity=0.50] (137.32,224.08) circle (  1.21);

\path[draw=drawColor,draw opacity=0.50,line width= 0.4pt,line join=round,line cap=round,fill=fillColor,fill opacity=0.50] (138.00,226.76) circle (  1.21);

\path[draw=drawColor,draw opacity=0.50,line width= 0.4pt,line join=round,line cap=round,fill=fillColor,fill opacity=0.50] (140.71,208.38) circle (  1.21);

\path[draw=drawColor,draw opacity=0.50,line width= 0.4pt,line join=round,line cap=round,fill=fillColor,fill opacity=0.50] (143.42,223.71) circle (  1.21);

\path[draw=drawColor,draw opacity=0.50,line width= 0.4pt,line join=round,line cap=round,fill=fillColor,fill opacity=0.50] (146.13,234.80) circle (  1.21);

\path[draw=drawColor,draw opacity=0.50,line width= 0.4pt,line join=round,line cap=round,fill=fillColor,fill opacity=0.50] (146.81,215.67) circle (  1.21);

\path[draw=drawColor,draw opacity=0.50,line width= 0.4pt,line join=round,line cap=round,fill=fillColor,fill opacity=0.50] (149.52,229.05) circle (  1.21);

\path[draw=drawColor,draw opacity=0.50,line width= 0.4pt,line join=round,line cap=round,fill=fillColor,fill opacity=0.50] (149.52,227.14) circle (  1.21);

\path[draw=drawColor,draw opacity=0.50,line width= 0.4pt,line join=round,line cap=round,fill=fillColor,fill opacity=0.50] (152.24,226.76) circle (  1.21);

\path[draw=drawColor,draw opacity=0.50,line width= 0.4pt,line join=round,line cap=round,fill=fillColor,fill opacity=0.50] (154.27,217.93) circle (  1.21);

\path[draw=drawColor,draw opacity=0.50,line width= 0.4pt,line join=round,line cap=round,fill=fillColor,fill opacity=0.50] (155.63,226.39) circle (  1.21);

\path[draw=drawColor,draw opacity=0.50,line width= 0.4pt,line join=round,line cap=round,fill=fillColor,fill opacity=0.50] (157.66,213.73) circle (  1.21);

\path[draw=drawColor,draw opacity=0.50,line width= 0.4pt,line join=round,line cap=round,fill=fillColor,fill opacity=0.50] (159.02,234.80) circle (  1.21);

\path[draw=drawColor,draw opacity=0.50,line width= 0.4pt,line join=round,line cap=round,fill=fillColor,fill opacity=0.50] (160.37,208.78) circle (  1.21);

\path[draw=drawColor,draw opacity=0.50,line width= 0.4pt,line join=round,line cap=round,fill=fillColor,fill opacity=0.50] (160.37,197.66) circle (  1.21);

\path[draw=drawColor,draw opacity=0.50,line width= 0.4pt,line join=round,line cap=round,fill=fillColor,fill opacity=0.50] (161.05,233.25) circle (  1.21);

\path[draw=drawColor,draw opacity=0.50,line width= 0.4pt,line join=round,line cap=round,fill=fillColor,fill opacity=0.50] (161.73,223.31) circle (  1.21);

\path[draw=drawColor,draw opacity=0.50,line width= 0.4pt,line join=round,line cap=round,fill=fillColor,fill opacity=0.50] (161.73,222.53) circle (  1.21);

\path[draw=drawColor,draw opacity=0.50,line width= 0.4pt,line join=round,line cap=round,fill=fillColor,fill opacity=0.50] (163.76,202.63) circle (  1.21);

\path[draw=drawColor,draw opacity=0.50,line width= 0.4pt,line join=round,line cap=round,fill=fillColor,fill opacity=0.50] (163.76,219.87) circle (  1.21);

\path[draw=drawColor,draw opacity=0.50,line width= 0.4pt,line join=round,line cap=round,fill=fillColor,fill opacity=0.50] (169.86,226.76) circle (  1.21);

\path[draw=drawColor,draw opacity=0.50,line width= 0.4pt,line join=round,line cap=round,fill=fillColor,fill opacity=0.50] (169.86,216.41) circle (  1.21);

\path[draw=drawColor,draw opacity=0.50,line width= 0.4pt,line join=round,line cap=round,fill=fillColor,fill opacity=0.50] (173.93,214.90) circle (  1.21);

\path[draw=drawColor,draw opacity=0.50,line width= 0.4pt,line join=round,line cap=round,fill=fillColor,fill opacity=0.50] (174.61,212.98) circle (  1.21);

\path[draw=drawColor,draw opacity=0.50,line width= 0.4pt,line join=round,line cap=round,fill=fillColor,fill opacity=0.50] (176.64,207.21) circle (  1.21);

\path[draw=drawColor,draw opacity=0.50,line width= 0.4pt,line join=round,line cap=round,fill=fillColor,fill opacity=0.50] (178.00,209.55) circle (  1.21);

\path[draw=drawColor,draw opacity=0.50,line width= 0.4pt,line join=round,line cap=round,fill=fillColor,fill opacity=0.50] (182.07,222.16) circle (  1.21);

\path[draw=drawColor,draw opacity=0.50,line width= 0.4pt,line join=round,line cap=round,fill=fillColor,fill opacity=0.50] (182.07,210.30) circle (  1.21);

\path[draw=drawColor,draw opacity=0.50,line width= 0.4pt,line join=round,line cap=round,fill=fillColor,fill opacity=0.50] (184.78,221.76) circle (  1.21);

\path[draw=drawColor,draw opacity=0.50,line width= 0.4pt,line join=round,line cap=round,fill=fillColor,fill opacity=0.50] (186.14,213.36) circle (  1.21);

\path[draw=drawColor,draw opacity=0.50,line width= 0.4pt,line join=round,line cap=round,fill=fillColor,fill opacity=0.50] (186.14,210.30) circle (  1.21);

\path[draw=drawColor,draw opacity=0.50,line width= 0.4pt,line join=round,line cap=round,fill=fillColor,fill opacity=0.50] (186.81,217.56) circle (  1.21);

\path[draw=drawColor,draw opacity=0.50,line width= 0.4pt,line join=round,line cap=round,fill=fillColor,fill opacity=0.50] (190.20,212.98) circle (  1.21);

\path[draw=drawColor,draw opacity=0.50,line width= 0.4pt,line join=round,line cap=round,fill=fillColor,fill opacity=0.50] (191.56,213.36) circle (  1.21);

\path[draw=drawColor,draw opacity=0.50,line width= 0.4pt,line join=round,line cap=round,fill=fillColor,fill opacity=0.50] (193.59,216.41) circle (  1.21);

\path[draw=drawColor,draw opacity=0.50,line width= 0.4pt,line join=round,line cap=round,fill=fillColor,fill opacity=0.50] (199.02,216.41) circle (  1.21);

\path[draw=drawColor,draw opacity=0.50,line width= 0.4pt,line join=round,line cap=round,fill=fillColor,fill opacity=0.50] (203.09,205.69) circle (  1.21);

\path[draw=drawColor,draw opacity=0.50,line width= 0.4pt,line join=round,line cap=round,fill=fillColor,fill opacity=0.50] (203.09,209.15) circle (  1.21);

\path[draw=drawColor,draw opacity=0.50,line width= 0.4pt,line join=round,line cap=round,fill=fillColor,fill opacity=0.50] (206.48,209.55) circle (  1.21);

\path[draw=drawColor,draw opacity=0.50,line width= 0.4pt,line join=round,line cap=round,fill=fillColor,fill opacity=0.50] (212.58,213.36) circle (  1.21);

\path[draw=drawColor,draw opacity=0.50,line width= 0.4pt,line join=round,line cap=round,fill=fillColor,fill opacity=0.50] (217.33,216.41) circle (  1.21);

\path[draw=drawColor,draw opacity=0.50,line width= 0.4pt,line join=round,line cap=round,fill=fillColor,fill opacity=0.50] (221.39,209.15) circle (  1.21);

\path[draw=drawColor,draw opacity=0.50,line width= 0.4pt,line join=round,line cap=round,fill=fillColor,fill opacity=0.50] (227.50,212.58) circle (  1.21);

\path[draw=drawColor,draw opacity=0.50,line width= 0.4pt,line join=round,line cap=round,fill=fillColor,fill opacity=0.50] (227.50,216.41) circle (  1.21);

\path[draw=drawColor,draw opacity=0.50,line width= 0.4pt,line join=round,line cap=round,fill=fillColor,fill opacity=0.50] (228.85,212.58) circle (  1.21);

\path[draw=drawColor,draw opacity=0.50,line width= 0.4pt,line join=round,line cap=round,fill=fillColor,fill opacity=0.50] (236.31,216.41) circle (  1.21);
\definecolor{drawColor}{RGB}{55,126,184}

\path[draw=drawColor,line width= 1.1pt,line join=round] ( 49.18,213.11) --
	( 49.85,213.07) --
	( 51.89,212.96) --
	( 53.24,212.87) --
	( 54.60,212.79) --
	( 62.06,212.64) --
	( 63.41,212.68) --
	( 64.09,212.70) --
	( 67.48,212.77) --
	( 68.84,212.75) --
	( 70.87,212.64) --
	( 70.87,212.64) --
	( 73.59,212.44) --
	( 74.94,212.38) --
	( 75.62,212.36) --
	( 76.98,212.39) --
	( 78.33,212.51) --
	( 79.69,212.69) --
	( 85.79,213.01) --
	( 87.15,212.47) --
	( 87.82,212.06) --
	( 90.54,209.34) --
	( 90.54,209.34) --
	( 90.54,209.34) --
	( 90.54,209.34) --
	( 90.54,209.34) --
	( 90.54,209.34) --
	( 91.21,208.40) --
	( 93.25,205.03) --
	( 93.25,205.03) --
	( 93.25,205.03) --
	( 93.25,205.03) --
	( 93.93,203.77) --
	( 93.93,203.77) --
	( 94.60,202.47) --
	( 94.60,202.47) --
	( 95.28,201.13) --
	( 95.28,201.13) --
	( 95.96,199.78) --
	( 95.96,199.78) --
	( 95.96,199.78) --
	( 96.64,198.43) --
	( 96.64,198.43) --
	( 97.32,197.10) --
	( 97.99,195.80) --
	( 97.99,195.80) --
	( 97.99,195.80) --
	(100.71,191.13) --
	(100.71,191.13) --
	(100.71,191.13) --
	(100.71,191.13) --
	(101.38,190.14) --
	(101.38,190.14) --
	(104.10,186.89) --
	(104.10,186.89) --
	(106.13,185.14) --
	(106.81,184.66) --
	(106.81,184.66) --
	(107.49,184.22) --
	(109.52,183.15) --
	(110.20,182.86) --
	(112.91,182.13) --
	(113.59,182.07) --
	(114.94,182.16) --
	(115.62,182.31) --
	(119.69,185.12) --
	(120.37,185.92) --
	(122.40,188.78) --
	(123.08,189.88) --
	(123.08,189.88) --
	(124.44,192.21) --
	(125.79,194.68) --
	(125.79,194.68) --
	(127.15,197.21) --
	(127.15,197.21) --
	(127.83,198.47) --
	(129.18,200.96) --
	(129.86,202.17) --
	(129.86,202.17) --
	(130.54,203.35) --
	(132.57,206.74) --
	(133.25,207.80) --
	(133.25,207.80) --
	(133.25,207.80) --
	(134.61,209.84) --
	(136.64,212.66) --
	(137.32,213.54) --
	(137.32,213.54) --
	(138.00,214.38) --
	(140.71,217.37) --
	(143.42,219.64) --
	(146.13,221.06) --
	(146.81,221.27) --
	(149.52,221.61) --
	(149.52,221.61) --
	(152.24,221.30) --
	(154.27,220.79) --
	(155.63,220.38) --
	(157.66,219.69) --
	(159.02,219.20) --
	(160.37,218.70) --
	(160.37,218.70) --
	(161.05,218.44) --
	(161.73,218.19) --
	(161.73,218.19) --
	(163.76,217.42) --
	(163.76,217.42) --
	(169.86,215.34) --
	(169.86,215.34) --
	(173.93,214.49) --
	(174.61,214.40) --
	(176.64,214.22) --
	(178.00,214.18) --
	(182.07,214.30) --
	(182.07,214.30) --
	(184.78,214.53) --
	(186.14,214.64) --
	(186.14,214.64) --
	(186.81,214.68) --
	(190.20,214.69) --
	(191.56,214.54) --
	(193.59,214.09) --
	(199.02,211.91) --
	(203.09,210.50) --
	(203.09,210.50) --
	(206.48,210.36) --
	(212.58,212.00) --
	(217.33,212.86) --
	(221.39,212.86) --
	(227.50,213.42) --
	(227.50,213.42) --
	(228.85,213.74) --
	(236.31,215.03);

\path[draw=drawColor,line width= 0.7pt,line join=round] ( 49.18,211.94) --
	( 49.85,211.95) --
	( 51.89,211.95) --
	( 53.24,211.93) --
	( 54.60,211.90) --
	( 62.06,211.93) --
	( 63.41,211.98) --
	( 64.09,212.01) --
	( 67.48,212.06) --
	( 68.84,212.01) --
	( 70.87,211.84) --
	( 70.87,211.84) --
	( 73.59,211.51) --
	( 74.94,211.35) --
	( 75.62,211.28) --
	( 76.98,211.17) --
	( 78.33,211.13) --
	( 79.69,211.12) --
	( 85.79,209.96) --
	( 87.15,208.86) --
	( 87.82,208.12) --
	( 90.54,203.71) --
	( 90.54,203.71) --
	( 90.54,203.71) --
	( 90.54,203.71) --
	( 90.54,203.71) --
	( 90.54,203.71) --
	( 91.21,202.24) --
	( 93.25,197.03) --
	( 93.25,197.03) --
	( 93.25,197.03) --
	( 93.25,197.03) --
	( 93.93,195.08) --
	( 93.93,195.08) --
	( 94.60,193.06) --
	( 94.60,193.06) --
	( 95.28,190.99) --
	( 95.28,190.99) --
	( 95.96,188.90) --
	( 95.96,188.90) --
	( 95.96,188.90) --
	( 96.64,186.83) --
	( 96.64,186.83) --
	( 97.32,184.80) --
	( 97.99,182.84) --
	( 97.99,182.84) --
	( 97.99,182.84) --
	(100.71,176.17) --
	(100.71,176.17) --
	(100.71,176.17) --
	(100.71,176.17) --
	(101.38,174.87) --
	(101.38,174.87) --
	(104.10,171.20) --
	(104.10,171.20) --
	(106.13,169.76) --
	(106.81,169.45) --
	(106.81,169.45) --
	(107.49,169.21) --
	(109.52,168.75) --
	(110.20,168.67) --
	(112.91,168.60) --
	(113.59,168.67) --
	(114.94,168.95) --
	(115.62,169.17) --
	(119.69,171.92) --
	(120.37,172.63) --
	(122.40,175.11) --
	(123.08,176.03) --
	(123.08,176.03) --
	(124.44,177.99) --
	(125.79,180.02) --
	(125.79,180.02) --
	(127.15,182.08) --
	(127.15,182.08) --
	(127.83,183.10) --
	(129.18,185.10) --
	(129.86,186.08) --
	(129.86,186.08) --
	(130.54,187.04) --
	(132.57,189.82) --
	(133.25,190.71) --
	(133.25,190.71) --
	(133.25,190.71) --
	(134.61,192.43) --
	(136.64,194.89) --
	(137.32,195.67) --
	(137.32,195.67) --
	(138.00,196.42) --
	(140.71,199.15) --
	(143.42,201.25) --
	(146.13,202.48) --
	(146.81,202.64) --
	(149.52,202.64) --
	(149.52,202.64) --
	(152.24,201.72) --
	(154.27,200.57) --
	(155.63,199.67) --
	(157.66,198.29) --
	(159.02,197.45) --
	(160.37,196.76) --
	(160.37,196.76) --
	(161.05,196.49) --
	(161.73,196.28) --
	(161.73,196.28) --
	(163.76,196.06) --
	(163.76,196.06) --
	(169.86,198.17) --
	(169.86,198.17) --
	(173.93,199.94) --
	(174.61,200.20) --
	(176.64,200.93) --
	(178.00,201.42) --
	(182.07,203.12) --
	(182.07,203.12) --
	(184.78,204.39) --
	(186.14,204.98) --
	(186.14,204.98) --
	(186.81,205.25) --
	(190.20,206.15) --
	(191.56,206.23) --
	(193.59,206.01) --
	(199.02,204.10) --
	(203.09,202.89) --
	(203.09,202.89) --
	(206.48,202.94) --
	(212.58,204.68) --
	(217.33,205.59) --
	(221.39,205.81) --
	(227.50,206.80) --
	(227.50,206.80) --
	(228.85,207.17) --
	(236.31,208.23);
\definecolor{fillColor}{RGB}{55,126,184}

\path[fill=fillColor,fill opacity=0.30] ( 49.18,213.49) --
	( 49.85,213.45) --
	( 51.89,213.29) --
	( 53.24,213.19) --
	( 54.60,213.09) --
	( 62.06,212.87) --
	( 63.41,212.91) --
	( 64.09,212.93) --
	( 67.48,213.01) --
	( 68.84,212.99) --
	( 70.87,212.90) --
	( 70.87,212.90) --
	( 73.59,212.75) --
	( 74.94,212.72) --
	( 75.62,212.73) --
	( 76.98,212.80) --
	( 78.33,212.96) --
	( 79.69,213.21) --
	( 85.79,214.02) --
	( 87.15,213.67) --
	( 87.82,213.36) --
	( 90.54,211.21) --
	( 90.54,211.21) --
	( 90.54,211.21) --
	( 90.54,211.21) --
	( 90.54,211.21) --
	( 90.54,211.21) --
	( 91.21,210.44) --
	( 93.25,207.69) --
	( 93.25,207.69) --
	( 93.25,207.69) --
	( 93.25,207.69) --
	( 93.93,206.66) --
	( 93.93,206.66) --
	( 94.60,205.59) --
	( 94.60,205.59) --
	( 95.28,204.49) --
	( 95.28,204.49) --
	( 95.96,203.39) --
	( 95.96,203.39) --
	( 95.96,203.39) --
	( 96.64,202.28) --
	( 96.64,202.28) --
	( 97.32,201.18) --
	( 97.99,200.10) --
	( 97.99,200.10) --
	( 97.99,200.10) --
	(100.71,196.10) --
	(100.71,196.10) --
	(100.71,196.10) --
	(100.71,196.10) --
	(101.38,195.20) --
	(101.38,195.20) --
	(104.10,192.09) --
	(104.10,192.09) --
	(106.13,190.24) --
	(106.81,189.70) --
	(106.81,189.70) --
	(107.49,189.20) --
	(109.52,187.92) --
	(110.20,187.57) --
	(112.91,186.61) --
	(113.59,186.51) --
	(114.94,186.54) --
	(115.62,186.67) --
	(119.69,189.50) --
	(120.37,190.32) --
	(122.40,193.32) --
	(123.08,194.47) --
	(123.08,194.47) --
	(124.44,196.93) --
	(125.79,199.55) --
	(125.79,199.55) --
	(127.15,202.23) --
	(127.15,202.23) --
	(127.83,203.57) --
	(129.18,206.21) --
	(129.86,207.50) --
	(129.86,207.50) --
	(130.54,208.77) --
	(132.57,212.35) --
	(133.25,213.48) --
	(133.25,213.48) --
	(133.25,213.48) --
	(134.61,215.61) --
	(136.64,218.56) --
	(137.32,219.46) --
	(137.32,219.46) --
	(138.00,220.33) --
	(140.71,223.41) --
	(143.42,225.74) --
	(146.13,227.22) --
	(146.81,227.45) --
	(149.52,227.91) --
	(149.52,227.91) --
	(152.24,227.80) --
	(154.27,227.50) --
	(155.63,227.24) --
	(157.66,226.78) --
	(159.02,226.41) --
	(160.37,225.97) --
	(160.37,225.97) --
	(161.05,225.72) --
	(161.73,225.45) --
	(161.73,225.45) --
	(163.76,224.50) --
	(163.76,224.50) --
	(169.86,221.03) --
	(169.86,221.03) --
	(173.93,219.31) --
	(174.61,219.11) --
	(176.64,218.63) --
	(178.00,218.41) --
	(182.07,218.01) --
	(182.07,218.01) --
	(184.78,217.89) --
	(186.14,217.84) --
	(186.14,217.84) --
	(186.81,217.81) --
	(190.20,217.53) --
	(191.56,217.30) --
	(193.59,216.77) --
	(199.02,214.49) --
	(203.09,213.02) --
	(203.09,213.02) --
	(206.48,212.82) --
	(212.58,214.42) --
	(217.33,215.27) --
	(221.39,215.20) --
	(227.50,215.61) --
	(227.50,215.61) --
	(228.85,215.92) --
	(236.31,217.29) --
	(236.31,212.71) --
	(228.85,211.49) --
	(227.50,211.15) --
	(227.50,211.15) --
	(221.39,210.45) --
	(217.33,210.37) --
	(212.58,209.50) --
	(206.48,207.82) --
	(203.09,207.90) --
	(203.09,207.90) --
	(199.02,209.24) --
	(193.59,211.33) --
	(191.56,211.70) --
	(190.20,211.77) --
	(186.81,211.46) --
	(186.14,211.34) --
	(186.14,211.34) --
	(184.78,211.06) --
	(182.07,210.48) --
	(182.07,210.48) --
	(178.00,209.82) --
	(176.64,209.68) --
	(174.61,209.54) --
	(173.93,209.52) --
	(169.86,209.47) --
	(169.86,209.47) --
	(163.76,210.12) --
	(163.76,210.12) --
	(161.73,210.70) --
	(161.73,210.70) --
	(161.05,210.94) --
	(160.37,211.20) --
	(160.37,211.20) --
	(159.02,211.76) --
	(157.66,212.37) --
	(155.63,213.30) --
	(154.27,213.88) --
	(152.24,214.61) --
	(149.52,215.13) --
	(149.52,215.13) --
	(146.81,214.90) --
	(146.13,214.71) --
	(143.42,213.35) --
	(140.71,211.14) --
	(138.00,208.24) --
	(137.32,207.43) --
	(137.32,207.43) --
	(136.64,206.59) --
	(134.61,203.89) --
	(133.25,201.96) --
	(133.25,201.96) --
	(133.25,201.96) --
	(132.57,200.95) --
	(130.54,197.78) --
	(129.86,196.67) --
	(129.86,196.67) --
	(129.18,195.54) --
	(127.83,193.22) --
	(127.15,192.04) --
	(127.15,192.04) --
	(125.79,189.67) --
	(125.79,189.67) --
	(124.44,187.35) --
	(123.08,185.14) --
	(123.08,185.14) --
	(122.40,184.11) --
	(120.37,181.37) --
	(119.69,180.61) --
	(115.62,177.82) --
	(114.94,177.64) --
	(113.59,177.49) --
	(112.91,177.50) --
	(110.20,178.01) --
	(109.52,178.23) --
	(107.49,179.09) --
	(106.81,179.46) --
	(106.81,179.46) --
	(106.13,179.88) --
	(104.10,181.52) --
	(104.10,181.52) --
	(101.38,184.92) --
	(101.38,184.92) --
	(100.71,186.02) --
	(100.71,186.02) --
	(100.71,186.02) --
	(100.71,186.02) --
	( 97.99,191.37) --
	( 97.99,191.37) --
	( 97.99,191.37) --
	( 97.32,192.90) --
	( 96.64,194.47) --
	( 96.64,194.47) --
	( 95.96,196.06) --
	( 95.96,196.06) --
	( 95.96,196.06) --
	( 95.28,197.66) --
	( 95.28,197.66) --
	( 94.60,199.25) --
	( 94.60,199.25) --
	( 93.93,200.80) --
	( 93.93,200.80) --
	( 93.25,202.30) --
	( 93.25,202.30) --
	( 93.25,202.30) --
	( 93.25,202.30) --
	( 91.21,206.29) --
	( 90.54,207.42) --
	( 90.54,207.42) --
	( 90.54,207.42) --
	( 90.54,207.42) --
	( 90.54,207.42) --
	( 90.54,207.42) --
	( 87.82,210.71) --
	( 87.15,211.24) --
	( 85.79,211.97) --
	( 79.69,212.15) --
	( 78.33,212.03) --
	( 76.98,211.98) --
	( 75.62,211.99) --
	( 74.94,212.03) --
	( 73.59,212.12) --
	( 70.87,212.37) --
	( 70.87,212.37) --
	( 68.84,212.49) --
	( 67.48,212.53) --
	( 64.09,212.46) --
	( 63.41,212.44) --
	( 62.06,212.39) --
	( 54.60,212.49) --
	( 53.24,212.55) --
	( 51.89,212.61) --
	( 49.85,212.69) --
	( 49.18,212.71) --
	cycle;

\path[] ( 49.18,213.49) --
	( 49.85,213.45) --
	( 51.89,213.29) --
	( 53.24,213.19) --
	( 54.60,213.09) --
	( 62.06,212.87) --
	( 63.41,212.91) --
	( 64.09,212.93) --
	( 67.48,213.01) --
	( 68.84,212.99) --
	( 70.87,212.90) --
	( 70.87,212.90) --
	( 73.59,212.75) --
	( 74.94,212.72) --
	( 75.62,212.73) --
	( 76.98,212.80) --
	( 78.33,212.96) --
	( 79.69,213.21) --
	( 85.79,214.02) --
	( 87.15,213.67) --
	( 87.82,213.36) --
	( 90.54,211.21) --
	( 90.54,211.21) --
	( 90.54,211.21) --
	( 90.54,211.21) --
	( 90.54,211.21) --
	( 90.54,211.21) --
	( 91.21,210.44) --
	( 93.25,207.69) --
	( 93.25,207.69) --
	( 93.25,207.69) --
	( 93.25,207.69) --
	( 93.93,206.66) --
	( 93.93,206.66) --
	( 94.60,205.59) --
	( 94.60,205.59) --
	( 95.28,204.49) --
	( 95.28,204.49) --
	( 95.96,203.39) --
	( 95.96,203.39) --
	( 95.96,203.39) --
	( 96.64,202.28) --
	( 96.64,202.28) --
	( 97.32,201.18) --
	( 97.99,200.10) --
	( 97.99,200.10) --
	( 97.99,200.10) --
	(100.71,196.10) --
	(100.71,196.10) --
	(100.71,196.10) --
	(100.71,196.10) --
	(101.38,195.20) --
	(101.38,195.20) --
	(104.10,192.09) --
	(104.10,192.09) --
	(106.13,190.24) --
	(106.81,189.70) --
	(106.81,189.70) --
	(107.49,189.20) --
	(109.52,187.92) --
	(110.20,187.57) --
	(112.91,186.61) --
	(113.59,186.51) --
	(114.94,186.54) --
	(115.62,186.67) --
	(119.69,189.50) --
	(120.37,190.32) --
	(122.40,193.32) --
	(123.08,194.47) --
	(123.08,194.47) --
	(124.44,196.93) --
	(125.79,199.55) --
	(125.79,199.55) --
	(127.15,202.23) --
	(127.15,202.23) --
	(127.83,203.57) --
	(129.18,206.21) --
	(129.86,207.50) --
	(129.86,207.50) --
	(130.54,208.77) --
	(132.57,212.35) --
	(133.25,213.48) --
	(133.25,213.48) --
	(133.25,213.48) --
	(134.61,215.61) --
	(136.64,218.56) --
	(137.32,219.46) --
	(137.32,219.46) --
	(138.00,220.33) --
	(140.71,223.41) --
	(143.42,225.74) --
	(146.13,227.22) --
	(146.81,227.45) --
	(149.52,227.91) --
	(149.52,227.91) --
	(152.24,227.80) --
	(154.27,227.50) --
	(155.63,227.24) --
	(157.66,226.78) --
	(159.02,226.41) --
	(160.37,225.97) --
	(160.37,225.97) --
	(161.05,225.72) --
	(161.73,225.45) --
	(161.73,225.45) --
	(163.76,224.50) --
	(163.76,224.50) --
	(169.86,221.03) --
	(169.86,221.03) --
	(173.93,219.31) --
	(174.61,219.11) --
	(176.64,218.63) --
	(178.00,218.41) --
	(182.07,218.01) --
	(182.07,218.01) --
	(184.78,217.89) --
	(186.14,217.84) --
	(186.14,217.84) --
	(186.81,217.81) --
	(190.20,217.53) --
	(191.56,217.30) --
	(193.59,216.77) --
	(199.02,214.49) --
	(203.09,213.02) --
	(203.09,213.02) --
	(206.48,212.82) --
	(212.58,214.42) --
	(217.33,215.27) --
	(221.39,215.20) --
	(227.50,215.61) --
	(227.50,215.61) --
	(228.85,215.92) --
	(236.31,217.29);

\path[] (236.31,212.71) --
	(228.85,211.49) --
	(227.50,211.15) --
	(227.50,211.15) --
	(221.39,210.45) --
	(217.33,210.37) --
	(212.58,209.50) --
	(206.48,207.82) --
	(203.09,207.90) --
	(203.09,207.90) --
	(199.02,209.24) --
	(193.59,211.33) --
	(191.56,211.70) --
	(190.20,211.77) --
	(186.81,211.46) --
	(186.14,211.34) --
	(186.14,211.34) --
	(184.78,211.06) --
	(182.07,210.48) --
	(182.07,210.48) --
	(178.00,209.82) --
	(176.64,209.68) --
	(174.61,209.54) --
	(173.93,209.52) --
	(169.86,209.47) --
	(169.86,209.47) --
	(163.76,210.12) --
	(163.76,210.12) --
	(161.73,210.70) --
	(161.73,210.70) --
	(161.05,210.94) --
	(160.37,211.20) --
	(160.37,211.20) --
	(159.02,211.76) --
	(157.66,212.37) --
	(155.63,213.30) --
	(154.27,213.88) --
	(152.24,214.61) --
	(149.52,215.13) --
	(149.52,215.13) --
	(146.81,214.90) --
	(146.13,214.71) --
	(143.42,213.35) --
	(140.71,211.14) --
	(138.00,208.24) --
	(137.32,207.43) --
	(137.32,207.43) --
	(136.64,206.59) --
	(134.61,203.89) --
	(133.25,201.96) --
	(133.25,201.96) --
	(133.25,201.96) --
	(132.57,200.95) --
	(130.54,197.78) --
	(129.86,196.67) --
	(129.86,196.67) --
	(129.18,195.54) --
	(127.83,193.22) --
	(127.15,192.04) --
	(127.15,192.04) --
	(125.79,189.67) --
	(125.79,189.67) --
	(124.44,187.35) --
	(123.08,185.14) --
	(123.08,185.14) --
	(122.40,184.11) --
	(120.37,181.37) --
	(119.69,180.61) --
	(115.62,177.82) --
	(114.94,177.64) --
	(113.59,177.49) --
	(112.91,177.50) --
	(110.20,178.01) --
	(109.52,178.23) --
	(107.49,179.09) --
	(106.81,179.46) --
	(106.81,179.46) --
	(106.13,179.88) --
	(104.10,181.52) --
	(104.10,181.52) --
	(101.38,184.92) --
	(101.38,184.92) --
	(100.71,186.02) --
	(100.71,186.02) --
	(100.71,186.02) --
	(100.71,186.02) --
	( 97.99,191.37) --
	( 97.99,191.37) --
	( 97.99,191.37) --
	( 97.32,192.90) --
	( 96.64,194.47) --
	( 96.64,194.47) --
	( 95.96,196.06) --
	( 95.96,196.06) --
	( 95.96,196.06) --
	( 95.28,197.66) --
	( 95.28,197.66) --
	( 94.60,199.25) --
	( 94.60,199.25) --
	( 93.93,200.80) --
	( 93.93,200.80) --
	( 93.25,202.30) --
	( 93.25,202.30) --
	( 93.25,202.30) --
	( 93.25,202.30) --
	( 91.21,206.29) --
	( 90.54,207.42) --
	( 90.54,207.42) --
	( 90.54,207.42) --
	( 90.54,207.42) --
	( 90.54,207.42) --
	( 90.54,207.42) --
	( 87.82,210.71) --
	( 87.15,211.24) --
	( 85.79,211.97) --
	( 79.69,212.15) --
	( 78.33,212.03) --
	( 76.98,211.98) --
	( 75.62,211.99) --
	( 74.94,212.03) --
	( 73.59,212.12) --
	( 70.87,212.37) --
	( 70.87,212.37) --
	( 68.84,212.49) --
	( 67.48,212.53) --
	( 64.09,212.46) --
	( 63.41,212.44) --
	( 62.06,212.39) --
	( 54.60,212.49) --
	( 53.24,212.55) --
	( 51.89,212.61) --
	( 49.85,212.69) --
	( 49.18,212.71);

\path[draw=drawColor,line width= 0.7pt,line join=round] ( 49.18,214.27) --
	( 49.85,214.19) --
	( 51.89,213.96) --
	( 53.24,213.82) --
	( 54.60,213.68) --
	( 62.06,213.35) --
	( 63.41,213.38) --
	( 64.09,213.40) --
	( 67.48,213.48) --
	( 68.84,213.48) --
	( 70.87,213.44) --
	( 70.87,213.44) --
	( 73.59,213.38) --
	( 74.94,213.41) --
	( 75.62,213.45) --
	( 76.98,213.61) --
	( 78.33,213.88) --
	( 79.69,214.26) --
	( 85.79,216.06) --
	( 87.15,216.08) --
	( 87.82,215.99) --
	( 90.54,214.96) --
	( 90.54,214.96) --
	( 90.54,214.96) --
	( 90.54,214.96) --
	( 90.54,214.96) --
	( 90.54,214.96) --
	( 91.21,214.55) --
	( 93.25,213.04) --
	( 93.25,213.04) --
	( 93.25,213.04) --
	( 93.25,213.04) --
	( 93.93,212.47) --
	( 93.93,212.47) --
	( 94.60,211.88) --
	( 94.60,211.88) --
	( 95.28,211.27) --
	( 95.28,211.27) --
	( 95.96,210.66) --
	( 95.96,210.66) --
	( 95.96,210.66) --
	( 96.64,210.04) --
	( 96.64,210.04) --
	( 97.32,209.41) --
	( 97.99,208.77) --
	( 97.99,208.77) --
	( 97.99,208.77) --
	(100.71,206.10) --
	(100.71,206.10) --
	(100.71,206.10) --
	(100.71,206.10) --
	(101.38,205.41) --
	(101.38,205.41) --
	(104.10,202.58) --
	(104.10,202.58) --
	(106.13,200.52) --
	(106.81,199.86) --
	(106.81,199.86) --
	(107.49,199.24) --
	(109.52,197.54) --
	(110.20,197.05) --
	(112.91,195.65) --
	(113.59,195.47) --
	(114.94,195.36) --
	(115.62,195.46) --
	(119.69,198.33) --
	(120.37,199.21) --
	(122.40,202.46) --
	(123.08,203.72) --
	(123.08,203.72) --
	(124.44,206.44) --
	(125.79,209.35) --
	(125.79,209.35) --
	(127.15,212.34) --
	(127.15,212.34) --
	(127.83,213.85) --
	(129.18,216.81) --
	(129.86,218.26) --
	(129.86,218.26) --
	(130.54,219.67) --
	(132.57,223.66) --
	(133.25,224.90) --
	(133.25,224.90) --
	(133.25,224.90) --
	(134.61,227.25) --
	(136.64,230.43) --
	(137.32,231.41) --
	(137.32,231.41) --
	(138.00,232.33) --
	(140.71,235.59) --
	(143.42,238.04) --
	(146.13,239.63) --
	(146.81,239.90) --
	(149.52,240.59) --
	(149.52,240.59) --
	(152.24,240.89) --
	(154.27,241.02) --
	(155.63,241.08) --
	(157.66,241.08) --
	(159.02,240.95) --
	(160.37,240.64) --
	(160.37,240.64) --
	(161.05,240.40) --
	(161.73,240.09) --
	(161.73,240.09) --
	(163.76,238.77) --
	(163.76,238.77) --
	(169.86,232.51) --
	(169.86,232.51) --
	(173.93,229.03) --
	(174.61,228.60) --
	(176.64,227.52) --
	(178.00,226.93) --
	(182.07,225.49) --
	(182.07,225.49) --
	(184.78,224.67) --
	(186.14,224.29) --
	(186.14,224.29) --
	(186.81,224.11) --
	(190.20,223.24) --
	(191.56,222.85) --
	(193.59,222.17) --
	(199.02,219.71) --
	(203.09,218.10) --
	(203.09,218.10) --
	(206.48,217.78) --
	(212.58,219.31) --
	(217.33,220.13) --
	(221.39,219.91) --
	(227.50,220.04) --
	(227.50,220.04) --
	(228.85,220.31) --
	(236.31,221.84);
\definecolor{drawColor}{RGB}{0,0,0}

\path[draw=drawColor,line width= 1.1pt,line join=round,line cap=round] ( 39.82,145.31) rectangle (245.67,258.53);
\end{scope}
\begin{scope}
\definecolor{drawColor}{RGB}{0,0,0}

\node[text=drawColor,anchor=base east,inner sep=0pt, outer sep=0pt, scale=  0.84] at ( 33.32,153.28) {-200};

\node[text=drawColor,anchor=base east,inner sep=0pt, outer sep=0pt, scale=  0.84] at ( 33.32,181.87) {-100};

\node[text=drawColor,anchor=base east,inner sep=0pt, outer sep=0pt, scale=  0.84] at ( 33.32,210.46) {0};

\node[text=drawColor,anchor=base east,inner sep=0pt, outer sep=0pt, scale=  0.84] at ( 33.32,239.05) {100};
\end{scope}
\begin{scope}
\definecolor{drawColor}{RGB}{0,0,0}

\path[draw=drawColor,line width= 0.6pt,line join=round] ( 36.32,156.18) --
	( 39.82,156.18);

\path[draw=drawColor,line width= 0.6pt,line join=round] ( 36.32,184.77) --
	( 39.82,184.77);

\path[draw=drawColor,line width= 0.6pt,line join=round] ( 36.32,213.36) --
	( 39.82,213.36);

\path[draw=drawColor,line width= 0.6pt,line join=round] ( 36.32,241.95) --
	( 39.82,241.95);
\end{scope}
\begin{scope}
\definecolor{drawColor}{RGB}{0,0,0}

\path[draw=drawColor,line width= 0.6pt,line join=round] ( 41.04,141.81) --
	( 41.04,145.31);

\path[draw=drawColor,line width= 0.6pt,line join=round] (108.84,141.81) --
	(108.84,145.31);

\path[draw=drawColor,line width= 0.6pt,line join=round] (176.64,141.81) --
	(176.64,145.31);

\path[draw=drawColor,line width= 0.6pt,line join=round] (244.45,141.81) --
	(244.45,145.31);
\end{scope}
\begin{scope}
\definecolor{drawColor}{RGB}{0,0,0}

\node[text=drawColor,anchor=base,inner sep=0pt, outer sep=0pt, scale=  0.84] at ( 41.04,133.03) {0};

\node[text=drawColor,anchor=base,inner sep=0pt, outer sep=0pt, scale=  0.84] at (108.84,133.03) {20};

\node[text=drawColor,anchor=base,inner sep=0pt, outer sep=0pt, scale=  0.84] at (176.64,133.03) {40};

\node[text=drawColor,anchor=base,inner sep=0pt, outer sep=0pt, scale=  0.84] at (244.45,133.03) {60};
\end{scope}
\begin{scope}
\definecolor{drawColor}{RGB}{0,0,0}

\node[text=drawColor,anchor=base,inner sep=0pt, outer sep=0pt, scale=  0.84] at (142.74,122.11) {Time (ms)};
\end{scope}
\begin{scope}
\definecolor{drawColor}{RGB}{0,0,0}

\node[text=drawColor,rotate= 90.00,anchor=base,inner sep=0pt, outer sep=0pt, scale=  0.84] at ( 12.79,201.92) {Acceleration (g)};
\end{scope}
\end{tikzpicture}

%% file: figs/covid_liksens.tex
% Created by tikzDevice version 0.12.3.1 on 2022-11-29 12:01:45
% !TEX encoding = UTF-8 Unicode
\begin{tikzpicture}[x=1pt,y=1pt]
\definecolor{fillColor}{RGB}{255,255,255}
\begin{scope}
\definecolor{drawColor}{RGB}{0,0,0}

\path[draw=drawColor,line width= 1.1pt,line join=round] ( 71.14,293.93) --
	( 71.72,295.95) --
	( 72.29,295.95) --
	( 72.86,295.95) --
	( 73.44,295.95) --
	( 74.01,295.95) --
	( 74.58,295.95) --
	( 75.16,295.95) --
	( 75.73,295.94) --
	( 76.30,295.93) --
	( 76.88,295.90) --
	( 77.45,295.85) --
	( 78.03,295.78) --
	( 78.60,295.69) --
	( 79.17,295.57) --
	( 79.75,295.43) --
	( 80.32,295.27) --
	( 80.89,295.12) --
	( 81.47,294.96) --
	( 82.04,294.83) --
	( 82.61,294.72) --
	( 83.19,294.69) --
	( 83.76,294.78) --
	( 84.33,294.99) --
	( 84.91,295.31) --
	( 85.48,295.83) --
	( 86.05,296.56) --
	( 86.63,297.47) --
	( 87.20,299.03) --
	( 87.77,300.52) --
	( 88.35,302.43) --
	( 88.92,304.10) --
	( 89.49,305.68) --
	( 90.07,306.87) --
	( 90.64,307.88) --
	( 91.22,308.63) --
	( 91.79,309.26) --
	( 92.36,309.98) --
	( 92.94,310.75) --
	( 93.51,311.52) --
	( 94.08,312.27) --
	( 94.66,312.90) --
	( 95.23,313.48) --
	( 95.80,314.03) --
	( 96.38,314.51) --
	( 96.95,314.90) --
	( 97.52,315.19) --
	( 98.10,315.31) --
	( 98.67,315.31) --
	( 99.24,315.18) --
	( 99.82,314.92) --
	(100.39,314.54) --
	(100.96,314.04) --
	(101.54,313.41) --
	(102.11,312.69) --
	(102.68,311.87) --
	(103.26,310.96) --
	(103.83,309.96) --
	(104.41,308.89) --
	(104.98,307.77) --
	(105.55,306.60) --
	(106.13,305.37) --
	(106.70,304.09) --
	(107.27,302.81) --
	(107.85,301.53) --
	(108.42,300.49) --
	(108.99,300.15) --
	(109.57,299.51) --
	(110.14,298.77) --
	(110.71,297.88) --
	(111.29,296.77) --
	(111.86,295.48) --
	(112.43,294.15) --
	(113.01,294.05) --
	(113.58,295.92) --
	(114.15,298.17) --
	(114.73,300.47) --
	(115.30,302.56) --
	(115.87,303.81) --
	(116.45,304.66) --
	(117.02,305.13) --
	(117.59,305.10) --
	(118.17,304.85) --
	(118.74,304.45) --
	(119.32,304.05) --
	(119.89,303.56) --
	(120.46,303.09) --
	(121.04,302.54) --
	(121.61,301.97) --
	(122.18,301.77) --
	(122.76,301.98) --
	(123.33,302.18) --
	(123.90,302.38) --
	(124.48,302.54) --
	(125.05,302.65) --
	(125.62,302.74) --
	(126.20,302.80) --
	(126.77,302.84) --
	(127.34,302.92) --
	(127.92,303.06);

\path[draw=drawColor,line width= 0.6pt,dash pattern=on 1pt off 3pt ,line join=round] ( 68.30,296.77) -- (130.76,296.77);
\definecolor{drawColor}{RGB}{0,0,0}

\path[draw=drawColor,draw opacity=0.50,line width= 0.6pt,line join=round] ( 82.61,290.52) -- ( 82.61,352.97);
\definecolor{drawColor}{RGB}{0,0,0}

\path[draw=drawColor,line width= 0.6pt,line join=round,line cap=round] ( 68.30,290.52) rectangle (130.76,352.97);
\end{scope}
\begin{scope}
\definecolor{drawColor}{RGB}{0,0,0}

\path[draw=drawColor,line width= 1.1pt,line join=round] ( 71.14,208.42) --
	( 71.72,231.79) --
	( 72.29,231.79) --
	( 72.86,231.79) --
	( 73.44,231.79) --
	( 74.01,231.79) --
	( 74.58,231.79) --
	( 75.16,231.79) --
	( 75.73,231.78) --
	( 76.30,231.76) --
	( 76.88,231.72) --
	( 77.45,231.64) --
	( 78.03,231.53) --
	( 78.60,231.38) --
	( 79.17,231.21) --
	( 79.75,230.99) --
	( 80.32,230.76) --
	( 80.89,230.50) --
	( 81.47,230.21) --
	( 82.04,229.91) --
	( 82.61,229.60) --
	( 83.19,229.28) --
	( 83.76,228.95) --
	( 84.33,228.60) --
	( 84.91,228.17) --
	( 85.48,227.73) --
	( 86.05,227.29) --
	( 86.63,226.81) --
	( 87.20,226.33) --
	( 87.77,225.84) --
	( 88.35,225.15) --
	( 88.92,224.40) --
	( 89.49,223.59) --
	( 90.07,222.72) --
	( 90.64,221.73) --
	( 91.22,220.59) --
	( 91.79,219.22) --
	( 92.36,217.54) --
	( 92.94,215.42) --
	( 93.51,213.69) --
	( 94.08,214.11) --
	( 94.66,214.46) --
	( 95.23,214.76) --
	( 95.80,215.05) --
	( 96.38,215.31) --
	( 96.95,215.54) --
	( 97.52,215.75) --
	( 98.10,216.12) --
	( 98.67,216.55) --
	( 99.24,216.92) --
	( 99.82,217.25) --
	(100.39,217.53) --
	(100.96,217.78) --
	(101.54,217.99) --
	(102.11,218.16) --
	(102.68,218.30) --
	(103.26,218.38) --
	(103.83,218.44) --
	(104.41,218.46) --
	(104.98,218.45) --
	(105.55,218.39) --
	(106.13,218.29) --
	(106.70,218.54) --
	(107.27,218.90) --
	(107.85,219.28) --
	(108.42,219.67) --
	(108.99,220.07) --
	(109.57,220.46) --
	(110.14,220.81) --
	(110.71,221.11) --
	(111.29,221.27) --
	(111.86,220.89) --
	(112.43,220.20) --
	(113.01,218.87) --
	(113.58,216.36) --
	(114.15,213.70) --
	(114.73,211.17) --
	(115.30,209.48) --
	(115.87,208.48) --
	(116.45,208.69) --
	(117.02,209.74) --
	(117.59,210.76) --
	(118.17,212.01) --
	(118.74,213.21) --
	(119.32,214.32) --
	(119.89,215.36) --
	(120.46,216.29) --
	(121.04,217.18) --
	(121.61,218.03) --
	(122.18,218.82) --
	(122.76,219.56) --
	(123.33,220.26) --
	(123.90,220.91) --
	(124.48,221.53) --
	(125.05,222.11) --
	(125.62,222.67) --
	(126.20,223.18) --
	(126.77,223.67) --
	(127.34,224.14) --
	(127.92,224.58);

\path[draw=drawColor,line width= 0.6pt,dash pattern=on 1pt off 3pt ,line join=round] ( 68.30,211.49) -- (130.76,211.49);
\definecolor{drawColor}{RGB}{0,0,0}

\path[draw=drawColor,draw opacity=0.50,line width= 0.6pt,line join=round] ( 87.20,205.24) -- ( 87.20,267.69);
\definecolor{drawColor}{RGB}{0,0,0}

\path[draw=drawColor,line width= 0.6pt,line join=round,line cap=round] ( 68.30,205.24) rectangle (130.76,267.69);
\end{scope}
\begin{scope}
\definecolor{drawColor}{RGB}{0,0,0}

\path[draw=drawColor,line width= 1.1pt,line join=round] ( 71.14,123.61) --
	( 71.72,135.47) --
	( 72.29,135.47) --
	( 72.86,135.47) --
	( 73.44,135.47) --
	( 74.01,135.47) --
	( 74.58,135.47) --
	( 75.16,135.47) --
	( 75.73,135.46) --
	( 76.30,135.43) --
	( 76.88,135.37) --
	( 77.45,135.26) --
	( 78.03,135.10) --
	( 78.60,134.89) --
	( 79.17,134.62) --
	( 79.75,134.28) --
	( 80.32,133.89) --
	( 80.89,133.44) --
	( 81.47,132.93) --
	( 82.04,132.36) --
	( 82.61,131.72) --
	( 83.19,131.01) --
	( 83.76,130.25) --
	( 84.33,129.44) --
	( 84.91,128.58) --
	( 85.48,127.83) --
	( 86.05,127.20) --
	( 86.63,126.96) --
	( 87.20,127.54) --
	( 87.77,129.88) --
	( 88.35,132.03) --
	( 88.92,133.62) --
	( 89.49,136.32) --
	( 90.07,139.11) --
	( 90.64,141.75) --
	( 91.22,144.14) --
	( 91.79,146.13) --
	( 92.36,147.64) --
	( 92.94,148.96) --
	( 93.51,150.12) --
	( 94.08,151.09) --
	( 94.66,151.84) --
	( 95.23,152.48) --
	( 95.80,152.98) --
	( 96.38,153.33) --
	( 96.95,153.50) --
	( 97.52,153.49) --
	( 98.10,153.29) --
	( 98.67,152.91) --
	( 99.24,152.36) --
	( 99.82,151.66) --
	(100.39,150.81) --
	(100.96,149.85) --
	(101.54,148.78) --
	(102.11,147.61) --
	(102.68,146.37) --
	(103.26,145.06) --
	(103.83,143.68) --
	(104.41,142.27) --
	(104.98,140.80) --
	(105.55,139.28) --
	(106.13,137.72) --
	(106.70,137.08) --
	(107.27,136.43) --
	(107.85,135.54) --
	(108.42,134.34) --
	(108.99,132.74) --
	(109.57,130.66) --
	(110.14,127.75) --
	(110.71,124.89) --
	(111.29,125.55) --
	(111.86,129.55) --
	(112.43,132.75) --
	(113.01,135.03) --
	(113.58,136.49) --
	(114.15,137.38) --
	(114.73,137.94) --
	(115.30,138.30) --
	(115.87,138.52) --
	(116.45,138.67) --
	(117.02,138.76) --
	(117.59,138.38) --
	(118.17,138.29) --
	(118.74,138.91) --
	(119.32,139.49) --
	(119.89,140.03) --
	(120.46,140.54) --
	(121.04,141.04) --
	(121.61,141.51) --
	(122.18,141.97) --
	(122.76,142.41) --
	(123.33,142.83) --
	(123.90,143.24) --
	(124.48,143.63) --
	(125.05,144.01) --
	(125.62,144.38) --
	(126.20,144.75) --
	(126.77,145.11) --
	(127.34,145.46) --
	(127.92,145.81);

\path[draw=drawColor,line width= 0.6pt,dash pattern=on 1pt off 3pt ,line join=round] ( 68.30,126.21) -- (130.76,126.21);
\definecolor{drawColor}{RGB}{0,0,0}

\path[draw=drawColor,draw opacity=0.50,line width= 0.6pt,line join=round] ( 80.89,119.95) -- ( 80.89,182.40);
\definecolor{drawColor}{RGB}{0,0,0}

\path[draw=drawColor,line width= 0.6pt,line join=round,line cap=round] ( 68.30,119.95) rectangle (130.76,182.40);
\end{scope}
\begin{scope}
\definecolor{drawColor}{RGB}{0,0,0}

\path[draw=drawColor,line width= 1.1pt,line join=round] ( 71.14, 38.17) --
	( 71.72, 54.35) --
	( 72.29, 54.35) --
	( 72.86, 54.35) --
	( 73.44, 54.35) --
	( 74.01, 54.35) --
	( 74.58, 54.35) --
	( 75.16, 54.35) --
	( 75.73, 54.34) --
	( 76.30, 54.32) --
	( 76.88, 54.27) --
	( 77.45, 54.18) --
	( 78.03, 54.05) --
	( 78.60, 53.88) --
	( 79.17, 53.67) --
	( 79.75, 53.42) --
	( 80.32, 53.12) --
	( 80.89, 52.77) --
	( 81.47, 52.37) --
	( 82.04, 51.95) --
	( 82.61, 51.48) --
	( 83.19, 50.97) --
	( 83.76, 50.43) --
	( 84.33, 49.87) --
	( 84.91, 49.30) --
	( 85.48, 48.74) --
	( 86.05, 48.15) --
	( 86.63, 47.56) --
	( 87.20, 46.94) --
	( 87.77, 46.36) --
	( 88.35, 45.93) --
	( 88.92, 46.12) --
	( 89.49, 47.39) --
	( 90.07, 49.75) --
	( 90.64, 52.73) --
	( 91.22, 55.93) --
	( 91.79, 58.97) --
	( 92.36, 61.81) --
	( 92.94, 63.70) --
	( 93.51, 65.34) --
	( 94.08, 66.58) --
	( 94.66, 67.64) --
	( 95.23, 68.58) --
	( 95.80, 69.38) --
	( 96.38, 69.96) --
	( 96.95, 70.41) --
	( 97.52, 70.73) --
	( 98.10, 70.91) --
	( 98.67, 70.92) --
	( 99.24, 70.77) --
	( 99.82, 70.43) --
	(100.39, 69.92) --
	(100.96, 69.25) --
	(101.54, 68.44) --
	(102.11, 67.48) --
	(102.68, 66.41) --
	(103.26, 65.23) --
	(103.83, 63.96) --
	(104.41, 62.62) --
	(104.98, 61.20) --
	(105.55, 59.72) --
	(106.13, 58.19) --
	(106.70, 56.60) --
	(107.27, 54.96) --
	(107.85, 53.27) --
	(108.42, 52.11) --
	(108.99, 52.48) --
	(109.57, 52.44) --
	(110.14, 51.71) --
	(110.71, 50.67) --
	(111.29, 49.18) --
	(111.86, 46.55) --
	(112.43, 43.02) --
	(113.01, 39.42) --
	(113.58, 39.80) --
	(114.15, 43.64) --
	(114.73, 47.27) --
	(115.30, 50.15) --
	(115.87, 51.71) --
	(116.45, 52.68) --
	(117.02, 53.27) --
	(117.59, 53.64) --
	(118.17, 53.86) --
	(118.74, 53.98) --
	(119.32, 53.68) --
	(119.89, 52.83) --
	(120.46, 52.36) --
	(121.04, 53.06) --
	(121.61, 53.73) --
	(122.18, 54.37) --
	(122.76, 54.97) --
	(123.33, 55.55) --
	(123.90, 56.12) --
	(124.48, 56.66) --
	(125.05, 57.18) --
	(125.62, 57.69) --
	(126.20, 58.17) --
	(126.77, 58.64) --
	(127.34, 59.09) --
	(127.92, 59.54);

\path[draw=drawColor,line width= 0.6pt,dash pattern=on 1pt off 3pt ,line join=round] ( 68.30, 40.92) -- (130.76, 40.92);
\definecolor{drawColor}{RGB}{0,0,0}

\path[draw=drawColor,draw opacity=0.50,line width= 0.6pt,line join=round] ( 82.61, 34.67) -- ( 82.61, 97.12);
\definecolor{drawColor}{RGB}{0,0,0}

\path[draw=drawColor,line width= 0.6pt,line join=round,line cap=round] ( 68.30, 34.67) rectangle (130.76, 97.12);
\end{scope}
\begin{scope}
\definecolor{drawColor}{RGB}{0,0,0}

\path[draw=drawColor,line width= 1.1pt,line join=round] (140.59,295.26) --
	(141.17,329.35) --
	(141.74,329.35) --
	(142.32,329.35) --
	(142.89,329.35) --
	(143.46,329.35) --
	(144.04,329.35) --
	(144.61,329.35) --
	(145.18,329.34) --
	(145.76,329.31) --
	(146.33,329.24) --
	(146.90,329.14) --
	(147.48,328.98) --
	(148.05,328.76) --
	(148.62,328.50) --
	(149.20,328.19) --
	(149.77,327.83) --
	(150.34,327.43) --
	(150.92,326.99) --
	(151.49,326.52) --
	(152.06,326.01) --
	(152.64,325.48) --
	(153.21,324.92) --
	(153.78,324.34) --
	(154.36,323.74) --
	(154.93,323.12) --
	(155.51,322.49) --
	(156.08,321.83) --
	(156.65,321.17) --
	(157.23,320.43) --
	(157.80,319.49) --
	(158.37,318.47) --
	(158.95,317.21) --
	(159.52,315.70) --
	(160.09,314.04) --
	(160.67,312.22) --
	(161.24,310.19) --
	(161.81,307.93) --
	(162.39,305.50) --
	(162.96,303.06) --
	(163.53,301.87) --
	(164.11,304.49) --
	(164.68,307.29) --
	(165.25,309.76) --
	(165.83,311.85) --
	(166.40,314.86) --
	(166.97,317.97) --
	(167.55,320.54) --
	(168.12,322.00) --
	(168.70,323.16) --
	(169.27,324.11) --
	(169.84,324.90) --
	(170.42,325.57) --
	(170.99,326.13) --
	(171.56,326.60) --
	(172.14,326.99) --
	(172.71,327.30) --
	(173.28,327.53) --
	(173.86,327.68) --
	(174.43,327.74) --
	(175.00,327.54) --
	(175.58,327.25) --
	(176.15,326.87) --
	(176.72,326.42) --
	(177.30,325.88) --
	(177.87,325.25) --
	(178.44,324.53) --
	(179.02,323.72) --
	(179.59,322.83) --
	(180.16,321.85) --
	(180.74,320.80) --
	(181.31,319.67) --
	(181.88,318.44) --
	(182.46,317.50) --
	(183.03,316.87) --
	(183.61,315.96) --
	(184.18,314.67) --
	(184.75,312.57) --
	(185.33,309.74) --
	(185.90,306.29) --
	(186.47,302.26) --
	(187.05,299.07) --
	(187.62,296.04) --
	(188.19,295.73) --
	(188.77,298.92) --
	(189.34,301.81) --
	(189.91,303.89) --
	(190.49,305.52) --
	(191.06,306.81) --
	(191.63,307.85) --
	(192.21,308.95) --
	(192.78,310.54) --
	(193.35,312.00) --
	(193.93,313.33) --
	(194.50,314.57) --
	(195.07,315.72) --
	(195.65,316.79) --
	(196.22,317.80) --
	(196.80,318.74) --
	(197.37,319.64);

\path[draw=drawColor,line width= 0.6pt,dash pattern=on 1pt off 3pt ,line join=round] (137.76,296.77) -- (200.21,296.77);
\definecolor{drawColor}{RGB}{0,0,0}

\path[draw=drawColor,draw opacity=0.50,line width= 0.6pt,line join=round] (162.39,290.52) -- (162.39,352.97);
\definecolor{drawColor}{RGB}{0,0,0}

\path[draw=drawColor,line width= 0.6pt,line join=round,line cap=round] (137.76,290.52) rectangle (200.21,352.97);
\end{scope}
\begin{scope}
\definecolor{drawColor}{RGB}{0,0,0}

\path[draw=drawColor,line width= 1.1pt,line join=round] (140.59,208.16) --
	(141.17,243.76) --
	(141.74,243.76) --
	(142.32,243.76) --
	(142.89,243.76) --
	(143.46,243.76) --
	(144.04,243.76) --
	(144.61,243.76) --
	(145.18,243.75) --
	(145.76,243.72) --
	(146.33,243.64) --
	(146.90,243.52) --
	(147.48,243.34) --
	(148.05,243.11) --
	(148.62,242.81) --
	(149.20,242.46) --
	(149.77,242.07) --
	(150.34,241.63) --
	(150.92,241.15) --
	(151.49,240.63) --
	(152.06,240.08) --
	(152.64,239.46) --
	(153.21,238.82) --
	(153.78,238.16) --
	(154.36,237.47) --
	(154.93,236.76) --
	(155.51,236.02) --
	(156.08,235.26) --
	(156.65,234.41) --
	(157.23,233.42) --
	(157.80,232.36) --
	(158.37,231.22) --
	(158.95,229.99) --
	(159.52,228.64) --
	(160.09,227.16) --
	(160.67,225.51) --
	(161.24,223.38) --
	(161.81,220.95) --
	(162.39,218.43) --
	(162.96,216.03) --
	(163.53,215.45) --
	(164.11,218.25) --
	(164.68,220.56) --
	(165.25,222.33) --
	(165.83,223.81) --
	(166.40,224.54) --
	(166.97,225.92) --
	(167.55,227.07) --
	(168.12,227.96) --
	(168.70,228.69) --
	(169.27,229.31) --
	(169.84,229.83) --
	(170.42,230.26) --
	(170.99,230.62) --
	(171.56,230.91) --
	(172.14,231.13) --
	(172.71,231.28) --
	(173.28,231.36) --
	(173.86,231.36) --
	(174.43,231.28) --
	(175.00,231.11) --
	(175.58,230.86) --
	(176.15,230.52) --
	(176.72,230.11) --
	(177.30,229.62) --
	(177.87,229.05) --
	(178.44,228.65) --
	(179.02,229.52) --
	(179.59,230.41) --
	(180.16,231.32) --
	(180.74,232.23) --
	(181.31,232.41) --
	(181.88,231.91) --
	(182.46,230.36) --
	(183.03,228.01) --
	(183.61,224.34) --
	(184.18,219.92) --
	(184.75,215.55) --
	(185.33,212.82) --
	(185.90,210.41) --
	(186.47,211.48) --
	(187.05,213.40) --
	(187.62,215.30) --
	(188.19,217.02) --
	(188.77,218.36) --
	(189.34,219.47) --
	(189.91,219.95) --
	(190.49,221.49) --
	(191.06,222.97) --
	(191.63,224.34) --
	(192.21,225.59) --
	(192.78,226.75) --
	(193.35,227.82) --
	(193.93,228.82) --
	(194.50,229.76) --
	(195.07,230.64) --
	(195.65,231.47) --
	(196.22,232.27) --
	(196.80,233.02) --
	(197.37,233.75);

\path[draw=drawColor,line width= 0.6pt,dash pattern=on 1pt off 3pt ,line join=round] (137.76,211.49) -- (200.21,211.49);
\definecolor{drawColor}{RGB}{0,0,0}

\path[draw=drawColor,draw opacity=0.50,line width= 0.6pt,line join=round] (160.67,205.24) -- (160.67,267.69);
\definecolor{drawColor}{RGB}{0,0,0}

\path[draw=drawColor,line width= 0.6pt,line join=round,line cap=round] (137.76,205.24) rectangle (200.21,267.69);
\end{scope}
\begin{scope}
\definecolor{drawColor}{RGB}{0,0,0}

\path[draw=drawColor,line width= 1.1pt,line join=round] (140.59,124.26) --
	(141.17,143.92) --
	(141.74,143.92) --
	(142.32,143.92) --
	(142.89,143.92) --
	(143.46,143.92) --
	(144.04,143.92) --
	(144.61,143.92) --
	(145.18,143.91) --
	(145.76,143.89) --
	(146.33,143.84) --
	(146.90,143.76) --
	(147.48,143.64) --
	(148.05,143.49) --
	(148.62,143.29) --
	(149.20,143.06) --
	(149.77,142.81) --
	(150.34,142.52) --
	(150.92,142.21) --
	(151.49,141.89) --
	(152.06,141.55) --
	(152.64,141.20) --
	(153.21,140.85) --
	(153.78,140.50) --
	(154.36,140.12) --
	(154.93,139.65) --
	(155.51,139.18) --
	(156.08,138.69) --
	(156.65,138.19) --
	(157.23,137.67) --
	(157.80,137.13) --
	(158.37,136.53) --
	(158.95,135.97) --
	(159.52,135.30) --
	(160.09,134.64) --
	(160.67,133.82) --
	(161.24,132.59) --
	(161.81,131.08) --
	(162.39,129.81) --
	(162.96,128.65) --
	(163.53,127.50) --
	(164.11,126.88) --
	(164.68,126.64) --
	(165.25,126.70) --
	(165.83,126.95) --
	(166.40,127.26) --
	(166.97,127.70) --
	(167.55,128.16) --
	(168.12,128.51) --
	(168.70,128.86) --
	(169.27,129.27) --
	(169.84,129.66) --
	(170.42,130.00) --
	(170.99,130.31) --
	(171.56,130.59) --
	(172.14,130.82) --
	(172.71,130.99) --
	(173.28,131.11) --
	(173.86,131.19) --
	(174.43,131.23) --
	(175.00,131.23) --
	(175.58,131.19) --
	(176.15,131.13) --
	(176.72,131.02) --
	(177.30,130.87) --
	(177.87,130.77) --
	(178.44,131.27) --
	(179.02,131.82) --
	(179.59,132.43) --
	(180.16,133.09) --
	(180.74,133.82) --
	(181.31,134.59) --
	(181.88,134.78) --
	(182.46,134.87) --
	(183.03,134.87) --
	(183.61,134.68) --
	(184.18,133.41) --
	(184.75,131.55) --
	(185.33,129.14) --
	(185.90,126.43) --
	(186.47,124.03) --
	(187.05,123.98) --
	(187.62,124.83) --
	(188.19,125.81) --
	(188.77,126.81) --
	(189.34,127.78) --
	(189.91,128.69) --
	(190.49,129.58) --
	(191.06,130.44) --
	(191.63,131.28) --
	(192.21,132.10) --
	(192.78,132.95) --
	(193.35,133.76) --
	(193.93,134.55) --
	(194.50,135.30) --
	(195.07,136.06) --
	(195.65,136.82) --
	(196.22,137.54) --
	(196.80,138.27) --
	(197.37,138.98);

\path[draw=drawColor,line width= 0.6pt,dash pattern=on 1pt off 3pt ,line join=round] (137.76,126.21) -- (200.21,126.21);
\definecolor{drawColor}{RGB}{0,0,0}

\path[draw=drawColor,draw opacity=0.50,line width= 0.6pt,line join=round] (152.64,119.95) -- (152.64,182.40);
\definecolor{drawColor}{RGB}{0,0,0}

\path[draw=drawColor,line width= 0.6pt,line join=round,line cap=round] (137.76,119.95) rectangle (200.21,182.40);
\end{scope}
\begin{scope}
\definecolor{drawColor}{RGB}{0,0,0}

\path[draw=drawColor,line width= 1.1pt,line join=round] (140.59, 40.09) --
	(141.17, 58.06) --
	(141.74, 58.06) --
	(142.32, 58.06) --
	(142.89, 58.06) --
	(143.46, 58.06) --
	(144.04, 58.06) --
	(144.61, 58.06) --
	(145.18, 58.05) --
	(145.76, 58.02) --
	(146.33, 57.97) --
	(146.90, 57.88) --
	(147.48, 57.75) --
	(148.05, 57.58) --
	(148.62, 57.36) --
	(149.20, 57.11) --
	(149.77, 56.82) --
	(150.34, 56.51) --
	(150.92, 56.16) --
	(151.49, 55.79) --
	(152.06, 55.41) --
	(152.64, 55.01) --
	(153.21, 54.61) --
	(153.78, 54.22) --
	(154.36, 53.83) --
	(154.93, 53.34) --
	(155.51, 52.85) --
	(156.08, 52.36) --
	(156.65, 51.87) --
	(157.23, 51.21) --
	(157.80, 50.11) --
	(158.37, 48.87) --
	(158.95, 47.44) --
	(159.52, 45.73) --
	(160.09, 44.42) --
	(160.67, 45.56) --
	(161.24, 46.33) --
	(161.81, 47.39) --
	(162.39, 50.64) --
	(162.96, 53.33) --
	(163.53, 55.63) --
	(164.11, 57.60) --
	(164.68, 59.27) --
	(165.25, 60.39) --
	(165.83, 61.27) --
	(166.40, 62.01) --
	(166.97, 62.62) --
	(167.55, 63.11) --
	(168.12, 63.49) --
	(168.70, 63.70) --
	(169.27, 63.79) --
	(169.84, 63.76) --
	(170.42, 63.62) --
	(170.99, 63.38) --
	(171.56, 63.02) --
	(172.14, 62.55) --
	(172.71, 61.97) --
	(173.28, 61.31) --
	(173.86, 60.61) --
	(174.43, 59.83) --
	(175.00, 59.01) --
	(175.58, 58.14) --
	(176.15, 57.17) --
	(176.72, 56.12) --
	(177.30, 55.00) --
	(177.87, 53.79) --
	(178.44, 52.48) --
	(179.02, 51.05) --
	(179.59, 50.30) --
	(180.16, 49.53) --
	(180.74, 48.17) --
	(181.31, 45.53) --
	(181.88, 41.83) --
	(182.46, 39.18) --
	(183.03, 39.55) --
	(183.61, 42.42) --
	(184.18, 44.83) --
	(184.75, 46.81) --
	(185.33, 48.24) --
	(185.90, 48.88) --
	(186.47, 49.31) --
	(187.05, 49.60) --
	(187.62, 49.80) --
	(188.19, 50.14) --
	(188.77, 50.99) --
	(189.34, 51.77) --
	(189.91, 52.50) --
	(190.49, 53.18) --
	(191.06, 53.81) --
	(191.63, 54.40) --
	(192.21, 54.96) --
	(192.78, 55.49) --
	(193.35, 55.99) --
	(193.93, 56.47) --
	(194.50, 56.93) --
	(195.07, 57.37) --
	(195.65, 57.79) --
	(196.22, 58.19) --
	(196.80, 58.59) --
	(197.37, 58.96);

\path[draw=drawColor,line width= 0.6pt,dash pattern=on 1pt off 3pt ,line join=round] (137.76, 40.92) -- (200.21, 40.92);
\definecolor{drawColor}{RGB}{0,0,0}

\path[draw=drawColor,draw opacity=0.50,line width= 0.6pt,line join=round] (156.08, 34.67) -- (156.08, 97.12);
\definecolor{drawColor}{RGB}{0,0,0}

\path[draw=drawColor,line width= 0.6pt,line join=round,line cap=round] (137.76, 34.67) rectangle (200.21, 97.12);
\end{scope}
\begin{scope}
\definecolor{drawColor}{RGB}{0,0,0}

\path[draw=drawColor,line width= 1.1pt,line join=round] (210.05,293.36) --
	(210.62,343.48) --
	(211.19,343.48) --
	(211.77,343.48) --
	(212.34,343.48) --
	(212.91,343.48) --
	(213.49,343.48) --
	(214.06,343.48) --
	(214.63,343.46) --
	(215.21,343.40) --
	(215.78,343.26) --
	(216.35,343.01) --
	(216.93,342.66) --
	(217.50,342.20) --
	(218.07,341.62) --
	(218.65,340.94) --
	(219.22,340.16) --
	(219.80,339.30) --
	(220.37,338.36) --
	(220.94,337.36) --
	(221.52,336.31) --
	(222.09,335.21) --
	(222.66,334.06) --
	(223.24,332.82) --
	(223.81,331.45) --
	(224.38,330.00) --
	(224.96,328.46) --
	(225.53,326.79) --
	(226.10,324.94) --
	(226.68,322.76) --
	(227.25,319.78) --
	(227.82,316.14) --
	(228.40,311.59) --
	(228.97,305.91) --
	(229.54,299.29) --
	(230.12,296.12) --
	(230.69,302.65) --
	(231.26,310.12) --
	(231.84,316.77) --
	(232.41,321.83) --
	(232.99,326.05) --
	(233.56,329.67) --
	(234.13,332.84) --
	(234.71,335.67) --
	(235.28,338.22) --
	(235.85,340.54) --
	(236.43,342.66) --
	(237.00,344.58) --
	(237.57,346.31) --
	(238.15,347.51) --
	(238.72,348.53) --
	(239.29,349.33) --
	(239.87,349.87) --
	(240.44,350.13) --
	(241.01,350.09) --
	(241.59,349.75) --
	(242.16,349.10) --
	(242.73,348.16) --
	(243.31,346.97) --
	(243.88,345.54) --
	(244.45,343.90) --
	(245.03,342.08) --
	(245.60,340.12) --
	(246.17,338.03) --
	(246.75,335.81) --
	(247.32,333.50) --
	(247.90,331.09) --
	(248.47,328.61) --
	(249.04,326.05) --
	(249.62,324.64) --
	(250.19,326.27) --
	(250.76,326.66) --
	(251.34,326.57) --
	(251.91,325.14) --
	(252.48,321.19) --
	(253.06,315.79) --
	(253.63,308.69) --
	(254.20,299.90) --
	(254.78,296.75) --
	(255.35,305.26) --
	(255.92,311.91) --
	(256.50,316.14) --
	(257.07,319.09) --
	(257.64,321.26) --
	(258.22,322.16) --
	(258.79,322.24) --
	(259.36,322.17) --
	(259.94,322.02) --
	(260.51,321.79) --
	(261.09,323.51) --
	(261.66,325.12) --
	(262.23,326.62) --
	(262.81,328.04) --
	(263.38,329.38) --
	(263.95,330.66) --
	(264.53,331.89) --
	(265.10,333.06) --
	(265.67,334.18) --
	(266.25,335.25) --
	(266.82,336.27);

\path[draw=drawColor,line width= 0.6pt,dash pattern=on 1pt off 3pt ,line join=round] (207.21,296.77) -- (269.66,296.77);
\definecolor{drawColor}{RGB}{0,0,0}

\path[draw=drawColor,draw opacity=0.50,line width= 0.6pt,line join=round] (221.52,290.52) -- (221.52,352.97);
\definecolor{drawColor}{RGB}{0,0,0}

\path[draw=drawColor,line width= 0.6pt,line join=round,line cap=round] (207.21,290.52) rectangle (269.66,352.97);
\end{scope}
\begin{scope}
\definecolor{drawColor}{RGB}{0,0,0}

\path[draw=drawColor,line width= 1.1pt,line join=round] (210.05,208.51) --
	(210.62,215.38) --
	(211.19,215.38) --
	(211.77,215.38) --
	(212.34,215.38) --
	(212.91,215.38) --
	(213.49,215.38) --
	(214.06,215.38) --
	(214.63,215.38) --
	(215.21,215.37) --
	(215.78,215.35) --
	(216.35,215.32) --
	(216.93,215.28) --
	(217.50,215.23) --
	(218.07,215.18) --
	(218.65,215.13) --
	(219.22,215.08) --
	(219.80,215.06) --
	(220.37,215.10) --
	(220.94,215.19) --
	(221.52,215.30) --
	(222.09,215.47) --
	(222.66,215.70) --
	(223.24,215.97) --
	(223.81,216.38) --
	(224.38,216.91) --
	(224.96,217.45) --
	(225.53,218.00) --
	(226.10,218.51) --
	(226.68,219.20) --
	(227.25,220.16) --
	(227.82,221.47) --
	(228.40,222.75) --
	(228.97,223.93) --
	(229.54,225.16) --
	(230.12,226.44) --
	(230.69,227.74) --
	(231.26,228.87) --
	(231.84,229.96) --
	(232.41,231.02) --
	(232.99,232.03) --
	(233.56,232.92) --
	(234.13,233.79) --
	(234.71,234.62) --
	(235.28,235.34) --
	(235.85,235.96) --
	(236.43,236.47) --
	(237.00,236.84) --
	(237.57,237.06) --
	(238.15,237.11) --
	(238.72,236.99) --
	(239.29,236.68) --
	(239.87,236.22) --
	(240.44,235.59) --
	(241.01,234.81) --
	(241.59,233.90) --
	(242.16,232.89) --
	(242.73,231.78) --
	(243.31,230.57) --
	(243.88,229.26) --
	(244.45,227.95) --
	(245.03,226.64) --
	(245.60,225.27) --
	(246.17,223.84) --
	(246.75,222.34) --
	(247.32,220.80) --
	(247.90,219.22) --
	(248.47,217.61) --
	(249.04,216.05) --
	(249.62,214.47) --
	(250.19,212.92) --
	(250.76,211.40) --
	(251.34,209.94) --
	(251.91,209.27) --
	(252.48,209.12) --
	(253.06,209.54) --
	(253.63,210.43) --
	(254.20,211.52) --
	(254.78,212.73) --
	(255.35,214.01) --
	(255.92,215.14) --
	(256.50,216.09) --
	(257.07,216.92) --
	(257.64,217.66) --
	(258.22,218.27) --
	(258.79,218.76) --
	(259.36,218.90) --
	(259.94,218.89) --
	(260.51,218.79) --
	(261.09,218.37) --
	(261.66,217.94) --
	(262.23,217.51) --
	(262.81,217.11) --
	(263.38,216.40) --
	(263.95,215.74) --
	(264.53,215.16) --
	(265.10,214.64) --
	(265.67,214.18) --
	(266.25,213.77) --
	(266.82,213.41);

\path[draw=drawColor,line width= 0.6pt,dash pattern=on 1pt off 3pt ,line join=round] (207.21,211.49) -- (269.66,211.49);
\definecolor{drawColor}{RGB}{0,0,0}

\path[draw=drawColor,draw opacity=0.50,line width= 0.6pt,line join=round] (219.80,205.24) -- (219.80,267.69);
\definecolor{drawColor}{RGB}{0,0,0}

\path[draw=drawColor,line width= 0.6pt,line join=round,line cap=round] (207.21,205.24) rectangle (269.66,267.69);
\end{scope}
\begin{scope}
\definecolor{drawColor}{RGB}{0,0,0}

\path[draw=drawColor,line width= 1.1pt,line join=round] (210.05,124.72) --
	(210.62,155.31) --
	(211.19,155.31) --
	(211.77,155.31) --
	(212.34,155.31) --
	(212.91,155.31) --
	(213.49,155.31) --
	(214.06,155.31) --
	(214.63,155.30) --
	(215.21,155.26) --
	(215.78,155.18) --
	(216.35,155.04) --
	(216.93,154.83) --
	(217.50,154.56) --
	(218.07,154.22) --
	(218.65,153.81) --
	(219.22,153.34) --
	(219.80,152.81) --
	(220.37,152.23) --
	(220.94,151.60) --
	(221.52,150.92) --
	(222.09,150.20) --
	(222.66,149.42) --
	(223.24,148.60) --
	(223.81,147.72) --
	(224.38,146.78) --
	(224.96,145.75) --
	(225.53,144.63) --
	(226.10,143.19) --
	(226.68,141.52) --
	(227.25,139.59) --
	(227.82,137.35) --
	(228.40,134.78) --
	(228.97,131.96) --
	(229.54,129.10) --
	(230.12,127.38) --
	(230.69,128.11) --
	(231.26,131.09) --
	(231.84,134.92) --
	(232.41,138.96) --
	(232.99,142.88) --
	(233.56,146.22) --
	(234.13,148.70) --
	(234.71,150.85) --
	(235.28,152.70) --
	(235.85,154.30) --
	(236.43,155.67) --
	(237.00,156.84) --
	(237.57,157.53) --
	(238.15,158.04) --
	(238.72,158.36) --
	(239.29,158.50) --
	(239.87,158.45) --
	(240.44,158.19) --
	(241.01,157.75) --
	(241.59,157.13) --
	(242.16,156.36) --
	(242.73,155.43) --
	(243.31,154.37) --
	(243.88,153.20) --
	(244.45,151.92) --
	(245.03,150.54) --
	(245.60,149.07) --
	(246.17,147.53) --
	(246.75,145.92) --
	(247.32,144.22) --
	(247.90,142.43) --
	(248.47,141.62) --
	(249.04,141.75) --
	(249.62,141.56) --
	(250.19,140.86) --
	(250.76,138.32) --
	(251.34,134.05) --
	(251.91,128.45) --
	(252.48,123.59) --
	(253.06,127.27) --
	(253.63,132.02) --
	(254.20,135.83) --
	(254.78,138.33) --
	(255.35,139.65) --
	(255.92,140.11) --
	(256.50,140.33) --
	(257.07,140.41) --
	(257.64,140.46) --
	(258.22,141.81) --
	(258.79,143.04) --
	(259.36,144.17) --
	(259.94,145.23) --
	(260.51,146.24) --
	(261.09,147.18) --
	(261.66,148.07) --
	(262.23,148.92) --
	(262.81,149.73) --
	(263.38,150.51) --
	(263.95,151.25) --
	(264.53,151.96) --
	(265.10,152.64) --
	(265.67,153.29) --
	(266.25,153.92) --
	(266.82,154.53);

\path[draw=drawColor,line width= 0.6pt,dash pattern=on 1pt off 3pt ,line join=round] (207.21,126.21) -- (269.66,126.21);
\definecolor{drawColor}{RGB}{0,0,0}

\path[draw=drawColor,draw opacity=0.50,line width= 0.6pt,line join=round] (219.80,119.95) -- (219.80,182.40);
\definecolor{drawColor}{RGB}{0,0,0}

\path[draw=drawColor,line width= 0.6pt,line join=round,line cap=round] (207.21,119.95) rectangle (269.66,182.40);
\end{scope}
\begin{scope}
\definecolor{drawColor}{RGB}{0,0,0}

\path[draw=drawColor,line width= 1.1pt,line join=round] (279.50,293.42) --
	(280.07,346.36) --
	(280.64,346.36) --
	(281.22,346.36) --
	(281.79,346.36) --
	(282.36,346.36) --
	(282.94,346.36) --
	(283.51,346.36) --
	(284.09,346.34) --
	(284.66,346.27) --
	(285.23,346.12) --
	(285.81,345.86) --
	(286.38,345.49) --
	(286.95,344.99) --
	(287.53,344.38) --
	(288.10,343.66) --
	(288.67,342.85) --
	(289.25,341.95) --
	(289.82,340.99) --
	(290.39,339.97) --
	(290.97,338.90) --
	(291.54,337.81) --
	(292.11,336.68) --
	(292.69,335.54) --
	(293.26,334.39) --
	(293.83,333.23) --
	(294.41,332.06) --
	(294.98,330.89) --
	(295.55,329.22) --
	(296.13,327.43) --
	(296.70,325.50) --
	(297.28,323.37) --
	(297.85,321.00) --
	(298.42,318.28) --
	(299.00,315.11) --
	(299.57,311.36) --
	(300.14,306.31) --
	(300.72,301.14) --
	(301.29,303.81) --
	(301.86,306.88) --
	(302.44,309.06) --
	(303.01,310.70) --
	(303.58,311.90) --
	(304.16,312.79) --
	(304.73,313.67) --
	(305.30,314.86) --
	(305.88,315.80) --
	(306.45,316.54) --
	(307.02,317.10) --
	(307.60,317.50) --
	(308.17,317.74) --
	(308.74,317.83) --
	(309.32,317.78) --
	(309.89,317.61) --
	(310.46,318.16) --
	(311.04,319.14) --
	(311.61,320.17) --
	(312.19,321.23) --
	(312.76,322.01) --
	(313.33,321.83) --
	(313.91,321.37) --
	(314.48,320.54) --
	(315.05,319.20) --
	(315.63,317.19) --
	(316.20,314.36) --
	(316.77,310.49) --
	(317.35,305.27) --
	(317.92,300.37) --
	(318.49,296.68) --
	(319.07,294.40) --
	(319.64,296.27) --
	(320.21,298.67) --
	(320.79,300.84) --
	(321.36,303.47) --
	(321.93,306.09) --
	(322.51,308.50) --
	(323.08,310.71) --
	(323.65,312.74) --
	(324.23,314.61) --
	(324.80,316.32) --
	(325.38,317.91) --
	(325.95,319.38) --
	(326.52,320.74) --
	(327.10,322.02) --
	(327.67,323.22) --
	(328.24,324.35) --
	(328.82,325.42) --
	(329.39,326.43) --
	(329.96,327.39) --
	(330.54,328.31) --
	(331.11,329.19) --
	(331.68,330.03) --
	(332.26,330.83) --
	(332.83,331.60) --
	(333.40,332.33) --
	(333.98,333.03) --
	(334.55,333.69) --
	(335.12,334.33) --
	(335.70,334.96) --
	(336.27,335.56);

\path[draw=drawColor,line width= 0.6pt,dash pattern=on 1pt off 3pt ,line join=round] (276.66,296.77) -- (339.11,296.77);
\definecolor{drawColor}{RGB}{0,0,0}

\path[draw=drawColor,draw opacity=0.50,line width= 0.6pt,line join=round] (296.13,290.52) -- (296.13,352.97);
\definecolor{drawColor}{RGB}{0,0,0}

\path[draw=drawColor,line width= 0.6pt,line join=round,line cap=round] (276.66,290.52) rectangle (339.11,352.97);
\end{scope}
\begin{scope}
\definecolor{drawColor}{RGB}{0,0,0}

\path[draw=drawColor,line width= 1.1pt,line join=round] (279.50,208.95) --
	(280.07,257.49) --
	(280.64,257.49) --
	(281.22,257.49) --
	(281.79,257.49) --
	(282.36,257.49) --
	(282.94,257.49) --
	(283.51,257.49) --
	(284.09,257.47) --
	(284.66,257.39) --
	(285.23,257.23) --
	(285.81,256.95) --
	(286.38,256.55) --
	(286.95,256.01) --
	(287.53,255.34) --
	(288.10,254.56) --
	(288.67,253.67) --
	(289.25,252.62) --
	(289.82,251.47) --
	(290.39,250.22) --
	(290.97,248.89) --
	(291.54,247.48) --
	(292.11,245.99) --
	(292.69,244.41) --
	(293.26,242.73) --
	(293.83,240.94) --
	(294.41,238.99) --
	(294.98,236.74) --
	(295.55,234.14) --
	(296.13,230.89) --
	(296.70,226.69) --
	(297.28,221.40) --
	(297.85,215.72) --
	(298.42,213.56) --
	(299.00,217.99) --
	(299.57,224.90) --
	(300.14,231.54) --
	(300.72,236.84) --
	(301.29,241.10) --
	(301.86,244.62) --
	(302.44,247.53) --
	(303.01,249.06) --
	(303.58,250.22) --
	(304.16,251.06) --
	(304.73,251.59) --
	(305.30,251.83) --
	(305.88,251.77) --
	(306.45,251.42) --
	(307.02,250.81) --
	(307.60,249.96) --
	(308.17,248.90) --
	(308.74,247.66) --
	(309.32,246.25) --
	(309.89,244.71) --
	(310.46,243.03) --
	(311.04,241.23) --
	(311.61,239.31) --
	(312.19,237.53) --
	(312.76,238.06) --
	(313.33,238.38) --
	(313.91,237.49) --
	(314.48,235.41) --
	(315.05,232.50) --
	(315.63,228.45) --
	(316.20,222.87) --
	(316.77,215.68) --
	(317.35,208.84) --
	(317.92,213.54) --
	(318.49,219.00) --
	(319.07,223.34) --
	(319.64,225.83) --
	(320.21,227.56) --
	(320.79,228.99) --
	(321.36,231.43) --
	(321.93,233.63) --
	(322.51,235.63) --
	(323.08,237.46) --
	(323.65,239.14) --
	(324.23,240.69) --
	(324.80,242.13) --
	(325.38,243.49) --
	(325.95,244.76) --
	(326.52,245.97) --
	(327.10,247.12) --
	(327.67,248.22) --
	(328.24,249.28) --
	(328.82,250.29) --
	(329.39,251.27) --
	(329.96,252.20) --
	(330.54,253.10) --
	(331.11,253.96) --
	(331.68,254.79) --
	(332.26,255.59) --
	(332.83,256.36) --
	(333.40,257.11) --
	(333.98,257.85) --
	(334.55,258.58) --
	(335.12,259.30) --
	(335.70,260.01) --
	(336.27,260.73);

\path[draw=drawColor,line width= 0.6pt,dash pattern=on 1pt off 3pt ,line join=round] (276.66,211.49) -- (339.11,211.49);
\definecolor{drawColor}{RGB}{0,0,0}

\path[draw=drawColor,draw opacity=0.50,line width= 0.6pt,line join=round] (291.54,205.24) -- (291.54,267.69);
\definecolor{drawColor}{RGB}{0,0,0}

\path[draw=drawColor,line width= 0.6pt,line join=round,line cap=round] (276.66,205.24) rectangle (339.11,267.69);
\end{scope}
\begin{scope}
\definecolor{drawColor}{RGB}{0,0,0}

\path[draw=drawColor,line width= 1.1pt,line join=round] (279.50,122.88) --
	(280.07,137.97) --
	(280.64,137.97) --
	(281.22,137.97) --
	(281.79,137.97) --
	(282.36,137.97) --
	(282.94,137.97) --
	(283.51,137.97) --
	(284.09,137.97) --
	(284.66,137.96) --
	(285.23,137.95) --
	(285.81,137.94) --
	(286.38,137.92) --
	(286.95,137.90) --
	(287.53,137.89) --
	(288.10,137.87) --
	(288.67,137.86) --
	(289.25,137.86) --
	(289.82,137.89) --
	(290.39,137.94) --
	(290.97,138.03) --
	(291.54,138.17) --
	(292.11,138.39) --
	(292.69,138.61) --
	(293.26,138.81) --
	(293.83,139.10) --
	(294.41,139.45) --
	(294.98,139.85) --
	(295.55,140.22) --
	(296.13,140.61) --
	(296.70,141.02) --
	(297.28,141.49) --
	(297.85,142.07) --
	(298.42,142.77) --
	(299.00,143.53) --
	(299.57,144.13) --
	(300.14,144.45) --
	(300.72,144.84) --
	(301.29,145.07) --
	(301.86,144.80) --
	(302.44,144.70) --
	(303.01,144.76) --
	(303.58,144.92) --
	(304.16,145.19) --
	(304.73,145.50) --
	(305.30,145.84) --
	(305.88,146.18) --
	(306.45,146.51) --
	(307.02,146.82) --
	(307.60,147.08) --
	(308.17,147.27) --
	(308.74,147.37) --
	(309.32,147.37) --
	(309.89,147.27) --
	(310.46,147.04) --
	(311.04,146.70) --
	(311.61,146.24) --
	(312.19,145.64) --
	(312.76,144.94) --
	(313.33,144.14) --
	(313.91,143.24) --
	(314.48,142.28) --
	(315.05,141.23) --
	(315.63,140.13) --
	(316.20,138.98) --
	(316.77,137.80) --
	(317.35,136.58) --
	(317.92,135.32) --
	(318.49,134.07) --
	(319.07,132.78) --
	(319.64,131.41) --
	(320.21,130.05) --
	(320.79,128.73) --
	(321.36,127.54) --
	(321.93,126.45) --
	(322.51,125.79) --
	(323.08,126.11) --
	(323.65,126.53) --
	(324.23,126.90) --
	(324.80,127.25) --
	(325.38,127.74) --
	(325.95,128.24) --
	(326.52,128.66) --
	(327.10,129.11) --
	(327.67,129.49) --
	(328.24,129.76) --
	(328.82,130.01) --
	(329.39,130.25) --
	(329.96,130.43) --
	(330.54,130.48) --
	(331.11,130.37) --
	(331.68,130.19) --
	(332.26,129.70) --
	(332.83,129.21) --
	(333.40,128.73) --
	(333.98,128.26) --
	(334.55,127.76) --
	(335.12,127.32) --
	(335.70,126.94) --
	(336.27,126.63);

\path[draw=drawColor,line width= 0.6pt,dash pattern=on 1pt off 3pt ,line join=round] (276.66,126.21) -- (339.11,126.21);
\definecolor{drawColor}{RGB}{0,0,0}

\path[draw=drawColor,draw opacity=0.50,line width= 0.6pt,line join=round] (289.25,119.95) -- (289.25,182.40);
\definecolor{drawColor}{RGB}{0,0,0}

\path[draw=drawColor,line width= 0.6pt,line join=round,line cap=round] (276.66,119.95) rectangle (339.11,182.40);
\end{scope}
\begin{scope}
\definecolor{drawColor}{RGB}{0,0,0}

\node[text=drawColor,anchor=base,inner sep=0pt, outer sep=0pt, scale=  1.00] at ( 99.53,101.59) {Portugal};
\end{scope}
\begin{scope}
\definecolor{drawColor}{RGB}{0,0,0}

\node[text=drawColor,anchor=base,inner sep=0pt, outer sep=0pt, scale=  1.00] at (168.98,101.59) {Netherlands};
\end{scope}
\begin{scope}
\definecolor{drawColor}{RGB}{0,0,0}

\node[text=drawColor,anchor=base,inner sep=0pt, outer sep=0pt, scale=  1.00] at ( 99.53,186.87) {Austria};
\end{scope}
\begin{scope}
\definecolor{drawColor}{RGB}{0,0,0}

\node[text=drawColor,anchor=base,inner sep=0pt, outer sep=0pt, scale=  1.00] at (168.98,186.87) {Sweden};
\end{scope}
\begin{scope}
\definecolor{drawColor}{RGB}{0,0,0}

\node[text=drawColor,anchor=base,inner sep=0pt, outer sep=0pt, scale=  1.00] at (238.43,186.87) {Switzerland};
\end{scope}
\begin{scope}
\definecolor{drawColor}{RGB}{0,0,0}

\node[text=drawColor,anchor=base,inner sep=0pt, outer sep=0pt, scale=  1.00] at (307.88,186.87) {Greece};
\end{scope}
\begin{scope}
\definecolor{drawColor}{RGB}{0,0,0}

\node[text=drawColor,anchor=base,inner sep=0pt, outer sep=0pt, scale=  1.00] at ( 99.53,272.16) {United Kingdom};
\end{scope}
\begin{scope}
\definecolor{drawColor}{RGB}{0,0,0}

\node[text=drawColor,anchor=base,inner sep=0pt, outer sep=0pt, scale=  1.00] at (168.98,272.16) {France};
\end{scope}
\begin{scope}
\definecolor{drawColor}{RGB}{0,0,0}

\node[text=drawColor,anchor=base,inner sep=0pt, outer sep=0pt, scale=  1.00] at (238.43,272.16) {Norway};
\end{scope}
\begin{scope}
\definecolor{drawColor}{RGB}{0,0,0}

\node[text=drawColor,anchor=base,inner sep=0pt, outer sep=0pt, scale=  1.00] at (307.88,272.16) {Belgium};
\end{scope}
\begin{scope}
\definecolor{drawColor}{RGB}{0,0,0}

\node[text=drawColor,anchor=base,inner sep=0pt, outer sep=0pt, scale=  1.00] at ( 99.53,357.44) {Denmark};
\end{scope}
\begin{scope}
\definecolor{drawColor}{RGB}{0,0,0}

\node[text=drawColor,anchor=base,inner sep=0pt, outer sep=0pt, scale=  1.00] at (168.98,357.44) {Italy};
\end{scope}
\begin{scope}
\definecolor{drawColor}{RGB}{0,0,0}

\node[text=drawColor,anchor=base,inner sep=0pt, outer sep=0pt, scale=  1.00] at (238.43,357.44) {Germany};
\end{scope}
\begin{scope}
\definecolor{drawColor}{RGB}{0,0,0}

\node[text=drawColor,anchor=base,inner sep=0pt, outer sep=0pt, scale=  1.00] at (307.88,357.44) {Spain};
\end{scope}
\begin{scope}
\definecolor{drawColor}{RGB}{0,0,0}

\path[draw=drawColor,line width= 0.6pt,line join=round] ( 70.57, 31.17) --
	( 70.57, 34.67);

\path[draw=drawColor,line width= 0.6pt,line join=round] ( 84.91, 31.17) --
	( 84.91, 34.67);

\path[draw=drawColor,line width= 0.6pt,line join=round] ( 99.24, 31.17) --
	( 99.24, 34.67);

\path[draw=drawColor,line width= 0.6pt,line join=round] (113.58, 31.17) --
	(113.58, 34.67);

\path[draw=drawColor,line width= 0.6pt,line join=round] (127.92, 31.17) --
	(127.92, 34.67);
\end{scope}
\begin{scope}
\definecolor{drawColor}{RGB}{0,0,0}

\node[text=drawColor,anchor=base,inner sep=0pt, outer sep=0pt, scale=  1.00] at ( 70.57, 21.28) {0};

\node[text=drawColor,anchor=base,inner sep=0pt, outer sep=0pt, scale=  1.00] at ( 84.91, 21.28) {25};

\node[text=drawColor,anchor=base,inner sep=0pt, outer sep=0pt, scale=  1.00] at ( 99.24, 21.28) {50};

\node[text=drawColor,anchor=base,inner sep=0pt, outer sep=0pt, scale=  1.00] at (113.58, 21.28) {75};

\node[text=drawColor,anchor=base,inner sep=0pt, outer sep=0pt, scale=  1.00] at (127.92, 21.28) {100};
\end{scope}
\begin{scope}
\definecolor{drawColor}{RGB}{0,0,0}

\path[draw=drawColor,line width= 0.6pt,line join=round] (140.02, 31.17) --
	(140.02, 34.67);

\path[draw=drawColor,line width= 0.6pt,line join=round] (154.36, 31.17) --
	(154.36, 34.67);

\path[draw=drawColor,line width= 0.6pt,line join=round] (168.70, 31.17) --
	(168.70, 34.67);

\path[draw=drawColor,line width= 0.6pt,line join=round] (183.03, 31.17) --
	(183.03, 34.67);

\path[draw=drawColor,line width= 0.6pt,line join=round] (197.37, 31.17) --
	(197.37, 34.67);
\end{scope}
\begin{scope}
\definecolor{drawColor}{RGB}{0,0,0}

\node[text=drawColor,anchor=base,inner sep=0pt, outer sep=0pt, scale=  1.00] at (140.02, 21.28) {0};

\node[text=drawColor,anchor=base,inner sep=0pt, outer sep=0pt, scale=  1.00] at (154.36, 21.28) {25};

\node[text=drawColor,anchor=base,inner sep=0pt, outer sep=0pt, scale=  1.00] at (168.70, 21.28) {50};

\node[text=drawColor,anchor=base,inner sep=0pt, outer sep=0pt, scale=  1.00] at (183.03, 21.28) {75};

\node[text=drawColor,anchor=base,inner sep=0pt, outer sep=0pt, scale=  1.00] at (197.37, 21.28) {100};
\end{scope}
\begin{scope}
\definecolor{drawColor}{RGB}{0,0,0}

\path[draw=drawColor,line width= 0.6pt,line join=round] (209.47,116.45) --
	(209.47,119.95);

\path[draw=drawColor,line width= 0.6pt,line join=round] (223.81,116.45) --
	(223.81,119.95);

\path[draw=drawColor,line width= 0.6pt,line join=round] (238.15,116.45) --
	(238.15,119.95);

\path[draw=drawColor,line width= 0.6pt,line join=round] (252.48,116.45) --
	(252.48,119.95);

\path[draw=drawColor,line width= 0.6pt,line join=round] (266.82,116.45) --
	(266.82,119.95);
\end{scope}
\begin{scope}
\definecolor{drawColor}{RGB}{0,0,0}

\node[text=drawColor,anchor=base,inner sep=0pt, outer sep=0pt, scale=  1.00] at (209.47,106.56) {0};

\node[text=drawColor,anchor=base,inner sep=0pt, outer sep=0pt, scale=  1.00] at (223.81,106.56) {25};

\node[text=drawColor,anchor=base,inner sep=0pt, outer sep=0pt, scale=  1.00] at (238.15,106.56) {50};

\node[text=drawColor,anchor=base,inner sep=0pt, outer sep=0pt, scale=  1.00] at (252.48,106.56) {75};

\node[text=drawColor,anchor=base,inner sep=0pt, outer sep=0pt, scale=  1.00] at (266.82,106.56) {100};
\end{scope}
\begin{scope}
\definecolor{drawColor}{RGB}{0,0,0}

\path[draw=drawColor,line width= 0.6pt,line join=round] (278.92,116.45) --
	(278.92,119.95);

\path[draw=drawColor,line width= 0.6pt,line join=round] (293.26,116.45) --
	(293.26,119.95);

\path[draw=drawColor,line width= 0.6pt,line join=round] (307.60,116.45) --
	(307.60,119.95);

\path[draw=drawColor,line width= 0.6pt,line join=round] (321.93,116.45) --
	(321.93,119.95);

\path[draw=drawColor,line width= 0.6pt,line join=round] (336.27,116.45) --
	(336.27,119.95);
\end{scope}
\begin{scope}
\definecolor{drawColor}{RGB}{0,0,0}

\node[text=drawColor,anchor=base,inner sep=0pt, outer sep=0pt, scale=  1.00] at (278.92,106.56) {0};

\node[text=drawColor,anchor=base,inner sep=0pt, outer sep=0pt, scale=  1.00] at (293.26,106.56) {25};

\node[text=drawColor,anchor=base,inner sep=0pt, outer sep=0pt, scale=  1.00] at (307.60,106.56) {50};

\node[text=drawColor,anchor=base,inner sep=0pt, outer sep=0pt, scale=  1.00] at (321.93,106.56) {75};

\node[text=drawColor,anchor=base,inner sep=0pt, outer sep=0pt, scale=  1.00] at (336.27,106.56) {100};
\end{scope}
\begin{scope}
\definecolor{drawColor}{RGB}{0,0,0}

\node[text=drawColor,anchor=base east,inner sep=0pt, outer sep=0pt, scale=  1.00] at ( 61.80,288.63) {0.0};

\node[text=drawColor,anchor=base east,inner sep=0pt, outer sep=0pt, scale=  1.00] at ( 61.80,307.42) {0.2};

\node[text=drawColor,anchor=base east,inner sep=0pt, outer sep=0pt, scale=  1.00] at ( 61.80,326.22) {0.4};

\node[text=drawColor,anchor=base east,inner sep=0pt, outer sep=0pt, scale=  1.00] at ( 61.80,345.01) {0.6};
\end{scope}
\begin{scope}
\definecolor{drawColor}{RGB}{0,0,0}

\path[draw=drawColor,line width= 0.6pt,line join=round] ( 64.80,292.08) --
	( 68.30,292.08);

\path[draw=drawColor,line width= 0.6pt,line join=round] ( 64.80,310.87) --
	( 68.30,310.87);

\path[draw=drawColor,line width= 0.6pt,line join=round] ( 64.80,329.66) --
	( 68.30,329.66);

\path[draw=drawColor,line width= 0.6pt,line join=round] ( 64.80,348.45) --
	( 68.30,348.45);
\end{scope}
\begin{scope}
\definecolor{drawColor}{RGB}{0,0,0}

\node[text=drawColor,anchor=base east,inner sep=0pt, outer sep=0pt, scale=  1.00] at ( 61.80,203.35) {0.0};

\node[text=drawColor,anchor=base east,inner sep=0pt, outer sep=0pt, scale=  1.00] at ( 61.80,222.14) {0.2};

\node[text=drawColor,anchor=base east,inner sep=0pt, outer sep=0pt, scale=  1.00] at ( 61.80,240.93) {0.4};

\node[text=drawColor,anchor=base east,inner sep=0pt, outer sep=0pt, scale=  1.00] at ( 61.80,259.73) {0.6};
\end{scope}
\begin{scope}
\definecolor{drawColor}{RGB}{0,0,0}

\path[draw=drawColor,line width= 0.6pt,line join=round] ( 64.80,206.79) --
	( 68.30,206.79);

\path[draw=drawColor,line width= 0.6pt,line join=round] ( 64.80,225.58) --
	( 68.30,225.58);

\path[draw=drawColor,line width= 0.6pt,line join=round] ( 64.80,244.38) --
	( 68.30,244.38);

\path[draw=drawColor,line width= 0.6pt,line join=round] ( 64.80,263.17) --
	( 68.30,263.17);
\end{scope}
\begin{scope}
\definecolor{drawColor}{RGB}{0,0,0}

\node[text=drawColor,anchor=base east,inner sep=0pt, outer sep=0pt, scale=  1.00] at ( 61.80,118.06) {0.0};

\node[text=drawColor,anchor=base east,inner sep=0pt, outer sep=0pt, scale=  1.00] at ( 61.80,136.86) {0.2};

\node[text=drawColor,anchor=base east,inner sep=0pt, outer sep=0pt, scale=  1.00] at ( 61.80,155.65) {0.4};

\node[text=drawColor,anchor=base east,inner sep=0pt, outer sep=0pt, scale=  1.00] at ( 61.80,174.44) {0.6};
\end{scope}
\begin{scope}
\definecolor{drawColor}{RGB}{0,0,0}

\path[draw=drawColor,line width= 0.6pt,line join=round] ( 64.80,121.51) --
	( 68.30,121.51);

\path[draw=drawColor,line width= 0.6pt,line join=round] ( 64.80,140.30) --
	( 68.30,140.30);

\path[draw=drawColor,line width= 0.6pt,line join=round] ( 64.80,159.09) --
	( 68.30,159.09);

\path[draw=drawColor,line width= 0.6pt,line join=round] ( 64.80,177.89) --
	( 68.30,177.89);
\end{scope}
\begin{scope}
\definecolor{drawColor}{RGB}{0,0,0}

\node[text=drawColor,anchor=base east,inner sep=0pt, outer sep=0pt, scale=  1.00] at ( 61.80, 32.78) {0.0};

\node[text=drawColor,anchor=base east,inner sep=0pt, outer sep=0pt, scale=  1.00] at ( 61.80, 51.57) {0.2};

\node[text=drawColor,anchor=base east,inner sep=0pt, outer sep=0pt, scale=  1.00] at ( 61.80, 70.36) {0.4};

\node[text=drawColor,anchor=base east,inner sep=0pt, outer sep=0pt, scale=  1.00] at ( 61.80, 89.16) {0.6};
\end{scope}
\begin{scope}
\definecolor{drawColor}{RGB}{0,0,0}

\path[draw=drawColor,line width= 0.6pt,line join=round] ( 64.80, 36.22) --
	( 68.30, 36.22);

\path[draw=drawColor,line width= 0.6pt,line join=round] ( 64.80, 55.02) --
	( 68.30, 55.02);

\path[draw=drawColor,line width= 0.6pt,line join=round] ( 64.80, 73.81) --
	( 68.30, 73.81);

\path[draw=drawColor,line width= 0.6pt,line join=round] ( 64.80, 92.60) --
	( 68.30, 92.60);
\end{scope}
\begin{scope}
\definecolor{drawColor}{RGB}{0,0,0}

\node[text=drawColor,anchor=base,inner sep=0pt, outer sep=0pt, scale=  1.00] at (203.71,  8.94) {Day};
\end{scope}
\begin{scope}
\definecolor{drawColor}{RGB}{0,0,0}

\node[text=drawColor,rotate= 90.00,anchor=base,inner sep=0pt, outer sep=0pt, scale=  1.00] at ( 43.58,193.82) {Likelihood sensitivity ($D_{\text{CJS}})$};
\end{scope}
\end{tikzpicture}

%% file: references.bib
@inbook{spiegelhalterBayesianApproachesClinical,
author = {David J. Spiegelhalter and Keith R. Abrams and Jonathan P. Myles},
publisher = {John Wiley \& Sons, Ltd},
isbn = {9780470092606},
title = {Prior Distributions},
booktitle = {Bayesian Approaches to Clinical Trials and Health‐Care Evaluation},
chapter = {5},
pages = {139-180},
doi = {https://doi.org/10.1002/0470092602.ch5},
year = {2003}
}

@article{seegerGaussianProcessesMachine2004,
  title = {Gaussian Processes for Machine Learning},
  author = {Seeger, Matthias},
  date = {2004-04},
  year = {2004},
  journal = {International Journal of Neural Systems},
  shortjournal = {Int. J. Neur. Syst.},
  volume = {14},
  number = {02},
  pages = {69--106},
  issn = {0129-0657, 1793-6462},
  doi = {10.1142/S0129065704001899}
}

@article{vehtariRankNormalizationFoldingLocalization2021,
  title = {Rank-{{Normalization}}, {{Folding}}, and {{Localization}}: {{An Improved Rˆ}} for {{Assessing Convergence}} of {{MCMC}} (with {{Discussion}})},
  shorttitle = {Rank-{{Normalization}}, {{Folding}}, and {{Localization}}},
  author = {Vehtari, Aki and Gelman, Andrew and Simpson, Daniel and Carpenter, Bob and Bürkner, Paul-Christian},
  date = {2021-06-01},
  year = {2021},
  journal= {Bayesian Analysis},
  shortjournal = {Bayesian Anal.},
  volume = {16},
  number = {2},
  issn = {1936-0975},
  doi = {10.1214/20-BA1221},
  urldate = {2022-11-30}
}

@article{siegmundImportanceSamplingMonte1976,
  title = {Importance {{Sampling}} in the {{Monte Carlo Study}} of {{Sequential Tests}}},
  author = {Siegmund, D.},
  date = {1976-07-01},
  year = {1976},
  journal = {The Annals of Statistics},
  shortjournal = {Ann. Statist.},
  volume = {4},
  number = {4},
  issn = {0090-5364},
  doi = {10.1214/aos/1176343541},
  urldate = {2022-11-28}
}

@misc{carpenter100K10Years2022,
  title = {From 0 to {{100K}} in 10 Years: Nurturing Open-Source Community},
  author = {Carpenter, Bob},
  date = {2022-06},
  year = {2022},
  url = {https://www.youtube.com/watch?v=P9gDFHl-Hss}
}

@article{liuReproductiveNumberCOVID192020,
  title = {The Reproductive Number of {{COVID-19}} Is Higher Compared to {{SARS}} Coronavirus},
  author = {Liu, Ying and Gayle, Albert A and Wilder-Smith, Annelies and Rocklöv, Joacim},
  date = {2020-03-13},
  year = {2020},
  journal = {Journal of Travel Medicine},
  volume = {27},
  number = {2},
  pages = {taaa021},
  issn = {1708-8305},
  doi = {10.1093/jtm/taaa021},
  urldate = {2022-11-29},
  langid = {english}
}

@book{clydeIntroductionBayesianThinking,
  title = {An {{Introduction}} to {{Bayesian Thinking}}},
  author = {Clyde, Merlise and Çetinkaya-Rundel, Mine and Rundel, Colin and Banks, David and Chai, Christine and Huang, Lizzy},
  url = {https://statswithr.github.io/book/},
  urldate = {2022-11-25},
  date = {2022},
  year = {2022}
}

@article{goelInformationHyperparamtersHierarchical1981,
  title = {Information {{About Hyperparamters}} in {{Hierarchical Models}}},
  author = {Goel, Prem K. and {DeGroot}, Morris H.},
  date = {1981-03},
  year = {1981},
  journal = {Journal of the American Statistical Association},
  shortjournal = {Journal of the American Statistical Association},
  volume = {76},
  number = {373},
  eprint = {2287059},
  eprinttype = {jstor},
  pages = {140},
  issn = {01621459},
  doi = {10.2307/2287059}
}

@article{gelmanRsquaredBayesianRegression2019,
  title = {R-Squared for {{Bayesian Regression Models}}},
  author = {Gelman, Andrew and Goodrich, Ben and Gabry, Jonah and Vehtari, Aki},
  date = {2019-07-03},
  year = {2019},
  journal = {The American Statistician},
  shortjournal = {The American Statistician},
  volume = {73},
  number = {3},
  pages = {307--309},
  issn = {0003-1305, 1537-2731},
  doi = {10.1080/00031305.2018.1549100},
  urldate = {2022-06-09},
  langid = {english}
}

@article{flaxmanReport13Estimating2020,
  title = {Estimating the Effects of Non-Pharmaceutical Interventions on {{COVID-19}} in {{Europe}}},
  author = {Flaxman, Seth and Mishra, Swapnil and Gandy, Axel and Unwin, H. Juliette T. and Mellan, Thomas A. and Coupland, Helen and Whittaker, Charles and Zhu, Harrison and Berah, Tresnia and Eaton, Jeffrey W. and Monod, Mélodie and Ghani, Azra C. and Donnelly, Christl A. and Riley, Steven and Vollmer, Michaela A. C. and Ferguson, Neil M. and Okell, Lucy C. and Bhatt, Samir},
  date = {2020-08},
  year = {2020},
  journal = {Nature},
  volume = {584},
  number = {7820},
  pages = {257--261},
  publisher = {{Nature Publishing Group}},
  issn = {1476-4687},
  doi = {10.1038/s41586-020-2405-7},
  urldate = {2022-11-28},
  issue = {7820},
  langid = {english}
}

@software{Magnusson_posteriordb_a_set_2021,
author = {Magnusson, Måns and Bürkner, Paul and Vehtari, Aki},
month = {9},
url = {https://github.com/stan-dev/posteriordb},
title = {{posteriordb: a set of posteriors for Bayesian inference and probabilistic programming}},
version = {0.3},
year = {2021}
}

@article{zhangBayesianRegressionUsing2022,
  title = {Bayesian {{Regression Using}} a {{Prior}} on the {{Model Fit}}: {{The R2-D2 Shrinkage Prior}}},
  shorttitle = {Bayesian {{Regression Using}} a {{Prior}} on the {{Model Fit}}},
  author = {Zhang, Yan Dora and Naughton, Brian P. and Bondell, Howard D. and Reich, Brian J.},
  date = {2022-04-03},
  year = {2022},
  journal = {Journal of the American Statistical Association},
  shortjournal = {Journal of the American Statistical Association},
  volume = {117},
  number = {538},
  pages = {862--874},
  issn = {0162-1459, 1537-274X},
  doi = {10.1080/01621459.2020.1825449},
  urldate = {2022-06-09},
  langid = {english}
}

@article{christmannMeasuringOverlapBinary2001,
  title = {Measuring Overlap in Binary Regression},
  author = {Christmann, Andreas and Rousseeuw, Peter J},
  date = {2001-07},
  year = {2001},
  journal = {Computational Statistics \& Data Analysis},
  shortjournal = {Computational Statistics \& Data Analysis},
  volume = {37},
  number = {1},
  pages = {65--75},
  issn = {01679473},
  doi = {10.1016/S0167-9473(00)00063-3},
  urldate = {2021-02-10}
 }

@article{Capretto2022,
        title={Bambi: A Simple Interface for Fitting Bayesian Linear Models in Python},
        volume={103},
        doi={10.18637/jss.v103.i15},
        number={15},
        journal={Journal of Statistical Software},
        author={Capretto, Tomás and Piho, Camen and Kumar, Ravin and Westfall, Jacob and Yarkoni, Tal and Martin, Osvaldo A},
        year={2022},
        pages={1–29}
        }

@article{greco_robust_2008,
	title = {Robust likelihood functions in {{Bayesian}} inference},
	volume = {138},
	issn = {0378-3758},
	doi = {10.1016/j.jspi.2007.05.001},
	pages = {1258--1270},
	number = {5},
	journal = {Journal of Statistical Planning and Inference},
	shortjournal = {Journal of Statistical Planning and Inference},
	author = {Greco, Luca and Racugno, Walter and Ventura, Laura},
	urldate = {2022-08-09},
	date = {2008-05-01},
	year = {2008},
	langid = {english}
}

@article{agostinelli_weighted_2013,
	title = {A weighted strategy to handle likelihood uncertainty in {{Bayesian}} inference},
	volume = {28},
	year = {2013},
	rights = {Springer-Verlag Berlin Heidelberg 2013},
	issn = {0943-4062},
	doi = {10.1007/s00180-011-0301-1},
	pages = {319--339},
	number = {1},
	journal = {Computational Statistics},
	author = {Agostinelli, Claudio and Greco, Luca},
	urldate = {2022-08-09},
	date = {2013-02},
	note = {Num Pages: 319-339
Place: Heidelberg, Netherlands
Publisher: Springer Nature B.V.}
}

@InProceedings{rubin1988SIR,
  author    = {D. B. Rubin},
  title     = {Using the {SIR} algorithm to simulate posterior distributions},
  booktitle = {Bayesian Statistics},
  year      = {1988},
  editor    = {J. M. Bernardo and M. H. DeGroot and D. V. Lindley and A. F. M. Smith},
  volume    = {3},
  series    = {Bayesian Statistics},
  publisher = {Oxford University Press},
  date      = {1988},
}

@Article{ruitortmayol2022PracticalHilbert,
title = {Practical {{Hilbert}} space approximate {{Bayesian}}
		  {{Gaussian}} processes for probabilistic
		  programming},
author = {Gabriel Riutort-Mayol and Paul-Christian B\"urkner and
		  Michael R. Andersen and Arno Solin and Aki Vehtari},
journal = {Statistics and Computing},
year = {2022},
note = {in press}
}

@article{pymc3,
author = {Salvatier, J. and Wiecki, T. V. and Fonnesbeck, C.},
year = {2016},
title = {Probabilistic programming in {Python} using {PyMC3}},
journal = {PeerJ Computer Science},
volume = {2},
number = {e55},
doi = {10.7717/peerj-cs.55}
}

@article{lele2007DataCloning,
author = {Lele, Subhash R. and Dennis, Brian and Lutscher, Frithjof},
title = {Data cloning: easy maximum likelihood estimation for complex
		  ecological models using Bayesian Markov chain Monte
		  Carlo methods},
journal = {Ecology Letters},
volume = {10},
number = {7},
pages = {551-563},
doi = {https://doi.org/10.1111/j.1461-0248.2007.01047.x},
year = {2007}
}

@book{diggle2007Geostatistics,
  title={Model-based {G}eostatistics},
  author={Diggle, Peter J. and Ribeiro, Paulo J.},
  year={2007},
  publisher={Springer}
}

@article{solin2020HilbertSpace,
title = {{Hilbert} space methods for reduced-rank {Gaussian} process
		  regression},
author = {Arno Solin and Simo S\"arkk\"a},
year = {2020},
pages = {419-446},
doi = {10.1007/s11222-019-09886-w},
journal = {Statistics and Computing},
volume = {30}
}

@Article{mclust,
    title = {{mclust} 5: clustering, classification and density
		  estimation using {G}aussian finite mixture models},
    author = {Luca Scrucca and Michael Fop and T. Brendan Murphy and
		  Adrian E. Raftery},
    journal = {The {R} Journal},
    year = {2016},
    volume = {8},
    number = {1},
    pages = {289--317},
    doi = {10.32614/RJ-2016-021}
  }

@Book{mass,
    title = {Modern Applied Statistics with {S}},
    author = {W. N. Venables and B. D. Ripley},
    publisher = {Springer},
    edition = {Fourth},
    address = {New York},
    year = {2002},
    note = {ISBN 0-387-95457-0},
    url = {https://www.stats.ox.ac.uk/pub/MASS4/},
  }

@misc{gabry2020PriorDistributions,
title = {Prior distributions for {rstanarm} models},
author = {Jonah Gabry and Ben Goodrich},
year = {2020},
url = {https://mc-stan.org/rstanarm/articles/priors.html}
}

@article{gagnon2021robustness,
  title = {Robustness {{Against Conflicting Prior Information}} in {{Regression}}},
  author = {Gagnon, Philippe},
  date = {2022-01},
  year = {2022},
  journal = {Bayesian Analysis},
  volume = {-1},
  pages = {1--24},
  publisher = {{International Society for Bayesian Analysis}},
  issn = {1936-0975, 1931-6690},
  doi = {10.1214/22-BA1330},
  urldate = {2022-11-28},
  issue = {-1}
}

@article{Silverman1985,
  doi = {10.1111/j.2517-6161.1985.tb01327.x},
  year = {1985},
  month = sep,
  publisher = {Wiley},
  volume = {47},
  number = {1},
  pages = {1--21},
  author = {B. W. Silverman},
  title = {Some Aspects of the Spline Smoothing Approach to
		  Non-Parametric Regression Curve Fitting},
  journal = {Journal of the Royal Statistical Society: Series B
		  (Methodological)}
}

@article{allabadiMeasuringBayesianRobustness2021,
author = {Al Labadi, Luai and Asl, Forough Fazeli and Wang, Ce},
title = {Measuring {{Bayesian}} Robustness Using {{R\'enyi}}
		  Divergence},
journal = {Stats},
volume = {4},
year = {2021},
number = {2},
pages = {251--268},
doi = {10.3390/stats4020018}
}

@article{allabadiOptimalRobustnessResults2017,
  title = {Optimal Robustness Results for Relative Belief Inferences
		  and the Relationship to Prior-Data Conflict},
  author = {Al Labadi, Luai and Evans, Michael},
  date = {2017-09},
  year = {2017},
  journal = {Bayesian Analysis},
  journal = {Bayesian Analysis},
  shortjournal = {Bayesian Anal.},
  volume = {12},
  pages = {705--728},
  issn = {1936-0975},
  doi = {10.1214/16-BA1024},
  urldate = {2020-08-28},
  langid = {english},
  number = {3}
}

@Book{spatstat,
     title = {Spatial Point Patterns: Methodology and Applications
		  with {R}},
     author = {Adrian Baddeley and Ege Rubak and Rolf Turner},
     year = {2015},
     publisher = {Chapman and Hall/CRC Press},
     address = {London},
     isbn = {978-1-4822-1021-7},
   }

@manual{bengtssonMatrixStatsFunctionsThat2020,
  title = {{{matrixStats}}: {{Functions}} That Apply to Rows and
		  Columns of Matrices (and to Vectors)},
  author = {Bengtsson, Henrik},
  date = {2020},
  year = {2020},
  url = {https://CRAN.R-project.org/package=matrixStats}}

@article{bergerRobustBayesianAnalysis1990,
  title = {Robust {{Bayesian}} Analysis: Sensitivity to the Prior},
  shorttitle = {Robust {{Bayesian}} Analysis},
  author = {Berger, James O.},
  year = {1990},
  date = {1990-07},
  journal = {Journal of Statistical Planning and Inference},
  volume = {25},
  pages = {303--328},
  issn = {03783758},
  doi = {10.1016/0378-3758(90)90079-A},
  urldate = {2021-05-05},
  langid = {english},
  number = {3}
}

@incollection{bergerBayesianRobustness2000,
  title = {Bayesian {{Robustness}}},
  booktitle = {Robust {{Bayesian Analysis}}},
  author = {Berger, James O. and Insua, David Ríos and Ruggeri,
		  Fabrizio},
  editor = {Insua, David Ríos and Ruggeri, Fabrizio},
  date = {2000},
  year = {2000},
  volume = {152},
  pages = {1--32},
  publisher = {{Springer New York}},
  location = {{New York, NY}},
  doi = {10.1007/978-1-4612-1306-2_1},
  series = {Lecture {{Notes}} in {{Statistics}}}
}

@article{bergerOverviewRobustBayesian1994,
  title = {An Overview of Robust {{Bayesian}} Analysis},
  author = {Berger, James O. and Moreno, Elías and Pericchi, Luis Raul
		  and Bayarri, M. Jesús and Bernardo, José M. and
		  Cano, Juan A. and De la Horra, Julián and Martín,
		  Jacinto and Ríos-Insúa, David and Betrò, Bruno and
		  Dasgupta, A. and Gustafson, Paul and Wasserman,
		  Larry and Kadane, Joseph B. and Srinivasan, Cid and
		  Lavine, Michael and O'Hagan, Anthony and Polasek,
		  Wolfgang and Robert, Christian P. and Goutis,
		  Constantinos and Ruggeri, Fabrizio and Salinetti,
		  Gabriella and Sivaganesan, Siva},
  date = {1994-06},
  year = {1994},
  journal = {Test},
    journal = {Test},
  shortjournal = {Test},
  volume = {3},
  pages = {5--124},
  issn = {1133-0686, 1863-8260},
  doi = {10.1007/BF02562676},
  langid = {english},
  number = {1}
}

@article{besagBayesianComputationStochastic1995,
  title = {Bayesian {{Computation}} and {{Stochastic Systems}}},
  author = {Besag, Julian and Green, Peter and Higdon, David and
		  Mengersen, Kerrie},
  date = {1995},
  year = {1995},
  journal = {Statistical Science},
  journal = {Statistical Science},
  volume = {10},
  pages = {3--41},
  eprint = {2246224},
  eprinttype = {jstor},
  langid = {english},
  doi = {10.1214/ss/1177010123},
  number = {1}
}

@article{bornnEfficientComputationalApproach2010,
  title = {An Efficient Computational Approach for Prior Sensitivity
		  Analysis and Cross-Validation},
  author = {Bornn, Luke and Doucet, Arnaud and Gottardo, Raphael},
  date = {2010},
  year = {2010},
  journal = {Canadian Journal of Statistics},
  journal = {Canadian Journal of Statistics},
  shortjournal = {Can. J. Statistics},
  pages = {47--64},
  issn = {03195724, 1708945X},
  doi = {10.1002/cjs.10045},
  langid = {english}
}

@article{brownMCMCGeneralizedLinear2010,
  title = {{{MCMC}} for {{Generalized Linear Mixed Models}} with
		  {{glmmBUGS}}},
  author = {Brown, Patrick and Zhou, Lutong},
  date = {2010},
  year = {2010},
  journal = {The R Journal},
  journal = {The R Journal},
  shortjournal = {The R Journal},
  volume = {2},
  pages = {13},
  issn = {2073-4859},
  doi = {10.32614/RJ-2010-003},
  urldate = {2020-09-14},
  langid = {english},
  number = {1}
}

@article{burknerBrmsPackageBayesian2017,
  title = {{{brms}}: {{An R}} Package for {{Bayesian}} Multilevel
		  Models Using {{Stan}}},
  author = {Bürkner, Paul-Christian},
  date = {2017},
  year = {2017},
  journal = {Journal of Statistical Software},
  journal = {Journal of Statistical Software},
  volume = {80},
  pages = {1--28},
  doi = {10.18637/jss.v080.i01},
  encoding = {UTF-8},
  number = {1}
}

@manual{burknerPosteriorToolsWorking2020,
  title = {{posterior}: {{Tools}} for Working with Posterior
		  Distributions},
  author = {Bürkner, Paul-Christian and Gabry, Jonah and Kay, Matthew
		  and Vehtari, Aki},
  date = {2022},
  year = {2022},
  url = {https://mc-stan.org/posterior}
}

@article{canavosBayesianEstimationSensitivity1975,
  title = {Bayesian Estimation: {{A}} Sensitivity Analysis},
  shorttitle = {Bayesian Estimation},
  author = {Canavos, George C.},
  year = {1975},
  date = {1975-09},
  journal = {Naval Research Logistics Quarterly},
  journal = {Naval Research Logistics Quarterly},
  shortjournal = {Naval Research Logistics},
  volume = {22},
  pages = {543--552},
  issn = {00281441, 19319193},
  doi = {10.1002/nav.3800220310},
  langid = {english},
  number = {3}
}

@article{chaComprehensiveSurveyDistance2007,
  title = {Comprehensive {{Survey}} on {{Distance}}/{{Similarity
		  Measures}} between {{Probability Density
		  Functions}}},
  author = {Cha, Sung-Hyuk},
  year = {2007},
  date = {2007},
  journal = {International Journal of Mathematical Models and
		  Methods in Applied Sciences},
  journal = {International Journal of Mathematical Models and
		  Methods in Applied Sciences},
  issn = {1998-0140},
  langid = {english}
}

@article{depaoliImportancePriorSensitivity2020,
  title = {The {{Importance}} of {{Prior Sensitivity Analysis}} in
		  {{Bayesian Statistics}}: {{Demonstrations Using}} an
		  {{Interactive Shiny App}}},
  shorttitle = {The {{Importance}} of {{Prior Sensitivity Analysis}}
		  in {{Bayesian Statistics}}},
  author = {Depaoli, Sarah and Winter, Sonja D. and Visser, Marieke},
  date = {2020-11-24},
  year = {2020},
  journal = {Frontiers in Psychology},
  journal = {Frontiers in Psychology},
  shortjournal = {Front. Psychol.},
  volume = {11},
  pages = {608045},
  issn = {1664-1078},
  doi = {10.3389/fpsyg.2020.608045}
}

@article{drostPhilentropyInformationTheory2018,
  title = {Philentropy: {{Information Theory}} and {{Distance
		  Quantification}} with {{R}}},
  shorttitle = {Philentropy},
  author = {Drost, Hajk-Georg},
  year = {2018},
  date = {2018-06-11},
  journal = {Journal of Open Source Software},
    journal = {Journal of Open Source Software},
  shortjournal = {JOSS},
  volume = {3},
  pages = {765},
  issn = {2475-9066},
  doi = {10.21105/joss.00765},
  number = {26}
}

@article{evansWeakInformativityInformation2011,
  title = {Weak Informativity and the Information in One Prior
		  Relative to Another},
  author = {Evans, Michael and Jang, Gun Ho},
  year = {2011},
  date = {2011-08},
  journal = {Statistical Science},
  journal = {Statistical Science},
  shortjournal = {Statist. Sci.},
  volume = {26},
  pages = {423--439},
  issn = {0883-4237},
  doi = {10.1214/11-STS357},
  langid = {english},
  number = {3}
}

@article{evansCheckingPriordataConflict2006,
  title = {Checking for Prior-Data Conflict},
  author = {Evans, Michael and Moshonov, Hadas},
  date = {2006-12},
  year = {2006},
  journal = {Bayesian Analysis},
  journal = {Bayesian Analysis},
  shortjournal = {Bayesian Anal.},
  volume = {1},
  pages = {893--914},
  issn = {1936-0975},
  doi = {10.1214/06-BA129},
  langid = {english},
  number = {4}
}

@book{Flury1988,
  doi = {10.1007/978-94-009-1217-5},
  year = {1988},
  publisher = {Springer Netherlands},
  author = {Bernhard Flury and Hans Riedwyl},
  title = {Multivariate Statistics: A Practical Approach}
}

@book{gelmanRegressionOtherStories2020,
  title = {Regression and {{Other Stories}}},
  author = {Gelman, Andrew and Hill, Jennifer and Vehtari, Aki},
  date = {2020},
  year = {2020},
  publisher = {{Cambridge University Press}},
  location = {{S.l.}},
  isbn = {978-1-107-02398-7},
  langid = {english}
}

@article{gelmanPriorCanOften2017,
  title = {The Prior Can Often Only Be Understood in the Context of
		  the Likelihood},
  author = {Gelman, Andrew and Simpson, Daniel and Betancourt,
		  Michael},
  year = {2017},
  date = {2017-10-19},
  journal = {Entropy},
    journal = {Entropy},

  shortjournal = {Entropy},
  volume = {19},
  pages = {555},
  issn = {1099-4300},
  doi = {10.3390/e19100555},
  langid = {english},
  number = {10}
}

@article{giordanoCovariancesRobustnessVariational2018,
  title = {Covariances, robustness, and variational Bayes},
  author = {Giordano, Ryan and Broderick, Tamara and Jordan, Michael
		  I.},
  date = {2018},
  year = {2018},
  journal = {Journal of Machine Learning Research},
    journal = {Journal of Machine Learning Research},

  volume = {19},
  pages = {1--49},
  issn = {1533-7928},
  url = {http://jmlr.org/papers/v19/17-670.html},
  urldate = {2020-07-01},
  number = {51}
}

@misc{gelmanBayesianWorkflow2020,
  title = {Bayesian {{Workflow}}},
  author = {Gelman, Andrew and Vehtari, Aki and Simpson, Daniel and
		  Margossian, Charles C. and Carpenter, Bob and Yao,
		  Yuling and Kennedy, Lauren and Gabry, Jonah and
		  Bürkner, Paul-Christian and Modrák, Martin},
  date = {2020-11-03},
  year = {2020},
  url = {http://arxiv.org/abs/2011.01808},
  urldate = {2020-11-04},
  archiveprefix = {arXiv},
  eprint = {2011.01808},
  eprinttype = {arxiv},
  primaryclass = {stat}
}

@Misc{rstanarm,
    title = {{rstanarm}: {Bayesian} applied regression modeling via
		  {Stan}.},
    author = {Ben Goodrich and Jonah Gabry and Imad Ali and Sam
		  Brilleman},
    note = {R package version 2.21.1},
    year = {2020},
    url = {https://mc-stan.org/rstanarm},
  }

@article{grinsztajnBayesianWorkflowDisease2021,
  title = {Bayesian Workflow for Disease Transmission Modeling in
		  {{Stan}}},
  author = {Grinsztajn, Léo and Semenova, Elizaveta and Margossian,
		  Charles C. and Riou, Julien},
  date = {2021},
  year = {2021},
  journal = {Statistical Medicine},
    journal = {Statistical Medicine},

  volume = {40},
  number = {27},
  pages = {6209--6234},
  doi = {10.1002/sim.9164}
}

@incollection{gustafsonLocalRobustnessBayesian2000,
  title = {Local {{Robustness}} in {{Bayesian Analysis}}},
  booktitle = {Robust {{Bayesian Analysis}}},
  author = {Gustafson, Paul},
  editor = {Insua, David Ríos and Ruggeri, Fabrizio},
  date = {2000},
  year = {2000},
  volume = {152},
  pages = {71--88},
  publisher = {{Springer New York}},
  location = {{New York, NY}},
  doi = {10.1007/978-1-4612-1306-2_4},
  editorb = {Bickel, P. and Diggle, P. and Fienberg, S. and
		  Krickeberg, K. and Olkin, I. and Wermuth, N. and
		  Zeger, S.},
  series = {Lecture {{Notes}} in {{Statistics}}}
}

@article{heinzeVariableSelectionReview2018,
  title = {Variable Selection - {{A}} Review and Recommendations for
		  the Practicing Statistician},
  author = {Heinze, Georg and Wallisch, Christine and Dunkler,
		  Daniela},
  date = {2018-05},
  year = {2018},
  journal = {Biometrical Journal},
    journal = {Biometrical Journal},

  shortjournal = {Biom. J.},
  volume = {60},
  pages = {431--449},
  issn = {03233847},
  doi = {10.1002/bimj.201700067},
  langid = {english},
  number = {3}
}

@article{hillSensitivityBayesianAnalysis1994,
  title = {Sensitivity of a {{Bayesian}} Analysis to the Prior
		  Distribution},
  author = {Hill, S.D. and Spall, J.C.},
  year = {1994},
  journal = {IEEE Transactions on Systems, Man, and Cybernetics},
    journal = {IEEE Transactions on Systems, Man, and Cybernetics},

  shortjournal = {IEEE Trans. Syst., Man, Cybern.},
  volume = {24},
  pages = {216--221},
  issn = {00189472},
  doi = {10.1109/21.281421},
  number = {2}
}

@article{hoGlobalRobustBayesian2020,
  doi = {10.21144/wp20-07},
  year = {2020},
  month = jul,
  publisher = {Federal Reserve Bank of Richmond},
  volume = {20},
  number = {07},
  pages = {1--46},
  author = {Paul Ho},
  title = {Global Robust Bayesian Analysis in Large Models},
  journal = {Federal Reserve Bank of Richmond Working Papers}
}

@article{hunanyan2021quantification,
  title = {Quantification of {{Empirical Determinacy}}: {{The Impact}} of {{Likelihood Weighting}} on {{Posterior Location}} and {{Spread}} in {{Bayesian Meta-Analysis Estimated}} with {{JAGS}} and {{INLA}}},
  shorttitle = {Quantification of {{Empirical Determinacy}}},
  author = {Sona Hunanyan and Håvard Rue and Martyn Plummer and Małgorzata Roos},
  date = {2022-01-01},
  year = {2022},
  journal = {Bayesian Analysis},
  shortjournal = {Bayesian Anal.},
  volume = {-1},
  issn = {1936-0975},
  doi = {10.1214/22-BA1325},
  urldate = {2022-11-28},
  issue = {-1}
}

@article{jacobiAutomatedSensitivityAnalysis2018,
  title = {Automated {{Sensitivity Analysis}} for {{Bayesian
		  Inference}} via {{Markov Chain Monte Carlo}}:
		  {{Applications}} to {{Gibbs Sampling}}},
  shorttitle = {Automated {{Sensitivity Analysis}} for {{Bayesian
		  Inference}} via {{Markov Chain Monte Carlo}}},
  author = {Jacobi, Liana and Joshi, Mark and Zhu, Dan},
  date = {2018},
  year = {2018},
  journal = {SSRN Electronic Journal},
    journal = {SSRN Electronic Journal},

  shortjournal = {SSRN Journal},
  issn = {1556-5068},
  doi = {10.2139/ssrn.2984054},
  langid = {english}
}

@article{johnsonFittingPercentageBody1996,
  title = {Fitting {{Percentage}} of {{Body Fat}} to {{Simple Body
		  Measurements}}},
  author = {Johnson, Roger W.},
  date = {1996-03},
  year = {1996},
  journal = {Journal of Statistics Education},
    journal = {Journal of Statistics Education},

  shortjournal = {Journal of Statistics Education},
  volume = {4},
  pages = {6},
  issn = {1069-1898},
  doi = {10.1080/10691898.1996.11910505},
  langid = {english},
  number = {1}
}

@article{kessy2018Sphere,
author = {Agnan Kessy and Alex Lewin and Korbinian Strimmer},
title = {Optimal Whitening and Decorrelation},
journal = {The American Statistician},
volume = {72},
number = {4},
pages = {309-314},
year = {2018},
publisher = {Taylor & Francis},
doi = {10.1080/00031305.2016.1277159}
}

@Manual{kosmidis2021,
    title = {{detectseparation}: Detect and Check for Separation and
		  Infinite Maximum Likelihood
Estimates},
    author = {Ioannis Kosmidis and Dirk Schumacher},
    year = {2021},
    note = {R package version 0.2},
    url = {https://CRAN.R-project.org/package=detectseparation},
  }

@article{kurtekBayesianSensitivityAnalysis2015,
  title = {Bayesian Sensitivity Analysis with the {{Fisher}}-{{Rao}}
		  Metric},
  author = {Kurtek, Sebastian and Bharath, Karthik},
  date = {2015-09},
  year = {2015},
  journal = {Biometrika},
  shortjournal = {Biometrika},
  volume = {102},
  pages = {601--616},
  issn = {0006-3444, 1464-3510},
  doi = {10.1093/biomet/asv026},
  langid = {english},
  number = {3}
}

@article{lekHowChoiceDistance2019,
  title = {How the choice of distance measure influences the detection
		  of prior-data conflict},
  author = {{Lek} and {van de Schoot}},
  year = {2019},
  date = {2019-04-29},
  journal = {Entropy},
    journal = {Entropy},

  shortjournal = {Entropy},
  volume = {21},
  pages = {446},
  issn = {1099-4300},
  doi = {10.3390/e21050446},
  langid = {english},
  number = {5}
}

@article{linDivergenceMeasuresBased1991,
  title = {Divergence Measures Based on the {{Shannon}} Entropy},
  author = {Lin, J.},
  year = {1991},
  journal = {IEEE Transactions on Information Theory},
    journal = {IEEE Transactions on Information Theory},

  shortjournal = {IEEE Trans. Inform. Theory},
  volume = {37},
  pages = {145--151},
  issn = {00189448},
  doi = {10.1109/18.61115},
  number = {1}
}

@article{lopesConfrontingPriorConvictions2011,
  title = {Confronting {{Prior Convictions}}: {{On Issues}} of {{Prior
		  Sensitivity}} and {{Likelihood Robustness}} in
		  {{Bayesian Analysis}}},
  shorttitle = {Confronting {{Prior Convictions}}},
  author = {Lopes, Hedibert F. and Tobias, Justin L.},
  date = {2011-09},
  year = {2011},
  journal = {Annual Review of Economics},
  journal = {Annual Review of Economics},
  shortjournal = {Annu. Rev. Econ.},
  volume = {3},
  pages = {107--131},
  issn = {1941-1383, 1941-1391},
  doi = {10.1146/annurev-economics-111809-125134},
  langid = {english},
  number = {1}
}

@misc{maroufyLocalGlobalRobustness2015,
  title = {Local and Global Robustness in Conjugate {{Bayesian}}
		  Analysis},
  author = {Maroufy, Vahed and Marriott, Paul},
  date = {2015-08-31},
  year = {2015},
  url = {http://arxiv.org/abs/1508.07937},
  urldate = {2020-07-01},
  archiveprefix = {arXiv},
  eprint = {1508.07937},
  eprinttype = {arxiv},
  primaryclass = {stat}
}

@Manual{mccartanAdjustrStanModel2021,
     title = {adjustr: Stan Model Adjustments and Sensitivity Analyses
		  using Importance
 Sampling},
     author = {Cory McCartan},
     year = {2022},
     note = {R package version 0.1.2},
     url = {https://corymccartan.github.io/adjustr}
   }

@incollection{nguyenNonparametricJensenShannonDivergence2015,
  title = {Non-Parametric {{Jensen}}-{{Shannon Divergence}}},
  booktitle = {Machine {{Learning}} and {{Knowledge Discovery}} in
		  {{Databases}}},
  author = {Nguyen, Hoang-Vu and Vreeken, Jilles},
  editor = {Appice, Annalisa and Rodrigues, Pedro Pereira and Santos
		  Costa, Vítor and Gama, João and Jorge, Alípio and
		  Soares, Carlos},
  date = {2015},
  year = {2015},
  volume = {9285},
  pages = {173--189},
  publisher = {{Springer International Publishing}},
  location = {{Cham}},
  doi = {10.1007/978-3-319-23525-7_11},
  series = {Lecture {{Notes}} in {{Computer Science}}}
}

@article{nottCheckingPriorDataConflict2020,
  title = {Checking for {{Prior}}-{{Data Conflict Using
		  Prior}}-to-{{Posterior Divergences}}},
  author = {Nott, David J. and Wang, Xueou and Evans, Michael and
		  Englert, Berthold-Georg},
  date = {2020-05},
  year = {2020},
  journal = {Statistical Science},
  journal = {Statistical Science},
  shortjournal = {Statist. Sci.},
  volume = {35},
  pages = {234--253},
  issn = {0883-4237},
  doi = {10.1214/19-STS731},
  langid = {english},
  number = {2}
}

@article{nottUsingPriorExpansions2020,
  title = {Using {{Prior Expansions}} for {{Prior}}-{{Data Conflict
		  Checking}}},
  author = {Nott, David J. and Seah, Max and Al Labadi, Luai and
		  Evans, Michael and Ng, Hui Khoon and Englert,
		  Berthold-Georg},
  date = {2020-03},
  year = {2020},
  journal = {Bayesian Analysis},
    journal = {Bayesian Analysis},
  shortjournal = {Bayesian Anal.},
  issn = {1936-0975},
  doi = {10.1214/20-BA1204},
  langid = {english}
}

@incollection{ohaganHSSS,
title = {{HSSS} model criticism},
booktitle = {Highly Structured Stochastic Systems},
author = {Anthony O'Hagan},
editor = {Peter J. Green and Nils Lid Hjort and Sylvia Richardson},
chapter = 14,
pages = {423-444},
year = {2003},
series = {Oxford Statistcal Science Series},
volume = 27,
publisher = {Oxford University Press}
}

@article{ohaganBayesianHeavytailedModels2012,
  title = {Bayesian Heavy-Tailed Models and Conflict Resolution: {{A}}
		  Review},
  shorttitle = {Bayesian Heavy-Tailed Models and Conflict Resolution},
  author = {O'Hagan, Anthony and Pericchi, Luis},
  date = {2012-11-01},
  year = {2012},
  journal = {Brazilian Journal of Probability and Statistics},
    journal = {Brazilian Journal of Probability and Statistics},

  shortjournal = {Braz. J. Probab. Stat.},
  volume = {26},
  issn = {0103-0752},
  doi = {10.1214/11-BJPS164},
  number = {4}
}

@article{oneillImportanceSamplingBayesian2009,
  title = {Importance Sampling for {{Bayesian}} Sensitivity Analysis},
  author = {O'Neill, B.},
  date = {2009-02},
  year = {2009},
  journal = {International Journal of Approximate Reasoning},
  journal = {International Journal of Approximate Reasoning},
  shortjournal = {International Journal of Approximate Reasoning},
  volume = {50},
  pages = {270--278},
  issn = {0888613X},
  doi = {10.1016/j.ijar.2008.03.015},
  langid = {english},
  number = {2}
}

@article{paananenImplicitlyAdaptiveImportance2021,
  title = {Implicitly Adaptive Importance Sampling},
  author = {Paananen, Topi and Piironen, Juho and Bürkner,
		  Paul-Christian and Vehtari, Aki},
  year = {2021},
  date = {2021-02-09},
  journal = {Statistics and Computing},
  journal = {Statistics and Computing},
  shortjournal = {Stat Comput},
  volume = {31},
  number = {16},
  issn = {1573-1375},
  doi = {10.1007/s11222-020-09982-2}
}

@InProceedings{paananen2019VariableSelection,
  author    = {Paananen, Topi and Piironen, Juho and Andersen, Michael Riis and Vehtari, Aki},
  title     = {Variable selection for Gaussian processes via sensitivity analysis of the posterior predictive distribution},
  booktitle = {Proceedings of the Twenty-Second International Conference on Artificial Intelligence and Statistics},
  year      = {2019},
  editor    = {Chaudhuri, Kamalika and Sugiyama, Masashi},
  volume    = {89},
  series    = {Proceedings of Machine Learning Research},
  pages     = {1743--1752},
  publisher = {PMLR},
  url       = {https://proceedings.mlr.press/v89/paananen19a.html},
}

@article{pavoneUsingReferenceModels2020,
  title = {Using Reference Models in Variable Selection},
  author = {Pavone, Federico and Piironen, Juho and Bürkner, Paul-Christian and Vehtari, Aki},
  date = {2022-05-14},
  year = {2022},
  journal = {Computational Statistics},
  shortjournal = {Comput Stat},
  issn = {0943-4062, 1613-9658},
  doi = {10.1007/s00180-022-01231-6},
  urldate = {2022-11-28},
  langid = {english}
}

@article{poirierRevisingBeliefsNonidentified1998,
  doi = {10.1017/s0266466698144043},
  year = {1998},
  month = aug,
  publisher = {Cambridge University Press ({CUP})},
  volume = {14},
  number = {4},
  pages = {483--509},
  author = {Dale J. Poirier},
  title = {Revising beliefs in nonidentified models},
  journal = {Econometric Theory}
}

@article{perezMCMCbasedLocalParametric2006,
  title = {{{MCMC}}-Based Local Parametric Sensitivity Estimations},
  author = {Pérez, C.J. and Martín, J. and Rufo, M.J.},
  year = {2006},
  date = {2006-11},
  journal = {Computational Statistics \& Data Analysis},
    journal = {Computational Statistics \& Data Analysis},

  shortjournal = {Computational Statistics \& Data Analysis},
  volume = {51},
  pages = {823--835},
  issn = {01679473},
  doi = {10.1016/j.csda.2005.09.005},
  langid = {english},
  number = {2}
}

@manual{rcoreteamLanguageEnvironmentStatistical2020,
  title = {R: {{A}} Language and Environment for Statistical
		  Computing},
  author = {{R Core Team}},
  year = {2022},
  date = {2022},
  location = {{Vienna, Austria}},
  url = {https://www.R-project.org/},
  organization = {{R Foundation for Statistical Computing}}
}

@article{reimherrPriorSampleSize2020,
  title = {Prior Sample Size Extensions for Assessing Prior Impact and Prior‐likelihood Discordance},
  author = {Reimherr, Matthew and Meng, Xiao‐Li and Nicolae, Dan L.},
  date = {2021-07},
  year = {2021},
  journal = {Journal of the Royal Statistical Society: Series B (Statistical Methodology)},
  shortjournal = {J R Stat Soc Series B},
  volume = {83},
  number = {3},
  pages = {413--437},
  issn = {1369-7412, 1467-9868},
  doi = {10.1111/rssb.12414},
  urldate = {2022-11-28},
  langid = {english}
}

@book{robert2004,
  doi = {10.1007/978-1-4757-4145-2},
  year = {2004},
  publisher = {Springer New York},
  author = {Christian P. Robert and George Casella},
  title = {{Monte Carlo} Statistical Methods}
}

@article{Roos2021,
  doi = {10.1002/bimj.202000193},
  year = {2021},
  month = aug,
  publisher = {Wiley},
  author = {Ma{\l}gorzata Roos and Sona Hunanyan and Haakon Bakka and
		  H{\aa}vard Rue},
  title = {Sensitivity and identification quantification by a relative
		  latent model complexity perturbation in Bayesian
		  meta-analysis},
  journal = {Biometrical Journal}
}

@article{roosSensitivityAnalysisBayesian2015,
  title = {Sensitivity {{Analysis}} for {{Bayesian Hierarchical
		  Models}}},
  author = {Roos, Małgorzata and Martins, Thiago G. and Held, Leonhard
		  and Rue, Håvard},
  year = {2015},
  date = {2015-06},
    journal = {Bayesian Analysis},

  journal = {Bayesian Analysis},
  shortjournal = {Bayesian Anal.},
  volume = {10},
  pages = {321--349},
  issn = {1936-0975},
  doi = {10.1214/14-BA909},
  langid = {english},
  number = {2}
}

@article{sailynoja2021graphical,
      title={Graphical Test for Discrete Uniformity and its
		  Applications in Goodness of Fit Evaluation and
		  Multiple Sample Comparison},
      author={Teemu S\"{a}ilynoja and Paul-Christian B\"{u}rkner and
		  Aki Vehtari},
      year={2022},
      journal = {Statistics and Computing},
      volume = {32},
      number = {32},
      doi={10.1007/s11222-022-10090-6}
}

@article{schadPrincipledBayesianWorkflow2021,
issn = {1082-989X},
journal = {Psychological methods},
pages = {103--126},
volume = {26},
publisher = {AMER PSYCHOLOGICAL ASSOC},
number = {1},
year = {2021},
title = {Toward a Principled Bayesian Workflow in Cognitive Science},
copyright = {2020, American Psychological Association},
language = {eng},
doi = {10.1037/met0000275},
address = {WASHINGTON},
author = {Schad, Daniel J and Betancourt, Michael and Vasishth,
		  Shravan},
}

@article{sivaganesanRobustBayesianDiagnostics1993,
  title = {Robust {{Bayesian}} Diagnostics},
  author = {Sivaganesan, Siva},
  year = {1993},
  date = {1993-05},
  journal = {Journal of Statistical Planning and Inference},
    journal = {Journal of Statistical Planning and Inference},

  shortjournal = {Journal of Statistical Planning and Inference},
  volume = {35},
  pages = {171--188},
  issn = {03783758},
  doi = {10.1016/0378-3758(93)90043-6},
  langid = {english},
  number = {2}
}

@article{skeneBayesianModellingSensitivity1986,
  title = {Bayesian {{Modelling}} and {{Sensitivity Analysis}}},
  author = {Skene, A. M. and Shaw, J. E. H. and Lee, T. D.},
  date = {1986},
  year = {1986},
  journal = {The Statistician},
    journal = {The Statistician},

  shortjournal = {The Statistician},
  volume = {35},
  pages = {281},
  issn = {00390526},
  doi = {10.2307/2987533},
  eprint = {10.2307/2987533},
  eprinttype = {jstor},
  number = {2}
}

@manual{standevelopmentteamStanModellingLanguage2021,
  title = {Stan {{Modelling Language Users Guide}} and {{Reference
		  Manual}}},
  author = {{Stan Development Team}},
  date = {2021},
  year = {2021},
  url = {https://mc-stan.org},
  version = {2.26}
}

@article{tsaiInfluenceMeasuresRobust2011,
  title = {Influence Measures and Robust Estimators of Dependence in
		  Multivariate Extremes},
  author = {Tsai, Yu-Ling and Murdoch, Duncan J. and Dupuis, Debbie
		  J.},
  date = {2011-12},
  year = {2011},
  journal = {Extremes},
    journal = {Extremes},

  shortjournal = {Extremes},
  volume = {14},
  pages = {343--363},
  issn = {1386-1999, 1572-915X},
  doi = {10.1007/s10687-010-0114-6},
  langid = {english},
  number = {4}
}

@article{vandeschootSystematicReviewBayesian2017,
  title = {A Systematic Review of {{Bayesian}} Articles in Psychology:
		  {{The}} Last 25 Years.},
  shorttitle = {A Systematic Review of {{Bayesian}} Articles in
		  Psychology},
  author = {van de Schoot, Rens and Winter, Sonja D. and Ryan, Oisín
		  and Zondervan-Zwijnenburg, Mariëlle and Depaoli,
		  Sarah},
  date = {2017-06},
  year = {2017},
  journal = {Psychological Methods},
    journal = {Psychological Methods},

  shortjournal = {Psychological Methods},
  volume = {22},
  pages = {217--239},
  issn = {1939-1463, 1082-989X},
  doi = {10.1037/met0000100},
  langid = {english},
  number = {2}
}

@manual{vehtariLooEfficientLeaveoneout2020,
  title = {{loo}: {{Efficient}} Leave-One-out Cross-Validation and
		  {{WAIC}} for {{Bayesian}} Models},
  author = {Vehtari, Aki and Gabry, Jonah and Magnusson, M\aans and Yao,
		  Yuling and Bürkner, Paul-Christian and Paananen,
		  Topi and Gelman, Andrew},
  date = {2020},
  year = {2020},
  url = {https://mc-stan.org/loo}
}

@misc{vehtariParetoSmoothedImportance2019,
  title = {Pareto {{Smoothed Importance Sampling}}},
  author = {Vehtari, Aki and Simpson, Daniel and Gelman, Andrew and
		  Yao, Yuling and Gabry, Jonah},
  date = {2022},
  year = {2022},
  url = {http://arxiv.org/abs/1507.02646},
  urldate = {2020-07-01},
  archiveprefix = {arXiv},
  eprint = {1507.02646},
  eprinttype = {arxiv},
  primaryclass = {stat}
}

@incollection{walter2009,
  doi = {10.1007/978-3-7908-2413-1_4},
  year = {2009},
  month = dec,
  publisher = {Physica-Verlag {HD}},
  pages = {59--78},
  editor = {Kneib, T and Tutz, G},
  author = {Gero Walter and Thomas Augustin},
  title = {Bayesian Linear Regression --- Different Conjugate Models
		  and Their (In)Sensitivity to Prior-Data Conflict},
  booktitle = {Statistical Modelling and Regression Structures}
}

@article{watsonApproximateModelsRobust2016,
  title = {Approximate {{Models}} and {{Robust Decisions}}},
  author = {Watson, James and Holmes, Chris},
  year = {2016},
  date = {2016-11},
  journal = {Statistical Science},
    journal = {Statistical Science},

  shortjournal = {Statist. Sci.},
  volume = {31},
  pages = {465--489},
  issn = {0883-4237},
  doi = {10.1214/16-STS592},
  langid = {english},
  number = {4}
}

@book{wickhamGgplot2ElegantGraphics2016,
  title = {{ggplot2}: {{Elegant}} Graphics for Data Analysis},
  author = {Wickham, Hadley},
  year = {2016},
  date = {2016},
  publisher = {{Springer-Verlag New York}},
  url = {https://ggplot2.tidyverse.org},
  isbn = {978-3-319-24277-4}
}

@InProceedings{paananen2021RSense,
  author    = {Paananen, Topi and Andersen, Michael Riis and Vehtari, Aki},
  title     = {Uncertainty-aware sensitivity analysis using R\'enyi divergences},
  booktitle = {Proceedings of the Thirty-Seventh Conference on Uncertainty in Artificial Intelligence},
  year      = {2021},
  editor    = {de Campos, Cassio and Maathuis, Marloes H.},
  volume    = {161},
  series    = {Proceedings of Machine Learning Research},
  pages     = {1185--1194},
  publisher = {PMLR},
  url       = {https://proceedings.mlr.press/v161/paananen21a.html},
}
